\documentclass[lettersize,journal]{IEEEtran}
\usepackage{amsmath,amsfonts}
\usepackage{algorithmic}
\usepackage{algorithm}
\usepackage{array}
\usepackage{textcomp}
\usepackage{stfloats}
\usepackage{url}
\usepackage{verbatim}
\usepackage{graphicx}
\usepackage{cite}
\usepackage{amssymb}
\usepackage{amsthm}
\usepackage{bm}
\usepackage{subfigure}
\usepackage{multirow}
\usepackage{hyperref}
\usepackage{cleveref}
\usepackage{gensymb}
\usepackage{booktabs}
\usepackage{soul}
\usepackage{color}
\usepackage{forest}
\usepackage{enumitem}

\usepackage{makecell}
\usepackage{pifont}

\definecolor{BlueGreen}{rgb}{0.643, 0.859, 0.871}
\definecolor{bounding}{rgb}{0.855, 0.392, 0.357}
\definecolor{Turquoise}{rgb}{0.251, 0.878, 0.816}
\definecolor{0}{rgb}{0.92, 0.3, 0.26}
\definecolor{1}{rgb}{0.961, 0.851, 0.471}
\definecolor{1_1}{rgb}{1.00, 0.933, 0.698}
\definecolor{2}{rgb}{0.522, 0.784, 0.953}

\newcommand{\revise}[1]{\textcolor{black}{#1}}

\newtheorem{theorem}{Theorem}

\newtheorem{remark}{Remark}
\newtheorem{proposition}{Proposition}

\usepackage[most]{tcolorbox}
\newtcolorbox[auto counter]{takeaway}[1][]{
    colback=gray!20, 
    colframe=gray!80,
    boxrule=0.5mm, 
    breakable, 
    enhanced, 
    before upper={\textbf{Takeaway~\thetcbcounter:}~},
    #1
}

\definecolor{customblue}{RGB}{43,105,175}

\begin{document}

\title{Exploring the Vulnerabilities of Federated Learning: A Deep Dive into Gradient Inversion Attacks}

\author{Pengxin Guo*, Runxi Wang*, Shuang Zeng, Jinjing Zhu, Haoning Jiang, Yanran Wang, Yuyin Zhou, Feifei Wang, Hui Xiong,~\IEEEmembership{Fellow, IEEE}, and Liangqiong Qu \\

Project page: \href{https://pengxin-guo.github.io/FLPrivacy}{\color{blue}pengxin-guo.github.io/FLPrivacy}
\thanks{Pengxin Guo, Runxi Wang, and Liangqiong Qu are with the School of Computing and Data Science, The University of Hong Kong, Hong Kong 999077, China (e-mail: \{guopx, u3637153\}@connect.hku.hk, liangqqu@hku.hk).}
\thanks{Shuang Zeng is with the Department of Mathematics, The University of Hong Kong, Hong Kong 999077, China (e-mail: zengsh9@connect.hku.hk).}
\thanks{Jinjing Zhu and Hui Xiong are with the Thrust of Artificial Intelligence, The Hong Kong University of Science and Technology (Guangzhou), Guangzhou 511458, China (email: jzhu706@connect.hkust-gz.edu.cn, xionghui@ust.hk).}
\thanks{Haoning Jiang is with the Department of Electronic and Electrical Engineering, Southern University of Science and Technology, Shenzhen 518055, China (e-mail: 12210308@mail.sustech.edu.cn).}
\thanks{Yanran Wang is with the Department of Biomedical Data Science, Stanford University, Stanford, CA 94305, USA (e-mail: joycewyr@stanford.edu).}
\thanks{Yuyin Zhou is with the Department of Computer Science and Engineering, University of California, Santa Cruz, CA 95064, USA (e-mail: yzhou284@ucsc.edu).}
\thanks{Feifei Wang is with the Department of Electrical and Electronic Engineering, The University of Hong Kong, Hong Kong 999077, China, and also with the Materials Innovation Institute for Life Sciences and Energy (MILES), HKU-SIRI, Shenzhen 518055, China (e-mail: ffwang@eee.hku.hk).}
\thanks{Pengxin Guo and Runxi Wang contributed equally to this work.}
\thanks{Corresponding author: Liangqiong Qu.}}

\markboth{Journal of \LaTeX\ Class Files,~Vol.~14, No.~8, August~2021}%
{Shell \MakeLowercase{\textit{et al.}}: A Sample Article Using IEEEtran.cls for IEEE Journals}

\IEEEpubid{0000--0000/00\$00.00~\copyright~2021 IEEE}

\maketitle

\begin{abstract}
  Federated Learning (FL) has emerged as a promising privacy-preserving collaborative model training paradigm without sharing raw data. However, recent studies have revealed that private information can still be leaked through shared gradient information and attacked by Gradient Inversion Attacks (GIA). While many GIA methods have been proposed, a detailed analysis, evaluation, and summary of these methods are still lacking. Although various survey papers summarize existing privacy attacks in FL, few studies have conducted extensive experiments to unveil the effectiveness of GIA and their associated limiting factors in this context. To fill this gap, we first undertake a systematic review of GIA and categorize existing methods into three types, i.e., \textit{optimization-based} GIA (OP-GIA), \textit{generation-based} GIA (GEN-GIA), and  \textit{analytics-based} GIA (ANA-GIA). Then, we comprehensively analyze and evaluate the three types of GIA in FL, providing insights into the factors that influence their performance, practicality, and potential threats. Our findings indicate that OP-GIA is the most practical attack setting despite its unsatisfactory performance, while GEN-GIA has many dependencies and ANA-GIA is easily detectable, making them both impractical. Finally, we offer a three-stage defense pipeline to users when designing FL frameworks and protocols for better privacy protection and share some future research directions from the perspectives of attackers and defenders that we believe should be pursued. We hope that our study can help researchers design more robust FL frameworks to defend against these attacks.
\end{abstract}


\section{Introduction} \label{sec:intro}
\IEEEPARstart{F}{ederated} Learning (FL) \cite{mcmahan2017communication,gupta2018distributed,zhang2024flhetbench,guo2025exploring} is a framework that enables multiple clients to collaboratively train a model without the need to disclose their private local data. 
The distinctive features of FL make it particularly suitable for developing machine learning models in privacy sensitive scenarios such as healthcare \cite{sadilek2021privacy,yan2023label,zeng2024tackling} and financial services \cite{long2020federated,chatterjee2023federated}. 
Although FL is designed to protect data privacy by only sharing model gradients, several studies 
have shown that attackers can still extract sensitive information about private training data through Gradient Inversion Attacks (GIA) \cite{zhu2019deep,huang2021evaluating}, Membership Inference Attacks (MIA) \cite{nasr2019comprehensive,zari2021efficient}, or Property Inference Attacks (PIA) \cite{wang2022poisoning,luo2021feature}. Among these privacy attacks, GIA, which focuses on reconstructing input data via either an \textit{honest-but-curious} server that follows the FL protocol but is interested in uncovering input data \cite{zhu2019deep,geiping2020inverting,huang2021evaluating}, or a \textit{malicious} server that may modify the model architecture or parameters sent to the user \cite{fowl2022robbing,boenisch2023curious,zhao2024loki}, has emerged as the most powerful and is the focus of this work.

\IEEEpubidadjcol

Many GIA methods have been proposed, but \textbf{a detailed analysis, evaluation, and summary of these methods is still lacking}. Although there are various surveys \cite{mothukuri2021survey,kumar2023impact,zhang2021survey,zhang2022survey,wen2023survey,liu2024vertical,shi2024dealing,carletti2025sok} summarizing existing privacy attacks in FL, few studies have conducted extensive experiments to reveal the effectiveness of GIA and their associated limiting factors in this context. Among the limited research on this topic, the works \cite{huang2021evaluating,du2025sok,baglin2024fedlad} are the most closely related to ours. 
However, these methods focus solely on analyzing partial GIA methods, leaving the applicability of their findings to other types of GIA uncertain.

To fill this gap, we aim to \textbf{conduct a comprehensive study of GIA, including a literature review, categorization of existing methods, in-depth analysis, and extensive evaluation}. Specifically, we first undertake a systematic review of current GIA methods and divide them into three categories (Section \ref{sec:ana_all_gia}): (1) \textit{Optimization-based} GIA (\textbf{OP-GIA}): OP-GIA works by minimizing the distance between the received gradients and the gradients computed from dummy data \cite{zhu2019deep,zhao2020idlg,geiping2020inverting,yin2021see,hatamizadeh2022gradvit,kariyappa2023cocktail,yue2023gradient,ye2024gradient}. (2) \textit{Generation-based} GIA (\textbf{GEN-GIA}): GEN-GIA utilizes a generator to reconstruct input data  \cite{jeon2021gradient,li2022auditing,fang2023gifd,ren2022grnn,xu2022cgir, zhang2023generative, sotthiwat2024generative,wu2023learning, xue2023fast, gu2024federated}.
(3) \textit{Analytics-based} GIA (\textbf{ANA-GIA}): ANA-GIA aims to recover input data in closed form using a \textit{malicious} server \cite{fowl2022robbing,zhao2023resource,zhao2024loki,pasquini2022eluding,wen2022fishing,boenisch2023curious,wang2024maximum,shi2025scale}.
Additionally, based on the different optimization components of the GEN-GIA methods, they can be further categorized into (i) optimizing latent vector $\bm{z}$ \cite{jeon2021gradient,li2022auditing,fang2023gifd}, (ii) optimizing generator’s parameters $\bm{W}$ \cite{ren2022grnn,xu2022cgir,zhang2023generative, sotthiwat2024generative}, and (iii) training an inversion generation model using an auxiliary dataset \cite{wu2023learning, xue2023fast}. Based on the type of modification, the ANA-GIA methods can be further categorized into (i) manipulating model architecture \cite{fowl2022robbing,zhao2023resource,zhao2024loki}, and (ii) manipulating model parameters \cite{pasquini2022eluding,wen2022fishing,boenisch2023curious,wang2024maximum,shi2025scale}. \revise{Concurrent work \cite{carletti2025sok} also classifies existing GIA methods into Optimization-based, Generative Model-based, and Analytic-based attacks, further validating the rationale behind our classification.}

{\setlength{\tabcolsep}{3pt}
\begin{table*} [t]
\centering
  \caption{Comparison of different types of GIA methods in terms of influence factors, reconstruction results, and extra reliance of each type of method. Influence factors include batch size, image resolution, model training state, the number of the same labels in one batch data, network architecture, and practical FedAvg with multiple local training steps. Reconstruction results include whether the reconstruction results are the original inputs and the visual quality of the reconstruction results.}
\resizebox{\linewidth}{!}{
  \begin{tabular}{cc cccccc cc c}
    \toprule
    \multirow{2}{*}{Taxonomy} & &  \multicolumn{6}{c}{Influence Factors} & \multicolumn{2}{c}{Reconstruction Results} & \multirow{2}{*}{Extra Reliance}\\
    \cmidrule(lr){3-8} \cmidrule(lr){9-10} 
    & & \makecell[c]{Batch \\ Size} & \makecell[c]{Image \\ Resolution} & \makecell[c]{\# Same \\ Label} & \makecell[c]{Model \\ Training State} & \makecell[c]{Network \\ Architecture} & \makecell[c]{Practical \\ FedAvg} & Original Inputs? & Visual Quality  \\
    \midrule
    OP-GIA & - & \checkmark & \checkmark & \checkmark & \checkmark & \checkmark & \checkmark & Yes & Low & No \\
    \midrule
    \multirow{3}{*}{GEN-GIA} & Opti. Lat. Vec. & \ding{55} & \ding{55} & \ding{55} & \ding{55} & \ding{55} & \ding{55} & No & High & Trained Generator \\
    & Opti. Gen. Para. & \checkmark & \checkmark & \checkmark & \checkmark & \checkmark & \checkmark & Yes & Middle & Sigmoid Activation \\
    & Train. Inv. Mod. & \checkmark & \checkmark & \checkmark & \ding{55} & \checkmark & \checkmark & Yes & Low & Auxiliary Dataset \\
    \midrule
    \multirow{2}{*}{ANA-GIA} & Manip. Mod. Arch. & \ding{55} & \ding{55} & \ding{55} & \ding{55} & \ding{55} & \ding{55} & Yes & High & Malicious Server \\
    & Manip. Mod. Para. & \ding{55} & \checkmark & \ding{55} & \checkmark & \checkmark & \ding{55} & Yes & Middle & Malicious Server \\
    \bottomrule    
  \end{tabular}
}
\label{tab:summary_results}
\end{table*}
\vspace{-0.1 cm}
}

Moreover, we conduct a comprehensive analysis and evaluation of the three types of GIA in FL, with the goal of providing actionable insights into the factors that influence their performance, practicality, and potential risks. Our study is structured to address the following research questions, progressing from foundational factors to a comparative analysis of threats, and finally focusing on strategies for efficient fine-tuning and their implications:

\textbf{R1.} \textbf{\textit{What are the crucial factors that impact the performance of different GIA methods and their associated underlying mechanisms?}}

\textbf{R2.} \textbf{\textit{Among all types of GIA with both \textit{honest-but-curious} and \textit{malicious} adversaries, which type is the most practical and poses the greatest threat to FL?}}

\textbf{R3.} \textbf{\textit{With Parameter-Efficient Fine-Tuning (PEFT) technologies being widely used in FL to fine-tune foundation models \cite{sun2024improving,yang2024dual,guo2025selective,zheng2025fedvlmbench}, what is the potential privacy leakage of FL under PEFT?}} 

To answer the first two research questions, we conduct extensive experiments on the three types of GIA to uncover the factors that influence GIA performance (Section \ref{sec:eva}).
Our results, summarized in Table \ref{tab:summary_results}, reveal that: 

\textbf{(I)} \textbf{\textit{OP-GIA is the most practical attack setting, but the performance is not satisfactory.}} Specifically, OP-GIA relies on minimal assumptions (i.e., NO extra reliance), making it the most practical attack setting among the three types of GIA. However, it is influenced by common FL training parameters, making it challenging to achieve satisfactory attack performance.
Moreover, practical FedAvg \cite{mcmahan2017communication}, where clients train the model locally for multiple iterations before sending updates, itself has the ability to resist OP-GIA (Section \ref{sec:opt_based_gia_result}).

\textbf{(II)} \textbf{\textit{GEN-GIA has many dependencies, making it pose a minimal threat to FL.}}
Specifically, some GEN-GIA methods (i.e., optimizing latent vector $\bm{z}$) can only achieve semantic-level recovery and heavily rely on the pre-trained generator. Other GEN-GIA methods (i.e., optimizing generator's parameters $\bm{W}$ and training an inversion model) can perform pixel-level attacks, but they have strong dependencies, such as reliance on the Sigmoid function and an auxiliary dataset (Section \ref{sec:exp_gan_based_gia}).

\textbf{(III)} \textbf{\textit{ANA-GIA can achieve satisfactory attack performance but is easily detected and defended against by clients.}} 
Specifically, while ANA-GIA can achieve satisfactory performance by manipulating model architecture or parameters, the modifications it makes to the network structure make it easily detectable and defendable by clients, thus rendering it impractical (Section \ref{sec:exp_ana_based_gia}).

For the research question \textbf{R3}, we construct the attack methods for FL under PEFT (Section \ref{sec:attack_peft}) and evaluate the privacy leakage (Section \ref{sec:exp_attack_peft}). Our experimental results reveal that:

\textbf{(IV)} \textbf{\textit{Attackers can breach privacy on low-resolution images but fail with high-resolution ones under PEFT.}} Specifically, as shown in Figure \ref{fig:lora_all}, the attackers achieve relatively good performance on CIFAR-10, CIFAR-100, and CelebA with a small resolution but perform poorly on ImageNet with a large resolution (Section \ref{sec:exp_attack_peft}).

Based on our experimental findings, we provide a three-stage defense pipeline for users when designing FL frameworks and protocols for better privacy protection: (1) avoid the Sigmoid activation function and use more complicated network architectures during network design, (2) adopt larger batch sizes and multi-step local training in the local training protocol, and (3) implement client-side validation to check for any potential malicious modification to the model architecture and parameters upon receiving the model from the server. By following this pipeline, users can better protect their data privacy when using FL without worrying about being attacked by current GIA methods.

We summarize our contributions as follows.
\begin{itemize}
    \item We undertake a systematic review of GIA and categorize existing methods into three types: \textit{optimization-based} GIA (OP-GIA), \textit{generation-based} GIA (GEN-GIA), and \textit{analytics-based} GIA (ANA-GIA) (Section \ref{sec:ana_all_gia}). We also provide a public repository to continually track developments in this fast-evolving field: \href{https://github.com/Pengxin-Guo/Awesome-Gradient-Inversion-Attacks}{\color{blue}Awesome-Gradient-Inversion-Attacks}.
    \item We introduce an error bound analysis (Theorem \ref{theo:opt_based_gia}) for data reconstruction in OP-GIA, which, for the first time, theoretically proves that the OP-GIA performance is linearly related to the square root of the batch size and image resolution. Furthermore, we propose a gradient similarity proposition (Proposition \ref{prop:comprae_different_model}) to investigate the impact of complex FL training parameters, such as model training state and label distributions, on OP-GIA performance (Section \ref{sec:opt_based_gia_method}).
    \item We conduct extensive experiments on the three types of GIA to uncover the factors influencing their performance and many key interesting findings are provided and summarized (Section \ref{sec:eva}). Based on our experimental findings, we provide a three-stage defense pipeline for users when designing FL frameworks and protocols for better privacy protection (Section \ref{sec:outlook}).
    \item We further investigate the privacy leakage in FL with PEFT in this work, revealing that attackers can breach privacy on low-resolution images but fail with high-resolution ones under PEFT (Sections \ref{sec:attack_peft} \& \ref{sec:exp_attack_peft}).
\end{itemize}

\section{Analysis of Gradient Inversion Attacks} \label{sec:ana_all_gia}

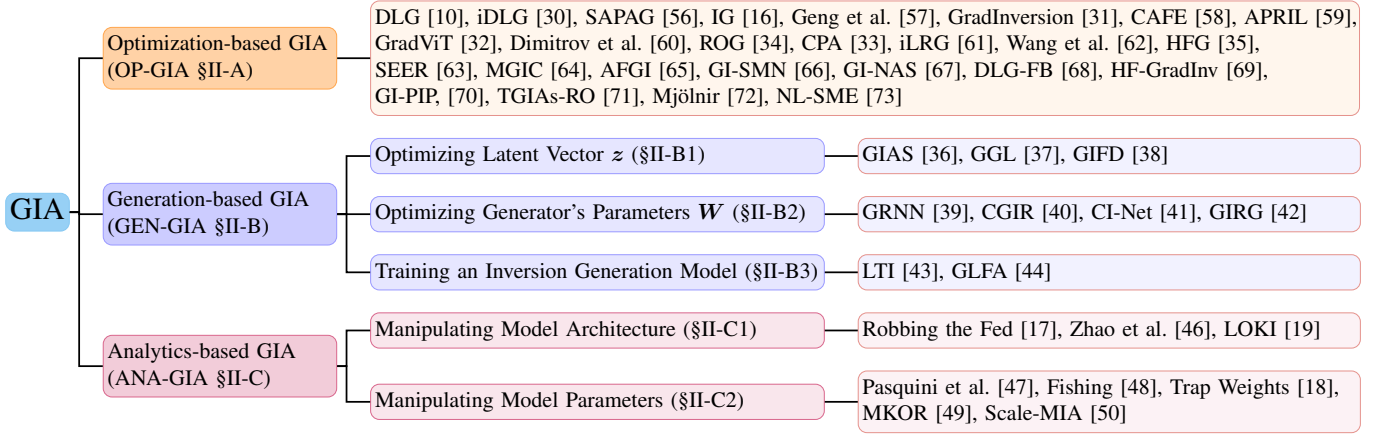
\begin{figure*}[t]
\centering
\resizebox{1\linewidth}{!}{
    \begin{forest}
  for tree={
    grow=0, 
    s sep=10pt, 
    l sep=15pt, 
    anchor=west,
    parent anchor=east,
    child anchor=west,
    edge={draw=black, thick}, 
    rectangle,
    draw,
    rounded corners,
    align=left,
    before typesetting nodes={inner sep=2pt},
    edge path={
          \noexpand\path [draw, \forestoption{edge}]
          (!u.parent anchor) -- ++(1.5mm,0) |- (.child anchor) \forestoption{edge label};},
        ver/.style={rotate=90, child anchor=north, parent anchor=south, anchor=center},
  },
  where level=1{text width=10.2em}{},
  where level=2{text width=20.2em}{},
  where level=3{text width=22.4em}{},
  [\Large GIA, color=2!100, fill=2!85, text=black,
    [Analytics-based GIA \\ (ANA-GIA \S \ref{sec:ANA_gia}), color=purple!60, fill=purple!20, text=black
      [Manipulating Model Parameters (\S \ref{sec:ana_gia_manip_para}), color=purple!60, fill=purple!10, text=black, 
        [{Pasquini et al. \cite{pasquini2022eluding}, Fishing~\cite{wen2022fishing}, Trap Weights \cite{boenisch2023curious}, \\ MKOR \cite{wang2024maximum}, \revise{Scale-MIA \cite{shi2025scale}}}, color=bounding!70, fill=purple!5, text=black,
        ]
      ]
      [Manipulating Model Architecture (\S \ref{sec:ana_gia_manip_arch}), color=purple!60, fill=purple!10, text=black, 
        [{Robbing the Fed~\cite{fowl2022robbing}, \revise{Zhao et al. \cite{zhao2023resource},} LOKI \cite{zhao2024loki}}, color=bounding!70, fill=purple!5, text=black,
        ]
      ]
    ]
    [Generation-based GIA \\ (GEN-GIA \S \ref{sec:gen_gia}), color=blue!40, fill=blue!20, text=black
        [Training an Inversion Generation Model (\S \ref{sec:gen_gia_inversion}), color=blue!40, fill=blue!10, text=black, 
            [{LTI \cite{wu2023learning}, GLFA \cite{xue2023fast}}, color=bounding!70, fill=blue!5, text=black, 
            ]
        ]
        [Optimizing Generator's Parameters $\bm{W}$ (\S \ref{sec:gen_gia_opt_w}), color=blue!40, fill=blue!10, text=black, 
            [{GRNN \cite{ren2022grnn}, \revise{CGIR \cite{xu2022cgir},} CI-Net~\cite{zhang2023generative}, GIRG \cite{sotthiwat2024generative}}, color=bounding!70, fill=blue!5, text=black, 
            ]
        ]
        [Optimizing Latent Vector $\bm{z}$ (\S \ref{sec:gen_gia_opt_z}), color=blue!40, fill=blue!10, text=black, 
            [{GIAS~\cite{jeon2021gradient}, GGL~\cite{li2022auditing}, GIFD \cite{fang2023gifd}},  color=bounding!70, fill=blue!5, text=black,
            ]
        ]
    ]
    [Optimization-based GIA \\ (OP-GIA \S \ref{sec:opt_based_gia_method}), color=orange!70, fill=orange!35, text=black
        [{DLG~\cite{zhu2019deep}, iDLG~\cite{zhao2020idlg}, SAPAG \cite{wang2020sapag}, 
        IG~\cite{geiping2020inverting}, 
        Geng et al. \cite{geng2021towards}, GradInversion~\cite{yin2021see}, CAFE \cite{jin2021cafe}, APRIL \cite{lu2022april}, \\  GradViT \cite{hatamizadeh2022gradvit}, Dimitrov et al. \cite{dimitrov2022data},  ROG \cite{yue2023gradient}, CPA~\cite{kariyappa2023cocktail}, iLRG \cite{ma2023instance}, Wang et al. \cite{wang2024towards}, HFG \cite{ye2024gradient}, \\ SEER \cite{garov2024hiding}, MGIC \cite{liu2024mgic}, AFGI \cite{liu2024raf}, GI-SMN \cite{qian2024gi}, GI-NAS \cite{yu2025gi}, DLG-FB \cite{leite2024federated}, \revise{HF-GradInv \cite{ye2024high},} \\ \revise{GI-PIP, \cite{sun2024gi},} TGIAs-RO \cite{li2025temporal}, \revise{Mjölnir \cite{liu2025mjolnir}}, \revise{NL-SME \cite{xia2025non}}}, color=bounding!70, fill=orange!7, text=black, text width=44.5em
        ]
    ]
  ]
    \end{forest}
}
\vskip -0.1in
    \caption{Taxonomy of existing GIA methods. The existing GIA methods can be divided into three types: optimization-based GIA (OP-GIA), which works by minimizing the distance between received gradients and gradients computed from dummy data; generation-based GIA (GEN-GIA), which utilizes a generator to reconstruct input data; and analytics-based GIA (ANA-GIA), which aims to recover input data in closed form. Moreover, GEN-GIA can be further divided into three categories: optimizing the latent vector $\bm{z}$, optimizing the generator’s parameters $\bm{W}$, and training an inversion generation model. ANA-GIA can be further divided into two categories: manipulating model architecture and manipulating model parameters.}
    \label{fig:overview}
\end{figure*}

Gradient Inversion Attacks (GIA) are tailored attacks intended for shared gradients in FL \cite{zhu2019deep}. They aim to reconstruct input data using the shared gradients, typically by a potential adversary seeking to reconstruct clients' private data. Adversaries are classified as \textit{honest-but-curious}\cite{goldreich2009foundations}, following the FL training protocol while trying to recover private data, or \textit{malicious}\cite{fowl2022robbing}, modifying model architecture or parameters to worsen privacy leakage. We divide existing GIA methods into three categories: \textit{optimization-based} GIA (\textbf{OP-GIA}), \textit{generation-based} GIA (\textbf{GEN-GIA}), and \textit{analytics-based} GIA (\textbf{ANA-GIA}), as illustrated in Figure \ref{fig:overview}.

\subsection{Optimization-based GIA} \label{sec:opt_based_gia_method}

\textit{Optimization-based} GIA (OP-GIA) typically operates under the threat model of an \textit{honest-but-curious} server to recover private data from clients \cite{zhu2019deep}. It aims to reconstruct input data by initializing random dummy data and minimizing the distance between the received gradients and those computed from the dummy data, a process also known as \textit{gradient matching} \cite{zhu2019deep}. 
Formally, given a neural network parameters $\theta$ and the gradients $\nabla_{\theta} \mathcal{L} (\bm{x}^*, \bm{y}^*)$ computed with a private data batch $(\bm{x}^*, \bm{y}^*)$, the attacker randomly initializes $(\bm{x}, \bm{y})$ and aims to solve the following optimization problem:
\begin{equation}
    {\arg\min_{\bm{x},\bm{y}}}\mathcal{D}(\nabla_{\theta}\mathcal{L}(\bm{x}^{*}, \bm{y}^{*}), \nabla_{\theta}\mathcal{L}(\bm{x}, \bm{y})) + \lambda\omega(\bm{x}),
    \label{eq:GIA}
\end{equation}
where $\mathcal{D}(\cdot, \cdot)$ denotes a distance metric, $\omega(\cdot)$ is a regularization term, and $\lambda$ is a regularization coefficient. 
Since simultaneously optimizing the input $\bm{x}$ and the label $\bm{y}$ in Eq. (\ref{eq:GIA}) is challenging \cite{zhu2019deep,zhao2020idlg,ma2023instance}, this has led to the development of methods that restore the label first, followed by input optimization \cite{zhao2020idlg,zhu2021r,yin2021see,ma2023instance,wang2024towards} \footnote{Label restoration is not the focus of this work, and we assume that label information is obtainable \cite{geiping2020inverting, zhao2020idlg, yin2021see, ma2023instance}.}. 
Then, the whole procedure of OP-GIA can be summarized in Algorithm \ref{algo:opt_based_gia}.

\renewcommand{\algorithmicrequire}{\textbf{Input:}}
\renewcommand{\algorithmicensure}{\textbf{Output:}}
\begin{algorithm}[t]\small
\label{alg1}
\caption{Optimization-based GIA}
\label{algo:opt_based_gia}
\begin{algorithmic}[1]
\REQUIRE Leaked gradients $\nabla_{\theta}\mathcal{L}(\bm{x}^{*}, \bm{y}^{*})$, model weight $\theta$, loss function $\mathcal{L}$, distance metric $\mathcal{D}(\cdot, \cdot)$, regulation term $\omega(\cdot)$, regularization coefficient $\lambda$, optimization learning rate $\eta$, and optimization iterations $I$.
\ENSURE Restored batch data $(\hat{\bm{x}}, \hat{\bm{y}})$.
\STATE Label Restoration: $\nabla_{\theta}\mathcal{L}(\bm{x}^{*}, \bm{y}^{*}) \rightarrow \hat{\bm{y}}$;
\STATE Initialize $\hat{\bm{x}_0}$;
\FOR{$i=1$ {\bfseries to} $I$}
\STATE $f(\bm{\hat{x}}_{i-1})=\mathcal{D}(\nabla_{\theta}\mathcal{L}(\bm{x}^{*}, \bm{y}^{*}), \nabla_{\theta}\mathcal{L}(\hat{\bm{x}}_{i-1}, \hat{\bm{y}})) + \lambda\omega(\hat{\bm{x}}_{i-1})$;
\STATE $\hat{\bm{x}}_{i}=\hat{\bm{x}}_{i-1}-\eta \nabla_{\hat{\bm{x}}_{i-1}}f (\hat{\bm{x}}_{i-1})$;
\ENDFOR
\STATE $\hat{\bm{x}} = \hat{\bm{x}}_I$;
\STATE \textbf{return} $(\hat{\bm{x}}, \hat{\bm{y}})$.
\end{algorithmic}
\end{algorithm}

Since the introduction of GIA in \cite{zhu2019deep}, various methods have been proposed to enhance attack performance by designing more powerful distance metrics or incorporating more complex regularization terms. For example, DLG \cite{zhu2019deep} uses $\ell_2$-distance as $\mathcal{D}(\cdot, \cdot)$ and does not use a regularization term. IG \cite{geiping2020inverting} adopts cosine similarity as $\mathcal{D}(\cdot, \cdot)$ and the Total Variance (TV) as $\omega(\cdot)$. GradInversion \cite{yin2021see} adopts $\ell_2$-distance as $\mathcal{D}(\cdot, \cdot)$ and divides $\omega(\cdot)$ into four terms, TV, $\ell_2$-distance, a Batch Normalization (BN) prior on the input $x$, and a group consistency regularization term. \revise{GI-PIP \cite{sun2024gi} utilizes cosine similarity as $\mathcal{D}(\cdot, \cdot)$ and divides $\omega(\cdot)$ into two terms: TV and Anomaly Score (AS).} Moreover, CPA \cite{kariyappa2023cocktail} uses Independent Component Analysis (ICA) to reconstruct features and aid further reconstruction. ROG \cite{yue2023gradient} employs an encoder to reduce input resolution, shrinking the original optimization space. HFG \cite{ye2024gradient} implements GIA stepwise to enhance reconstruction. 

Despite the progress, recent studies reveal that OP-GIA is often influenced by various FL training parameters, such as image resolution and batch size \cite{zhu2019deep,huang2021evaluating}. 
However, a comprehensive theoretical understanding detailing the reasons behind OP-GIA's susceptibility to these parameters as well as the extent of their impact (whether linear, exponential, or otherwise), remains unclear. 
To fill this gap, 
we provide the following theoretical analysis (the proof is provided in Section I-A in the Supplementary Material) obtained by Algorithm \ref{algo:opt_based_gia}, aiming to shed light on the relationship between FL parameters and attack performance.

\begin{theorem} \label{theo:opt_based_gia}
    If $f$ is $\mu$ strong convex and $L$-smooth, choose step-size $\eta \leq \sqrt{\frac{2}{\mu+L}}$, then Algorithm \ref{algo:opt_based_gia} obtains $\hat{\bm{x}}$ satisfying the following convergence guarantees:
    \begin{equation*}
        \|\hat{\bm{x}}-\bm{x}^*\|_2 \leq (1-\frac{\mu}{\mu+L})^{I}\|\hat{\bm{x}}_0-\bm{x}^*\|_2 + \frac{\sqrt{2BCHW(\mu+L)}}{\mu} \kappa,
    \end{equation*}
where $C, H, W$ denote the image resolution, $B$ is the batch size, and $\kappa$ is the upper bound of $\|\nabla_{\hat{\bm{x}}} f({\hat{\bm{x}}}) - \frac{1}{T}\sum_{t=1}^T \nabla_{\hat{\bm{x}}} f_t({\hat{\bm{x}}})\|_2$ for $T \geq 2$, where $f_t$ is constructed based on model weights $\theta_t$ at $t$ temporal. Here we assume there are $T$ communication rounds in FL, and $\theta_t$ denotes the model weights at time $t$.
\end{theorem}

\begin{remark}
    Intuitively, Theorem \ref{theo:opt_based_gia} provides a convergence analysis and details the quantitative relationship between the reconstruction error upper bound and both image resolution and batch size. Specifically, the reconstruction error is linearly related to the square root of the batch size and image resolution. Therefore, as image resolution and batch size increase, the attack performance deteriorates, which is also demonstrated in Section \ref{sec:opt_based_gia_result} in our experiments and other works \cite{zhu2019deep,geiping2020inverting,li2025temporal}. 
\end{remark}

\begin{proposition} \label{prop:comprae_different_model}
  For any two model weights $\theta_{t_1}$ and $\theta_{t_2}$, if the leaked gradients of different batch data on $\theta_{t_1}$ are more similar than those on $\theta_{t_2}$, i.e., 
  the cardinality of the set 
  $\{\bm{x}^{*,j}: \mathcal{D}(\nabla_{\theta_{t_1}}\mathcal{L}(\bm{x}^{*,i}, \bm{y}^{*,i}), \nabla_{\theta_{t_1}}\mathcal{L}(\bm{x}^{*,j}, \bm{y}^{*,j}) < \epsilon \}$ 
  is greater than the cardinality of the set 
  $\{\bm{x}^{*,j}: \mathcal{D}(\nabla_{\theta_{t_2}}\mathcal{L}(\bm{x}^{*,i}, \bm{y}^{*,i}), \nabla_{\theta_{t_2}}\mathcal{L}(\bm{x}^{*,j}, \bm{y}^{*,j}) < \epsilon \}$
  for any $i$ and $\epsilon > 0$, then recovering the input data using the leaked gradients by Algorithm \ref{algo:opt_based_gia} on $\theta_{t_1}$ is harder than on $\theta_{t_2}$. 
\end{proposition}

\begin{remark}
Proposition \ref{prop:comprae_different_model} states that for two different models, if the leaked gradients of different batch data on one model are more similar than those on another, then recovering input data using the former's leaked gradients is much harder. This implies that, in addition to the batch size and image resolution, the performance of OP-GIA is also influenced by the model training state and the label distribution on each batch size. Specifically, a well-trained model tends to exhibit more similarity in the gradients of different data points compared to an untrained model, resulting in a worse attack performance on the well-trained model. Furthermore, when a batch contains a higher number of identical labels, the gradients become more similar to those of a single image from that particular class, further degrading attack performance. These findings are also validated in Section \ref{sec:opt_based_gia_result}.  
\end{remark}

In summary, Theorem \ref{theo:opt_based_gia} and Proposition \ref{prop:comprae_different_model} offer valuable insights into the factors affecting the performance of OP-GIA. Theorem \ref{theo:opt_based_gia} lays a strong theoretical foundation for understanding the impact of critical training parameters, such as batch size and image resolution, on a given fixed model. In contrast, Proposition \ref{prop:comprae_different_model}  delves deeper into the influence of more complex and general parameters on attack performance by comparing the shared gradients of two distinct models. This insight is particularly valuable as it reveals the interplay between the model training state and attack performance, enabling researchers to identify potential vulnerabilities and devise appropriate countermeasures. Together, these theoretical analyses serve as a powerful tool for guiding future work on OP-GIA methods, allowing researchers to evaluate the effectiveness and limitations of their models' attack performance more comprehensively. 

\subsection{Generation-based GIA}
\label{sec:gen_gia}

Another type of GIA utilizes a generator to reconstruct the input data, which we refer to as \textit{generation-based} GIA (GEN-GIA) \cite{jeon2021gradient,li2022auditing,fang2023gifd,ren2022grnn,xu2022cgir,zhang2023generative,sotthiwat2024generative,wu2023learning,xue2023fast,gu2024federated}. Differing from OP-GIA, which directly optimizes the input data, this type of method uses a generator to produce the input data. Based on the different optimization components of these methods, we divide them into the following categories: 1) optimizing the latent vector $\bm{z}$; 2) optimizing the generator's parameters $\bm{W}$; and 3) training an inversion generation model using an auxiliary dataset.

\subsubsection{Optimizing Latent Vector $z$} \label{sec:gen_gia_opt_z}

Due to the large search space complicating the optimization process in OP-GIA methods, optimizing the latent vector $\bm{z}$ of a pre-trained generator is proposed to mitigate this issue, as it significantly reduces the search space \cite{jeon2021gradient,li2022auditing,fang2023gifd}. Formally, the optimizing process can be illustrated as:

{\small 
\begin{equation} \label{eq:gan_based_gia}
    {\arg\min_{\bm{z}}}\mathcal{D}(\nabla_{\theta}\mathcal{L}(\bm{x}^{*}, \bm{y}^{*}), \nabla_{\theta}\mathcal{L}(G(\bm{z}, \bm{y}^*), \bm{y}^*)) + \lambda\omega(G(\bm{z,\bm{y}^*})),
\end{equation}
}where $\bm{z}$ denotes the latent vector fed into the generator and the dimension is usually small, $G$ is a pre-trained generator. Here they \cite{jeon2021gradient,li2022auditing,fang2023gifd} assume the label $\bm{y}$ can be accurately recovered by other method \cite{zhao2020idlg,yin2021see,dang2021revealing,ma2023instance}. After reconstructing $\hat{\bm{z}}$ using Eq.~(\ref{eq:gan_based_gia}), it is fed into the pre-trained generator $G$ to obtain reconstruction images $\hat{\bm{x}}$ via $\hat{\bm{x}} = G(\hat{\bm{z}}, \bm{y}^*)$. Despite its promise, there are two limitations for this type of method that cannot be ignored. First, the data distribution of the recovered data should be similar to the data that the generator $G$ was pretrained on \cite{li2022auditing}. Moreover, since the generator $G$ is pre-trained and fixed, only semantically similar images can be reconstructed \cite{li2022auditing, jeon2021gradient,fang2023gifd}, as illustrated in Section \ref{sec:exp_gan_based_gia}.

\subsubsection{Optimizing Generator's Parameters $W$} \label{sec:gen_gia_opt_w}

Since fixing the generator's parameters can only produce semantically similar images, some works attempt to optimize the generator's parameters $\bm{W}$ to achieve pixel-level attacks \cite{ren2022grnn,xu2022cgir,zhang2023generative,sotthiwat2024generative}, resulting in the following optimization problem:

{\footnotesize
\begin{equation} \label{eq:gan_based_gia_para}
    {\arg\min_{\bm{W}}}\mathcal{D}(\nabla_{\theta}\mathcal{L}(\bm{x}^{*}, \bm{y}^{*}), \nabla_{\theta}\mathcal{L}(G(\bm{W};\bm{z}, \bm{y}^*), \bm{y}^*)) + \lambda\omega(G(\bm{W};\bm{z}, \bm{y}^*)),
\end{equation}
}where $\bm{W}$ denotes the parameters of the generator, and $\bm{z}$ denotes the randomly initialized latent vector that remains fixed during training. Differing from the optimization of the latent vector $\bm{z}$ in Eq. (\ref{eq:gan_based_gia}), the generator's parameters are optimized in Eq. (\ref{eq:gan_based_gia_para}), allowing for pixel-level similar image reconstruction, as demonstrated in Section \ref{sec:exp_gan_based_gia}. However, such methods heavily rely on the activation functions of the target model \cite{zhang2023generative}. Specifically, this approach only works when the target model adopts the Sigmoid activation function and fails with other activation functions, as validated in Section \ref{sec:exp_gan_based_gia}. This strong dependence limits its practicality.

\subsubsection{Training an Inversion Generation Model} \label{sec:gen_gia_inversion}

Unlike the previous two types of GEN-GIA methods that rely on gradient matching, another approach to utilizing the generator model is to train an inversion generation model to generate the input data \cite{wu2023learning, xue2023fast}. This involves optimizing the model's parameters to learn the mapping from a given set of gradients or other information to reconstruct the original input data. For example, Wu et al. \cite{wu2023learning} propose Learning to Invert (LTI), which trains an inversion model to reconstruct training samples from their gradients with the assistance of an auxiliary dataset. Xue et al. \cite{xue2023fast} introduce FGLA, which first extracts the features of each sample in a batch and then directly generates the user data based on a trained generator. Both methods can recover the input images directly based on the given gradients by solving an inverse problem. However, these methods require an auxiliary dataset to train the inversion model, which poses a significant limitation.

\subsection{Analytics-based GIA}
\label{sec:ANA_gia}
\begin{figure}[h]
    \centering
\includegraphics[width=0.9\linewidth]{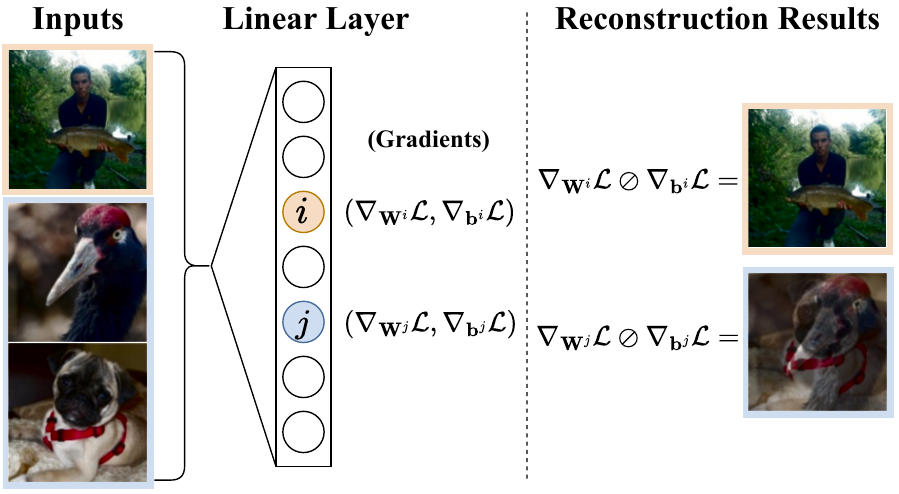}
\vskip -0.1in
    \caption{Reconstruction results of the linear layer. Neuron $i$ is activated by a single image, resulting in an accurate reconstruction, while neuron $j$, activated by two images, leads to a reconstruction that is a combination of both images.}
    \label{fig:ana_explanation}
\end{figure}

In contrast to the OP-GIA and GEN-GIA that recover input data using a gradient matching or inversion function, \textit{analytics-based} GIA (ANA-GIA) aims to reconstruct the input data in closed form \cite{phong2017privacy,zhu2021r,fowl2022robbing,zhao2024loki,wen2022fishing,boenisch2023curious,wang2024maximum,pasquini2022eluding,dimitrov2024spear}.
These approaches leverage the characteristic of the linear layer, where the linear composition of the input can be calculated based on the gradients of the weight and bias \cite{phong2017privacy}. Specifically, consider a linear layer $\bm{y}= \bm{W} \bm{x} + \bm{b},$
where $\bm{W}$ is a weight matrix, $\bm{b}$ is a bias, and $\bm{x}$ is the layer's input. 
When only a single image $\bm{x}^i$ activates a neuron $i$, the input $\bm{x}^i$ can be directly computed as 
$\bm{x}^i= \nabla_{\bm{W}^i}\mathcal{L} \oslash  \nabla_{\bm{b}^i}\mathcal{L}.$   
Here $\mathcal{L}$ is the loss function, $\oslash$ denotes entry-wise division, $i$ is the activated neuron, $\bm{x}^i$ is the input that activates neuron $i$, and $\nabla_{\bm{W}^i}\mathcal{L}$, $\nabla_{\bm{b}^i}\mathcal{L}$ are the weight gradient and bias gradient of the neuron respectively. If the linear layer is at the beginning of a network, the reconstructed data will be the original input. However, exact reconstruction only occurs when a single data sample activates a neuron. If multiple inputs activate the same neuron, the reconstruction becomes a combination of all activating images, as illustrated in Figure \ref{fig:ana_explanation}.

To address this issue, a \textit{malicious} adversary is employed to manipulate model architecture or parameters, ensuring that each neuron is activated by only a single sample whenever possible, thereby facilitating such attacks \cite{fowl2022robbing,pasquini2022eluding,boenisch2023curious,zhao2024loki,wang2024maximum,wen2022fishing}. Based on the type of modification, these ANA-GIA methods can be categorized into 1) manipulating model architecture \cite{fowl2022robbing,zhao2023resource,zhao2024loki} and 2) manipulating model parameters \cite{pasquini2022eluding,boenisch2023curious,wang2024maximum,wen2022fishing,shi2025scale}.

\subsubsection{Manipulating Model Architecture} \label{sec:ana_gia_manip_arch}
Fowl et al. \cite{fowl2022robbing} first introduce Robbing the Fed which inserts a specifically designed linear layer at the network's outset to exactly recover the input data, called ``binning''. 
This layer, formulated as $M(\bm{x}) = \text{ReLU}(\bm{W}_* \bm{x} + \bm{b}_*)$, uses weights to measure a known continuous cumulative density function (CDF) of the input data (e.g., image brightness), with each neuron's bias acting as a cutoff. 
The goal of this method is to ensure that only one input activates each ``bin'', where the ``bin'' is defined as the neuron with the largest cutoff that is activated. With well-designed weights $\bm{W}_*$ and biases $\bm{b}_*$, the original input $\bm{x}^i$ can be reconstructed as:
\begin{equation} \label{eq:robbing_the_fed}
    \bm{x}^i = \left(\nabla_{\bm{W}_*^i} \mathcal{L}-\nabla_{\bm{W}_*^{i+1}} \mathcal{L}\right) \oslash \left(\nabla_{\bm{b}_*^i} \mathcal{L}-\nabla_{\bm{b}_*^{i+1}} \mathcal{L}\right),
\end{equation}
where $\bm{W}^i_*$ is the $i$-th row of $\bm{W}_*$, $i$ is the activated bin and $i + 1$ is the bin with the next higher cutoff bias. 
The proposition 1 in \cite{fowl2022robbing} (also provided in Section II in the Supplementary Material) 
shows that the number of exactly recovered data
relates to the relationship between number of print bins $k$ and batch size $B$. However, if we choose a relatively large number of bins, the influence of batch size becomes minimal, as shown in Table \ref{tab:robbing_the_fed} in our experiments. \revise{Zhao et al. \cite{zhao2023resource} discuss a fundamental issue in prior work \cite{fowl2022robbing} on privacy attacks against FL with secure aggregation, highlighting that treating the aggregate update as a single large batch incurs unnecessary overheads, and they demonstrate that viewing the update as the aggregation of multiple individual updates allows the application of sparsity to alleviate resource overhead.}
Later, Zhao et al. \cite{zhao2024loki} propose LOKI, which inserts an attack module at the start of a model. The module consists of a convolutional layer followed by two fully connected layers, designed to perform successful attacks even under secure aggregation. LOKI sends customized convolutional kernels to each client as an identity mapping set, allowing the server to separate weight gradients from the clients despite the use of secure aggregation. The server then uses these weight gradients to reconstruct the original data points. Although these methods can achieve exact recovery, they involve modifications to the model architecture, making them easily detectable and defensible by clients \cite{garov2024hiding}.

\subsubsection{Manipulating Model Parameters} \label{sec:ana_gia_manip_para}

Since modifying model architecture is easily detectable, some works propose manipulating only the model parameters to facilitate such attacks \cite{pasquini2022eluding,boenisch2023curious,wang2024maximum,wen2022fishing,shi2025scale}. These methods can be understood as attacks based on \textit{gradient sparsity}. In this context, while the aggregation protocol nominally runs as intended, all but one of the data points return a zero gradient for certain model parameters, allowing the averaging process to still produce these entries directly \cite{wen2022fishing}. Formally, for a batch of data $\bm{x} = \{\bm{x}^1, \bm{x}^2, ..., \bm{x}^B\}$ and the corresponding labels $\bm{y} = \{\bm{y}^1, \bm{y}^2, ..., \bm{y}^B\}$, the gradients of the loss function $\mathcal{L}$ with respect to the model parameters $\theta$ on this batched data are:
\begin{equation} \label{eq:fishing}
\begin{aligned}
    \nabla_\theta\mathcal{L}
    \left(\bm{x}, \bm{y}\right)
    &= \frac{1}{B}
    \sum_{k=1}^B
    \nabla_\theta\mathcal{L}
    \left(\bm{x}^k, \bm{y}^k\right) \\
    &= \frac{1}{B}
    \left( \nabla_\theta\mathcal{L}
    \left(\bm{x}^t, \bm{y}^t\right) + 
    \sum_{\substack{{k}=1, {k} \neq {t}}}^{B}
    \nabla_\theta\mathcal{L}
    \left(\bm{x}^k, \bm{y}^k\right)
    \right).
\end{aligned}
\end{equation}
By deliberately modifying the model parameters $\theta$, the \textit{malicious} server can make $\sum_{\substack{{k}=1, {k} \neq {t}}}^{B}
    \nabla_\theta\mathcal{L}
    \left(\bm{x}^k, \bm{y}^k\right) \to 0$. Thus, it can achieve:
\begin{equation} \label{eq:fishing-result}
    \nabla_\theta\mathcal{L}
    \left(\bm{x}, \bm{y}\right)
    \approx 
    \frac{1}{B}
    \nabla_\theta\mathcal{L}
    \left(\bm{x}^t, \bm{y}^t\right).
\end{equation}
As a result, the gradient of a single data point is ``isolated'' from the large batch, breaking the aggregation.
For example, Pasquini et al. \cite{pasquini2022eluding} propose distributing different model weights to different users to tamper with the updates so that their aggregation will leak information about the update of a target user. Wen et al. \cite{wen2022fishing} introduce Fishing, in which the \textit{malicious} server artificially decreases the confidence of the network's predictions for the target class,  significantly increasing the contribution of the gradient information calculated from data belonging to the selected class. Boenisch et al. \cite{boenisch2023curious} propose Trap Weights, which re-scale components in the model's weight matrix to extract individual training data points from a chosen subset of participating users. \revise{Shi et al. \cite{shi2025scale} introduce Scale-MIA, which decomposes the reconstruction task into a two-step process involving the reconstruction of latent space representations, leveraging specially crafted linear layers, followed by the use of a fine-tuned generative decoder to reconstruct the entire input batch.} These methods share a similar idea: by manipulating model weights sent to the client to isolate a data point from batch data. However, they can only reconstruct the original input data when the first layer of the target model is a fully connected layer; otherwise, they can only isolate the gradients of a single data point and must be combined with OP-GIA methods to continue recovering \cite{boenisch2023curious,wen2022fishing}. Moreover, due to the modifications to the model parameters, they can also be detected by clients \cite{garov2024hiding}.

\subsection{Attacks under Parameter-Efficient Fine-Tuning} \label{sec:attack_peft}

The aforementioned works have primarily focused on the traditional scenario where clients share the gradients of the entire model with the server. In this section, we consider a more prevalent situation where clients fine-tune a foundation model using Parameter-Efficient Fine-Tuning (PEFT) technologies, such as Low-Rank Adaptation (LoRA) \cite{hu2022lora}. These approaches involve sharing only the gradients of a small subset of parameters with the server, and there is a lack of research on privacy leakage analysis in these scenarios. To fill this gap, we will explore these issues in this work.

When a client model is fine-tuned using LoRA, we can adopt the following optimization problem to recover the input data:

{\footnotesize
\begin{equation} \label{eq:LoRA-GIA}
   {\arg\min_{\bm{W}}}\mathcal{D}(\nabla_{\theta_{\bm{L}}}\mathcal{L}(\bm{x}^{*}, \bm{y}^{*}), \nabla_{\theta_{\bm{L}}}\mathcal{L}(G(\bm{W};\bm{z}), \bm{y}^*)) + \lambda\omega(G(\bm{W};\bm{z})),
\end{equation}
}where $G$ denotes a generator parameterized by $\bm{W}$ as used in GEN-GIA (Section \ref{sec:gen_gia_opt_w}), $\theta_{\bm{L}}$ represents the parameters of the LoRA matrices, and $\omega(\cdot)$ is the regularization term. The regularization term consists of two components: 1) Total Variation, which enhances the quality of reconstructed images, and 2) Patch Consistency, which addresses the misalignment of patches caused by the Vision Transformer \cite{hatamizadeh2022gradvit}.

\subsection{Extension to Practical FedAvg}

The aforementioned works focused on reconstructing raw data from known gradients in ideal settings, rather than considering practical environments in production FL. Specifically, these studies often assume training with FedSGD, where clients compute a gradient update on a single local batch of data and then send it to the server. In contrast, real-world FL applications typically build on FedAvg \cite{mcmahan2017communication}, where clients train the model locally for multiple iterations before sending updates. Under FedAvg, reconstructing clients' private data becomes significantly more challenging  \cite{dimitrov2022data,wang2023more}, which is also demonstrated in our later evaluation section.

\section{Evaluation} \label{sec:eva}

In this section, we provide a comprehensive evaluation of privacy leakage in FL across various types of GIA to answer the following research questions:

\textbf{R1.} {\textit{What are the crucial factors that impact the performance of different types of GIA?}}

\textbf{R2.} {\textit{Among all types of GIA, which type is the most practical and poses the greatest threat to FL?}}

\textbf{R3.} {\textit{What's the privacy leakage of FL under PEFT?}} 

To answer the research question \textbf{R1}, we divide the influence factors into two types: data level and model level. At the data level, we investigate the influence of batch size, image resolution, and the number of the same labels within one batch on the performance of each type of GIA. At the model level, we explore the influence of model training state and network architectures on the performance of each type of GIA. For the research question \textbf{R2}, we utilize the additional reliance on each type of GIA to measure its practicality and combine this with the reconstruction results to demonstrate its threat to FL. For the research question \textbf{R3}, we test the privacy leakage of FL under LoRA fine-tuning. In the following sections, we examine each type of GIA separately in terms of their influence factors, practicality, and threats to FL. Then, we discuss them collectively and provide guidance for users on designing FL models and training protocols to enhance data privacy protection.

\subsection{Experimental Setup} \label{sec:exp_setup}

We evaluate the privacy leakage of FL in three nature image classification datasets: CIFAR-10 \cite{krizhevsky2009learning}, CIFAR-100 \cite{krizhevsky2009learning}, and ImageNet \cite{deng2009imagenet}, and a facial image dataset: CelebA\cite{liu2015deep}. For each dataset, we select a subset of 64 images to evaluate the attack performance. There are no same labels in the subset of CIFAR-100, ImageNet, and CelebA. For the CelebA dataset, we choose the images of the first 1,000 celebrities and assign the corresponding number of the celebrity as the label. The illustration of these selected subsets is shown in Figure III.1 in the Supplementary Material. We adopt ResNet-18 \cite{he2016deep} as the baseline network, except for the experiments comparing different model architectures and attacks under PEFT. \revise{Evaluation on other backbones, such as LeNet \cite{lecun2002gradient}, AlexNet \cite{krizhevsky2012imagenet}, VGG \cite{simonyan2014very}, and GoogLeNet \cite{szegedy2015going}, is provided in Section V of the Supplementary Material.} 

We experiment with both \textit{honest-but-curious} and \textit{malicious} adversary settings, and focus on image reconstruction tasks due to its widespread interest. For OP-GIA and GEN-GIA, the adversary is an \textit{honest-but-curious} server, while for ANA-GIA, the adversary is a \textit{malicious} server.

For the OP-GIA, we choose IG \cite{geiping2020inverting} as the attack method for evaluation. For optimization, we optimize the attack for 24,000 iterations using Adam optimizer \cite{kingma2015adam}, with an initial learning rate 0.1. The learning rate is decayed by a factor of 0.1 at 3/8, 5/8, 7/8 of the optimization. The coefficient of the TV regularization term is 1e-2 for CIFAR-10 and CIFAR-100, and 1e-6 for ImageNet.

For the GEN-GIA, we employ GGL \cite{li2022auditing}, which optimizes the latent vector $\bm{z}$, CI-Net \cite{zhang2023generative}, which optimizes the generator's parameters $\bm{W}$, and LTI \cite{wu2023learning}, which learns an inversion generation model to evaluate the privacy leakage in FL. For GGL, similar to the experimental settings in \cite{li2022auditing}, we use a pre-trained BigGAN \cite{brock2019large} as the generator and the gradient-free Covariance Matrix Adaptation Evolution Strategy (CMA-ES) \cite{hansen2016cma} as the optimizer. For CI-Net, following the settings in \cite{zhang2023generative}, we use the CI-Net as the generator and optimize the attack for 2000 iterations using the Adam optimizer with learning rate 1e-3. For LTI, following the settings in \cite{wu2023learning}, we adopt an MLP model as the generator and optimize the attack for 5000 epochs using the Adam optimizer with learning rate 1e-4. 

For the ANA-GIA, we choose Robbing the Fed \cite{fowl2022robbing}, which manipulates model architectures, and Fishing \cite{wen2022fishing}, which manipulates model parameters, as the attack method for evaluation. For Robbing the Fed, we modify the ResNet-18 to include an imprint module with a different number of bins in front to perform attacks. For Fishing, we modify the parameters of the ResNet-18 model, enabling the isolation of gradients and the execution of optimization-based attacks on a single image.

All experiments are conducted on NVIDIA  L40S and GeForce RTX 4090 GPUs. The peak signal to noise ratio (PSNR) \cite{hore2010image}, structural similarity (SSIM) \cite{wang2004image}, learned perceptual image patch similarity (LPIPS) \cite{zhang2018unreasonable}, \revise{Jaccard similarity (Jaccard) \cite{yue2023gradient}, and relative data leakage value (RDLV) \cite{hatamizadeh2023gradient}} are adopted as the metrics for evaluating the attack performance. Lower LPIPS, higher PSNR, SSIM, \revise{Jaccard, and RDLV} of reconstructed images indicate better attack performance.


\subsection{Optimization-based GIA} \label{sec:opt_based_gia_result}
\begin{figure}[h]
\centering
\subfigure[] { \label{fig:gia_all}
\includegraphics[width=0.46\columnwidth]{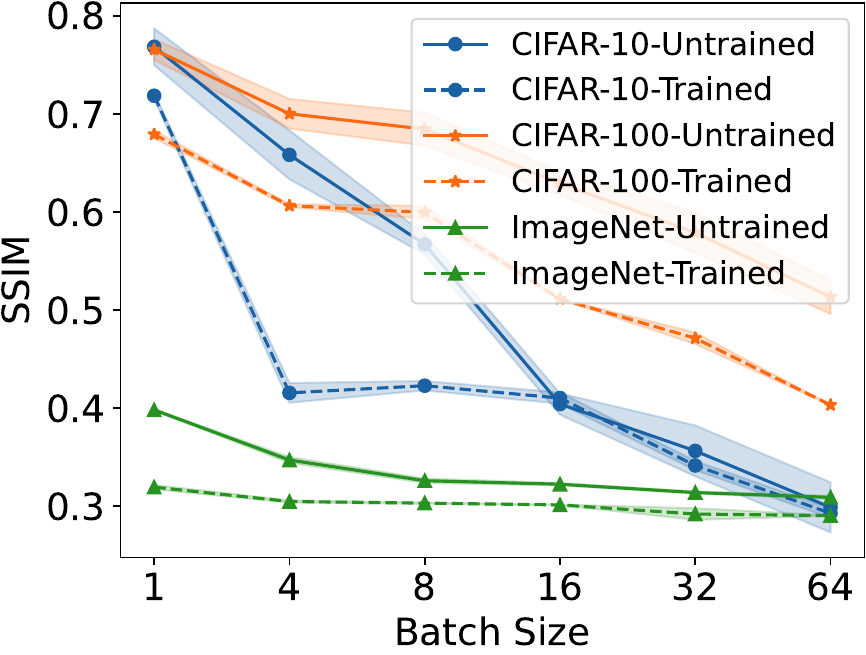}}   
\hfill
\subfigure[] { \label{fig:compare_arch} 
\includegraphics[width=0.46\columnwidth]{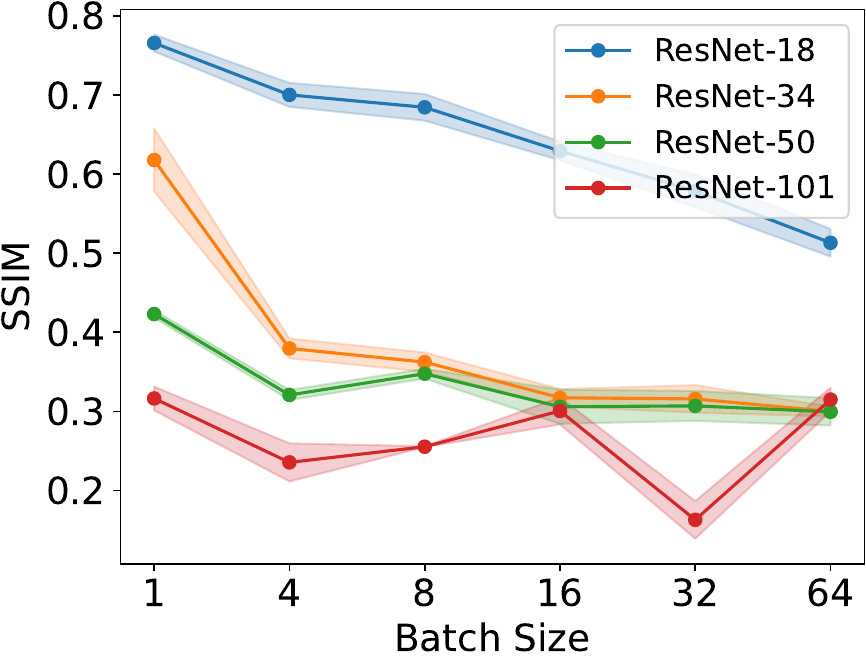}} 
\vskip -0.1in
    \caption{(a) Reconstruction results of IG evaluated on models in different training states on various datasets with different image resolutions and batch sizes. (b) Reconstruction results of IG with different network architectures on the CIFAR-100 dataset. The shaded region represents the standard deviation.
    These results show that a larger batch size, higher image resolution, more complicated network architecture, and better model training state lead to worse OP-GIA performance.
    }
\end{figure}

To validate the influence factors of OP-GIA, we evaluate the attack performance of IG \cite{geiping2020inverting} on various datasets with different image resolutions and batch sizes. Additionally, we conduct attacks using different network architectures and models at different training states. The reconstruction results are shown in Figures \ref{fig:gia_all} and \ref{fig:compare_arch}. 
Additional evaluation metrics, such as PSNR, LPIPS, \revise{Jaccard, and RDLV}, are provided in Figures IV.2 and IV.3 in the Supplementary Material. 
The visualization of reconstruction results are provided in Section IV-A1 in the Supplementary Material. 
Based on these results, we conclude that:

\begin{takeaway} \label{takeaway:op-gia}
    OP-GIA has no additional reliance, but its performance is not satisfactory and is affected by many factors.
    Larger batch size, higher image resolution, more complicated network architecture, better model training state, and more same labels in one batch lead to worse OP-GIA performance.
\end{takeaway}

\textit{\textbf{Larger batch size, worse OP-GIA performance.}} According to Figure \ref{fig:gia_all}, we can see that with the increase in batch size, the attack performance decreases, which is consistent with Theorem \ref{theo:opt_based_gia}. 
Intuitively, the size of the optimization problem in Eq. (\ref{eq:GIA}) is $\mathcal{O}(B \times d_{in})$, where $B$ is the batch size and $d_{in}$ denotes the dimensionality of the input data. The difficulty of performing this optimization scales with the dimensionality of the input, preventing OP-GIA from scaling to high-dimensional inputs. Therefore, as $B$ increases, the attack performance will decrease \cite{torn1989global,zhu2019deep,geiping2020inverting,li2025temporal}.

\textit{\textbf{Higher image resolution, worse OP-GIA performance.}} Comparing the attack performance on CIFAR-10/100 (with resolution $3 * 32 * 32$) and ImageNet (with resolution $3 * 224 * 224$) in Figure \ref{fig:gia_all}, we observe that higher image resolution leads to worse attack performance. This phenomenon is consistent with Theorem \ref{theo:opt_based_gia}, and the reasoning is similar to the influence of batch size. 
Moreover, the image resolution has a more significant impact than the batch size. As as shown in Theorem \ref{theo:opt_based_gia}, the reconstruction error for OP-GIA is linearly related to the square root of the batch size and image resolution. Since increasing the image resolution involves increasing both the height ($H$) and width ($W$), the reconstruction error can be considered to have a linear correlation with the image resolution. This suggests that the image resolution has a more significant impact than the batch size. 
Moreover, the experimental results on ImageNet with different image resolutions in Figure IV.4 in the Supplementary Material further support this point.

\textit{\textbf{More complicated network architecture, worse OP-GIA performance.}} From the results in Figure \ref{fig:compare_arch}, we can observe that as the model architecture becomes more complicated (i.e., transitioning from ResNet-18 to ResNet-101), the attack performance degrades. Notably, when the client adopts ResNet-50, the attack effectiveness against a batch size of 1 is even worse than using ResNet-18 with a batch size of 64. This is because more complicated network architectures can make the optimization process (as expressed in Eq. (\ref{eq:GIA})) more prone to getting trapped in local minima \cite{li2025temporal}. As a result, the attack performance decreases.

\begin{figure}[h]
    \centering
\includegraphics[width=0.8\linewidth]{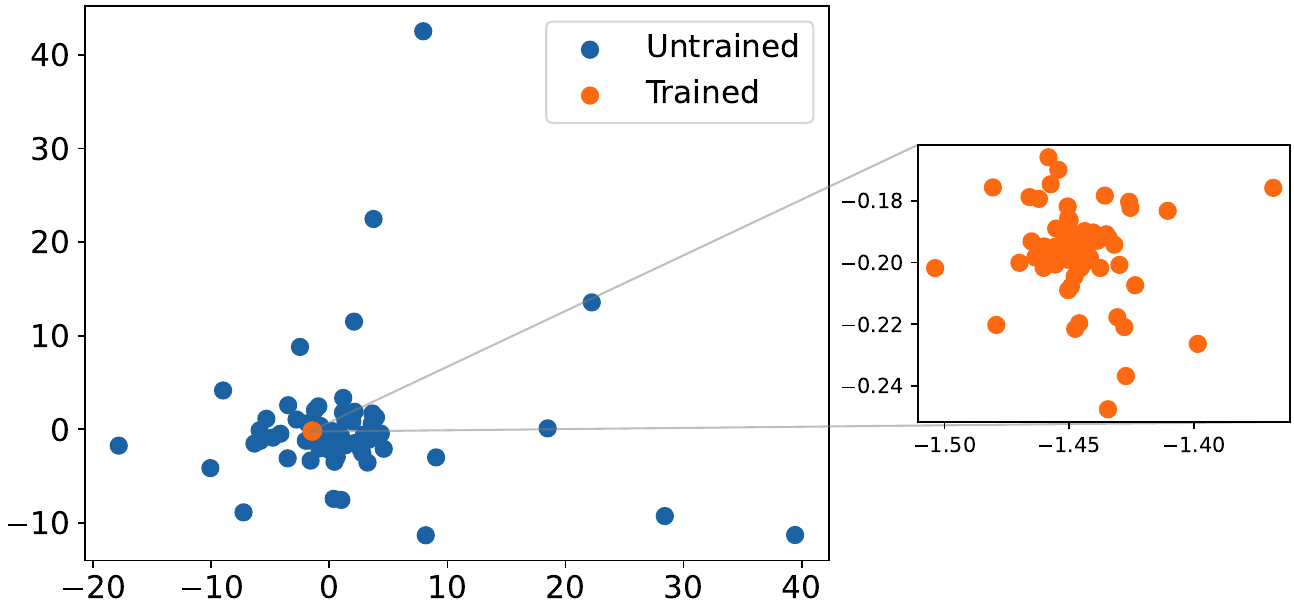}
\vskip -0.1in
    \caption{t-SNE visualization of gradients of different CIFAR-100 data points on untrained and trained models. It shows that the gradients are more similar for the trained model than the untrained model.}
    \label{fig:grad_compare}
\end{figure}

\textit{\textbf{Better model training state, worse OP-GIA performance.}} By comparing the attack performance on untrained and well-trained models in Figure \ref{fig:gia_all}, we observe that the attack performance is worse on the well-trained model. This is because the gradients tend to be more similar for well-trained models, as illustrated in Figure \ref{fig:grad_compare}, making it difficult to recover private data by comparing gradient values. This observation is consistent with Proposition \ref{prop:comprae_different_model}, which states that for two different models, if the leaked gradients of different batch data on one model are more similar than those on another, then recovering input data using the former's leaked gradients is much harder.

\begin{figure*}[t]
    \centering
    \includegraphics[width=1\linewidth]{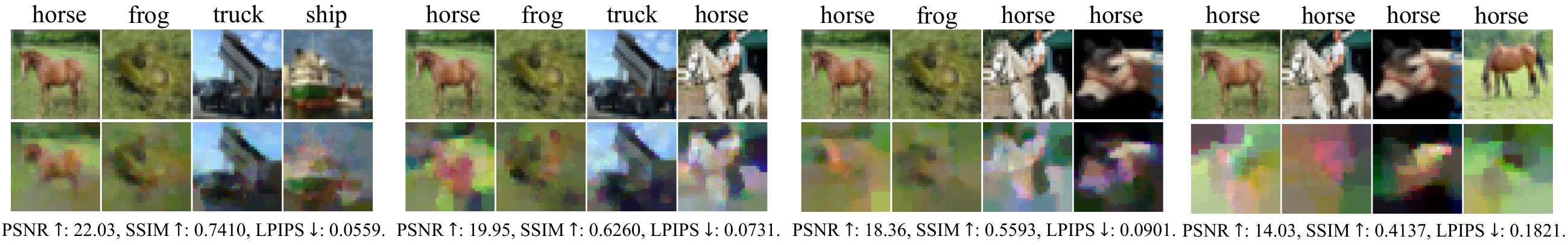}
    \vskip -0.1in
    \caption{Reconstruction results of IG on the CIFAR-10 dataset with a batch size of 4. From left to right, the number of images with the same label are 0, 2, 3, and 4. The first row represents the ground truth, while the second row shows the reconstruction results. These results indicate that more same labels in one batch lead to worse OP-GIA performance.}
\label{fig:compare_label_number_ig}
\end{figure*}

\textit{\textbf{More same labels in one batch, worse OP-GIA performance.}} 
We find that the attack performance on CIFAR-10 is inferior to that on CIFAR-100 when the batch size is large, as shown in Figure \ref{fig:gia_all}. Since both datasets share the same resolution, we hypothesize that this is due to the higher number of images with the same label in CIFAR-10. To validate this point, we compare the attack performance with different numbers of images having the same label within a batch, and the results are shown in Figure \ref{fig:compare_label_number_ig}. As illustrated in this figure, as the number of images with the same label increases, the attack performance decreases, which is consistent with our hypothesis.

To analyze the reasons for this finding, we compare the gradient relationships between these batch images and individual images, which are shown in Table \ref{tab:cos_sim}. $I_i$ denotes that there are $i$ identical labels in the batch data. This also corresponds to each batch images from left to right in Figure \ref{fig:compare_label_number_ig}. $I_i^j$ denotes the $j$-th image in $I_i$. From Table \ref{tab:cos_sim}, we can observe that as the number of images with the same label increases, the cosine similarity between the gradients of a single image and these batch images also increases (i.e., the first row in Table \ref{tab:cos_sim}). Furthermore, the cosine similarity between the gradients of individual images and the batch images with the same labels are all large (i.e., the last column in Table \ref{tab:cos_sim}). Therefore, we conclude that the reason for the difficulty in recovering batch images with the same labels comes from the similarity of gradients between the individual images and the batch images, which is consistent with the objective Eq. (\ref{eq:GIA}) and our gradient similarity  Proposition~\ref{prop:comprae_different_model}. Since the objective function is based on gradient matching, when the gradient between batch images and a single image is similar, it's hard to distinguish them, which makes recovering the input based on gradient matching more difficult. 

\begin{table} [h]
\centering
  \caption{Cosine similarity between pairwise gradients. $I_i$ denotes that there are $i$ identical labels in the batch data. $I_i^j$ denotes the $j$-th image in $I_i$. It shows that as the number of images with the same label increases, the cosine similarity between the gradients of a single image and these batch images also increases (i.e., the first row).}
\label{tab:cos_sim}
\resizebox{0.75\linewidth}{!}{
  \begin{tabular}{lcccc}
    \toprule
     & $I_1$ & $I_2$ & $I_3$ & $I_4$  \\
    \midrule
    $I_{1/2/3/4}^1$ & -0.0305 & 0.7443 & 0.9029 & 0.9541 \\
    $I_4^2$ & - & - & - & 0.9715 \\
    $I_4^3$ & - & - & - & 0.9623 \\
    $I_4^4$ & - & - & - & 0.9454 \\
    \bottomrule
  \end{tabular}
}
\end{table}

\paragraph{Extension to Practical FedAvg.}
\label{para:exp_pracical_fedavg_op_gia}

Previous evaluations are based on an ideal setting where training is performed using FedSGD where clients compute a gradient update on a single local batch of data, and then send it to the server. In this part, we evaluate the privacy leakage in a practical FedAvg scenario \cite{mcmahan2017communication} in which clients train the model locally for multiple iterations before sending the updated model back to the server. In this way, the server only observes the aggregates of the client's local updates.
The experimental results of attacking practical FedAvg on the CIFAR-100 dataset are provided in Table \ref{tab:attack_fedavg}. 
Strong Simulation means that the server knows the local training information (i.e., learning rate, batch size, epochs, etc.) that can simulate the training process at the server side. Weak Simulation denotes that the server does not know the local training information and adopts different training hyperparameters to simulate the local training process. No Simulation means that the server makes no simulation at the server side and considers the local updates as a FedSGD gradient to conduct attacks. From these results, we can conclude that \textbf{\textit{practical FedAvg itself has the ability to resist OP-GIA}}.


\begin{table}[h]
\centering
\caption{Reconstruction results of IG on the CIFAR-100 dataset under practical FedAvg. E denotes the number of local training epochs, and B represents the batch size. These results demonstrate that practical FedAvg itself has the ability to resist OP-GIA.}
\label{tab:attack_fedavg}
{\fontsize{22}{28}\selectfont
\resizebox{1\linewidth}{!}{
\begin{tabular}{cc ccc ccc ccc}
\toprule
& & \multicolumn{3}{c}{Strong Simulation} & \multicolumn{3}{c}{Weak Simulation} & \multicolumn{3}{c}{No Simulation} \\
\cmidrule(lr){3-5} \cmidrule(lr){6-8} \cmidrule(lr){9-11} 
E & B & PSNR $\uparrow$ & SSIM $\uparrow$ & LPIPS $\downarrow$ & PSNR $\uparrow$ & SSIM $\uparrow$ & LPIPS $\downarrow$ & PSNR $\uparrow$ & SSIM $\uparrow$ & LPIPS $\downarrow$ \\
\midrule
1 & 1 & 12.02 & 0.3031 & 0.2049 & 12.26 & 0.3971 & 0.1522 & 11.40 & 0.2301 & 0.2252 \\
1 & 8 & 14.11 & 0.5026 & 0.1295 & 11.75 & 0.3615 & 0.1681 & 11.53 & 0.1824 & 0.2512 \\
1 & 16 & 14.50 & 0.5050 & 0.1315 & 12.26 & 0.3971 & 0.1522 & 11.57 & 0.1503& 0.2751 \\
2 & 1 & 12.13 & 0.3653 & 0.1701 & 11.08 & 0.2672 & 0.2120 & 11.30 & 0.2011& 0.2372 \\
2 & 8 & 12.99 & 0.4252 & 0.1486 & 12.36 & 0.3871 & 0.1624 & 11.44 & 0.1573& 0.2877 \\
2 & 16 & 15.17 & 0.5261 & 0.1138 & 13.67 & 0.4502 & 0.1409 & 11.62 & 0.1375 & 0.2880 \\
5 & 1 & 11.51 & 0.3462 & 0.1816 & 10.93 & 0.2647 & 0.2363 & 11.48 & 0.1898 & 0.2482 \\
5 & 8 & 14.58 & 0.5210 & 0.1232 & 11.63 & 0.3321 & 0.1928 & 10.48 & 0.1666 & 0.3045 \\
5 & 16 & 14.85 & 0.5253 & 0.1137 & 12.22 & 0.3486 & 0.1702 & 11.03 & 0.1311 & 0.3281 \\
\bottomrule
\end{tabular}
}
}
\end{table}

Specifically, as shown in Table \ref{tab:attack_fedavg}, when the server does not have access to the training information on the client (i.e., weak simulation and no simulation), the attack performance is poor. Even when the server has knowledge of the local training information and can precisely simulate the training process on the server side, the reconstruction results are still unsatisfactory. This implies that practical FedAvg itself has the ability to resist OP-GIA.

\subsection{Generation-based GIA} \label{sec:exp_gan_based_gia}

GEN-GIA can be further divided into three categories: 1) optimizing the latent vector $\bm{z}$; 2) optimizing the generator's parameters $\bm{W}$; and 3) training an inversion generation model using an auxiliary dataset. In this section, we evaluate each type of method separately and then provide an overall observation.

\subsubsection{Optimizing Latent Vector $z$}
As mentioned above, by optimizing the input latent vector $\bm{z}$ in Eq. (\ref{eq:gan_based_gia}), GEN-GIA can only generate semantically similar images, not pixel-level reconstructions. This limitation makes numerical metrics such as PSNR, SSIM, and LPIPS unsuitable for evaluating the attack performance. Thus, we primarily rely on visual comparisons. We choose GGL \cite{li2022auditing} to evaluate the privacy leakage of FL under GEN-GIA when optimizing the latent vector $\bm{z}$. Partial reconstruction results are provided in Figure \ref{fig:ggl_vis_imagenet_cifar100}, while the complete results can be found in Section IV-B1 in the Supplementary Material. From these results we can conclude that \textbf{\textit{when optimizing latent vector $\bm{z}$, GEN-GIA can even generate semantically similar images when using random Gaussian noise instead of real gradients, as long as the label information is available, indicating that it is not affected by the factors influencing OP-GIA. However, it heavily relies on the pre-trained generator and only can achieve semantic-level recovery.}}

\begin{figure} [h]
    \centering
\includegraphics[width=1\linewidth]{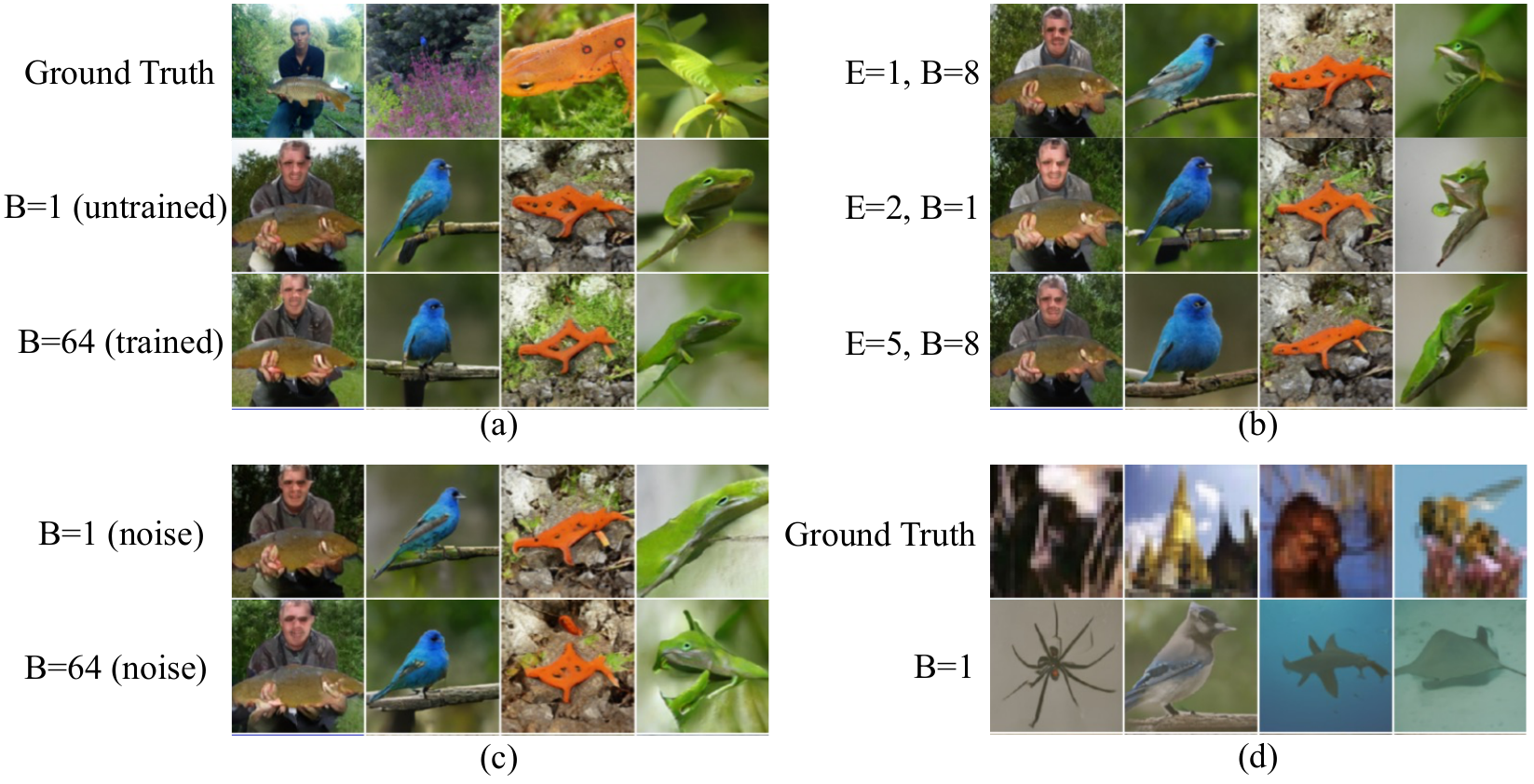}
\vskip -0.1in
    \caption{Reconstruction results of GGL. (a)-(c) Reconstruction results on the ImageNet dataset: (a) with different batch sizes and model training states; (b) under practical FedAvg scenario; (c) with random Gaussian noise. The ground truth for (b) and (c) is similar to (a) and is omitted. (d) Reconstruction results on the CIFAR-100 dataset with a batch size of one and an untrained model. These results show that when optimizing the latent vector $\bm{z}$, GEN-GIA can generate semantically similar images and is not affected by the factors influencing OP-GIA. However, it heavily relies on the pre-trained generator and only can achieve semantic-level recovery.}
\label{fig:ggl_vis_imagenet_cifar100}
\end{figure}

As illustrated in Figure \ref{fig:ggl_vis_imagenet_cifar100}, the reconstruction results of the GEN-GIA are not influenced by the factors affecting OP-GIA. This is likely because GEN-GIA only recovers a latent vector and then use this vector along with label information as input to a pre-trained generator to produce the reconstructed images via $\hat{\bm{x}} = G(\hat{\bm{z}}, \bm{y}^*)$. Consequently, the inferred label information and the pre-trained generator are crucial, while the obtained gradients are less important. To verify this hypothesis, we replace the real gradients in Eq. (\ref{eq:gan_based_gia}) with random Gaussian noise for gradient matching. As shown in Figure \ref{fig:ggl_vis_imagenet_cifar100}(c), even with random Gaussian noise, as long as the label information is available, it is still possible to reconstruct the images, supporting our hypothesis. Note that we assume the label information can be accurately recovered by other methods \cite{zhao2020idlg,yin2021see,dang2021revealing,ma2023instance}, which is consistent with the assumptions in most GEN-GIA methods \cite{jeon2021gradient,li2022auditing,fang2023gifd,ren2022grnn,zhang2023generative,sotthiwat2024generative}. Moreover, the recovery of label information is not the focus of this work.

Additionally, the results for the CIFAR-100 dataset in Figure \ref{fig:ggl_vis_imagenet_cifar100}(d) show that the reconstructed images are not even semantically similar to the input data. This suggests that GEN-GIA heavily relies on the pre-trained generator. In detail, the generator used in our experiment is pre-trained on the ImageNet dataset, which is dissimilar to CIFAR-100 data, resulting in poor reconstruction for CIFAR-100. This underscores the importance of the pre-trained generator for GEN-GIA.

Therefore, despite the advantages of GEN-GIA with optimizing latent vector $\bm{z}$ in not being influenced by many factors, it heavily relies on the pre-trained generator and can only generate data that is semantically similar, rather than reconstructing the original inputs. This implies that GEN-GIA, when optimizing the latent vector $\bm{z}$, suffers from significant constraints and poses little threat to FL.

\subsubsection{Optimizing Generator's Parameters $W$}

\begin{figure}[h]
\centering
\subfigure[] {
\includegraphics[width=0.45\linewidth]{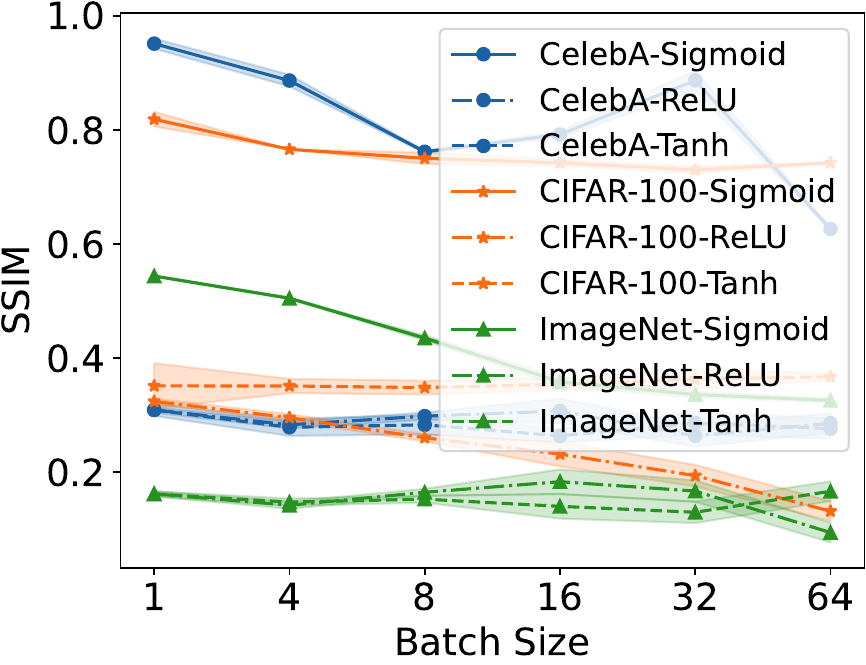}
\label{fig:CI-Net-ssim}
}
\subfigure[] {
\includegraphics[width=0.45\linewidth]{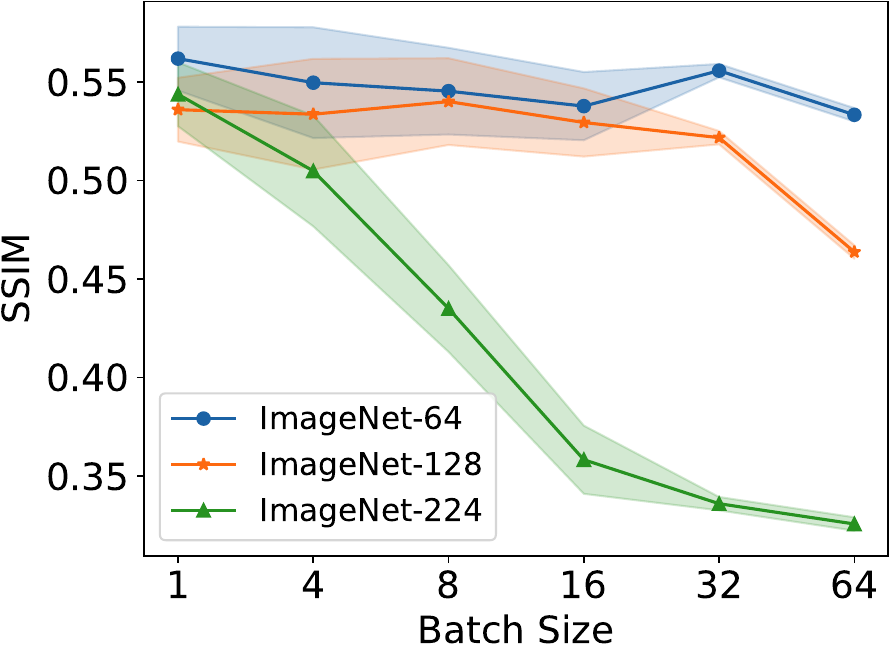}
\label{fig:CI-Net-bs}
}
\vskip -0.1in
    \caption{(a) Reconstruction results of CI-Net evaluated on ResNet-18 with different activation functions on various datasets with different batch sizes. (b) Reconstruction results of CI-Net on ImageNet with different resolutions under the Sigmoid activation function.
    These results show that GEN-GIA with optimizing the generator's parameters $\bm{W}$ is affected by the factors that influence OP-GIA. Moreover, it only works when the target model adopts the Sigmoid activation function and fails with other activation functions.
    }
    \label{fig:CI-Net-all}
\end{figure}

When optimizing the generator's parameters $\bm{W}$, GEN-GIA can reconstruct pixel-wise similar images, thereby enabling pixel-level attacks. We choose CI-Net \cite{zhang2023generative} to evaluate the privacy leakage of FL under GEN-GIA when optimizing the generator's parameters $\bm{W}$. The reconstruction results of CI-Net are provided in Figure \ref{fig:CI-Net-all}.
Additional evaluation metrics, such as PSNR, LPIPS, \revise{Jaccard, and RDLV}, are provided in Figures IV.12 and IV.13 in the Supplementary Material. More experimental results regarding varying numbers of samples that share the same label in one batch and other activation functions (e.g., Leaky-ReLU, RReLU, and GeLU) are shown in Figures IV.15 and IV.14 in the Supplementary Material. 
These results show that \textbf{\textit{when optimizing the generator's parameters $\bm{W}$, GEN-GIA can achieve pixel-level attacks, but is affected by the factors that influence OP-GIA. Moreover, it only works when the target model adopts the Sigmoid activation function and fails with other activation functions.}}

\begin{figure}[h]
\centering
\subfigure[] { \label{fig:distribution} 
\includegraphics[width=0.46\columnwidth]{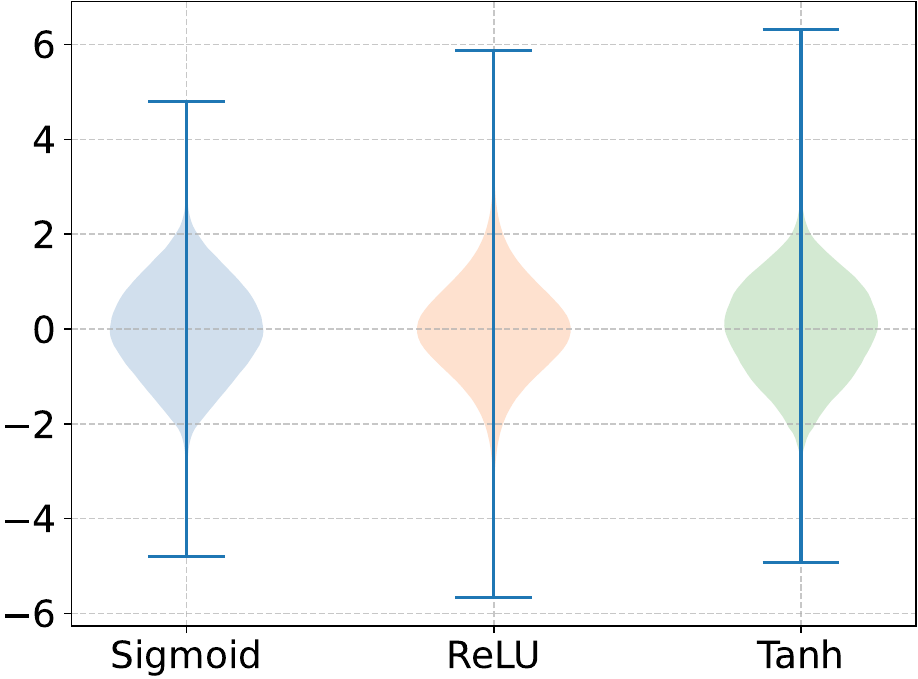}} 
\subfigure[] { \label{fig:sigmoid-tanh}
\includegraphics[width=0.46\columnwidth]{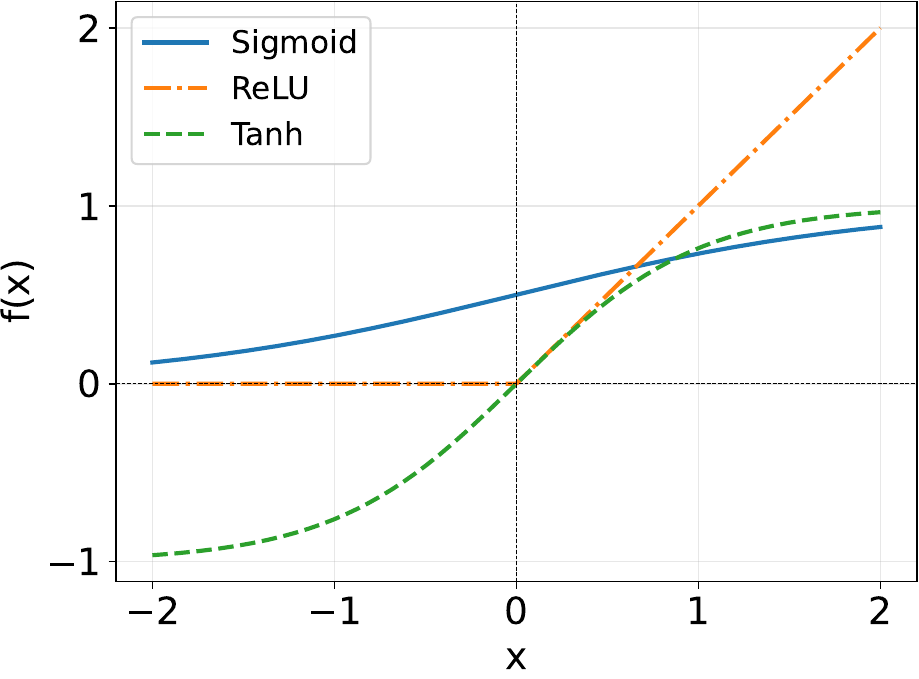}}   
\hfill
\vskip -0.1in
    \caption{(a) Distributions of inputs to different activation functions. It shows that the inputs mostly fall within the range of [-2, 2]. (b) Data transformation of different activation functions under the input range [-2, 2]. It shows that the Sigmoid activation function is approximately linear within this range, whereas other activation functions are not.}
\end{figure}

As shown in Figure \ref{fig:CI-Net-ssim}, CI-Net achieves satisfactory reconstruction results on the model with Sigmoid activation function, while it fails to reconstruct with other activation functions. To explore the reasons behind the vulnerability of the Sigmoid activation function, we start by analyzing the input to the activation function, as shown in Figure \ref{fig:distribution}. These results show that the inputs to the activation function mostly lie in the range $[-2, 2]$. Surprisingly, the Sigmoid activation function is approximately linear within this range, while other activation functions (e.g., ReLU and Tanh) are not, as illustrated in Figure \ref{fig:sigmoid-tanh}. Thus, we attribute the vulnerability of the Sigmoid activation function to its local linearity. Specifically, the data transformation under the Sigmoid activation function is linear, whereas it is non-linear for other activation functions, making it more vulnerable under the Sigmoid activation function.

Therefore, although GEN-GIA optimizes the generator’s parameters $\bm{W}$ can achieve pixel-level attacks, it only works when the target model adopts the Sigmoid activation function and fails with other activation functions. Since most current models rarely utilize Sigmoid activation functions, this type of attack poses little threat to FL.

\subsubsection{Training an Inversion Generation Model}

\begin{figure}[h]
\centering
\subfigure[] {
\includegraphics[width=0.45\linewidth]{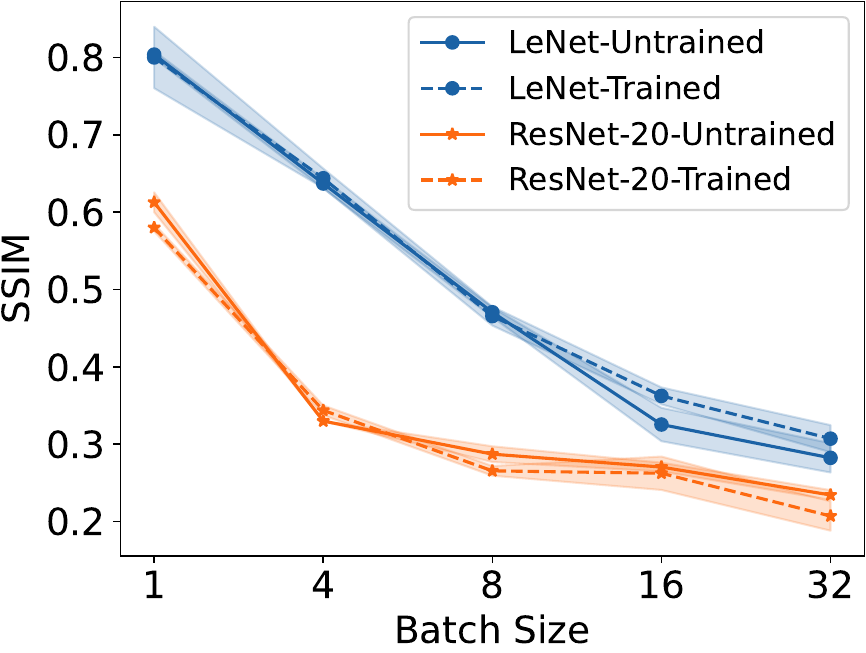}
\label{fig:LTI-model}
}
\subfigure[] {
\includegraphics[width=0.45\linewidth]{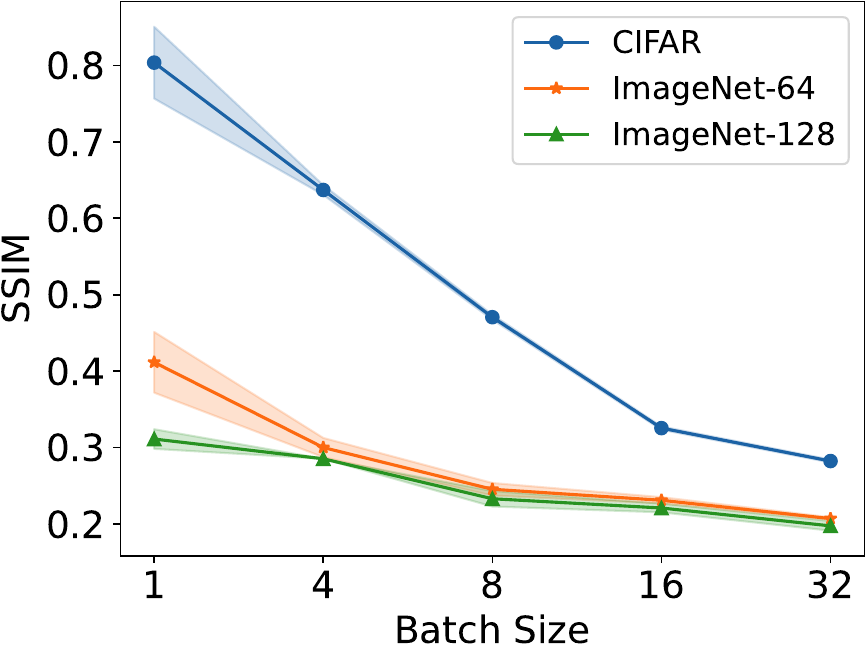}
\label{fig:LTI-resolution}
}
\vskip -0.1in
    \caption{(a) Reconstruction results of LTI evaluated on different models with different training states on CIFAR-10 with different batch sizes. (b) Reconstruction results of LTI on different datasets with different resolutions on LeNet.
    These results show that when training an inversion generation model, GEN-GIA can achieve pixel-level attacks but is influenced by most of the factors that affect OP-GIA, except for the model's training state. 
    }
    \label{fig:LTI-all}
\end{figure}

Here, we choose LTI \cite{wu2023learning} to evaluate the privacy leakage of FL under GEN-GIA with training an inversion generation model. 
Reconstruction results are shown in Figure \ref{fig:LTI-all}.
Additional evaluation metrics, such as PSNR, LPIPS, \revise{Jaccard, and RDLV}, are provided in Figures IV.16 and IV.17 in the Supplementary Material.
Reconstruction results for the varying numbers of samples that share the same label in one batch are provided in Figure IV.18 in the Supplementary Material.
From these results, we can conclude that \textbf{\textit{when training an inversion generation model, GEN-GIA can achieve pixel-level attacks but is influenced by most of the factors that affect OP-GIA, except for the model's training state. Moreover, such a paradigm relies on an auxiliary dataset with a data distribution similar to the local data to train the inversion model \cite{wu2023learning}, which is difficult to achieve in real-world applications.}}

In summary, combining all the experimental results of GEN-GIA, including optimizing of latent vector $\bm{z}$ and generator's parameters $\bm{W}$, and training an inversion generation model using an auxiliary dataset, we conclude that:
\begin{takeaway}
    GEN-GIA has many dependencies, which makes it pose a minimal threat to FL. Some GEN-GIA methods (i.e., optimizing latent vector $\bm{z}$) can only achieve semantic-level recovery and heavily rely on the pre-trained generator. Other GEN-GIA methods (i.e., optimizing generator's parameters $\bm{W}$ and training an inversion model) can perform pixel-level attacks, but they have strong dependencies, such as reliance on the Sigmoid function and an auxiliary dataset.
\end{takeaway}

\subsection{Analytics-based GIA}
\label{sec:exp_ana_based_gia}

The success of ANA-GIA relies on a \textit{malicious} server that alters the model architecture or the model parameters sent to the client. In this section, we use Robbing the Fed \cite{fowl2022robbing}, which manipulates model architecture, and Fishing \cite{wen2022fishing}, which manipulates model parameters, as examples to demonstrate the attack performance of ANA-GIA.

\subsubsection{Manipulating Model Architecture} Since ANA-GIA with manipulating model architecture can achieve exact recovery, resulting in reconstruction results that are exactly the same as the original images, we choose the number of reconstruction images as the metric here. The experimental results of Robbing the Fed \cite{fowl2022robbing} are illustrated in Table \ref{tab:robbing_the_fed}.
Reconstruction results for the varying numbers of samples that share the same label in one batch are provided in Figure IV.19 in the Supplementary Material.
These results show that \textbf{\textit{ANA-GIA, when manipulating model architecture, can achieve great attack performance irrespective of batch size, image resolution, model training state, or the number of same labels in one batch, provided it adopts a relatively large number of bins.}}

\begin{table} [h]
\centering
  \caption{The number of reconstruction images of Robbing the Fed with 1000 bins on different batch sizes. It shows that ANA-GIA, when manipulating model architecture, can achieve great attack performance irrespective of batch size, image resolution, or model training state, provided it adopts a relatively large number of bins.}
  \label{tab:robbing_the_fed}
\resizebox{0.7\linewidth}{!}{
  \begin{tabular}{c ccc}
    \toprule
     Batch Size & {CIFAR-10} & {CIFAR-100} & {ImageNet} \\
    \midrule
    1 & 64/64 & 64/64 & 64/64 \\
    8 & 63/64 & 64/64 & 64/64 \\
    32 & 60/64 & 61/64 & 64/64 \\
    64 & 60/64 & 60/64 & 61/64 \\
    \bottomrule
  \end{tabular}
}
\end{table}

As shown in Table \ref{tab:robbing_the_fed}, Robbing the Fed achieves impressive attack performance regardless of image resolution or model training state, which aligns with Proposition 1 in \cite{fowl2022robbing}. Furthermore, the influence of batch size is minimal if we choose a relatively large number of bins, i.e., 1000. This results in the batch size being smaller than the number of bins, thereby minimizing the batch size's influence. Additionally, unlike the OP-GIA which is affected by the number of same labels in a batch, Robbing the Fed is not influenced by this factor, as illustrated in Figure IV.19 in the Supplementary Material. Furthermore, in the case of practical FedAvg scenario \cite{mcmahan2017communication}, the authors further propose a sparse variant of Robbing the Fed utilizing a two-sided activation function (such as Hardtanh) and weight/bias scaling to maintain the same leakage as in FedSGD \cite{fowl2022robbing}.

\begin{figure}[h]
    \centering
\includegraphics[width=0.8\linewidth]{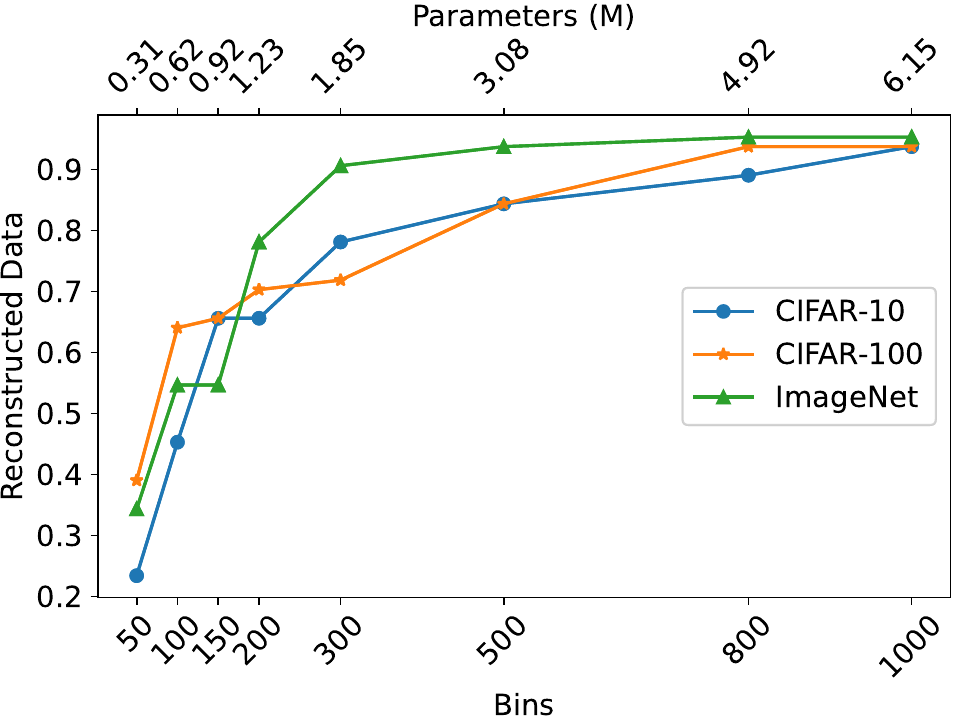}
\vskip -0.1in
    \caption{Proportion of exactly reconstructed images for a batch size of 64 with different numbers of bins, where the upper horizontal axis represents the number of additional parameters introduced. This indicates that there exists a trade-off between attack performance and model size overhead in ANA-GIA when manipulating model architecture. }
    \label{fig:robbing_the_fed}
\end{figure}

However, despite these advantages, \textbf{\textit{ANA-GIA with manipulating model architecture is easily detectable due to the modifications it makes to the network structure, rendering it impractical \cite{garov2024hiding}.}}
Moreover, introducing a linear layer can lead to storage and communication overhead. As shown in Figure \ref{fig:robbing_the_fed}, increasing the number of perfectly reconstructed images requires increasing the number of imprint bins, which, in turn, necessitates an increase in the number of parameters. From these results, we can see that there exists a trade-off between the attack performance and model size overhead in ANA-GIA when manipulating model architecture.

\subsubsection{Manipulating Model Parameters} 

\begin{figure}[h]
\centering
\subfigure[] { \label{fig:fish}
\includegraphics[width=0.46\columnwidth]{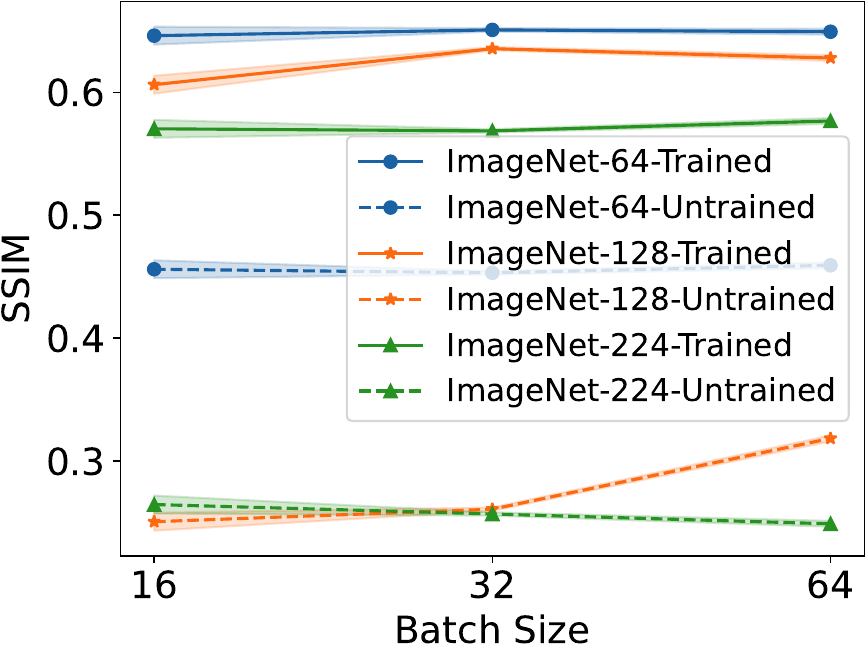}}   
\hfill
\subfigure[] { \label{fig:fish_arch}
\includegraphics[width=0.46\columnwidth]{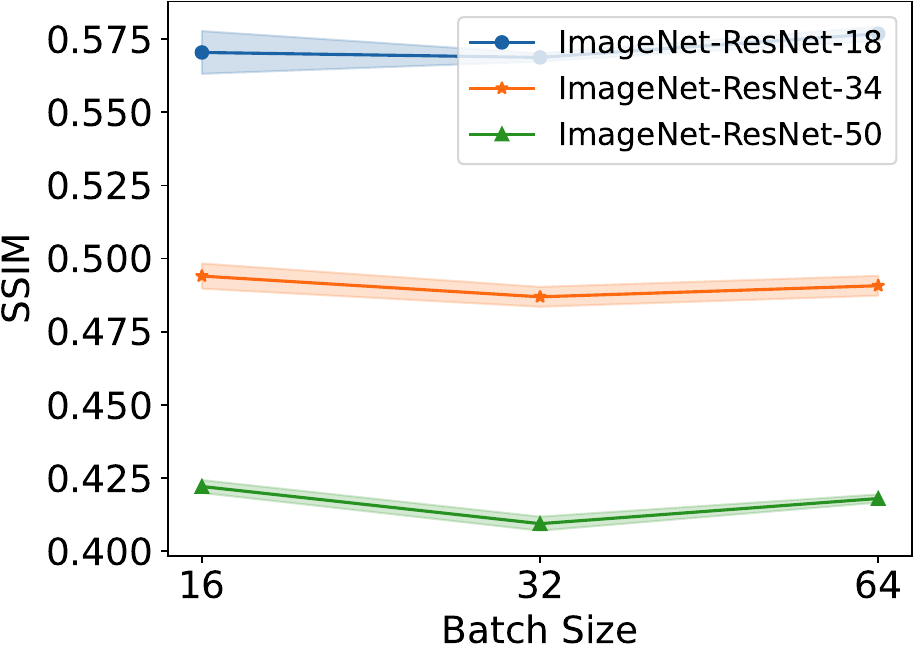}}
\vskip -0.1in
    \caption{Reconstruction results of Fishing on ImageNet with different (a) image resolutions and model training states, and (b) network architectures. These results show that the attack performance of ANA-GIA, which manipulates model parameters, is not affected by batch size but worsens with larger image resolutions, worse model training states, and more complicated model architectures.
    } 
    \label{fig:fish_all}
\end{figure}

As discussed in Section \ref{sec:ana_gia_manip_para}, ANA-GIA that manipulates model parameters can isolate the gradients of a single data point from a batch, making its performance independent of batch size, and the number of same labels in one batch. In this section, we utilize Fishing \cite{wen2022fishing} as the attack method on the ImageNet dataset to examine other impact factors, such as image resolution, model training state, and network architectures. We further utilize several large batch sizes to show that the attack performance is indeed not influenced by batch size. To test the influence of image resolutions, we resize the data in the ImageNet dataset to $64 * 64$ and $128 * 128$ to compare the performance with the original resolution of $224 * 224$. The experimental results of Fishing \cite{wen2022fishing} on the ImageNet dataset are shown in Figure \ref{fig:fish_all}. Additional evaluation metrics, such as PSNR, LPIPS, \revise{Jaccard, and RDLV}, are provided in Figures IV.20 and IV.21 in the Supplementary Material.

As the results in Figure \ref{fig:fish} show, \textbf{\textit{ANA-GIA that manipulates model parameters can achieve satisfactory attack performance regardless of batch size. However, performance decreases with increasing image resolution and from trained to untrained models.}} The impact of the model's training state has an inverse influence compared to OP-GIA, which may be because a well-trained model has better feature extraction ability, making it more compatible with the feature-fishing strategy used in the Fishing attack.
Moreover, as shown in Figure \ref{fig:fish_arch}, \textbf{\textit{the attack performance will also decrease with more complicated network architectures}}, which is similar to OP-GIA.
Additionally, under practical FedAvg scenario \cite{mcmahan2017communication}, performance is maintained because breaking the aggregation makes the batch size equivalent to 1, which eliminates the protection offered by practical FedAvg compared to OP-GIA \cite{wen2022fishing}.

Despite achieving satisfactory attack performance, \textbf{\textit{the modifications to the model parameters can also make it detectable and defensible by clients \cite{garov2024hiding}.}} Moreover, it can only isolate the gradients of a single data point from a batch and must be combined with OP-GIA methods to continue recovering \cite{boenisch2023curious,wen2022fishing}, making it not a purely analytics-based method.

Based on the aforementioned experimental results and the analysis of ANA-GIA involving the manipulation of model architecture and parameters, we conclude that:
\begin{takeaway}
    ANA-GIA can achieve satisfactory attack performance but is easily detected and defended against by clients.
\end{takeaway}

\subsection{Attacks under Parameter-Efficient Fine-Tuning} \label{sec:exp_attack_peft}

In the previous sections, we focused on the traditional scenario where clients share the gradients of the entire model with the server. In this section, we evaluate the potential privacy leakage under Parameter-Efficient Fine-Tuning (PEFT) for foundation models to answer the research question \textbf{R3}.


We choose pre-trained ViT \cite{dosovitskiy2020image} as the base models and utilize Low-Rank Adaptation (LoRA) \cite{hu2022lora} to fine-tune them. In this paradigm, the server receives only the gradients of the LoRA parameters. The reconstruction results using Eq. (\ref{eq:LoRA-GIA}) are shown in Figures \ref{fig:lora_all}, while the visualization results are provided in Section IV-D1 in the Supplementary Material.
Additional evaluation metrics, such as PSNR, LPIPS, \revise{Jaccard, and RDLV}, are provided in Figures IV.22 and IV.23 in the Supplementary Material. More experimental results on ImageNet with different image resolutions are shown in Figure IV.24 in the Supplementary Material. 
From these results, we can conclude that:

\begin{takeaway}
    Attackers can breach privacy on low-resolution images but fail with high-resolution ones under PEFT. Moreover, smaller pre-trained models are better at protecting privacy.
\end{takeaway}

\begin{figure}[h]
\centering
\subfigure[] { \label{fig:LoRA-all}
\includegraphics[width=0.45\columnwidth]{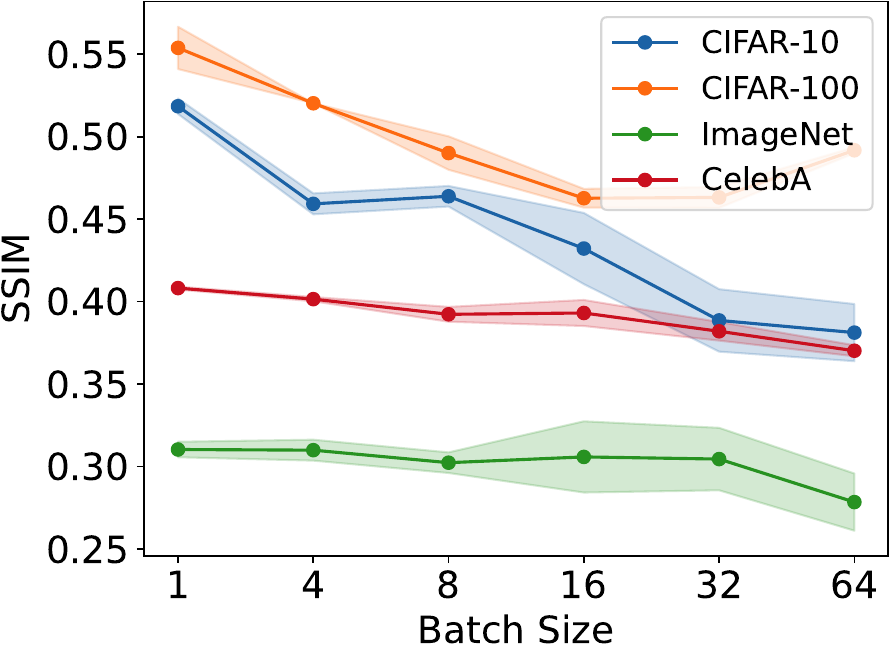}}   
\hfill
\subfigure[] { \label{fig:LoRA-model} 
\includegraphics[width=0.45\columnwidth]{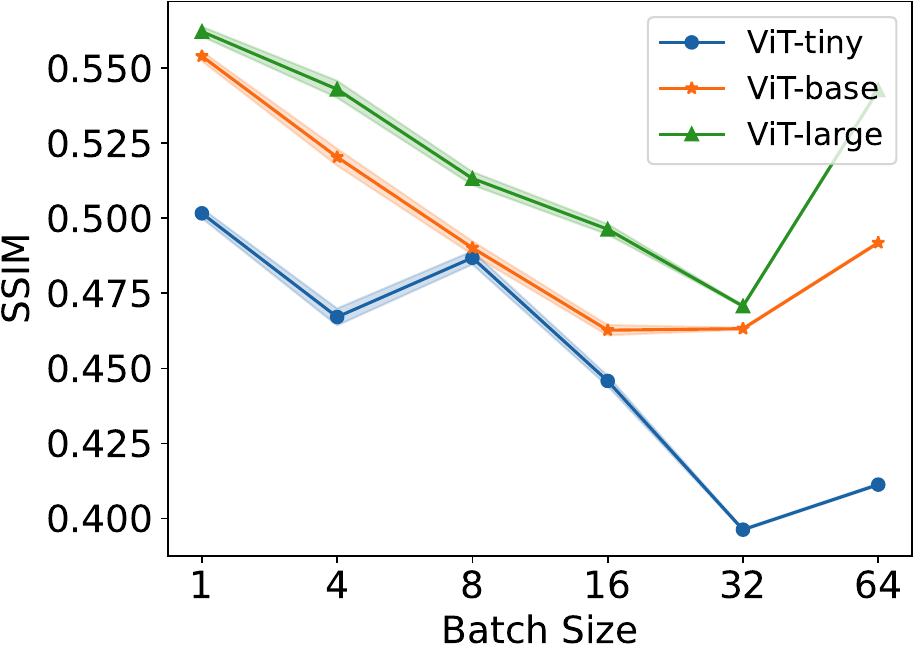}} 
\vskip -0.1in
    \caption{(a) Reconstruction results of Eq. (\ref{eq:LoRA-GIA}) evaluated on the ViT-base fine-tuned with LoRA on different datasets with different batch sizes. 
    (b) Reconstruction results of Eq. (\ref{eq:LoRA-GIA}) evaluated on different ViT architectures fine-tuned with LoRA on the CIFAR-100 dataset. These results show that attackers can breach privacy on low-resolution images but fail with high-resolution ones under PEFT. Moreover, smaller models are better at protecting privacy.
    }
    \label{fig:lora_all}
\end{figure}

Specifically, as shown in Figure \ref{fig:LoRA-all}, the attackers achieve relatively good performance on CIFAR-10, CIFAR-100, and CelebA with a small resolution but perform poorly on ImageNet with a large resolution. This indicates that attackers can breach privacy on low-resolution images but fail with high-resolution ones. The reconstruction results on ImageNet at different image resolutions, shown in Figure IV.24 in the Supplementary Material, further support this point. Moreover, we evaluate the privacy leakage across different ViT architectures fine-tuned by LoRA, with the results shown in Figure \ref{fig:LoRA-model}. According to these results, we find that smaller pre-trained models are better at protecting privacy. This may be because, with a smaller pre-trained model, the LoRA parameters are fewer, resulting in relatively small leaked gradients and less information leakage.

\section{Outlook}
\label{sec:outlook}

\subsection{Discussion}

From the evaluation, we understand the influence factors of each type of GIA method and their corresponding practicality and threat to FL. Here, we provide an overall discussion of all GIA methods and offer insights on how to defend against each type of method.

OP-GIA is the most practical setting since it has no additional reliance. However, its effectiveness is heavily influenced by many FL training factors, such as batch size, image resolution, the number of same labels in one batch, model training state, and network architectures.
To defend against these attacks, we can increase the batch size and choose more complicated network architectures. Furthermore, we find that practical FedAvg with multiple local training steps can inherently resist OP-GIA, indicating that the actual threat posed by OP-GIA to FL is limited. Therefore, it is recommended to perform multiple local training iterations when using the FL algorithm.

As for GEN-GIA, when optimizing the latent vector $\bm{z}$, it can achieve semantic-level recovery as long as the label information is provided. If users do not care about semantic privacy leakage, they do not need to be concerned about such attacks. When optimizing the generator's parameters $\bm{W}$, although it can achieve pixel-level attacks, it only works when the target model uses Sigmoid activation functions and fails with other activation functions. Users can simply choose different activation functions when designing local models to defend against this attack. The third variance of GEN-GIA, which involves training an inversion generation model, demands an auxiliary dataset with a distribution similar to the local data. This requirement is difficult to satisfy in real applications since the server may not know the distribution of the local data. Furthermore, even with a distributionally aligned auxiliary dataset, the performance of such attacks is unsatisfactory. In summary, these constraints--the generation of semantically similar results, specificity to the Sigmoid activation function, and the unrealistic requirement for an auxiliary dataset- significantly limit GEN-GIA's real-world threat to FL systems.

For ANA-GIA, although it can achieve satisfactory attack performance, both manipulating model architecture and parameters are easily detected by clients. The client only needs to conduct some simple checks when receiving the model sent from the server to defend against such attacks \cite{garov2024hiding}. 

Overall, current GIA methods all have their limitations, and if users use the FL training protocol carefully, the privacy leakage of local data can be minimized.

\subsection{Defensive Guidelines}
Based on our extensive evaluation and analysis, we propose actionable strategies to protect FL systems against GIA without using complicated defense mechanisms, which usually lead to privacy-utility trade-offs \cite{guo2025new}. \revise{Existing defense mechanisms against GIA can be broadly categorized into cryptographic methods, such as Secure Multi-party Computing (SMC) \cite{yao1982protocols, bonawitz2017practical, xu2022non} and Homomorphic Encryption (HE) \cite{gentry2009fully, zhang2020privacy, park2022privacy}, and perturbation-based methods like Differential Privacy (DP) \cite{dwork2006differential, geyer2017differentially, wei2021gradient, han2025ppfl}. Cryptographic solutions offer strong security guarantees by performing computations on encrypted data, but their high computational and communication overheads make them impractical for many large-scale deep learning applications \cite{bonawitz2017practical, zhang2020privacy}. DP-based methods, which involve adding calibrated noise to gradients, are more lightweight but often lead to a degradation in model performance, creating a difficult trade-off between privacy and utility \cite{geyer2017differentially, mcmahan2018learning}.
Beyond these theoretically grounded defenses, there are several empirical strategies, such as gradient pruning or masking \cite{zhu2019deep, huang2021evaluating, li2022auditing}, which zero out small-magnitude gradients; gradient perturbation \cite{wei2020framework, wang2024more, zhao2024loki}, which adds Gaussian noise to gradients; learning algorithm modification \cite{ye2024gradient, wu2024concealing}, which augments the training data with a designed vicinal distribution \cite{ye2024gradient} or obfuscates the gradients of the sensitive data with concealed samples \cite{wu2024concealing}; and model modification \cite{scheliga2022precode}, which enhances the local architecture of arbitrary models by inserting PRECODE, a PRivacy EnhanCing mODulE.
However, as demonstrated in their respective studies, most empirical methods also struggle with the inherent privacy-utility trade-off.
In contrast, our work proposes a set of practical guidelines derived from an extensive empirical analysis of GIA vulnerabilities. Rather than applying external defense strategies, our guidelines focus on strategically designing and optimizing the FL training protocol itself to enhance data privacy without compromising model performance. 
}

Specifically, the defender we are discussing here is how to design FL training protocols to safeguard local data privacy. Our recommendations are organized into a three-stage defense pipeline. First, for network design, users should avoid adopting the Sigmoid activation function, as it is particularly vulnerable to GEN-GIA. Moreover, employing more complicated network architectures can make many attack methods more challenging. Second, during the local training protocol, users are encouraged to use a larger batch size and train locally for multiple steps before sending model updates to the server, as this can also hinder attacks. Third, simple client-side validation should be adopted: when receiving the model from the server, users should verify the model architecture and parameters to avoid being attacked by ANA-GIA. By integrating these practices, users can better protect their data privacy when using FL without worrying about being attacked by current GIA methods. The above guidelines can be summarized into the following key takeaway tips:
\begin{takeaway}
    Three-stage defense pipeline: (1) avoid the Sigmoid activation function and use more complicated network architectures during network design, (2) adopt larger batch sizes and multi-step local training in the local training protocol, and (3) implement client-side validation to check for any potential malicious modification to the model architecture and parameters upon receiving the model from the server.
\end{takeaway}

\subsection{Implications for Attack Design}
While this work advocates for privacy protection, the experiments conducted may also provide some inspiration for improving attack strategies. Since attack and defense are interactive games that complement each other, a better attack method can promote the development of a better defense. For designing attack methods, researchers should not only consider attack performance but also practical applicability. For example, although current ANA-GIA methods can achieve high attack performance, modifications to the model architecture or parameters make them easily detected and defended against by clients, thus limiting their practicality \cite{garov2024hiding}. Moreover, current GEN-GIA methods depend on factors such as pre-trained generators, the Sigmoid activation function, and an auxiliary dataset, which limits their practicality. Thus, when designing attack methods in the future, researchers should prioritize their practicality and minimize reliance on external factors, similar to OP-GIA. Furthermore, researchers should pay more attention to practical scenarios like FedAvg, where clients train the model locally for multiple iterations before sending updates, rather than focusing solely on FedSGD.

\section{Conclusion}

In this work, we conduct a comprehensive study on various types of GIA in FL. We thoroughly examine and assess existing GIA methods, dividing them into three categories: \textit{optimization-based} GIA (OP-GIA), \textit{generation-based} GIA (GEN-GIA), and \textit{analytics-based} GIA (ANA-GIA).
Theoretically, we provide an error bound analysis for data reconstruction and a gradient similarity proposition in the context of OP-GIA. These theoretical analyses serve as powerful tools for guiding future work on OP-GIA methods, allowing researchers to evaluate the effectiveness and limitations of their models’ attack performance more comprehensively. 
Empirically, we conduct extensive evaluations, revealing that: (1) OP-GIA is the most practical attack setting, but its performance is not satisfactory; (2) GEN-GIA has many dependencies, posing a minimal threat to FL; (3) ANA-GIA can achieve satisfactory attack performance but is easily detected and defended against by clients.
Based on these experimental findings, we provide insights on how to defend against GIA in FL.
We hope our study can assist researchers in designing more robust FL frameworks to defend against these attacks.

\section*{Acknowledgments}

This work was supported by National Natural Science Foundation of China (62306253, T2522030), Early Career Scheme from the Research Grants Council of Hong Kong SAR (27207025, 27204623), and Guangdong Natural Science Fund-General Programme (2024A1515010233).

\bibliography{TPAMI_revision/main}
\bibliographystyle{IEEEtran}

\clearpage
\setcounter{section}{0}

\renewcommand{\thefigure}{\Roman{section}.\arabic{figure}}
\setcounter{figure}{0}

\renewcommand{\thetable}{\Roman{section}.\arabic{table}}
\setcounter{table}{0}

\renewcommand{\theequation}{\Roman{section}.\arabic{equation}}
\setcounter{equation}{0}

\twocolumn[
        \centering
        \Large
        \textbf{Exploring the Vulnerabilities of Federated Learning: A Deep Dive into Gradient Inversion Attacks}\\
        \vspace{0.5em}Supplementary Material \\
        \vspace{1.0em}
       ] 

\section{Theoretical Proof} \label{sec:theoretical_proof}

\subsection{Proof of Theorem 1} \label{app_sec:proof_theo_1}

\begin{proof}
In order to prove Theorem 1, we first assume that there exists $T$ temporal model weights $\theta_t$ and leaked gradients $\nabla_{\theta_t}\mathcal{L}(\bm{x}^{*}, \bm{y}^{*})$. Consequently, multiple $f_t(\cdot)$ can be constructed, i.e., $f_t(\bm{x}) = \mathcal{D}(\nabla_{\theta_t}\mathcal{L}(\bm{x}^{*}, \bm{y}^{*}), \nabla_{\theta_t}\mathcal{L}({\bm{x}}, \hat{\bm{y}})) + \lambda\omega({\bm{x}})$. This assumption is practical since clients usually share gradients with the server multiple times, and any of these leaked gradients can be used to conduct an attack. 

Then, according to Algorithm 1, we have:
\begin{equation} \label{eq:eq_1}
    \begin{aligned}
        &\| \hat{\bm{x}}_{i} - \bm{x}^* \|_2 \\
        = \ &\|\hat{\bm{x}}_{i-1}-\eta \nabla_{\hat{\bm{x}}_{i-1}}f (\hat{\bm{x}}_{i-1}) - \bm{x}^* \|_2 \\
        = \ &\|\hat{\bm{x}}_{i-1} - \eta \nabla_{\hat{\bm{x}}_{i-1}}f (\hat{\bm{x}}_{i-1}) - \bm{x}^* \\
        \ &- \eta \frac{1}{T}\sum_{t=1}^T \nabla_{\hat{\bm{x}}_{i-1}} f_t({\hat{\bm{x}}_{i-1}}) + \eta \frac{1}{T}\sum_{t=1}^T \nabla_{\hat{\bm{x}}_{i-1}} f_t({\hat{\bm{x}}_{i-1}}) \|_2 \\
        \leq \ &\|\hat{\bm{x}}_{i-1} - \bm{x}^* - \eta \frac{1}{T}\sum_{t=1}^T \nabla_{\hat{\bm{x}}_{i-1}} f_t({\hat{\bm{x}}_{i-1}}) \|_2 \\
        \ &+ \eta \| \frac{1}{T}\sum_{t=1}^T \nabla_{\hat{\bm{x}}_{i-1}} f_t({\hat{\bm{x}}_{i-1}}) - \nabla_{\hat{\bm{x}}_{i-1}}f (\hat{\bm{x}}_{i-1})\|_2 \\
        = \ &\|\hat{\bm{x}}_{i-1} - \bm{x}^* - \eta \nabla_{\hat{\bm{x}}_{i-1}} \bar{f}({\hat{\bm{x}}_{i-1}}) \|_2 \\
        \ &+ \eta \| \nabla_{\hat{\bm{x}}_{i-1}} \bar{f}({\hat{\bm{x}}_{i-1}}) - \nabla_{\hat{\bm{x}}_{i-1}}f (\hat{\bm{x}}_{i-1})\|_2, \\
    \end{aligned}
\end{equation}
where $\nabla_{\hat{\bm{x}}_{i-1}} \bar{f}({\hat{\bm{x}}_{i-1}}) = \frac{1}{T}\sum_{t=1}^T \nabla_{\hat{\bm{x}}_{i-1}} f_t({\hat{\bm{x}}_{i-1}})$.  

For the first term in Eq. (\ref{eq:eq_1}),
\begin{equation*}
    \begin{aligned}
        &\|\hat{\bm{x}}_{i-1} - \bm{x}^* - \eta \nabla_{\hat{\bm{x}}_{i-1}} \bar{f}({\hat{\bm{x}}_{i-1}})\|_2^2 \\
        = \ &\|\hat{\bm{x}}_{i-1} - \bm{x}^*\|_2^2 - 2\eta \langle \hat{\bm{x}}_{i-1} - \bm{x}^*, \nabla_{\hat{\bm{x}}_{i-1}} \bar{f}({\hat{\bm{x}}_{i-1}})\rangle \\
        \ &+ \eta^2 \| \nabla_{\hat{\bm{x}}_{i-1}} \bar{f}({\hat{\bm{x}}_{i-1}})\|_2^2. \\
    \end{aligned}
\end{equation*}

Since $f$ is $L-$smooth and $\mu$ strong convex, according to Lemma 3.11 in \cite{bubeck2015convex}, we obtain:
\begin{equation*}
\begin{aligned}
    & \langle \hat{\bm{x}}_{i-1} - \bm{x}^*, \nabla_{\hat{\bm{x}}_{i-1}} \bar{f}({\hat{\bm{x}}_{i-1}}) \rangle \\
    \geq \ &\frac{1}{\mu+L}\|\nabla_{\hat{\bm{x}}_{i-1}} \bar{f}({\hat{\bm{x}}_{i-1}})\|_2^2 + \frac{\mu L}{\mu+L}\|\hat{\bm{x}}_{i-1}-\bm{x}^*\|_2^2.
\end{aligned}
\end{equation*}

Let $\eta \leq \sqrt{\frac{2}{\mu+L}}$, then we get:
{\small
\begin{equation*}
    \begin{aligned}
    & \|\hat{\bm{x}}_{i-1} - \bm{x}^* - \eta \nabla_{\hat{\bm{x}}_{i-1}} \bar{f}({\hat{\bm{x}}_{i-1}})\|_2^2 \\
    \leq \ & (1-\frac{2 \mu}{\mu+L})\|\hat{\bm{x}}_{i-1}-\bm{x}^*\|_2^2 +(\eta^2 - \frac{2}{\mu+L})\|\nabla_{\hat{\bm{x}}_{i-1}} \bar{f}({\hat{\bm{x}}_{i-1}})\|_2^2 \\
    \leq \ & (1-\frac{2 \mu}{\mu+L})\|\hat{\bm{x}}_{i-1}-\bm{x}^*\|_2^2.
    \end{aligned}
\end{equation*}
}
Since $\sqrt{1-x} \leq 1 - \frac{x}{2}$, we get:
{\small
\begin{equation} \label{eq:eq_2}
    \|\hat{\bm{x}}_{i-1} - \bm{x}^* - \eta \nabla_{\hat{\bm{x}}_{i-1}} \bar{f}({\hat{\bm{x}}_{i-1}})\|_2 
    \leq  (1-\frac{\mu}{\mu+L})\|\hat{\bm{x}}_{i-1}-\bm{x}^*\|_2.
\end{equation}
}For the second term in Eq. (\ref{eq:eq_1}), since the dimension of $\nabla_{\bm{x}}f(\bm{x})$ is $B \times C \times H \times W$, where $B$ is the batch size, $C, H, W$ denote the image resolution, then we have:
{\small
\begin{equation} \label{eq:eq_3}
    \begin{aligned}
    & \| \nabla_{\hat{\bm{x}}_{i-1}} \bar{f}({\hat{\bm{x}}_{i-1}}) - \nabla_{\hat{\bm{x}}_{i-1}}f (\hat{\bm{x}}_{i-1})\|_2 \\
    = \ & \|\nabla_{\hat{\bm{x}}_{i-1}}f (\hat{\bm{x}}_{i-1}) - \nabla_{\hat{\bm{x}}_{i-1}} \bar{f}({\hat{\bm{x}}_{i-1}})\|_2 \\
    = \ & \sum_{j=1}^{B \times C \times H \times W} \sqrt{[\nabla_{\hat{\bm{x}}_{i-1}}f (\hat{\bm{x}}_{i-1}) - \nabla_{\hat{\bm{x}}_{i-1}} \bar{f}({\hat{\bm{x}}_{i-1}})]^2_{(j)}} \\
    \leq \ & \sqrt{BCHW} \sqrt{\sum_{j=1}^{B \times C \times H \times W}[\nabla_{\hat{\bm{x}}_{i-1}}f (\hat{\bm{x}}_{i-1}) - \nabla_{\hat{\bm{x}}_{i-1}} \bar{f}({\hat{\bm{x}}_{i-1}})]^2_{(j)}} \\
    = \ & \sqrt{BCHW} \kappa,
    \end{aligned}
\end{equation}
}where $\kappa$ is the upper bound of $\|\nabla_{\hat{\bm{x}}} f({\hat{\bm{x}}}) -\nabla_{\hat{\bm{x}}} \bar{f}({\hat{\bm{x}}})\|_2$, and the inequality is due to Cauchy–Schwarz inequality.

Combine Eqs. (\ref{eq:eq_1}), (\ref{eq:eq_2}), and (\ref{eq:eq_3}), and use $\eta \leq \sqrt{\frac{2}{\mu+L}}$, we have:
\begin{equation} \label{eq:eq_4}
    \| \hat{\bm{x}}_{i} - \bm{x}^* \|_2 
    \leq (1-\frac{\mu}{\mu+L})\|\hat{\bm{x}}_{i-1}-\bm{x}^*\|_2 + \sqrt{\frac{2BCHW}{\mu+L}} \kappa.
\end{equation}

Apply Eq. (\ref{eq:eq_4}) recursively for $i = 1, \ldots, I$, we obtain:
{\small
\begin{equation}
     \|\hat{\bm{x}}-\bm{x}^*\|_2 \leq (1-\frac{\mu}{\mu+L})^{I}\|\hat{\bm{x}}_0-\bm{x}^*\|_2 + \frac{\sqrt{2BCHW(\mu+L)}}{\mu} \kappa.
\end{equation}
}
\end{proof}

\subsection{Proof of Proposition 1}

\begin{proof}
Suppose the ground truth data $(\bm{x}^*, \bm{y}^*)$ and reconstruction results $(\hat{\bm{x}}, \hat{\bm{y}})$ obtained by Algorithm 1 satisfy: $\mathcal{D}(\nabla_{\theta}\mathcal{L}(\bm{x}^{*}, \bm{y}^{*}), \nabla_{\theta}\mathcal{L}(\hat{\bm{x}}, \hat{\bm{y}})) < \epsilon, \ \epsilon > 0$.
Then, the set of all possible reconstruction results obtained by Algorithm 1 for model parameters $\theta_{t_1}$ and $\theta_{t_2}$ can be written as:
\begin{equation*}
\begin{aligned}
\hat{A}_1 &= \{\hat{\bm{x}}: \mathcal{D}(\nabla_{\theta_{t_1}}\mathcal{L}(\bm{x}^{*}, \bm{y}^{*}), \nabla_{\theta_{t_1}}\mathcal{L}(\hat{\bm{x}}, \hat{\bm{y}})) < \epsilon \}, \\
\hat{A}_2 &= \{\hat{\bm{x}}: \mathcal{D}(\nabla_{\theta_{t_2}}\mathcal{L}(\bm{x}^{*}, \bm{y}^{*}), \nabla_{\theta_{t_2}}\mathcal{L}(\hat{\bm{x}}, \hat{\bm{y}})) < \epsilon \},
\end{aligned}
\end{equation*}
where $\hat{A}_1$ and $\hat{A}_2$ represent the sets of all possible reconstruction results obtained by Algorithm 1 for $\theta_{t_1}$ and $\theta_{t_2}$, respectively. Then, all the elements in these sets can be considered local minima obtained by Algorithm 1, while only one of them is the optimal solution and the others are not. Thus, the presence of more elements in the set indicates that the reconstruction task is more challenging \cite{ge2015escaping}.

According to the assumption that the cardinality of the set 
$\{\bm{x}^{*,j}: \mathcal{D}(\nabla_{\theta_{t_1}}\mathcal{L}(\bm{x}^{*,i}, \bm{y}^{*,i}), \nabla_{\theta_{t_1}}\mathcal{L}(\bm{x}^{*,j}, \bm{y}^{*,j}) < \epsilon \}$ 
is greater than the cardinality of the set 
$\{\bm{x}^{*,j}: \mathcal{D}(\nabla_{\theta_{t_2}}\mathcal{L}(\bm{x}^{*,i}, \bm{y}^{*,i}), \nabla_{\theta_{t_2}}\mathcal{L}(\bm{x}^{*,j}, \bm{y}^{*,j}) < \epsilon \}$
for any $i$ and $\epsilon > 0$, we have $|\hat{A}_{1}| > |\hat{A}_{2}|$, where $|A|$ denotes the cardinality of the set $A$. This means recovering the input data using the leaked gradients by Algorithm 1 on $\theta_{t_1}$ is harder than on $\theta_{t_2}$.

\end{proof}

\section{Proposition 1. in Robbing the Fed}
\label{app_sec:robbing_the_fed_prop}

\begin{proposition} \label{prop:robbing_the_fed} \cite{fowl2022robbing}
If the server knows the cumulative density function (assumed to be continuous) of some quantity associated with user data that can be measured with a linear function $h : \mathbb{R}^m \rightarrow \mathbb{R}$, then for a batch of size $B$ and a number of imprint bins $k > B > 2$, by using an appropriate combination of linear layer and ReLU activation, the server can expect to exactly recover 

{ \tiny
    \begin{equation*}
        \frac{1}{\left(\begin{array}{c}
k+B-1 \\
k-1
\end{array}\right)}\left[\sum_{i=1}^{B-2} i \cdot\left(\begin{array}{c}
k \\
i
\end{array}\right) \cdot\left(\sum_{j=1}^{\left\lfloor\frac{B-i}{2}\right\rfloor}\left(\begin{array}{c}
k-i \\
j
\end{array}\right)\left(\begin{array}{c}
B-i-j-1 \\
j-1
\end{array}\right)\right)\right]+ r(B, k)
    \end{equation*}
}samples of user data (where the data is in $\mathbb{R}^m$) perfectly, where $\left(\begin{array}{l}
k \\
i
\end{array}\right)$ denotes the number of ways to select the $i$ bins that have exactly 1 element, and $r(B, k)$ is a correction term.
\end{proposition}


\section{Visualization of Selected Subset}

\begin{figure}[h]
  \centering
  \subfigure[CIFAR-10.] {     
\includegraphics[width=0.46\linewidth]{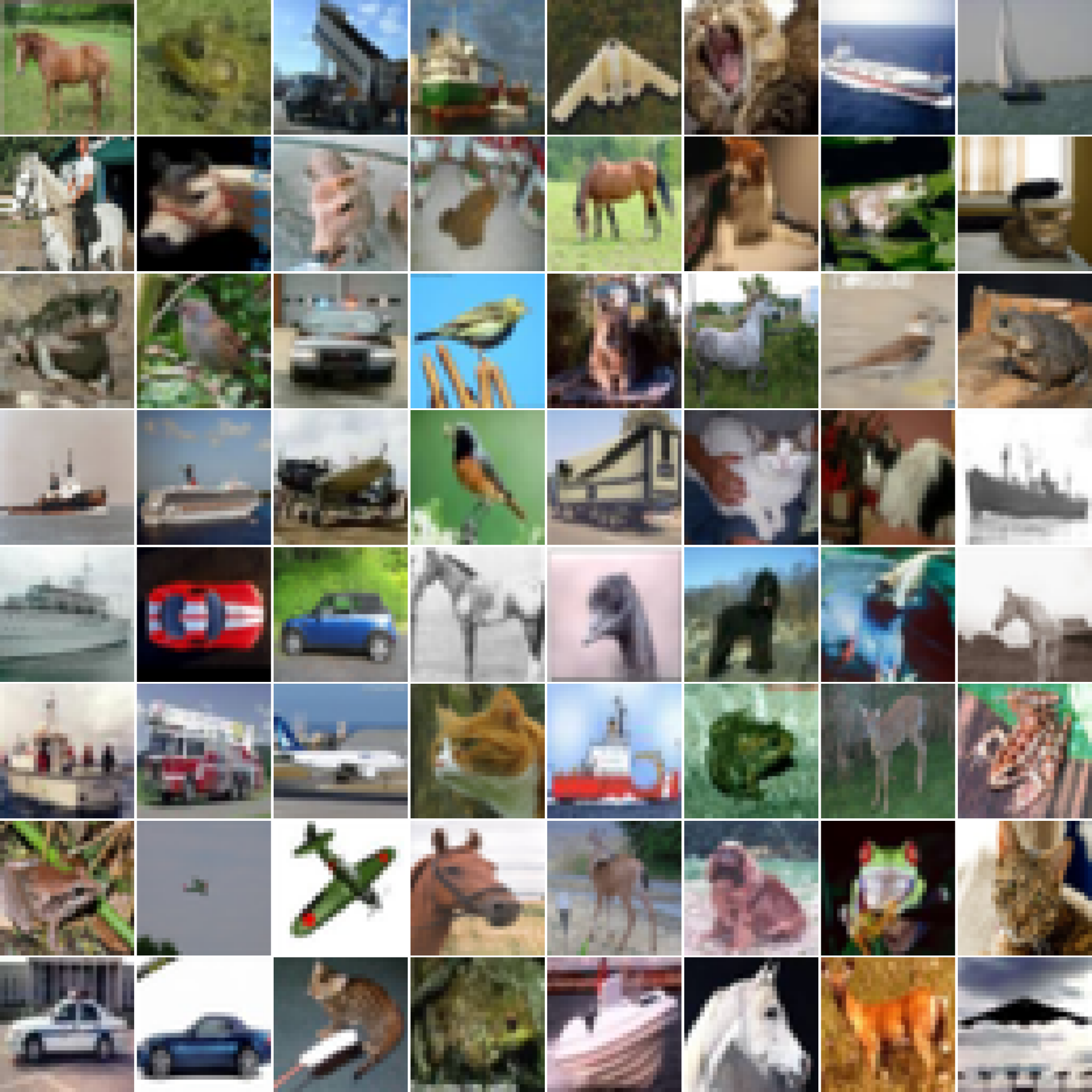}  
} 
    \subfigure[CIFAR-100.] {     
\includegraphics[width=0.46\linewidth]{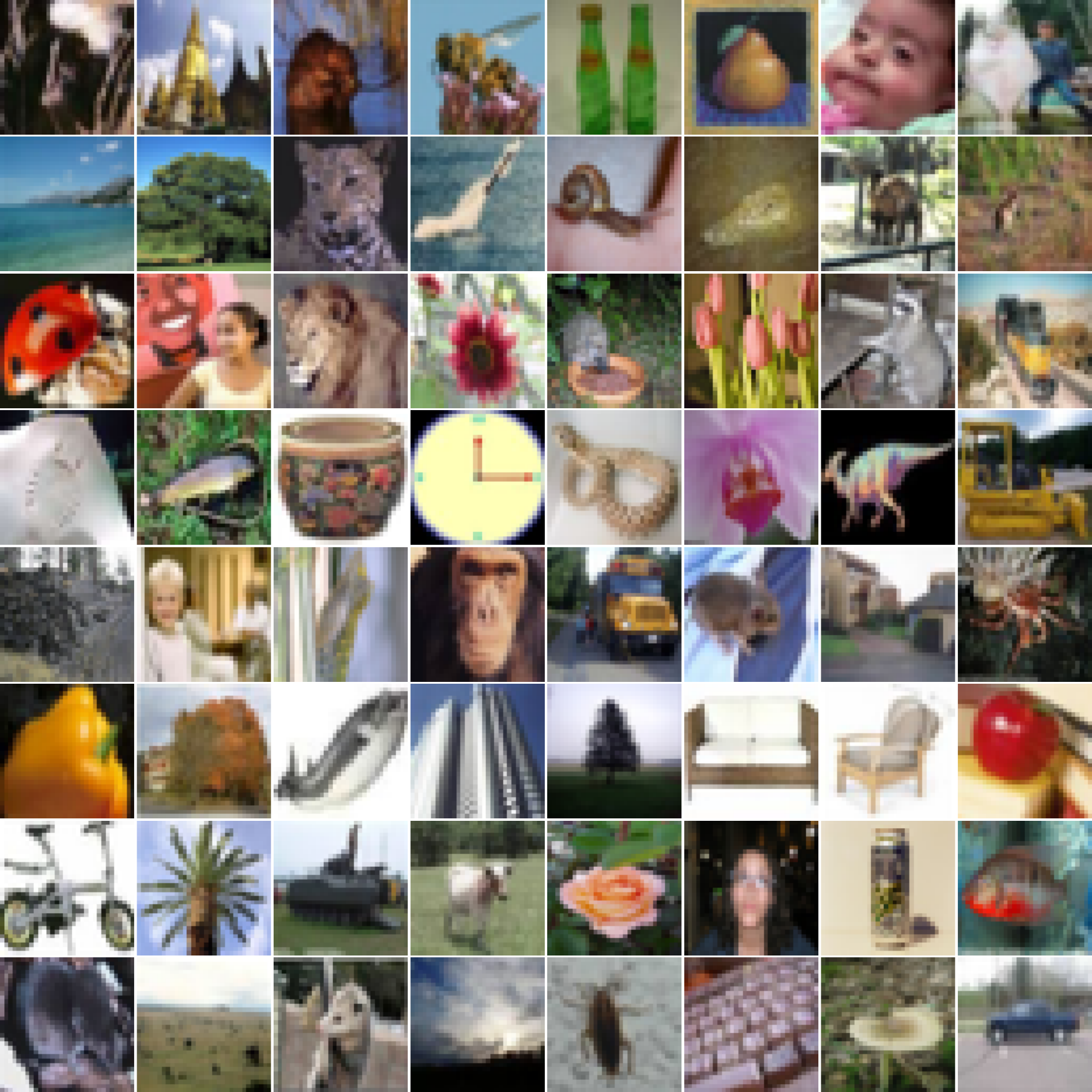}  
} 
    \subfigure[ImageNet.] {     
\includegraphics[width=0.46\linewidth]{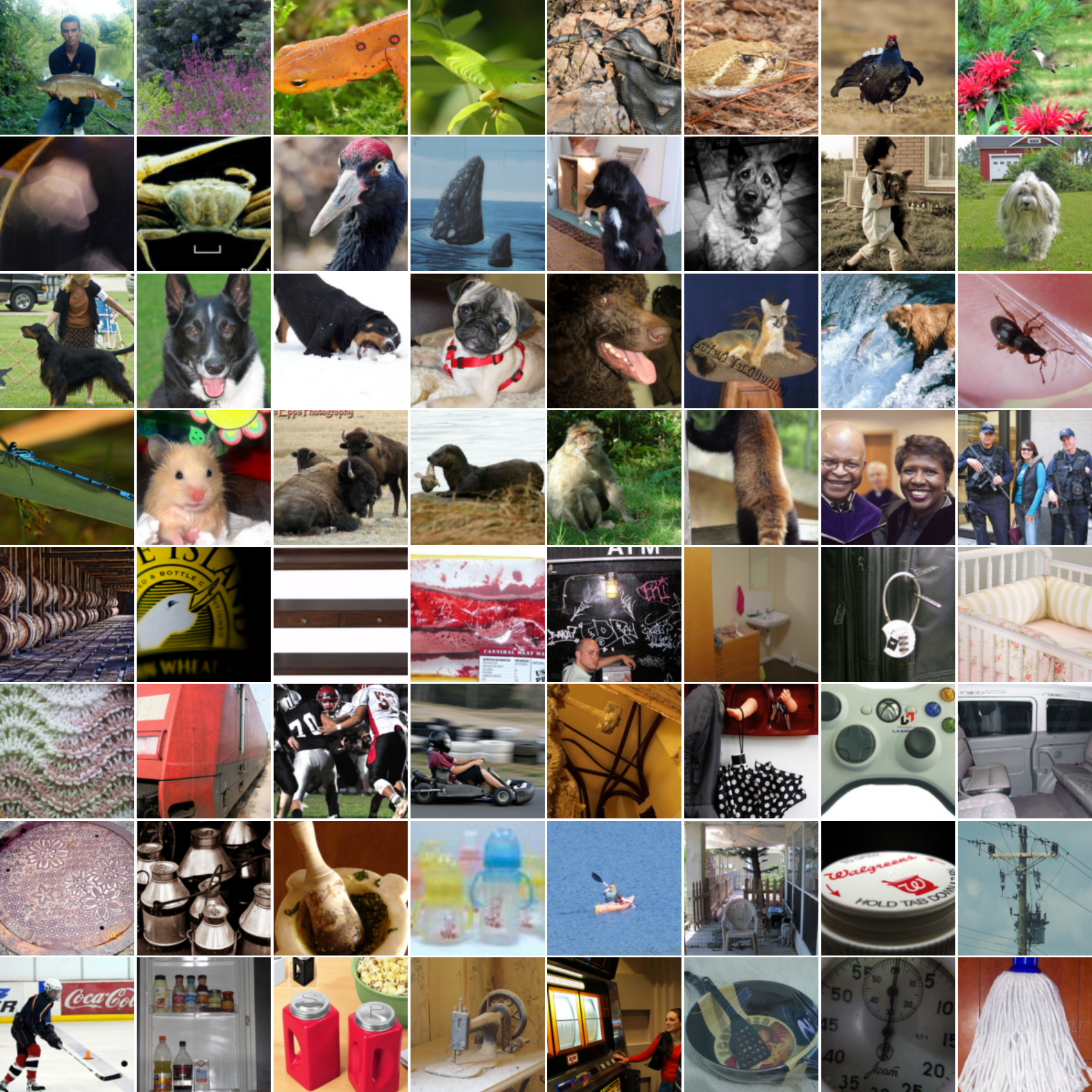}  
} 
    \subfigure[CelebA.] {     
\includegraphics[width=0.46\linewidth]{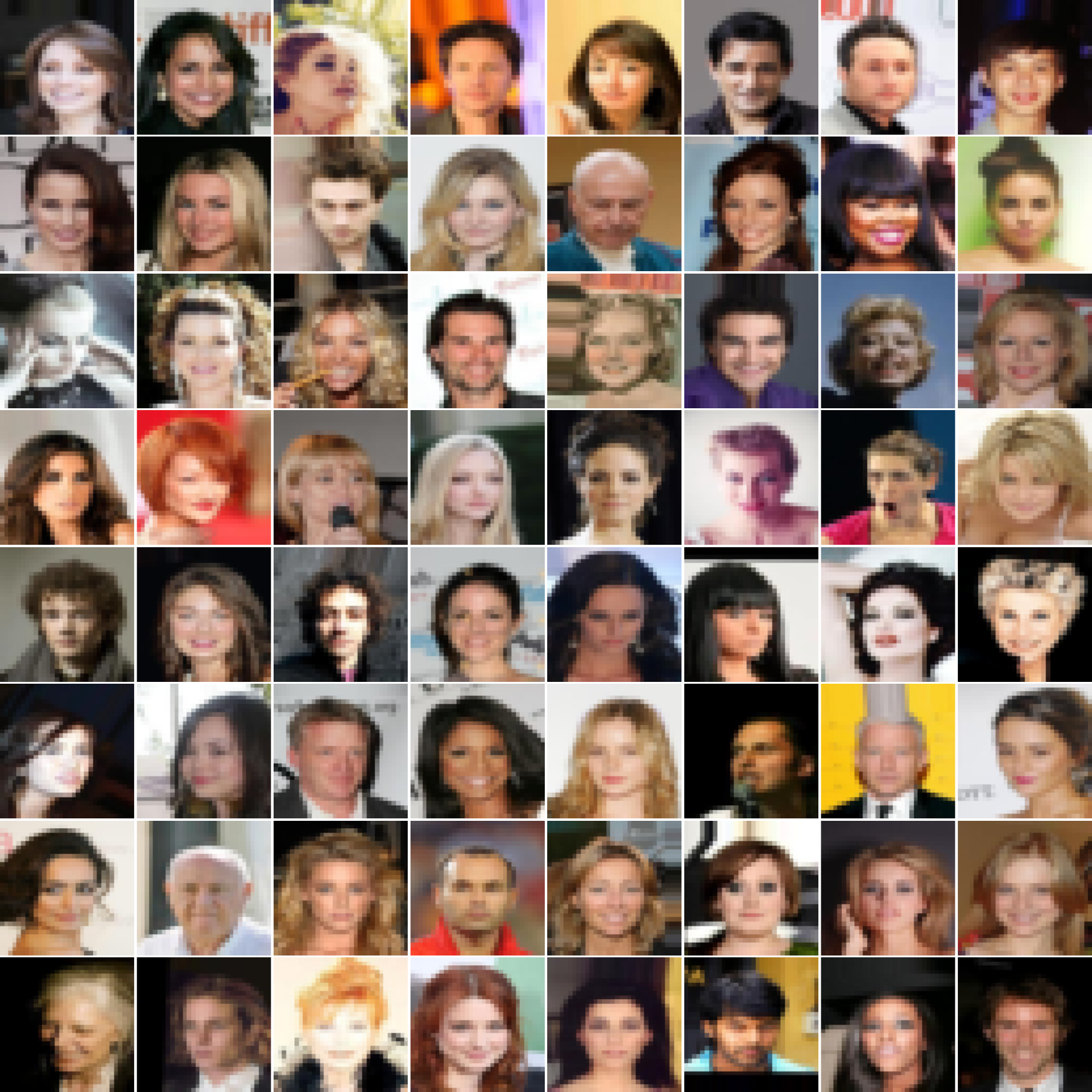}  
} 
  \caption{Visualization of each subset. We resize the images to $64*64$ for the CelebA dataset. The image resolutions for CIFAR-10 and CIFAR-100 are $32*32$, while they are $224*224$ for ImageNet.}
  \label{app_fig:vis_all_data}
\end{figure}

The illustration of these selected subsets is shown in Figure \ref{app_fig:vis_all_data}.

\section{More Experimental Results}

\subsection{Optimization-based GIA}

\begin{figure}[h]
  \centering
  \subfigure[PSNR $\uparrow$.] {     
\includegraphics[width=0.29\linewidth]{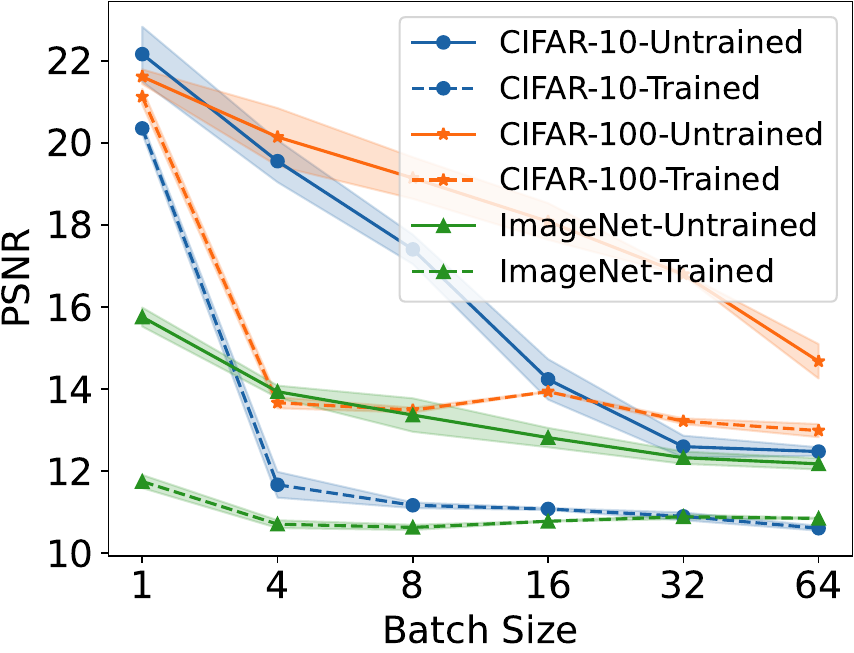}  
}   
\hfill
   \subfigure[SSIM $\uparrow$.] {    
\includegraphics[width=0.29\linewidth]{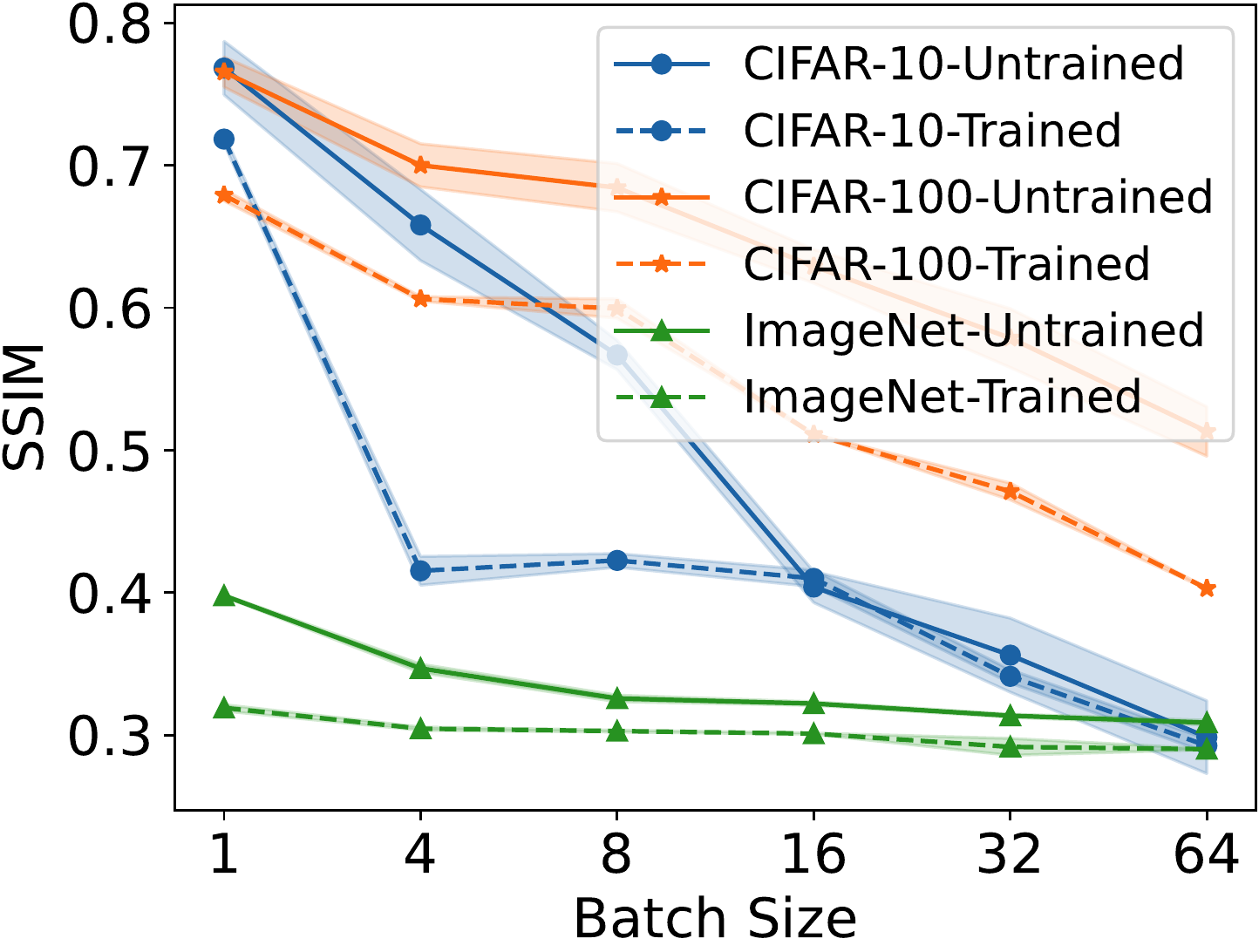}  
} 
\hfill
    \subfigure[LPIPS $\downarrow$.] {    
\includegraphics[width=0.29\linewidth]{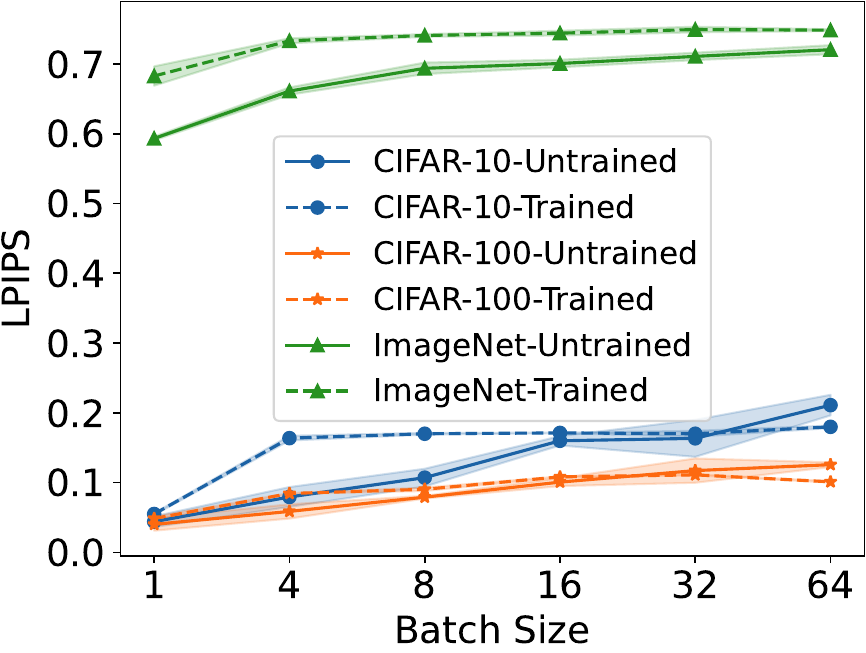} 
} 
\revise{
\hfill
    \subfigure[Jaccard $\uparrow$.] {    
\includegraphics[width=0.29\linewidth]{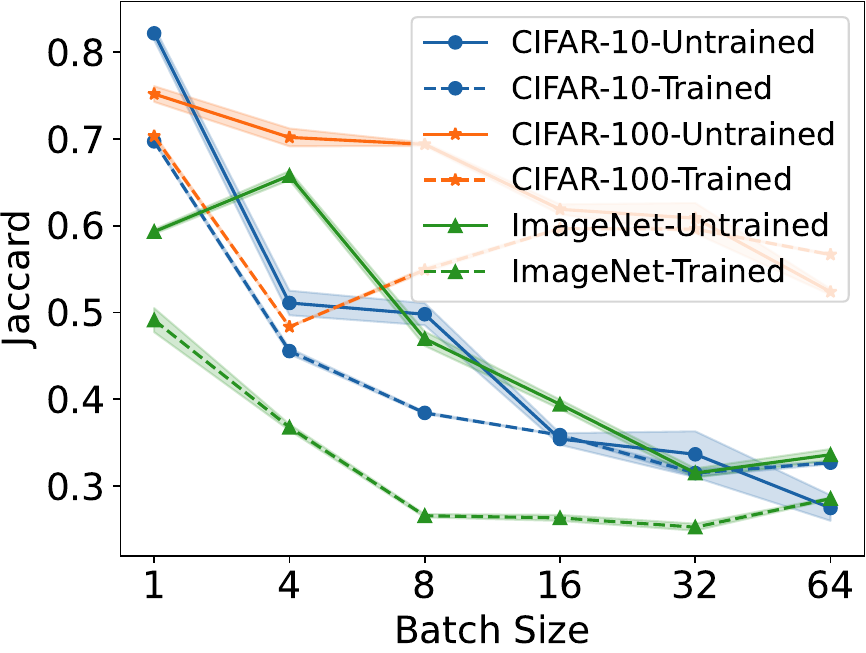} 
} 
    \subfigure[RDLV $\uparrow$.] {    
\includegraphics[width=0.29\linewidth]{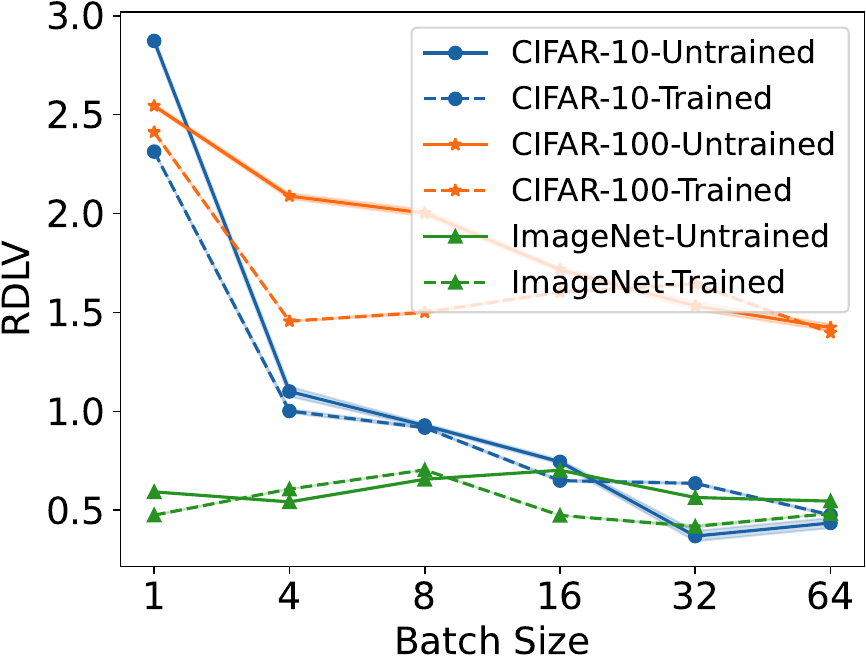} 
} 
}
  \caption{Reconstruction results of IG evaluated on models in different training states on various datasets with different image resolutions and batch sizes, where the shaded region represents the standard deviation. These results show that a larger batch size, higher image resolution, and better model training state lead to worse OP-GIA performance.
  }
  \label{app_fig:gia_all}
\end{figure}

\begin{figure}[h]
  \centering
  \subfigure[PSNR $\uparrow$.] {     
\includegraphics[width=0.29\linewidth]{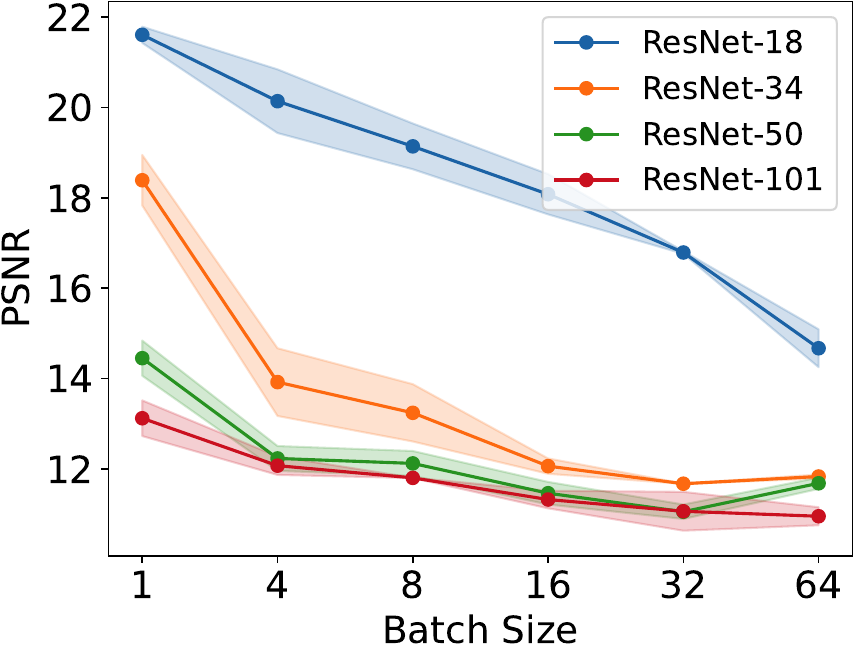}  
}   
\hfill
   \subfigure[SSIM $\uparrow$.] {    
\includegraphics[width=0.29\linewidth]{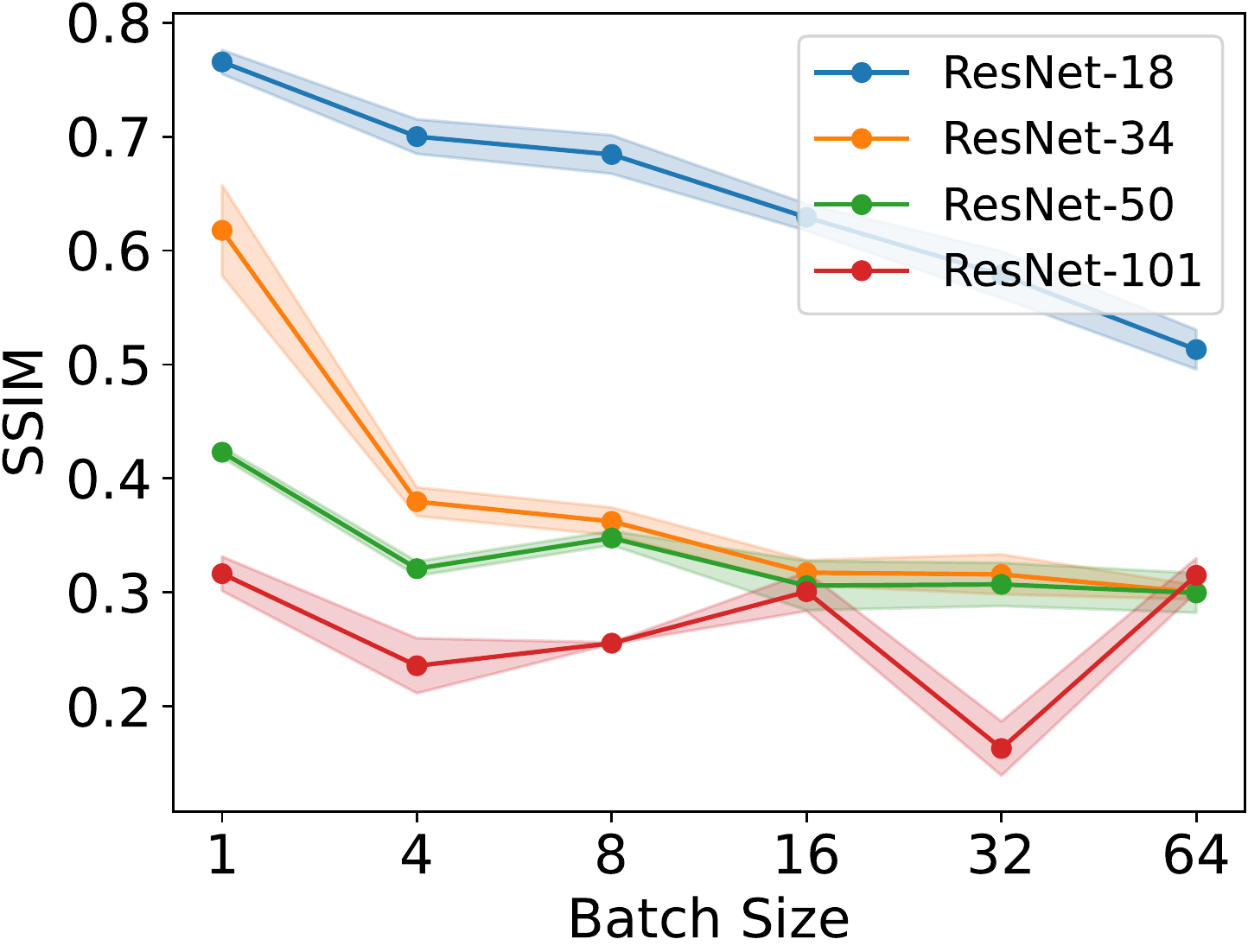}  
} 
\hfill
    \subfigure[LPIPS $\downarrow$.] {    
\includegraphics[width=0.29\linewidth]{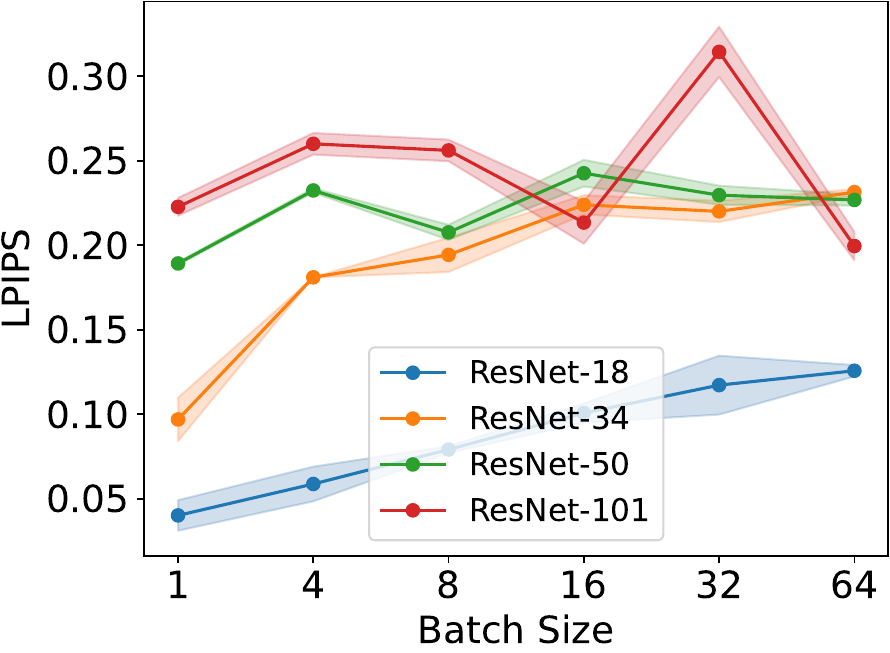}
}   
\revise{
\hfill 
    \subfigure[Jaccard $\uparrow$.] {    
\includegraphics[width=0.29\linewidth]{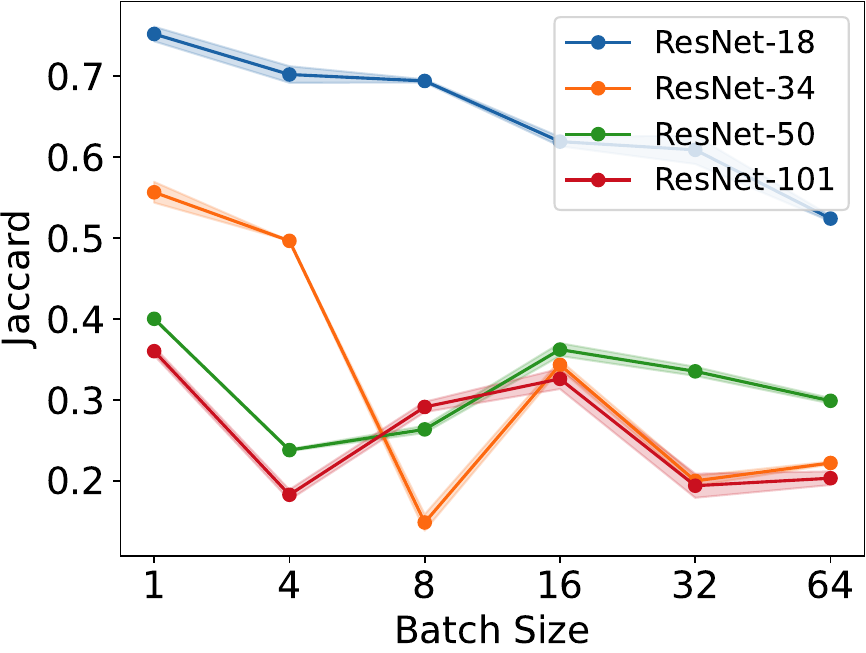}
}   
    \subfigure[RDLV $\uparrow$.] {    
\includegraphics[width=0.29\linewidth]{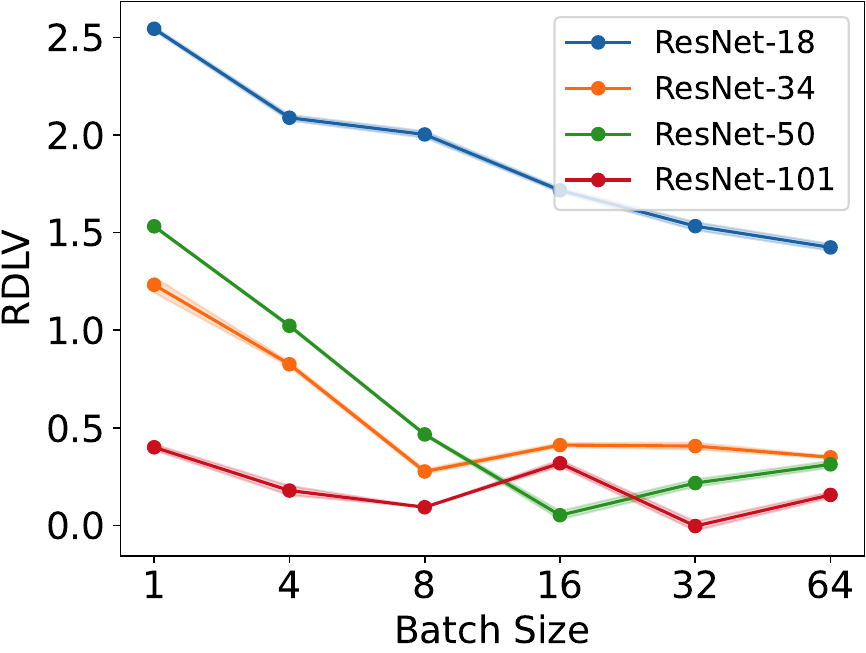}
}
}
  \caption{Reconstruction results of IG with different network architectures on the CIFAR-100 dataset, where the shaded region represents the standard deviation. These results show that more complicated network architecture lead to worse OP-GIA performance.
}
  \label{app_fig:compare_arch}
\end{figure}

The reconstruction results of IG with all evaluation metrics are shown in Figures \ref{app_fig:gia_all} and \ref{app_fig:compare_arch}. These results show that larger batch size, higher image resolution, more complicated network architecture, and better model training state lead to worse OP-GIA performance.

\begin{figure}[h]
  \centering
  \subfigure[PSNR $\uparrow$.] {     
\includegraphics[width=0.29\linewidth]{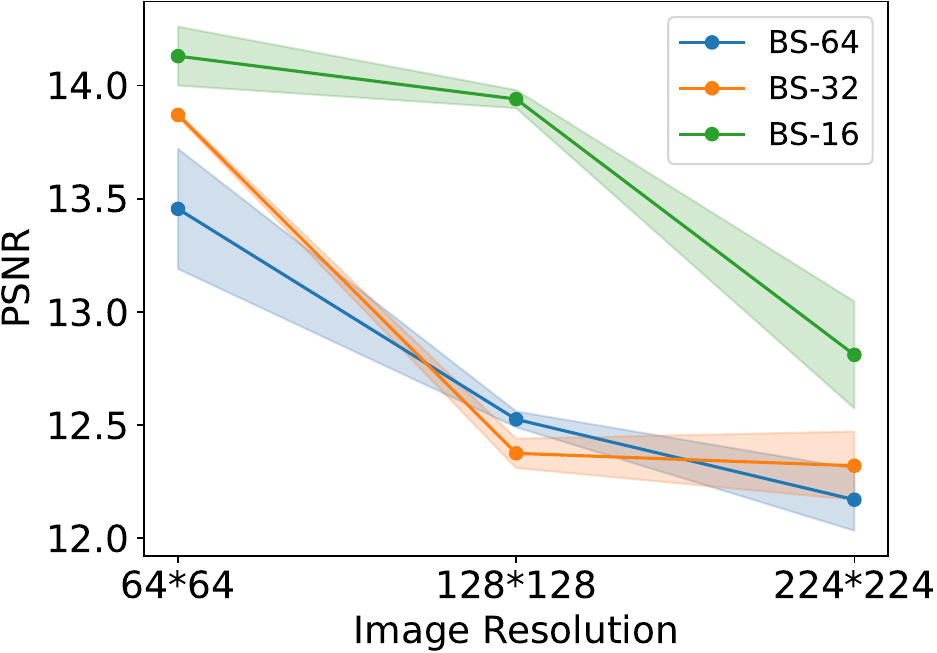}  
}   
\hfill
   \subfigure[SSIM $\uparrow$.] {    
\includegraphics[width=0.29\linewidth]{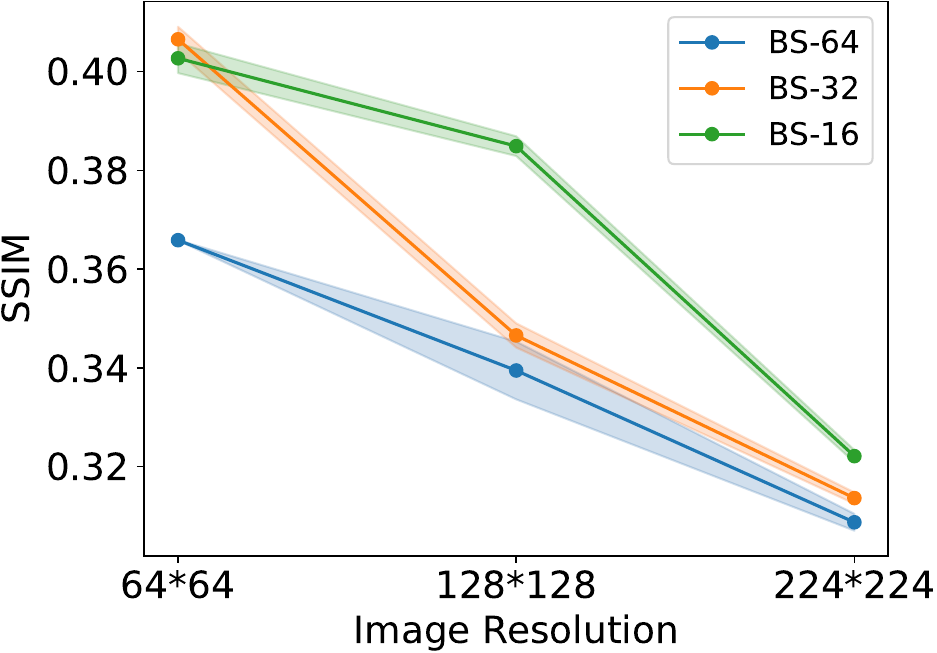}  
} 
\hfill
    \subfigure[LPIPS $\downarrow$.] {    
\includegraphics[width=0.29\linewidth]{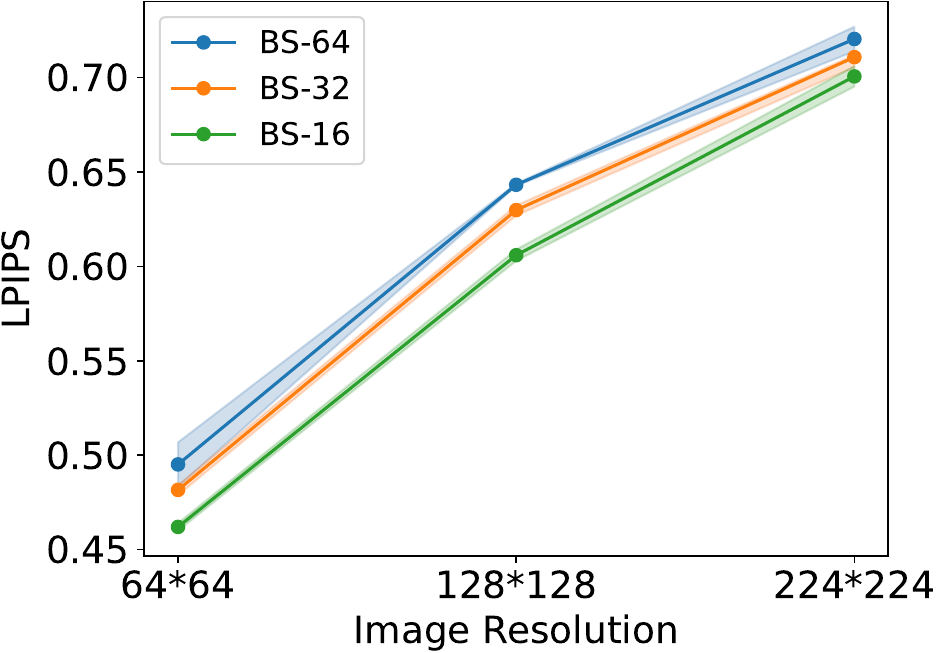}  
}
\revise{
\hfill 
    \subfigure[Jaccard $\uparrow$.] {    
\includegraphics[width=0.29\linewidth]{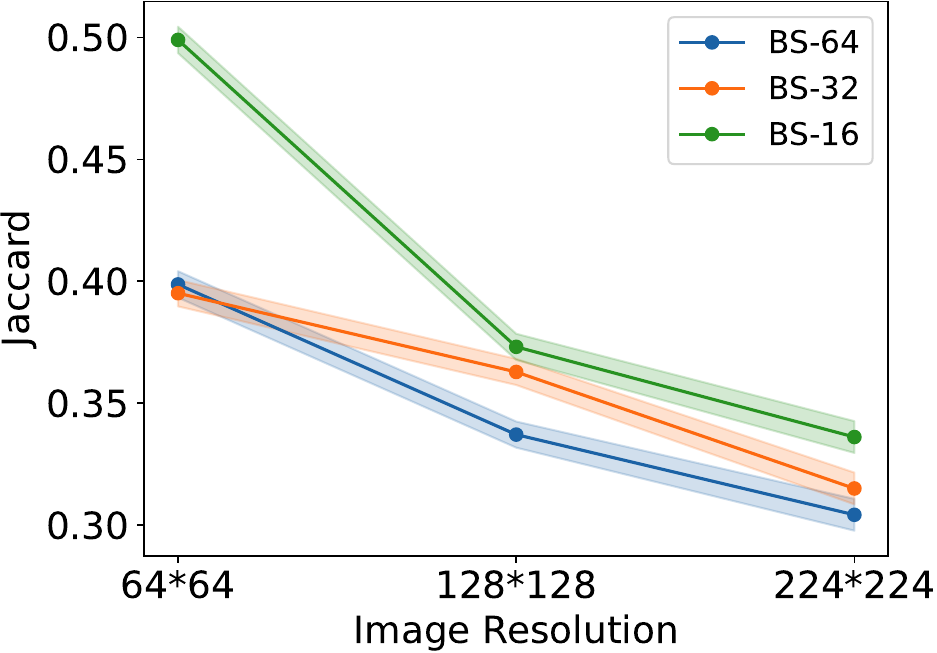}  
}
    \subfigure[RDLV $\uparrow$.] {    
\includegraphics[width=0.29\linewidth]{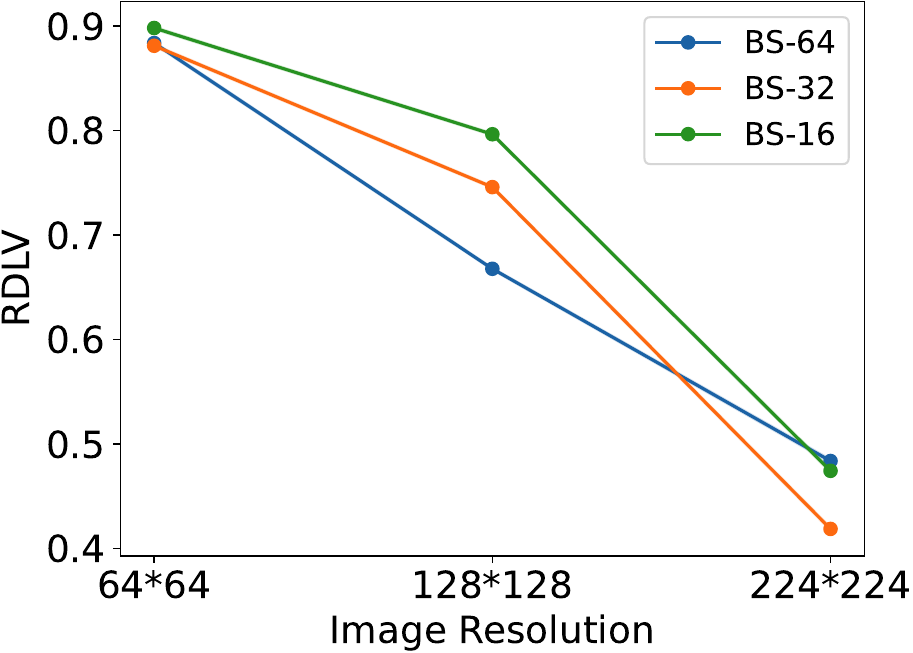}  
}
}
  \caption{Reconstruction results of IG with different image resolutions and batch sizes on the ImageNet dataset, where the shaded region represents the standard deviation. It shows that the image resolution has a more significant impact on the performance of OP-GIA than the batch size.
  }
  \label{app_fig:compare_resolution}
\end{figure}

The reconstruction results of IG on the ImageNet dataset with different resolutions \footnote{We resize the images to different resolutions.} and batch sizes are shown in Figure \ref{app_fig:compare_resolution}.
From these results, we can see that doubling the image resolution leads to a larger decrease in attack performance compared to doubling the batch size. This suggests that the image resolution has a more significant impact on the performance of OP-GIA than the batch size. 

\subsubsection{Visualization} \label{app_sec:op-gia_visua}
\begin{figure}[!h]
\centering
\subfigure[Batch size = 1. PSNR $\uparrow$: 22.03, SSIM $\uparrow$: 0.7410, LPIPS $\downarrow$: 0.0559.] {    
\includegraphics[width=0.29\linewidth]{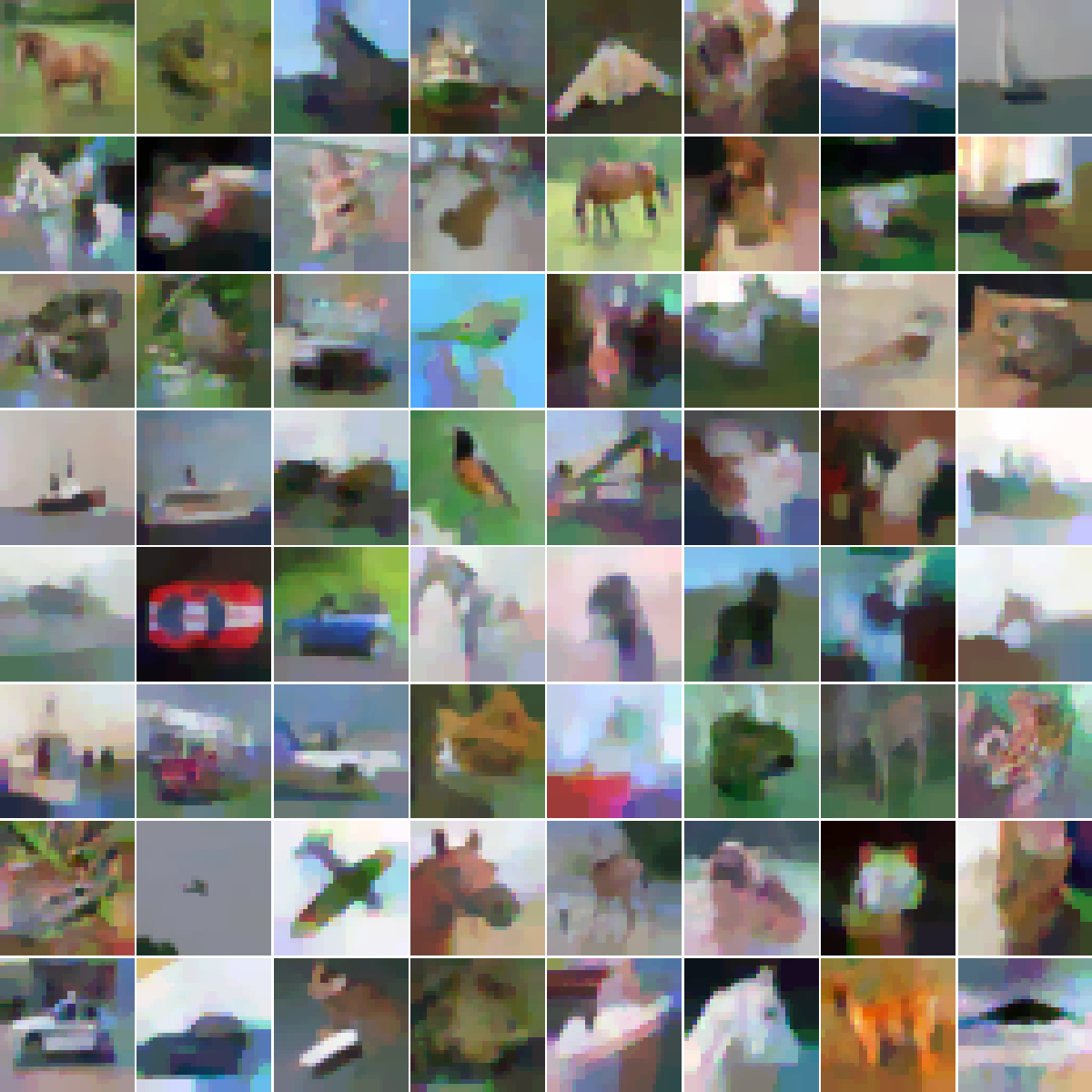}  
}   
\subfigure[Batch size = 32. PSNR $\uparrow$: 12.59, SSIM $\uparrow$: 0.3561, LPIPS $\downarrow$: 0.1634.] {    
\includegraphics[width=0.29\linewidth]{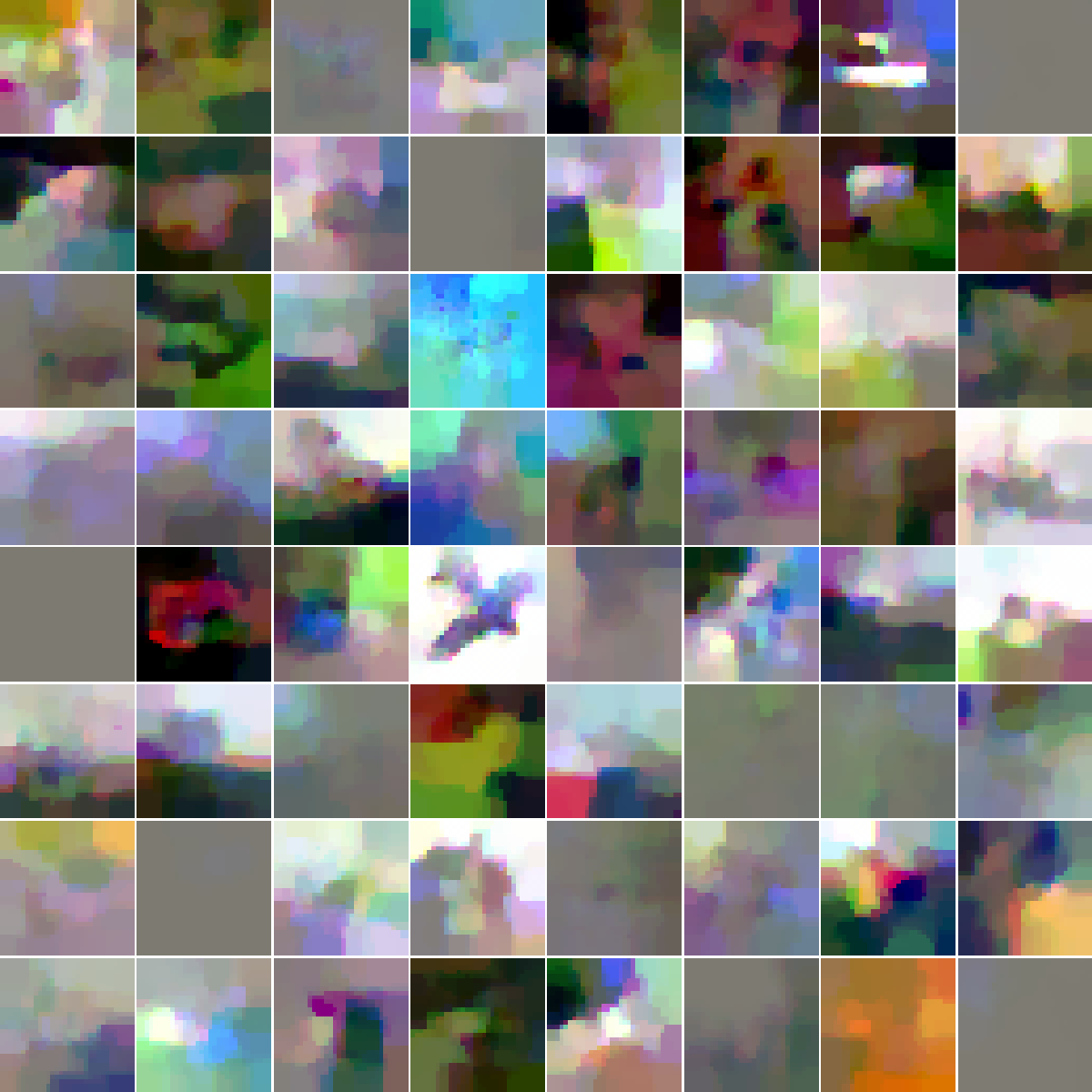}  
} 
\subfigure[Batch size = 64. PSNR $\uparrow$: 12.47, SSIM $\uparrow$: 0.2985, LPIPS $\downarrow$: 0.2110.] {     
\includegraphics[width=0.29\linewidth]{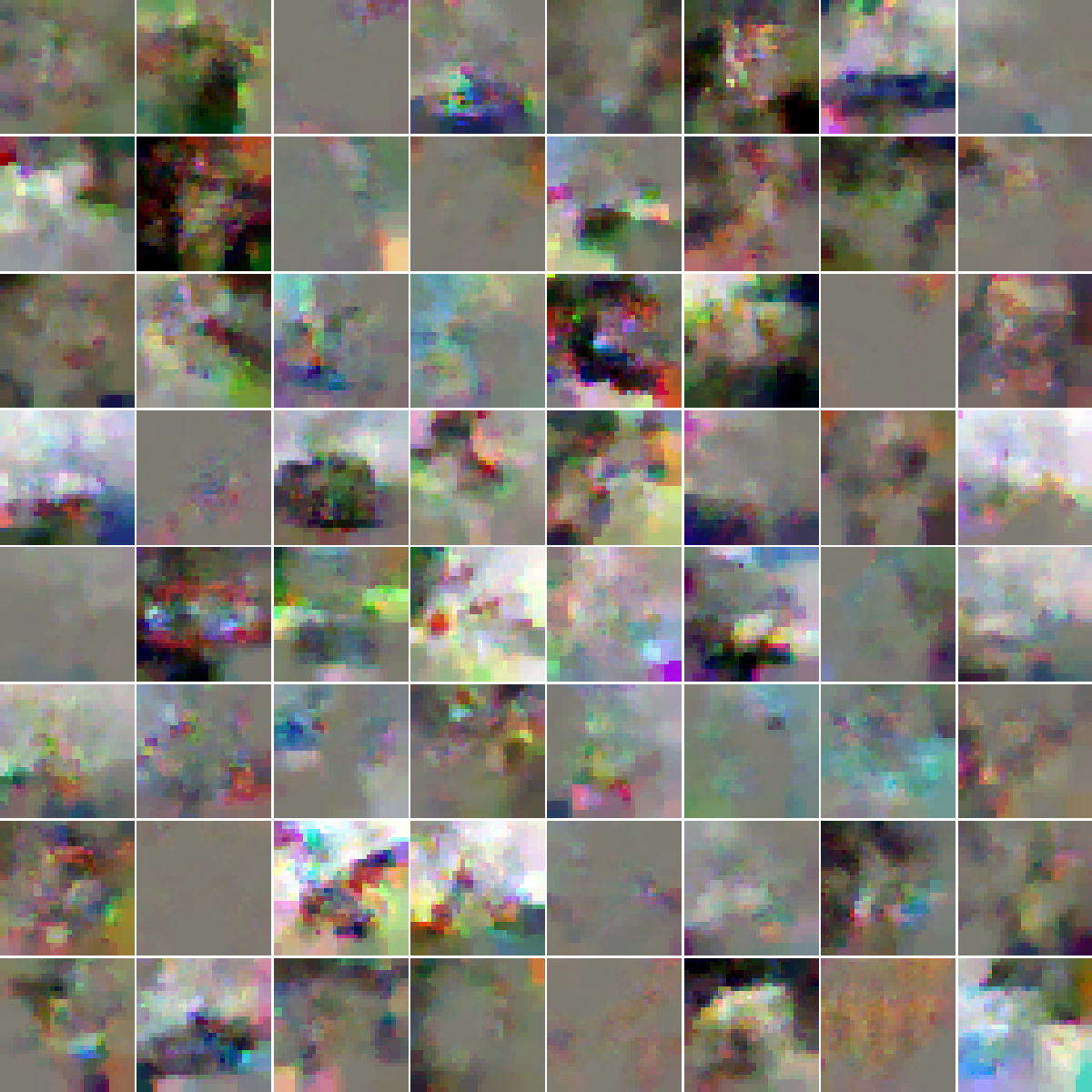}  
}   
\caption{Visualization of reconstruction results of IG on the CIFAR-10 dataset of untrained ResNet-18.}
  \label{app_fig:vis_cifar10}
\end{figure}

\begin{figure}[!h]
  \centering
    \subfigure[Batch size = 1. PSNR $\uparrow$: 21.39, SSIM $\uparrow$: 0.7448, LPIPS $\downarrow$: 0.0572.] {    
\includegraphics[width=0.29\linewidth]{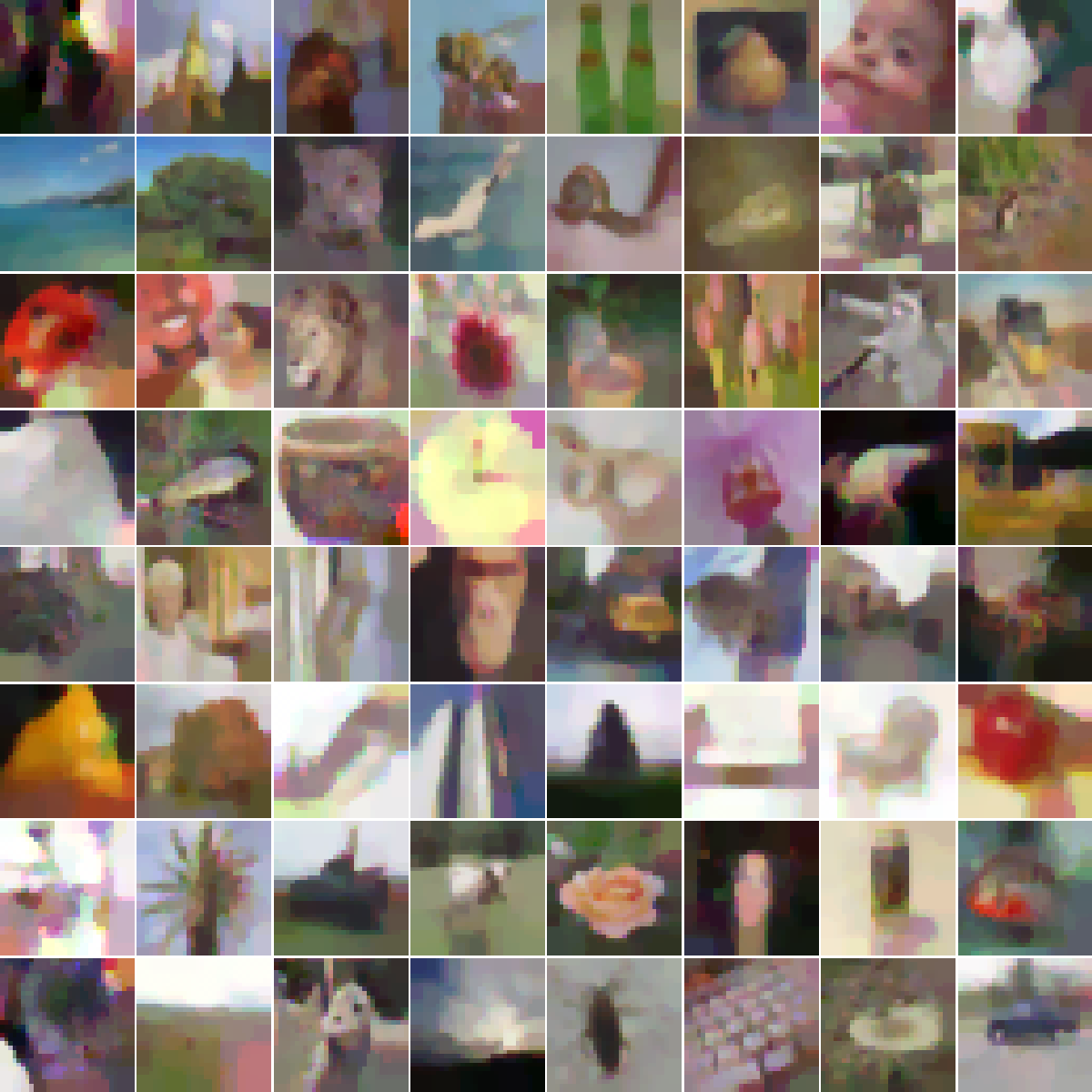}  
}   
   \subfigure[Batch size = 32. PSNR $\uparrow$: 16.72, SSIM $\uparrow$: 0.5787, LPIPS $\downarrow$: 0.1172.] {    
\includegraphics[width=0.29\linewidth]{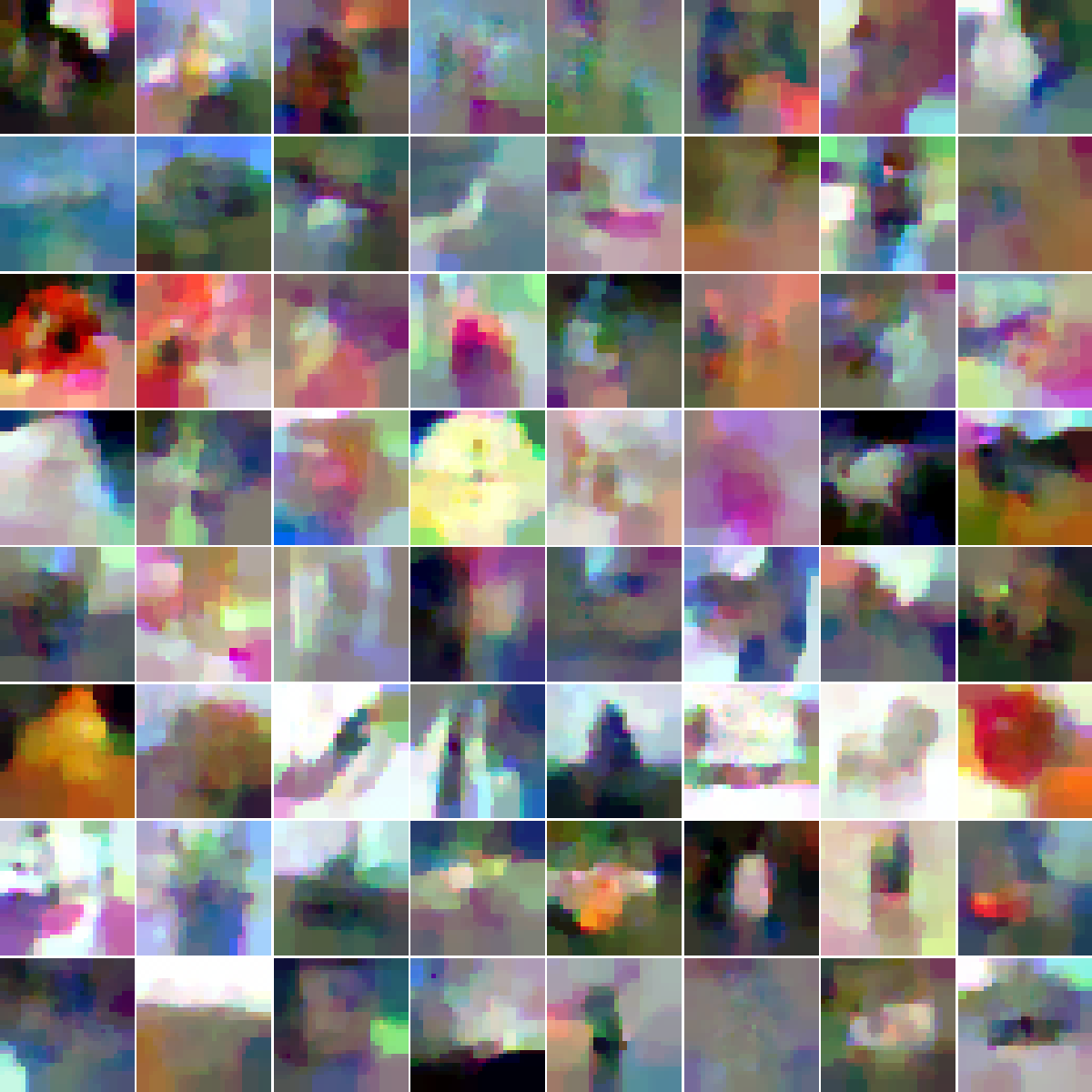}  
}  
   \subfigure[Batch size = 64. PSNR $\uparrow$: 14.99, SSIM $\uparrow$: 0.5131, LPIPS $\downarrow$: 0.1222.] {    
\includegraphics[width=0.29\linewidth]{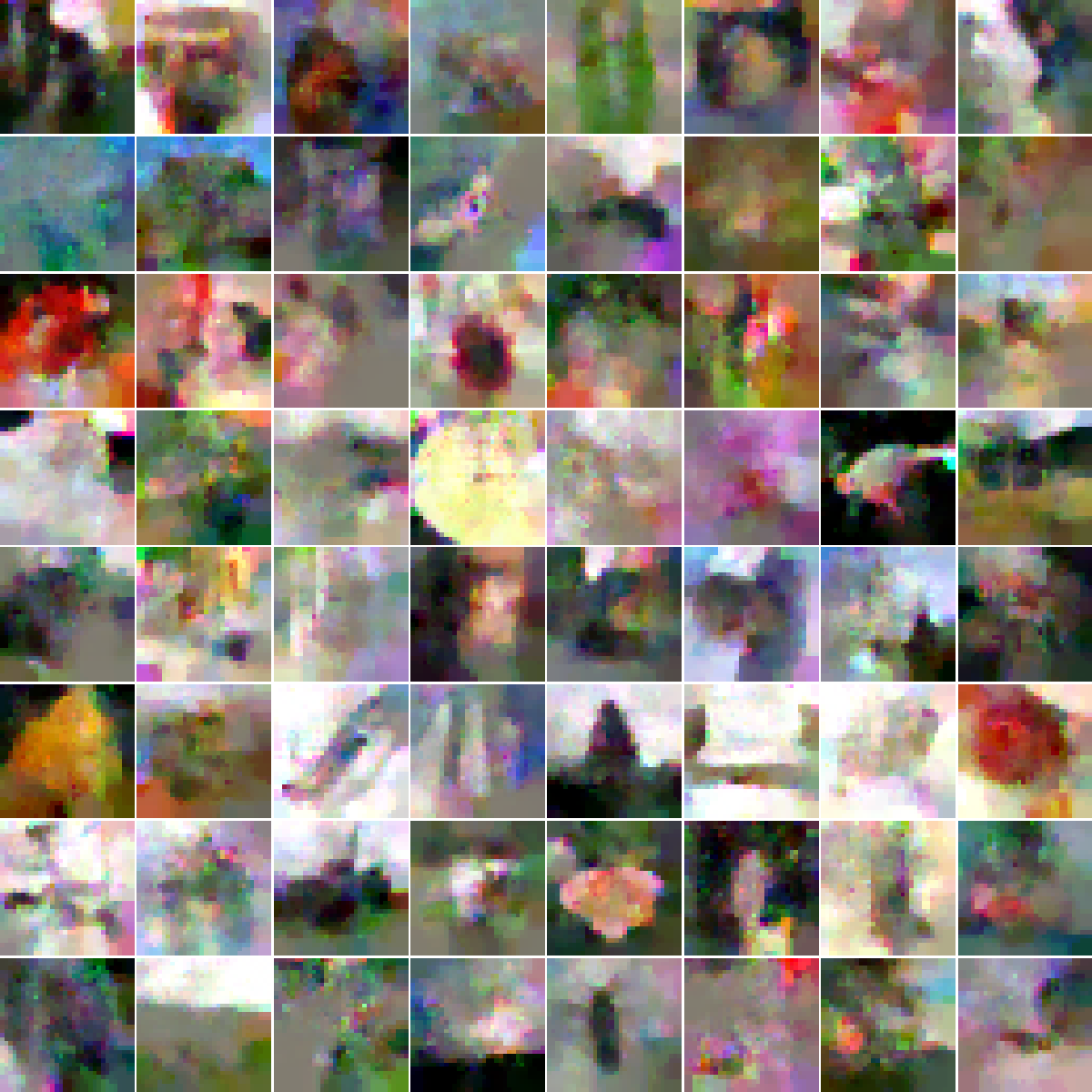}  
}   
  \caption{Visualization of reconstruction results of IG on the CIFAR-100 dataset of untrained ResNet-18.}
  \label{app_fig:vis_cifar100}
\end{figure}

\begin{figure}[!h]
  \centering
  \subfigure[Batch size = 1. PSNR $\uparrow$: 15.75, SSIM $\uparrow$: 0.3979, LPIPS $\downarrow$: 0.5933.] {    
\includegraphics[width=0.29\linewidth]{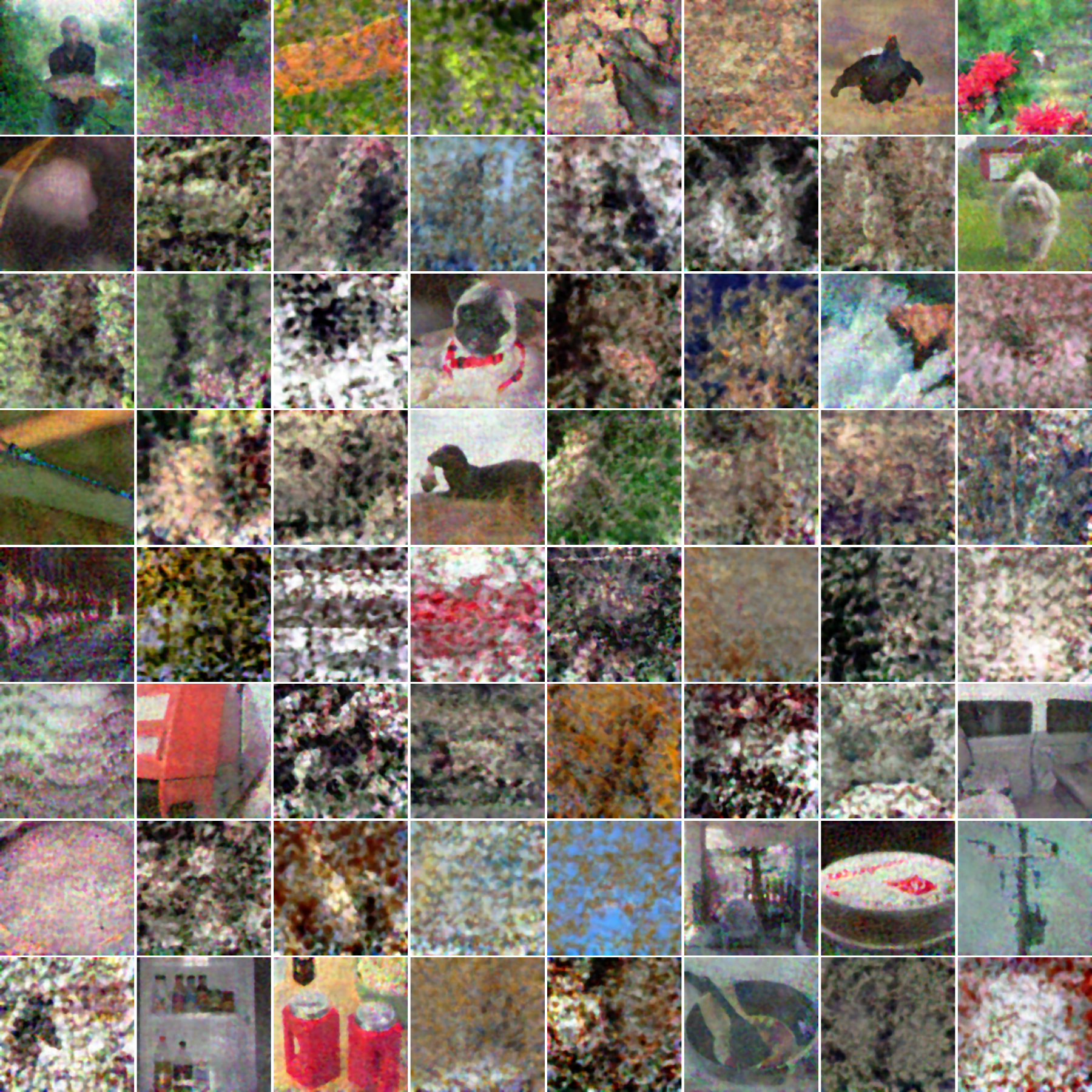}  
}   
  \subfigure[Batch size = 32. PSNR $\uparrow$: 12.32, SSIM $\uparrow$: 0.3136, LPIPS $\downarrow$: 0.7005.] {    
\includegraphics[width=0.29\linewidth]{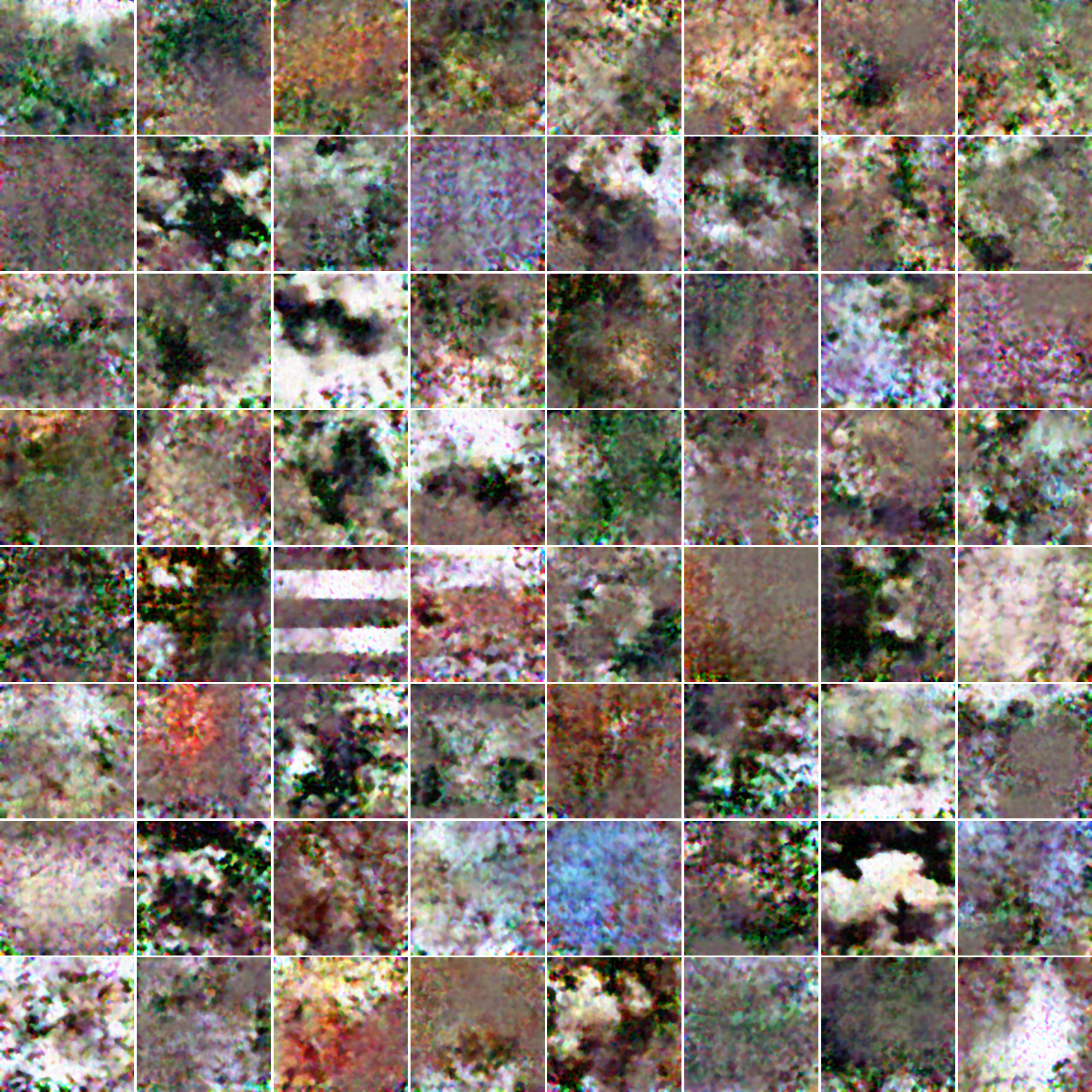}  
}   
  \subfigure[Batch size = 64.  PSNR $\uparrow$: 12.17, SSIM $\uparrow$: 0.3087, LPIPS $\downarrow$: 0.7204.] {     
\includegraphics[width=0.29\linewidth]{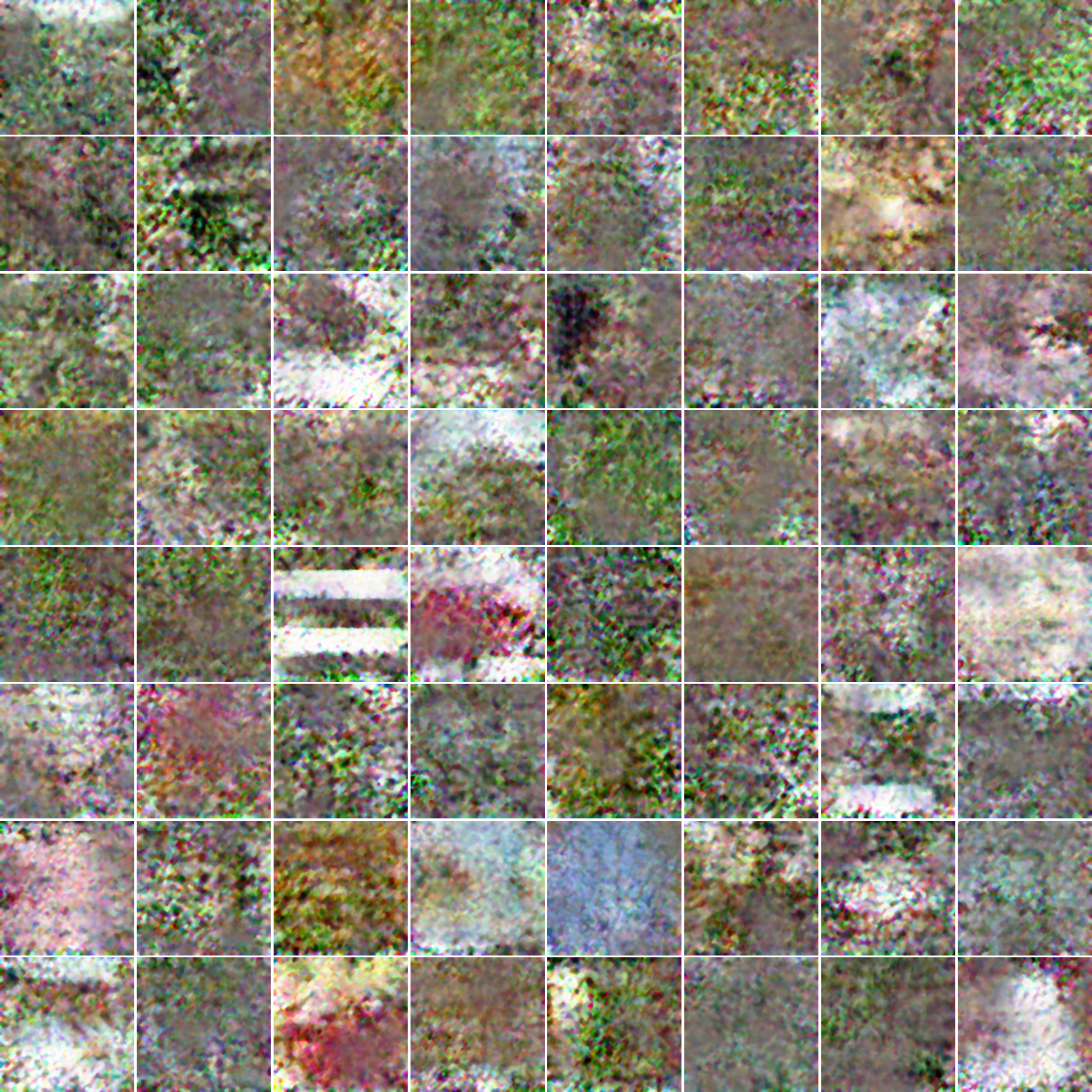}  
}   
  \caption{Visualization of reconstruction results of IG on the ImageNet dataset of untrained ResNet-18.}
  \label{app_fig:vis_imagenet}
\end{figure}

The visualization of reconstruction results of IG on the CIFAR-10, CIFAR-100, and ImageNet datasets are shown in Figures \ref{app_fig:vis_cifar10}, \ref{app_fig:vis_cifar100}, and \ref{app_fig:vis_imagenet}, respectively.

\subsection{Generation-based GIA} 

\subsubsection{Optimizing Latent Vector $z$} \label{sec:vis_gan_op_z}

\begin{figure}[h]
  \centering  
      \subfigure[Batch size = 1. 
      ] {     
\includegraphics[width=0.29\linewidth]{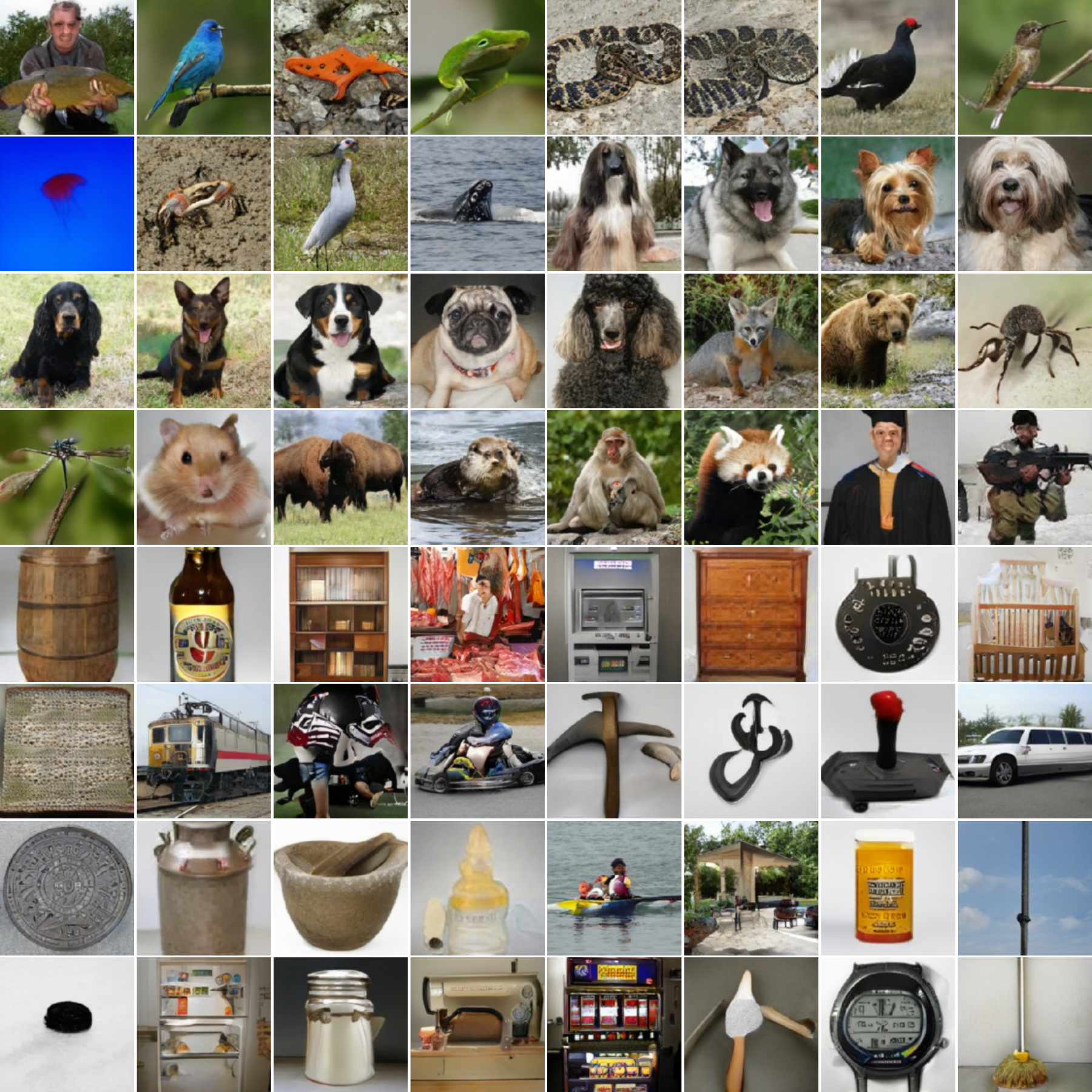}  
} 
   \subfigure[Batch size = 32. 
   ] {    
\includegraphics[width=0.29\linewidth]{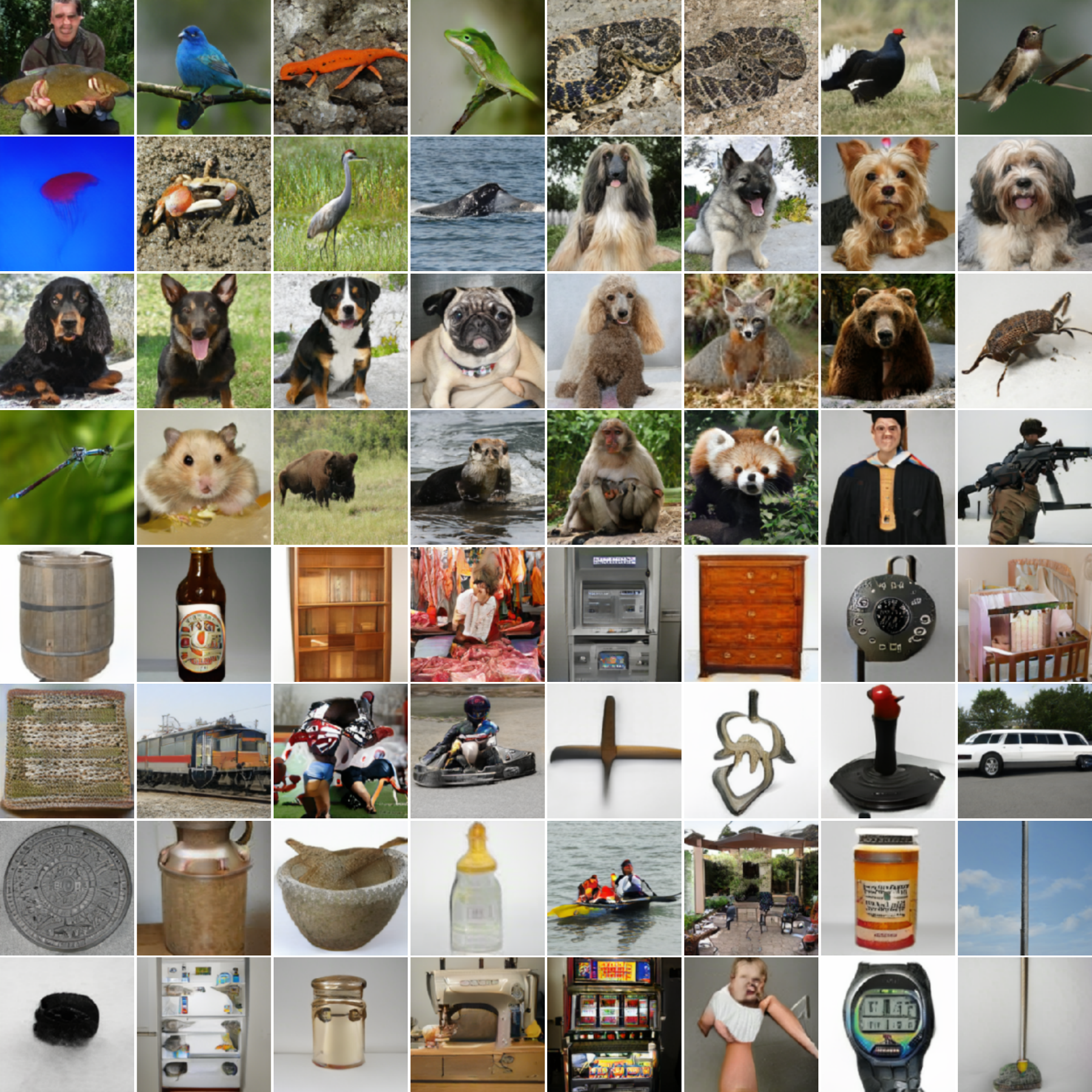}  
} 
  \subfigure[Batch size = 64. 
  ] {     
\includegraphics[width=0.29\linewidth]{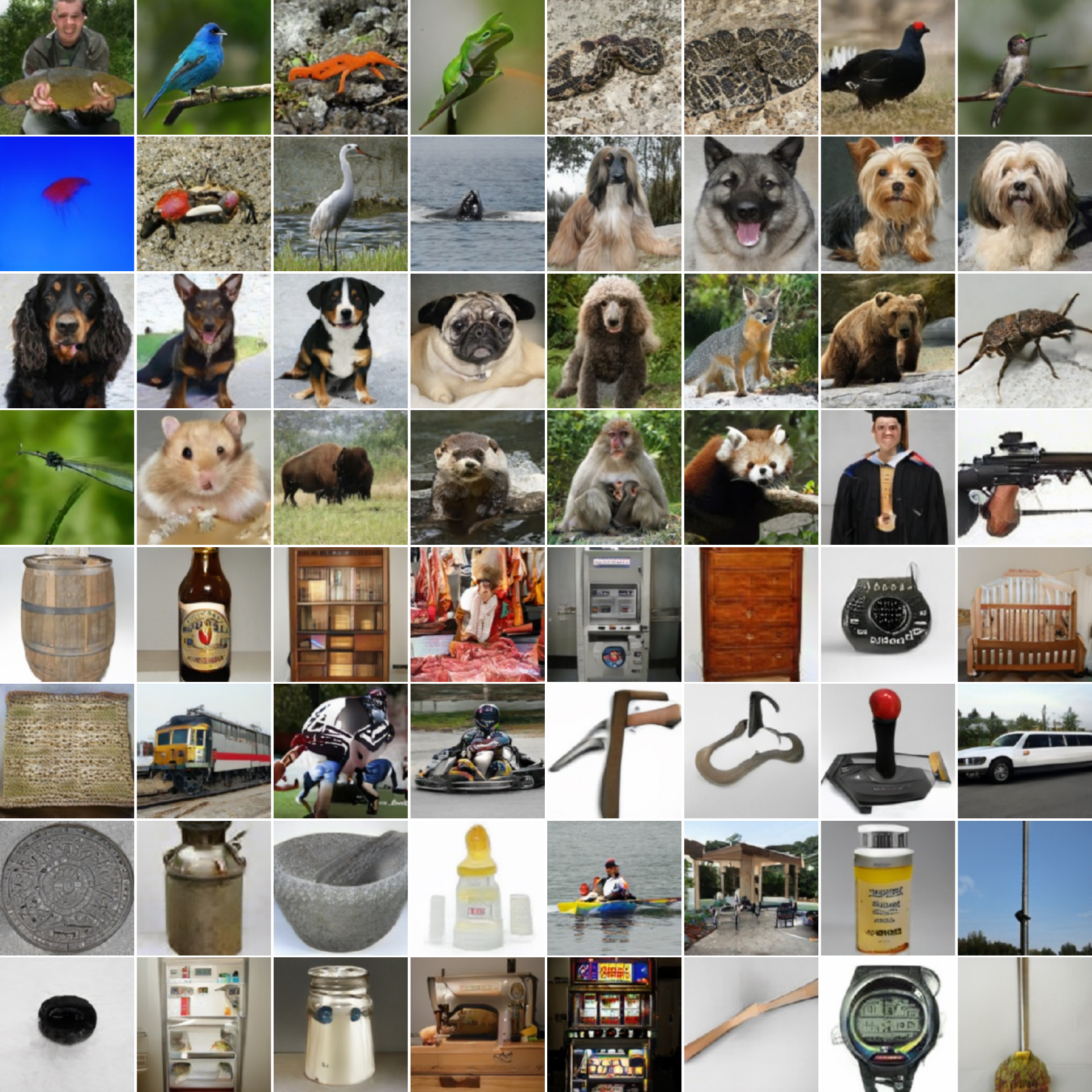}  
} 
  \caption{Reconstruction results of GGL on the ImageNet dataset of untrained ResNet-18. These results show that the attack performance of GGL is not affected by batch size.}
  \label{fig:vis_imagenet_gan}
\end{figure}

\begin{figure}[h]
  \centering  
    \subfigure[Batch size = 1. 
    ] {     
\includegraphics[width=0.29\linewidth]{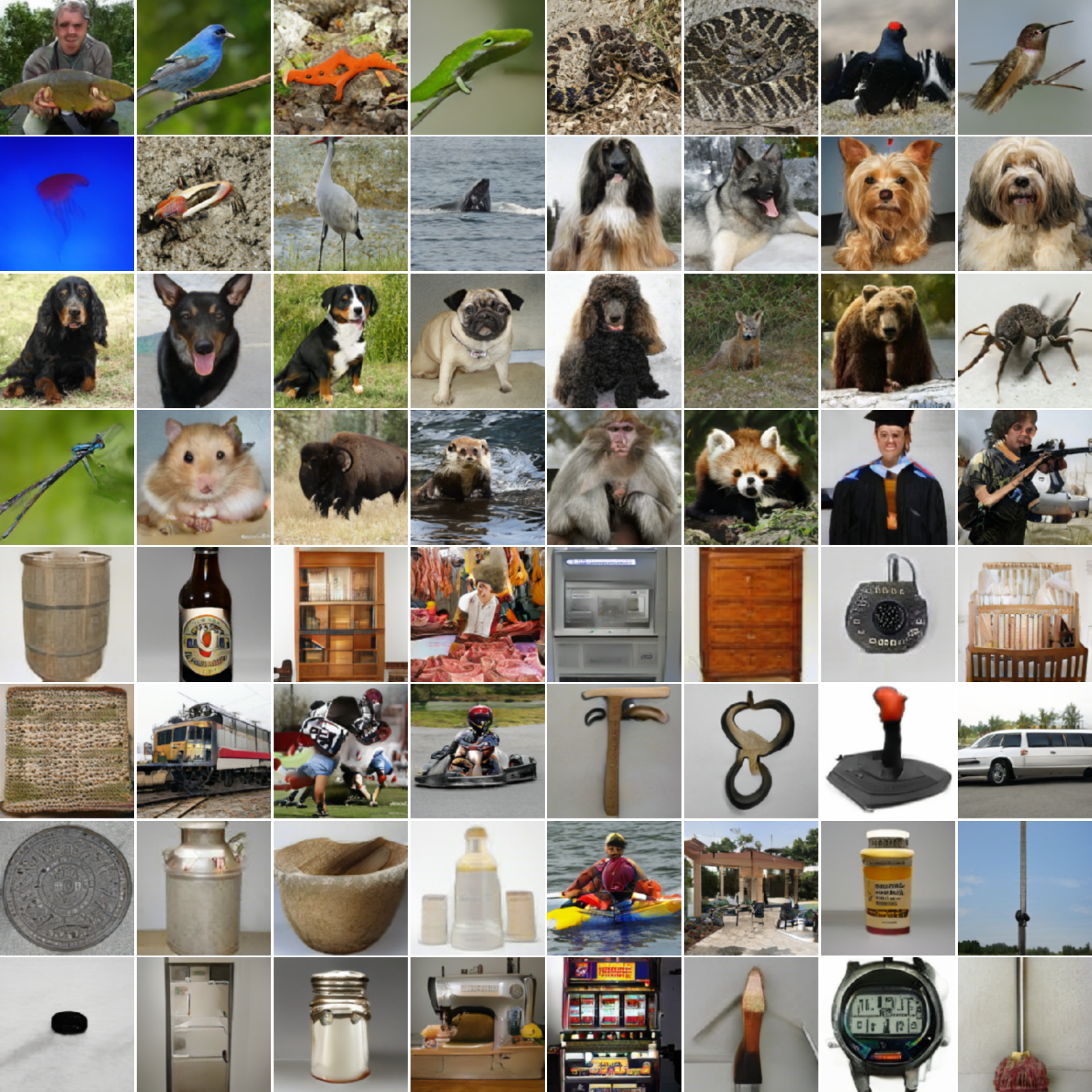}  
} 
      \subfigure[Batch size = 32. 
      ] {     
\includegraphics[width=0.29\linewidth]{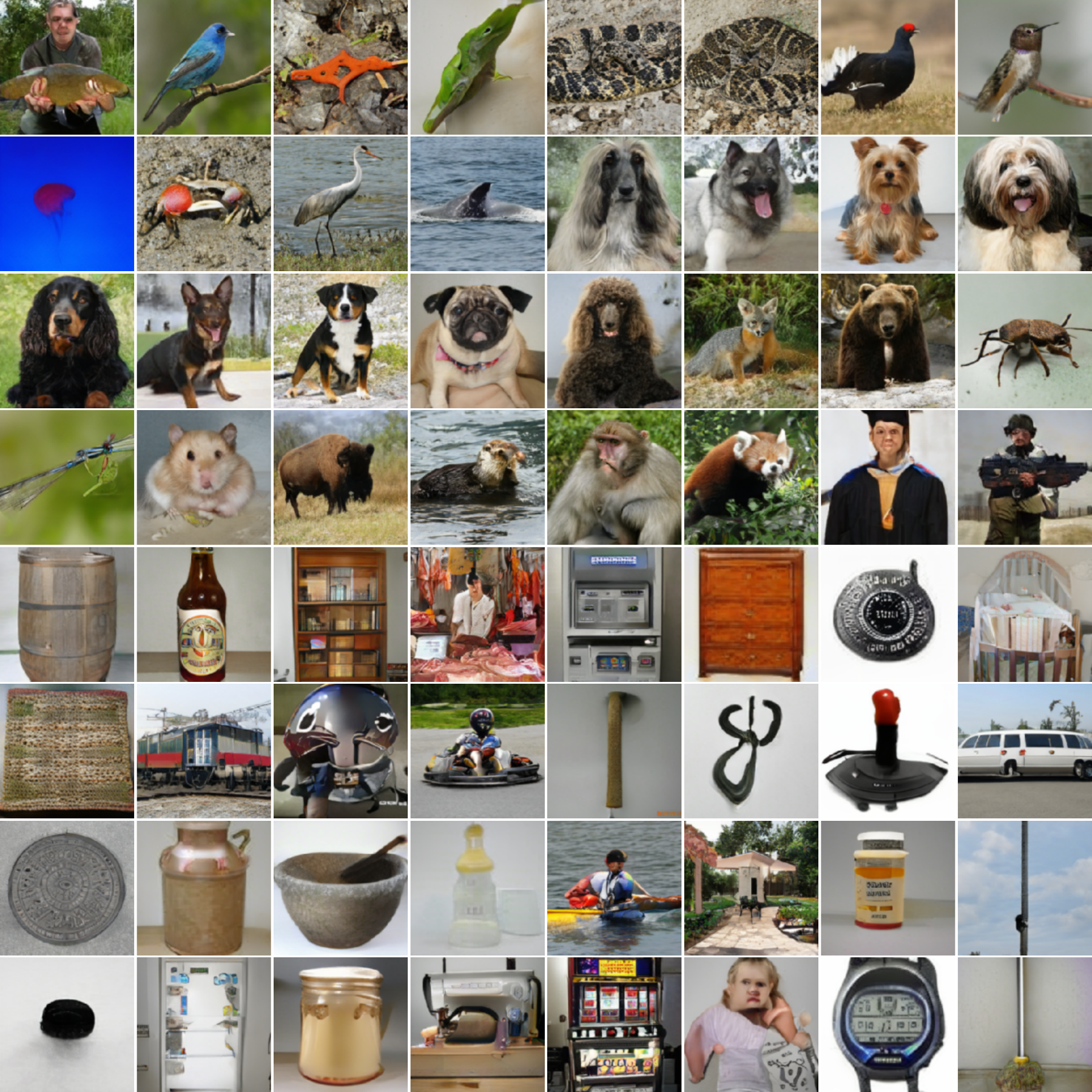}  
} 
   \subfigure[Batch size = 64. 
   ] {    
\includegraphics[width=0.29\linewidth]{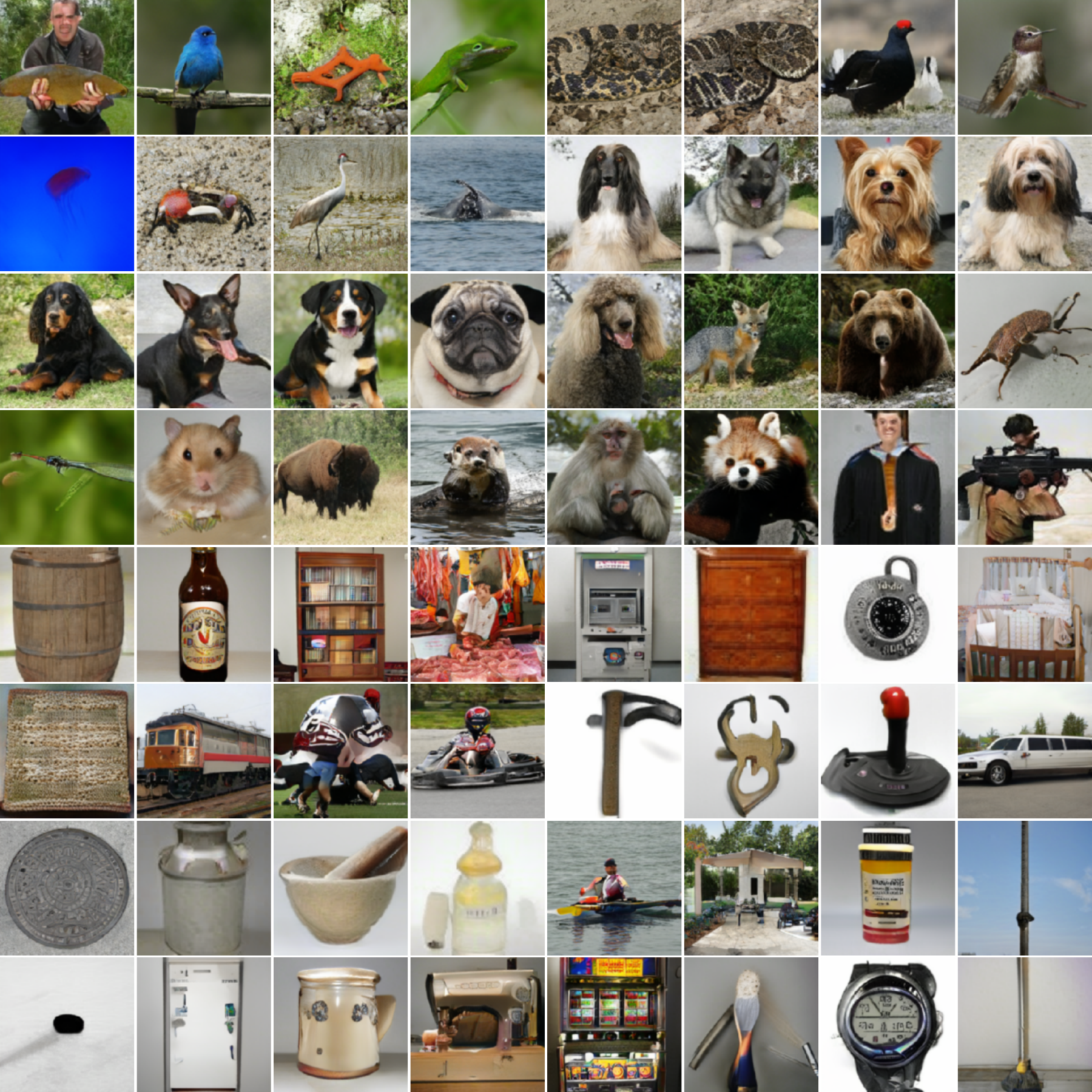}  
} 
   \caption{Reconstruction results of GGL on the ImageNet dataset of trained ResNet-18. These results show that the attack performance of GGL is not affected by model training states and batch size.}
  \label{fig:vis_imagenet_gan_trained_model}
\end{figure}

\begin{figure}[!h]
  \centering  
    \subfigure[Epoch = 1, batch size = 8. 
    ] {     
\includegraphics[width=0.29\linewidth]{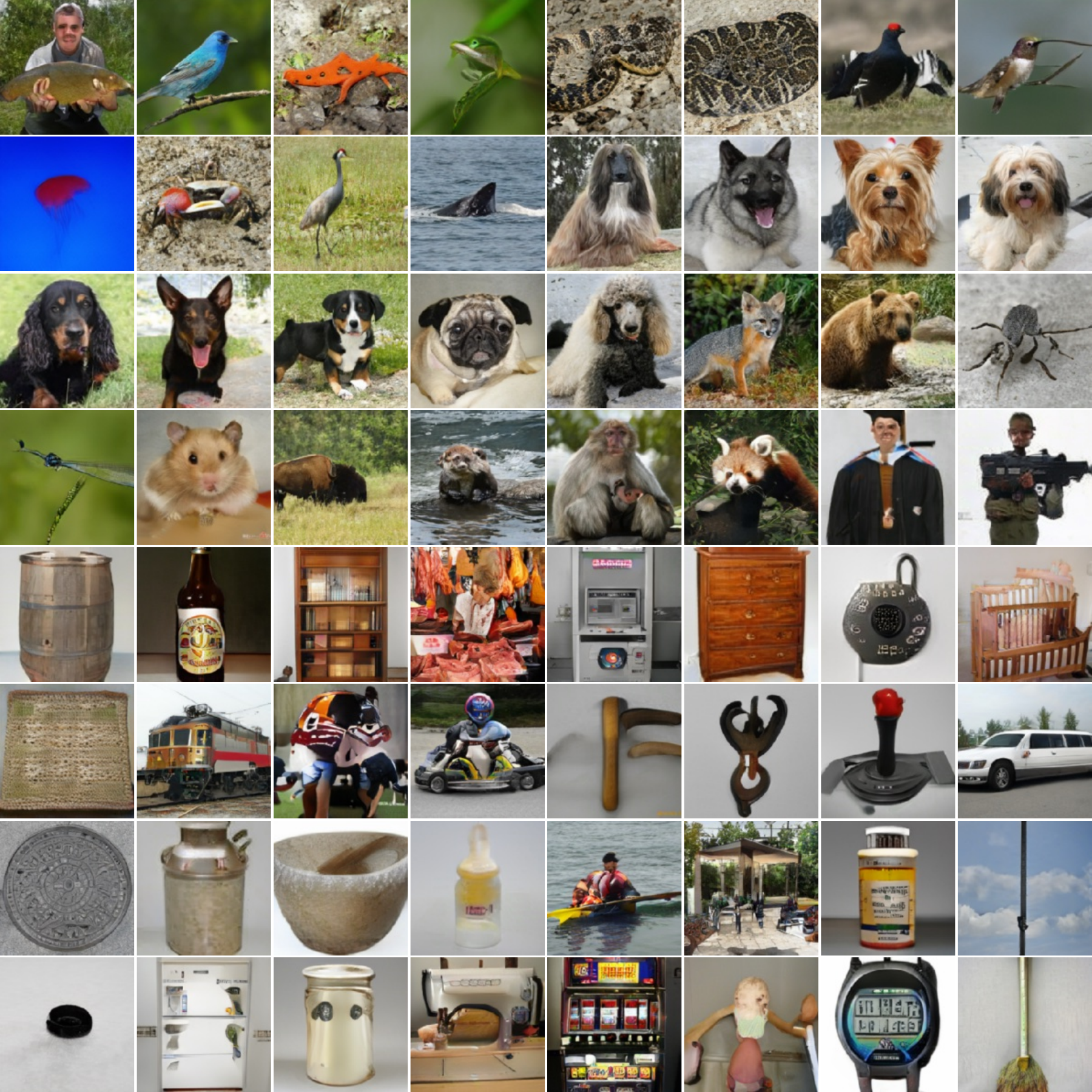}  
} 
      \subfigure[Epoch = 2, batch size = 1. 
      ] {     
\includegraphics[width=0.29\linewidth]{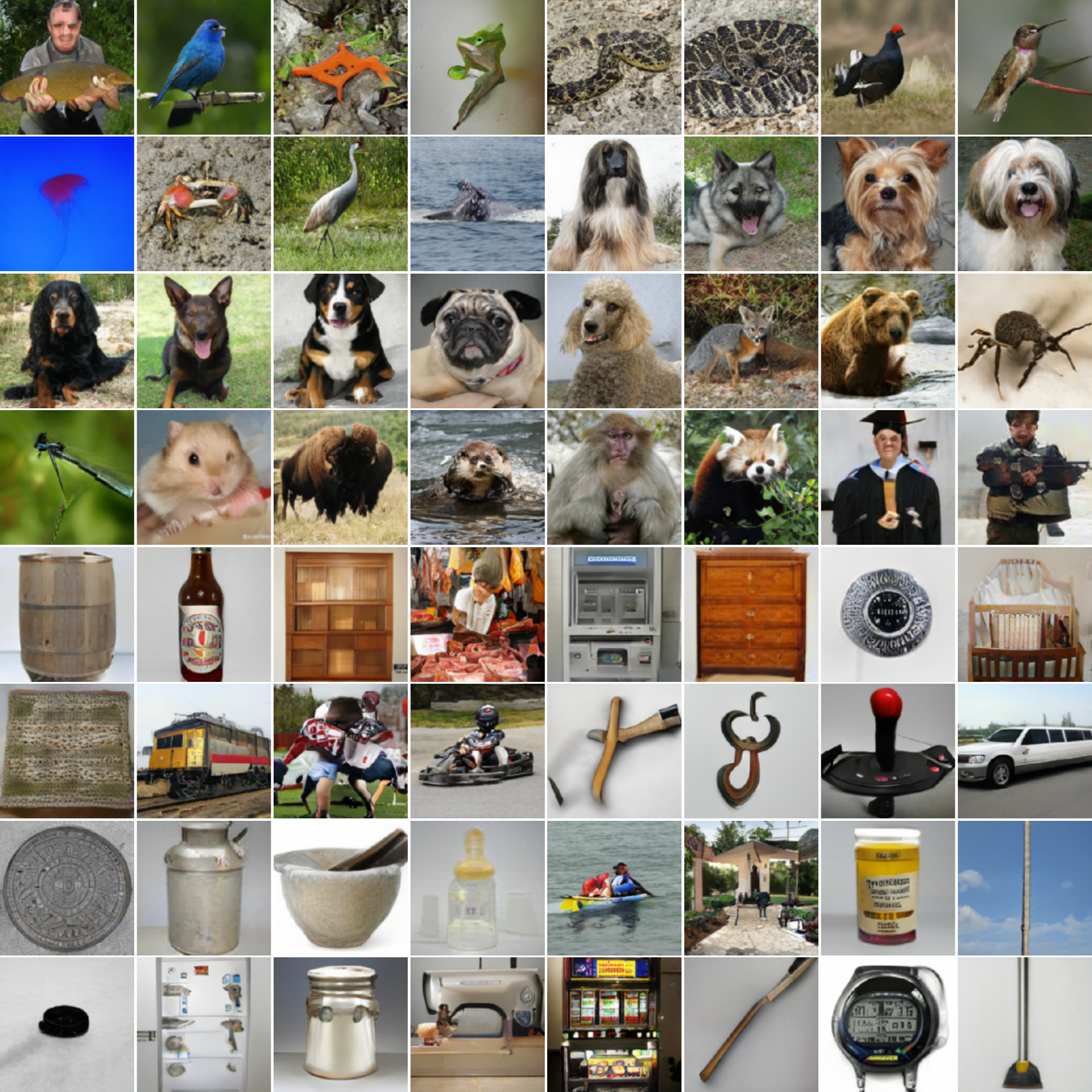}  
} 
   \subfigure[Epoch = 5, batch size = 8. 
   ] {    
\includegraphics[width=0.29\linewidth]{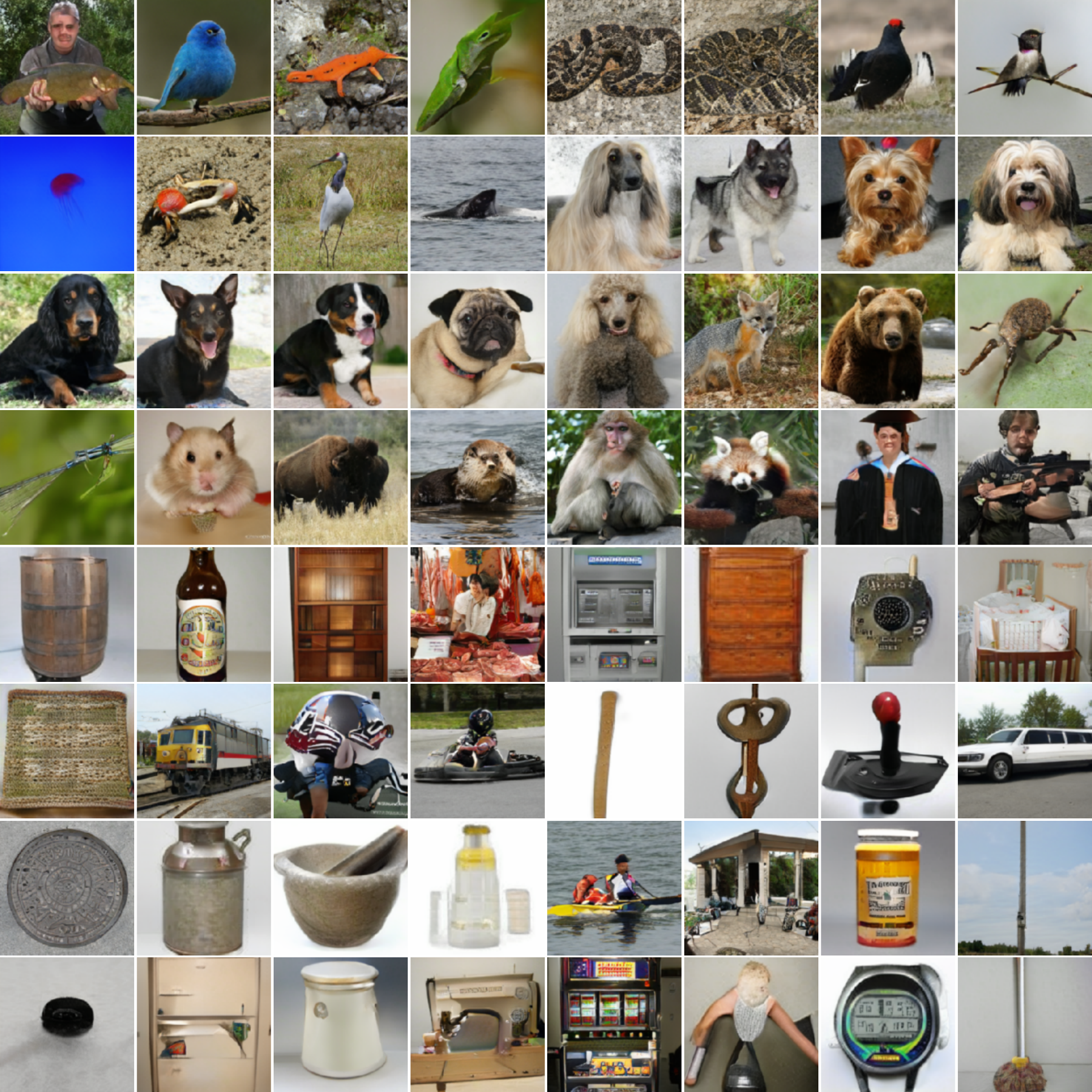}  
} 
   \caption{Reconstruction results of GGL on the ImageNet dataset of untrained ResNet-18 under practical FedAvg scenario. These results show that the attack performance of GGL is not affected by practical FedAvg scenario.}
  \label{fig:vis_imagenet_gan_fedavg}
\end{figure}

\begin{figure*}[t]
    \centering
    \includegraphics[width=1\linewidth]{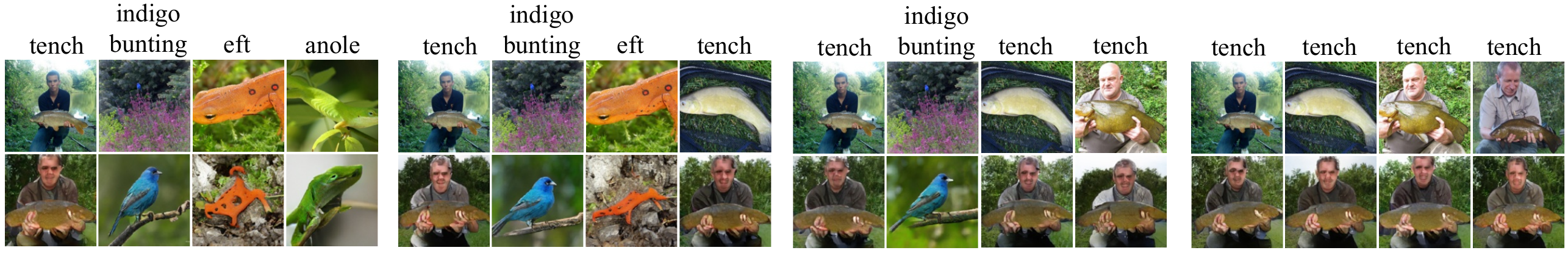}
    \caption{Reconstruction results of GGL on the ImageNet dataset with a batch size of 4. From left to right, the number of images with the same label is 0, 2, 3, and 4. The first row represents the ground truth, while the second row shows the reconstruction results. It shows that the reconstruction results of GGL are not affected by the number of images with the same label within one batch.}
    \label{app_fig:compare_label_number_ggl}
\end{figure*}

The reconstruction results of GGL are shown in Figures \ref{fig:vis_imagenet_gan}, \ref{fig:vis_imagenet_gan_trained_model}, \ref{fig:vis_imagenet_gan_fedavg}, and \ref{app_fig:compare_label_number_ggl}. 
These results show that when optimizing latent vector $\bm{z}$, GEN-GIA can even generate semantically similar images when using random Gaussian noise instead of real gradients, as long as the label information is available, indicating that it is not affected by the factors influencing OP-GIA. However, it heavily relies on the pre-trained generator and only can achieve semantic-level recovery.

\subsubsection{Optimizing Generator's Parameters $W$}

\begin{figure}[!h]
  \centering
  \subfigure[PSNR $\uparrow$.] {     
\includegraphics[width=0.29\linewidth]{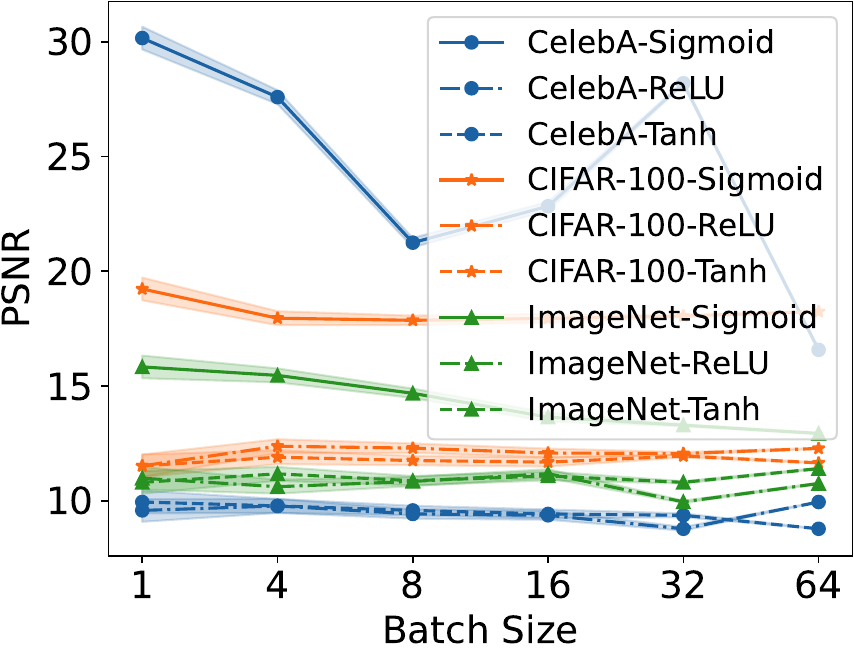}  
}   
\hfill
   \subfigure[SSIM $\uparrow$.] {    
\includegraphics[width=0.29\linewidth]{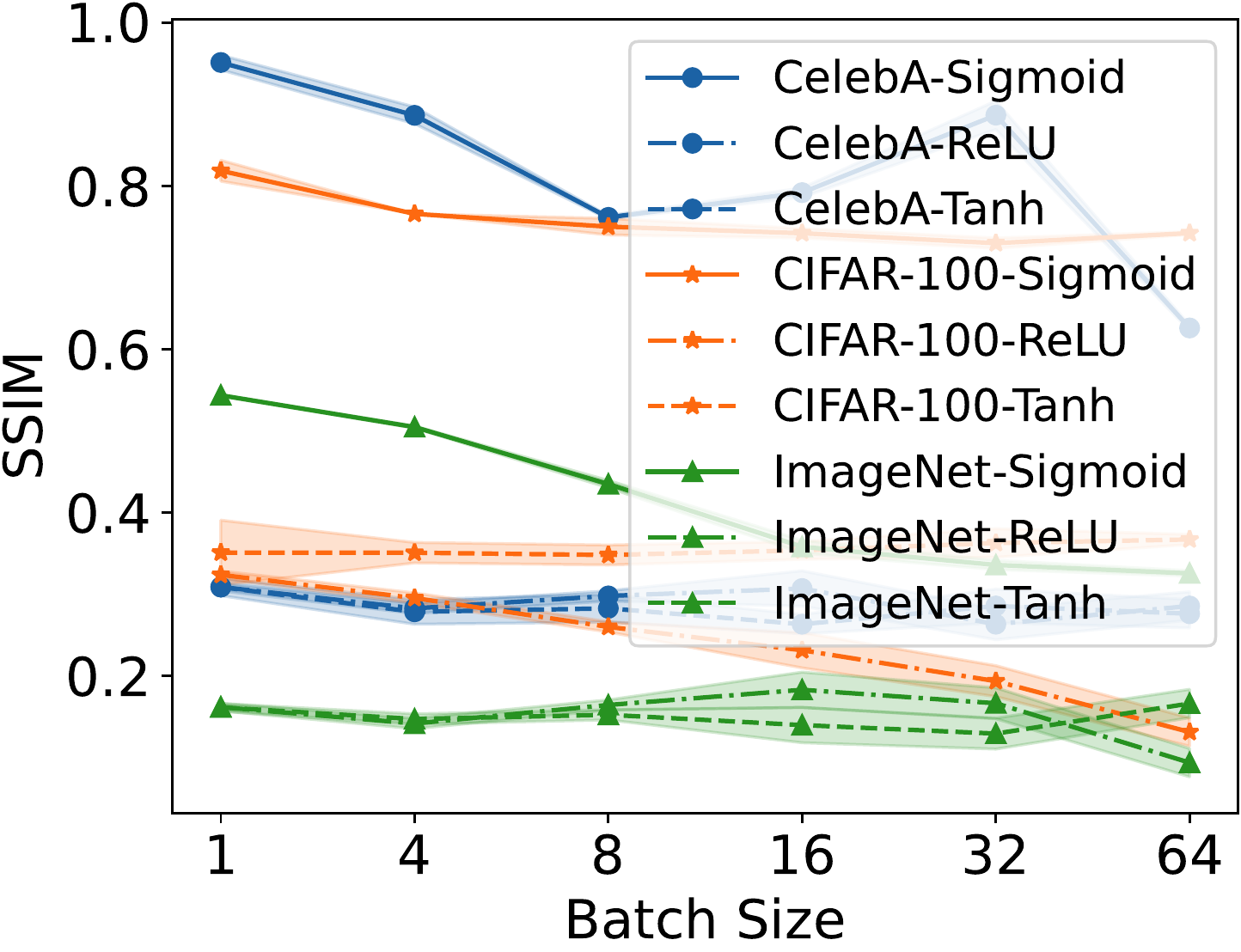}  
} 
\hfill
    \subfigure[LPIPS $\downarrow$.] {    
\includegraphics[width=0.29\linewidth]{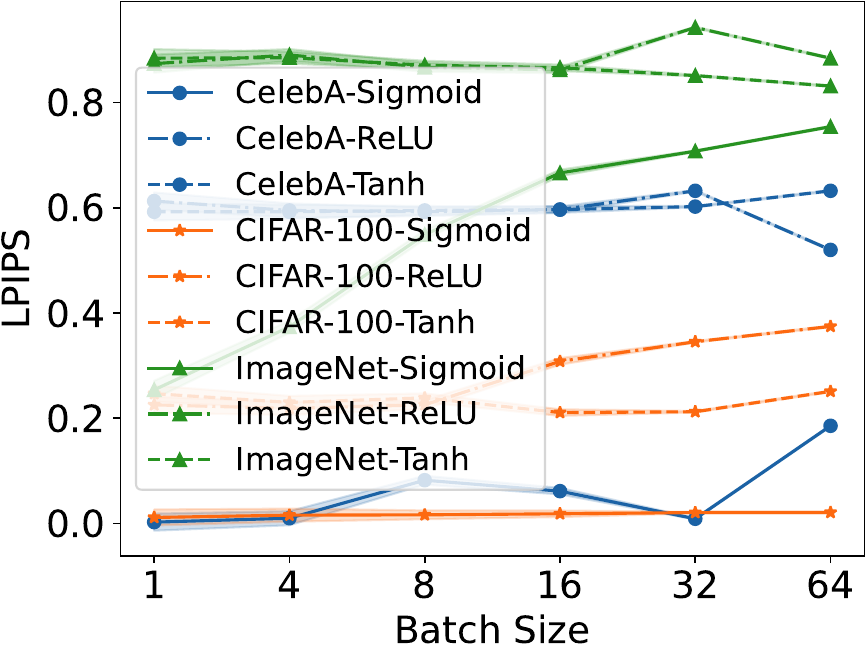}  
}   
\hfill 
\revise{
    \subfigure[Jaccard $\uparrow$.] {    
\includegraphics[width=0.29\linewidth]{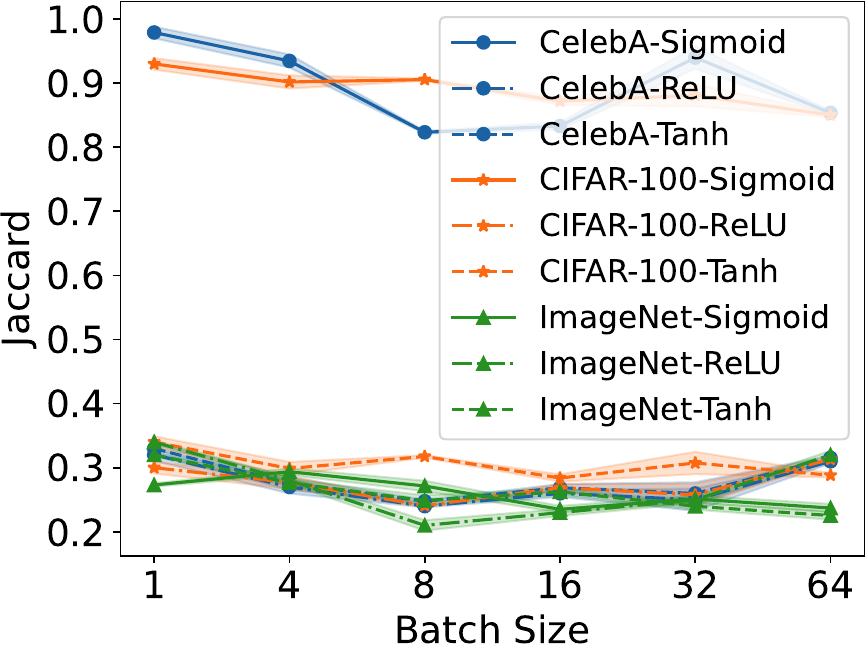}  
}   
    \subfigure[RDLV $\uparrow$.] {    
\includegraphics[width=0.29\linewidth]{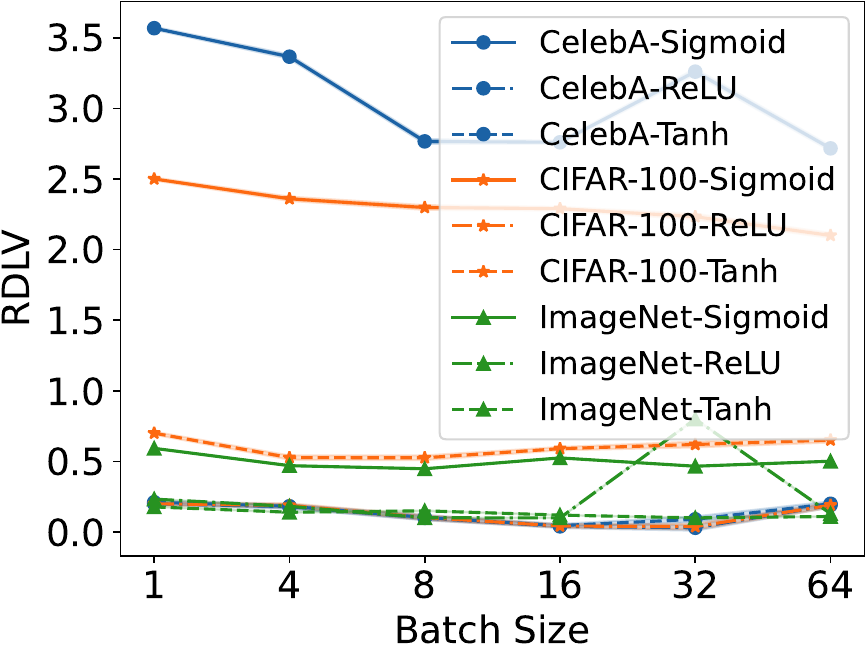}  
}
}
  \caption{Reconstruction results of CI-Net evaluated on ResNet-18 with different activation functions on various datasets with different batch sizes. These results show that GEN-GIA with optimizing the generator's parameters $\bm{W}$ is affected by the factors that influence OP-GIA. Moreover, it only works when the target model adopts the Sigmoid activation function and fails with other activation functions.
  }
  \label{app_fig:CI-Net-all}
\end{figure}

\begin{figure}[!h]
  \centering
  \subfigure[PSNR $\uparrow$.] {     
\includegraphics[width=0.29\linewidth]{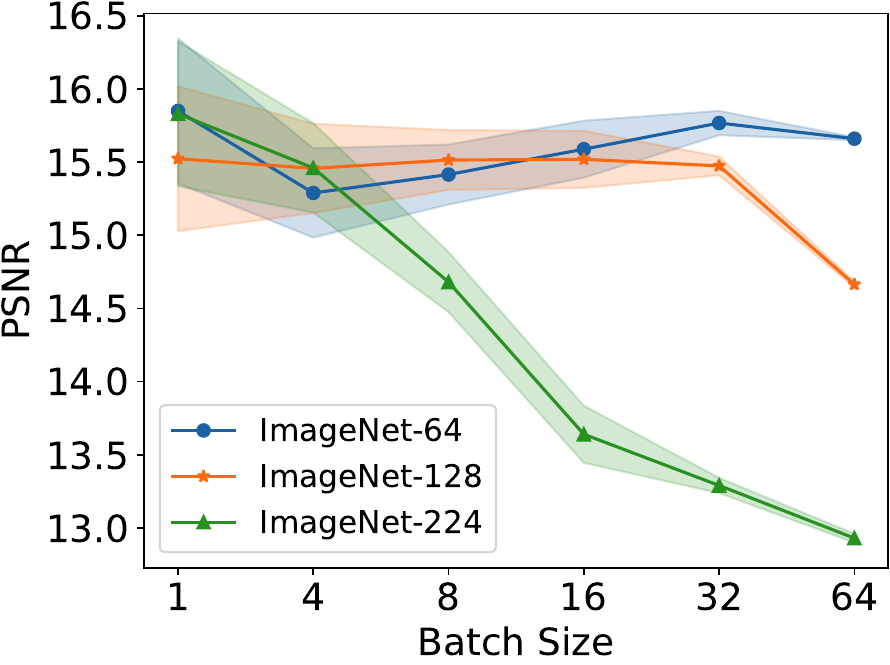}  
}   
\hfill
   \subfigure[SSIM $\uparrow$.] {    
\includegraphics[width=0.29\linewidth]{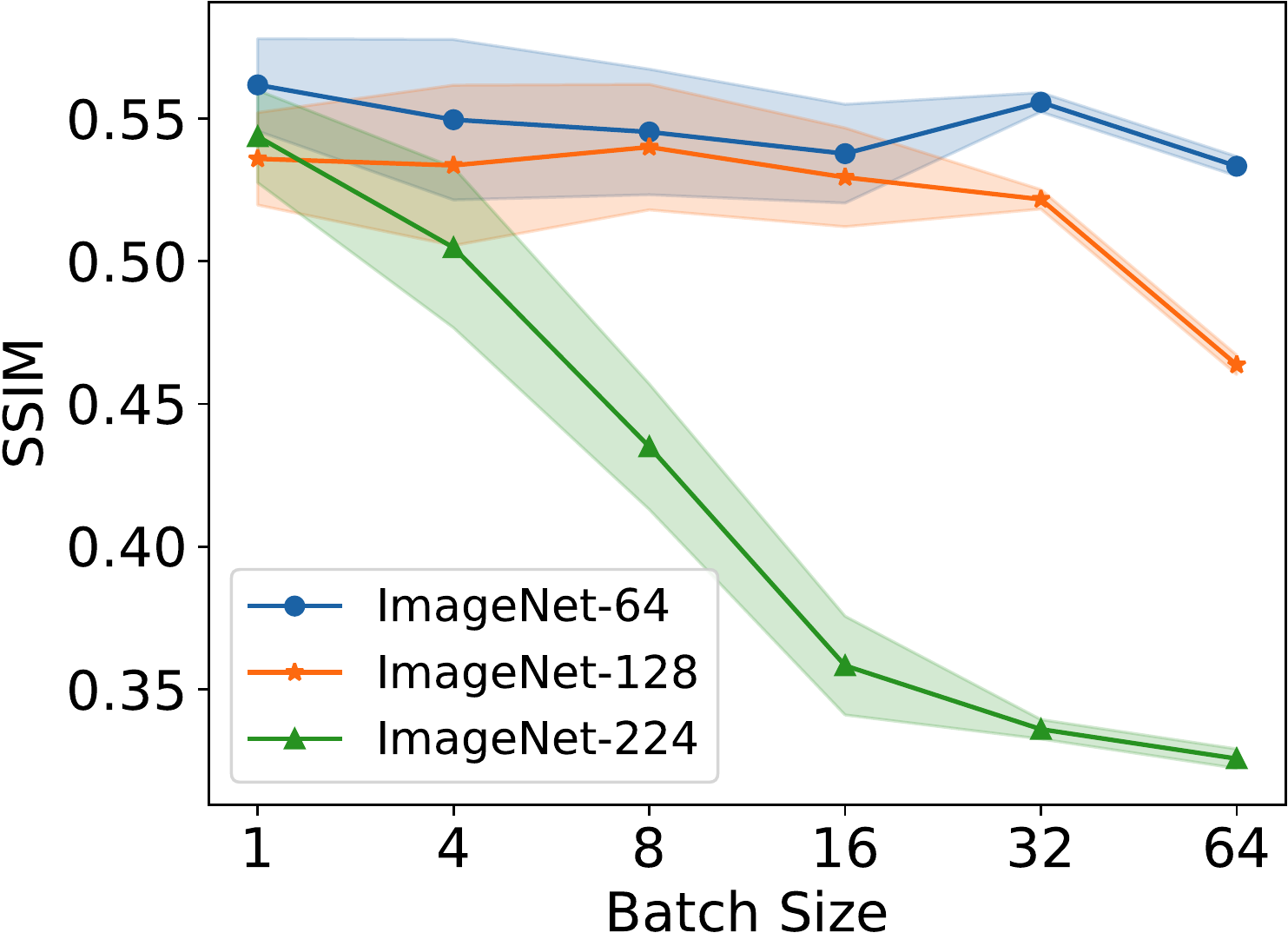}  
} 
\hfill
    \subfigure[LPIPS $\downarrow$.] {    
\includegraphics[width=0.29\linewidth]{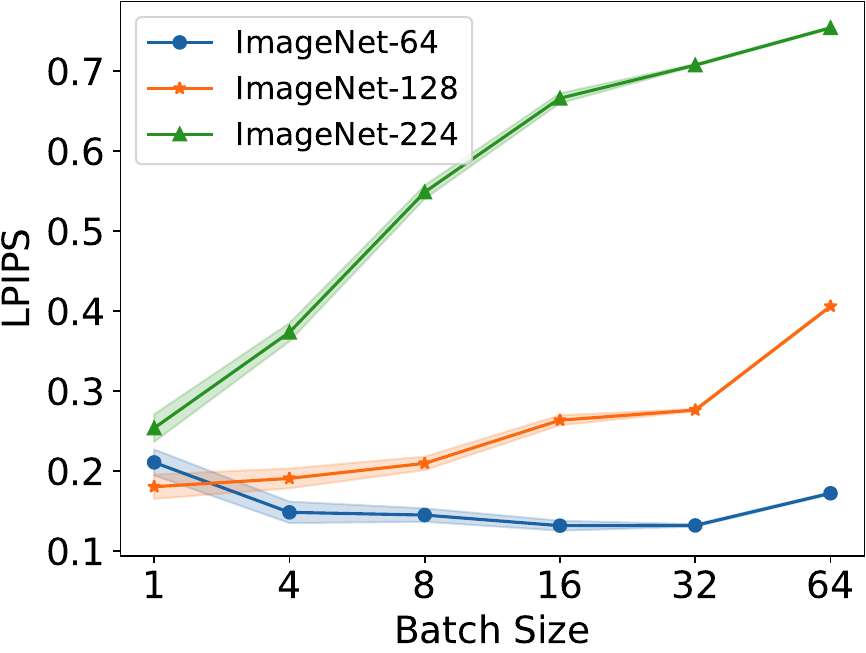}  
}   
\hfill 
\revise{
    \subfigure[Jaccard $\uparrow$.] {    
\includegraphics[width=0.29\linewidth]{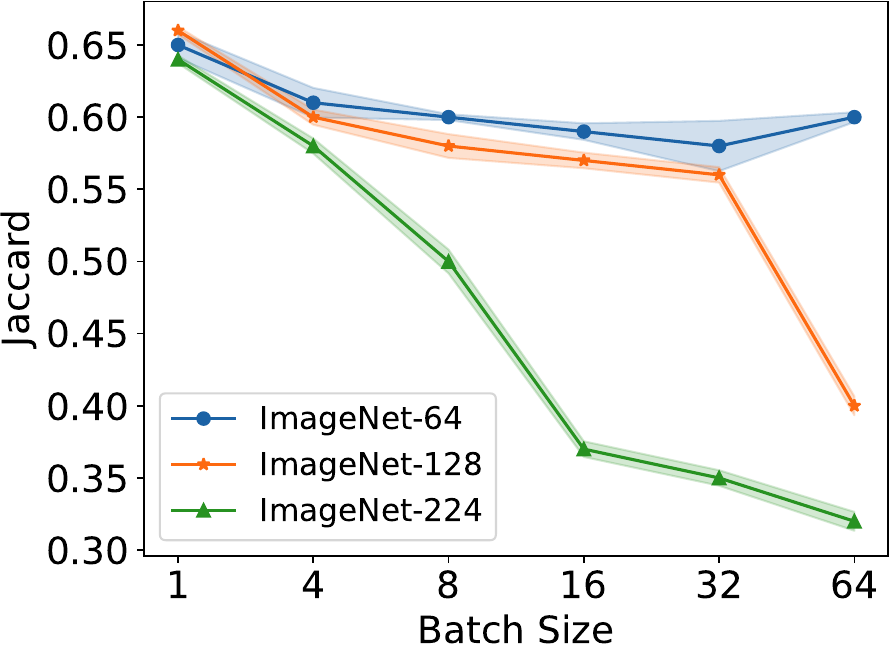}  
}   
    \subfigure[RDLV $\uparrow$.] {    
\includegraphics[width=0.29\linewidth]{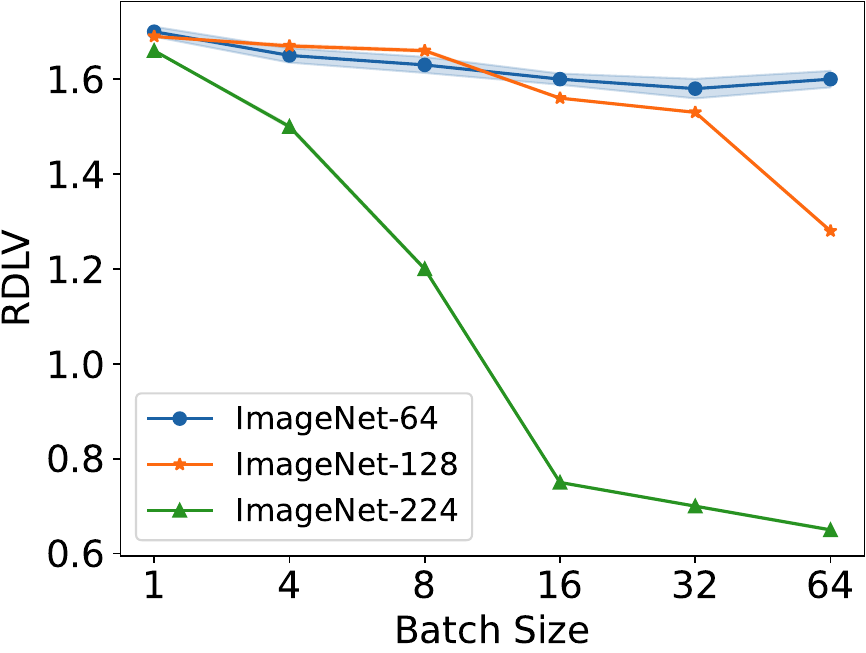}  
}
}
  \caption{Reconstruction results of CI-Net on ImageNet with different resolutions under the Sigmoid activation function. These results show that images with lower resolution are relatively easier to reconstruct for CI-Net.
  }
  \label{app_fig:CI-Net-bs-all}
\end{figure}

\begin{figure}[!h]
  \centering
  \subfigure[PSNR $\uparrow$.] {     
\includegraphics[width=0.29\linewidth]{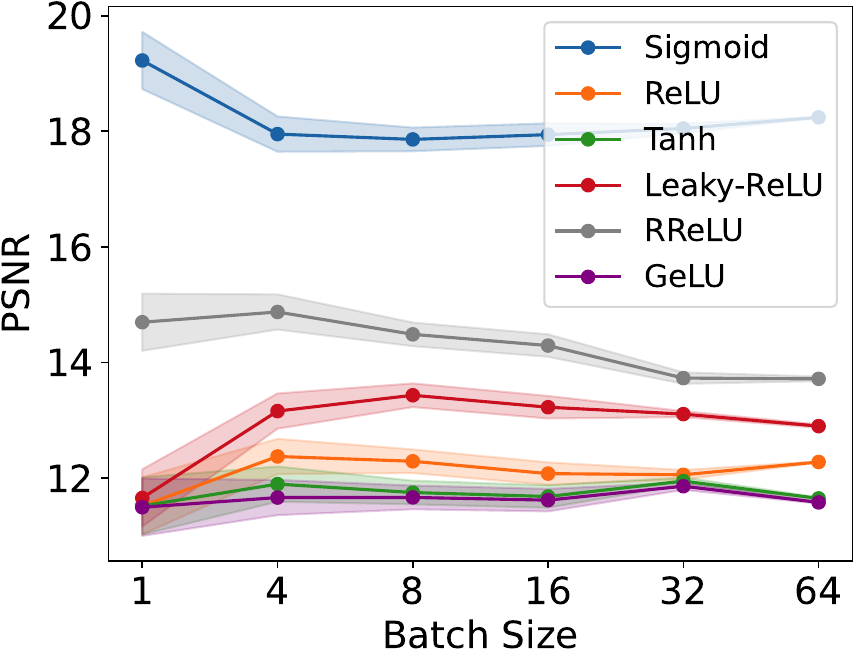}  
}   
\hfill
   \subfigure[SSIM $\uparrow$.] {    
\includegraphics[width=0.29\linewidth]{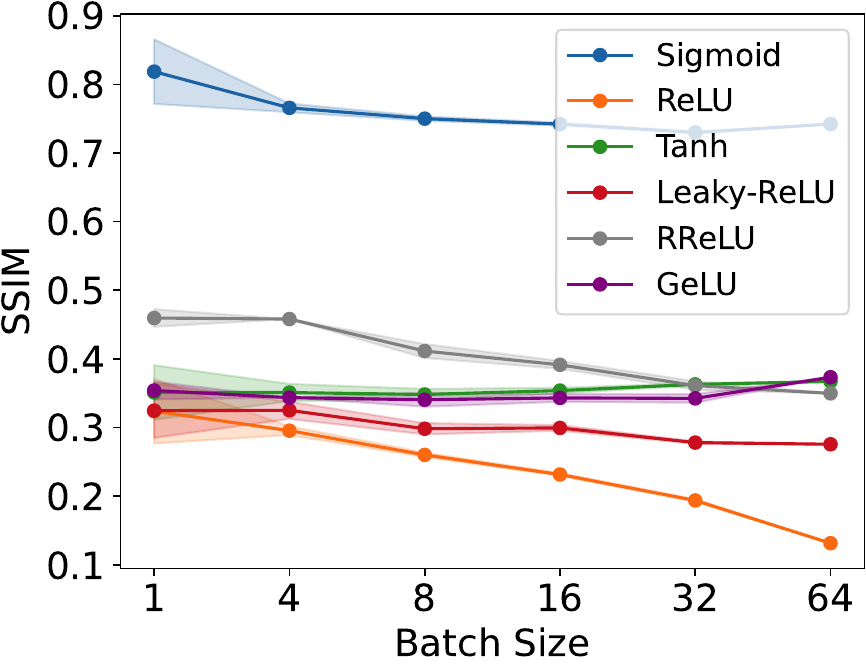}  
} 
\hfill
    \subfigure[LPIPS $\downarrow$.] {    
\includegraphics[width=0.29\linewidth]{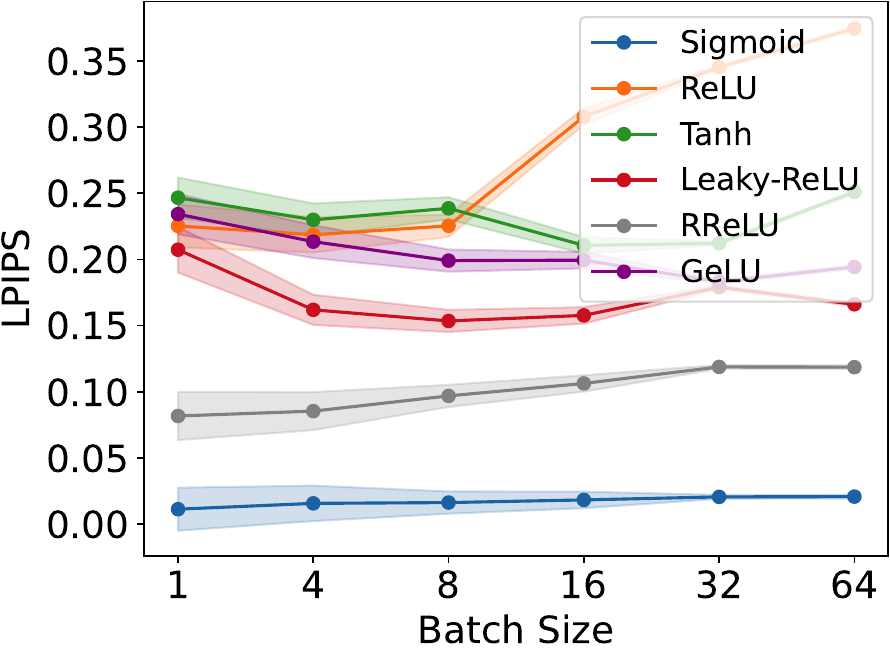}  
}   
\hfill 
\revise{
    \subfigure[Jaccard $\uparrow$.] {    
\includegraphics[width=0.29\linewidth]{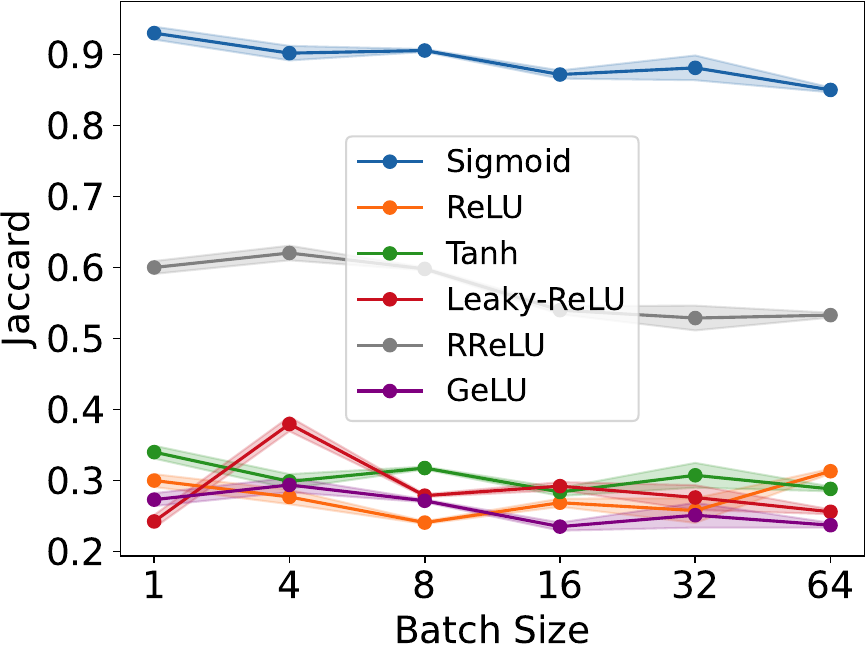}  
}   
    \subfigure[RDLV $\uparrow$.] {    
\includegraphics[width=0.29\linewidth]{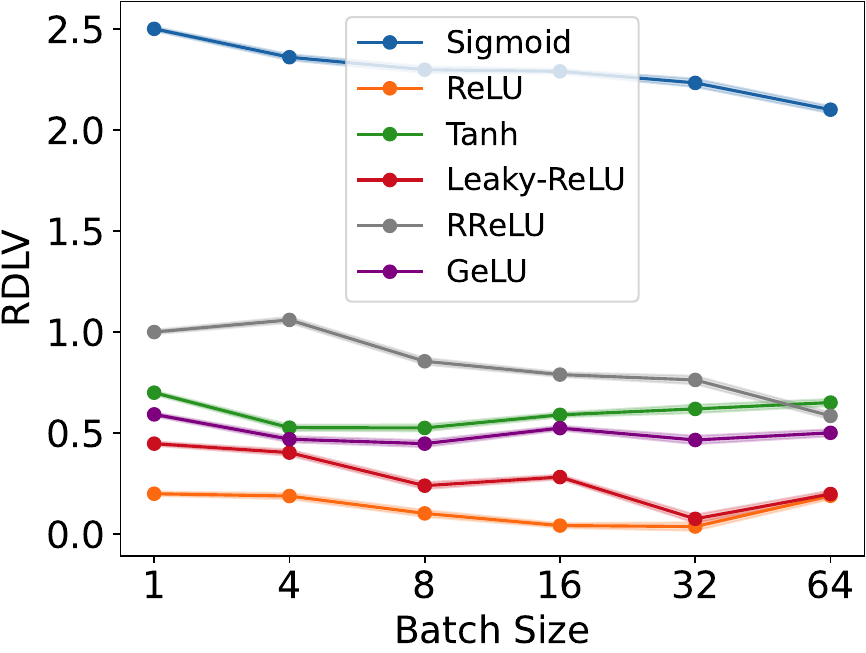}  
}
}
  \caption{Reconstruction results of CI-Net on CIFAR-100 under various activation functions. These results further show that when optimizing the generator's parameters $\bm{W}$, GEN-GIA only succeeds when the target model adopts the Sigmoid activation function.
  }
  \label{app_fig:CI-Net-acts-all}
\end{figure}

\begin{figure}[!h]
  \centering
  \subfigure[No same label.] {     
\includegraphics[width=0.29\linewidth]{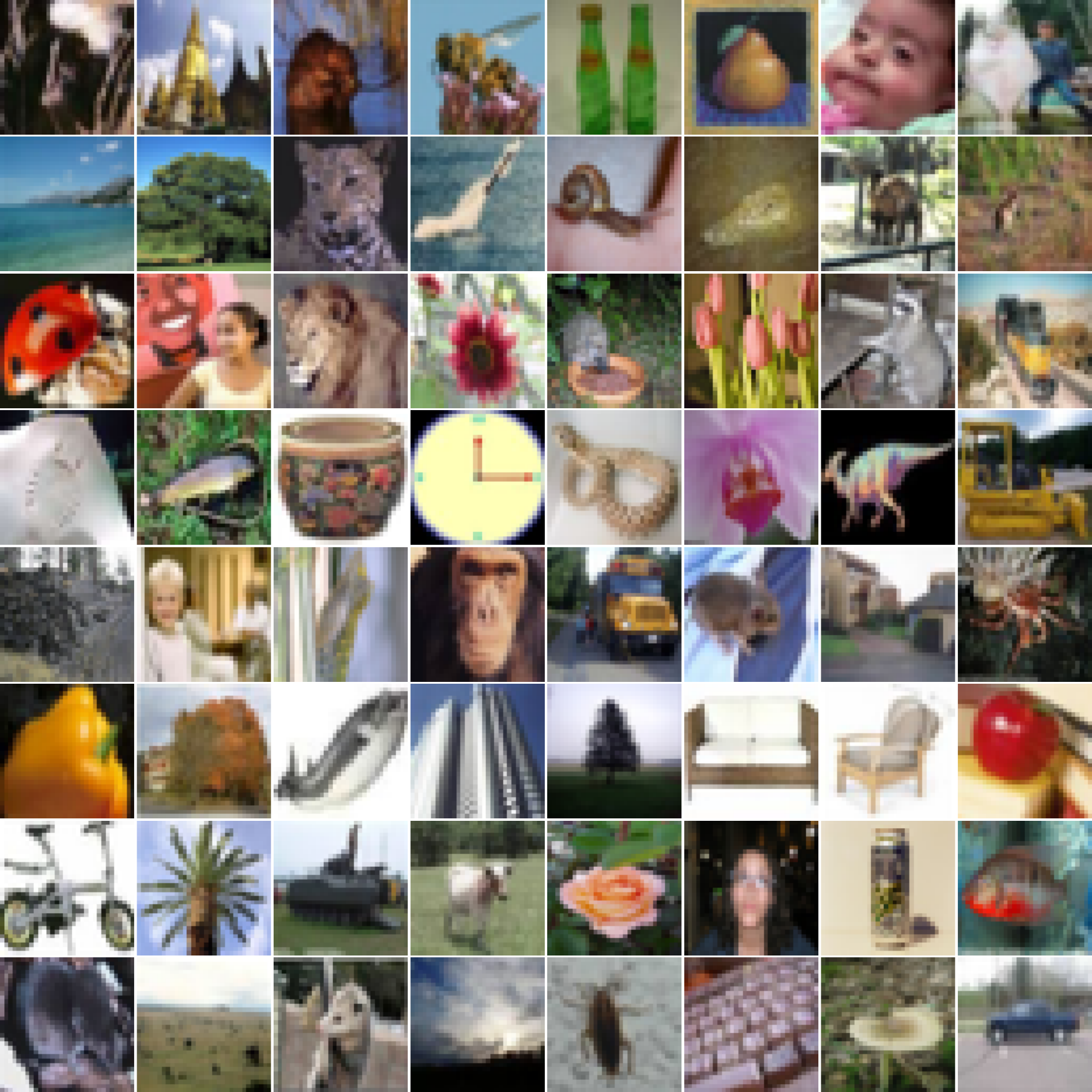}  
}   
\hfill
   \subfigure[Randomly selected.] {    
\includegraphics[width=0.29\linewidth]{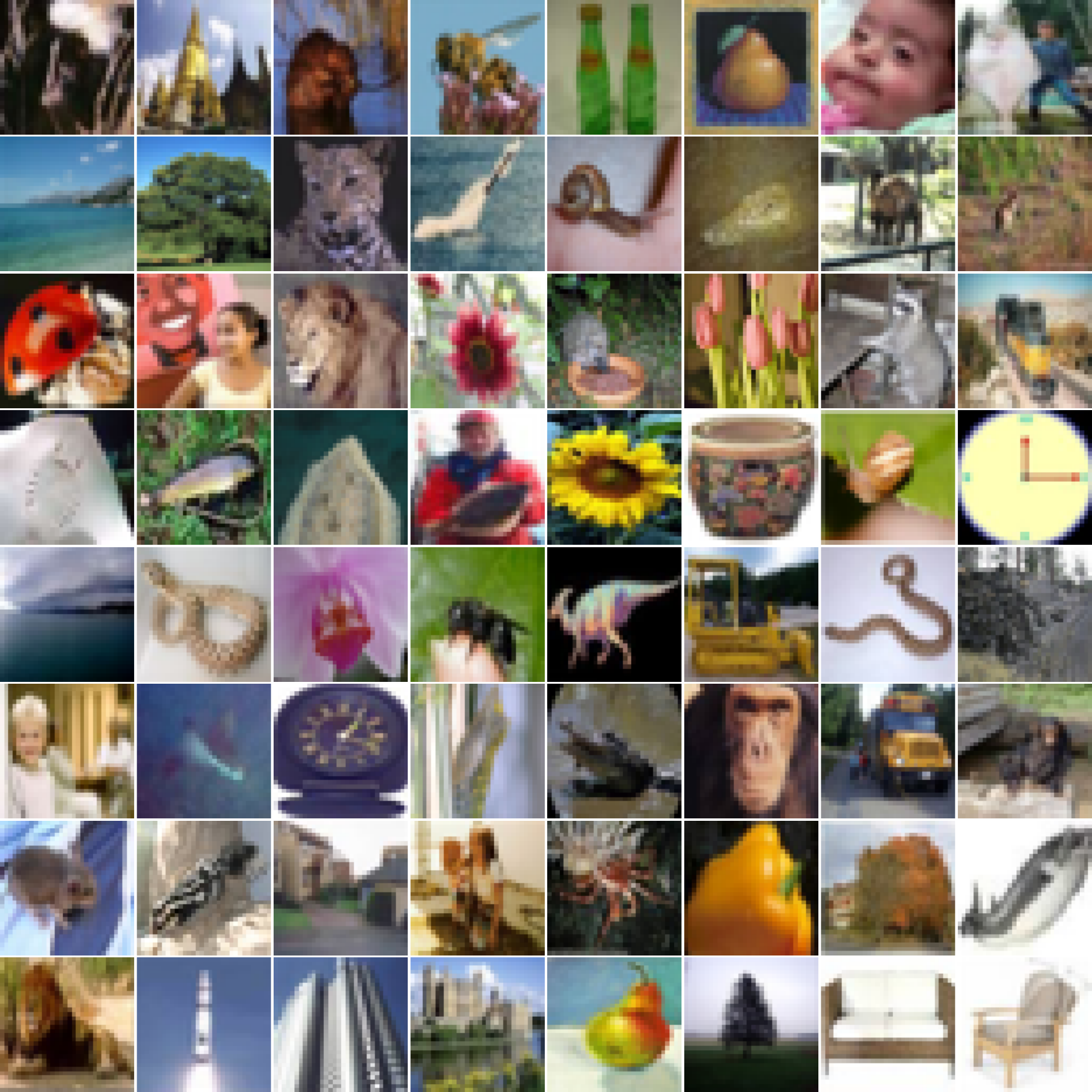}  
} 
\hfill
    \subfigure[All same label.] {    
\includegraphics[width=0.29\linewidth]{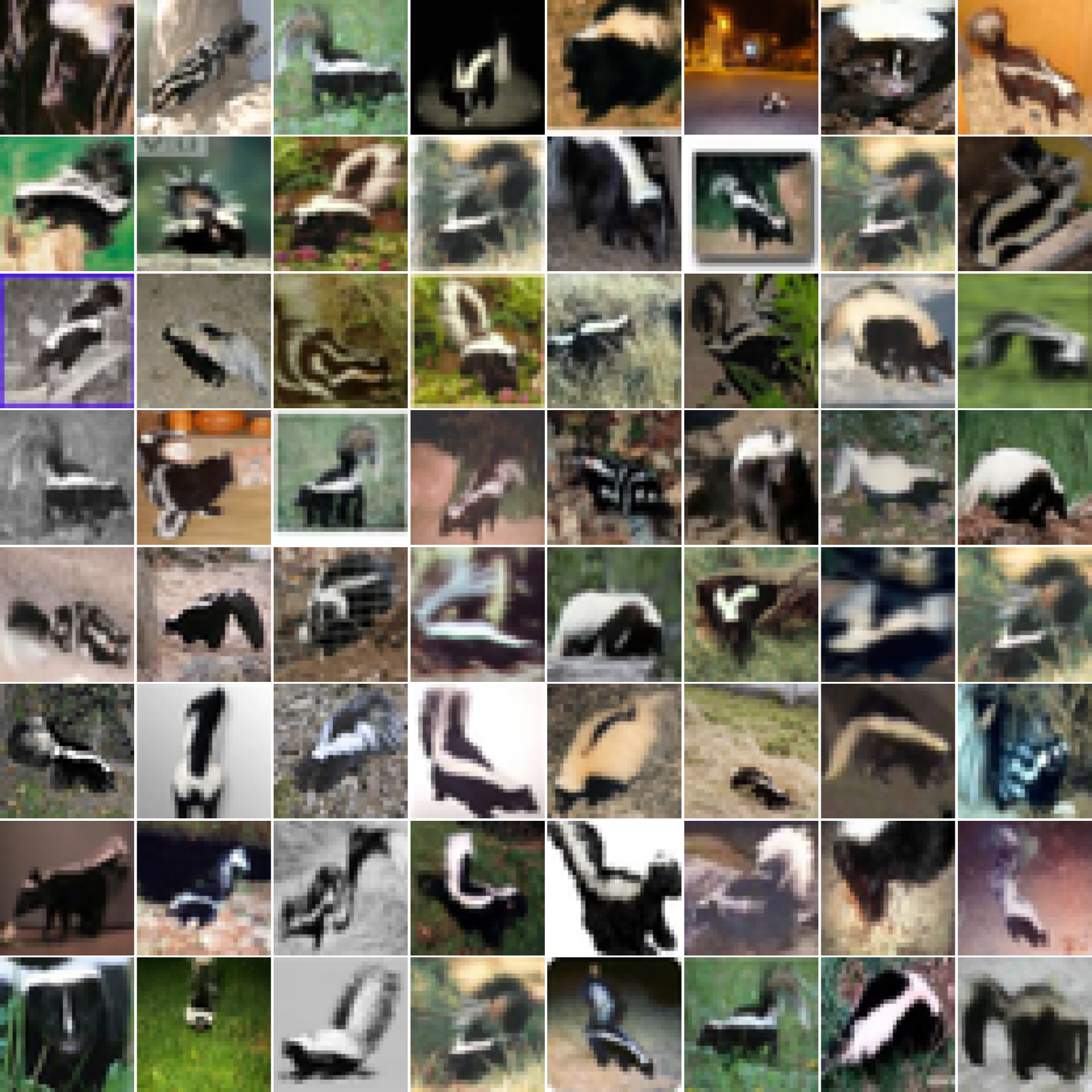}  
}   

  \subfigure[No same label. PSNR $\uparrow$: 18.15, SSIM $\uparrow$: 0.7556, LPIPS $\downarrow$: 0.0184.] {     
\includegraphics[width=0.29\linewidth]{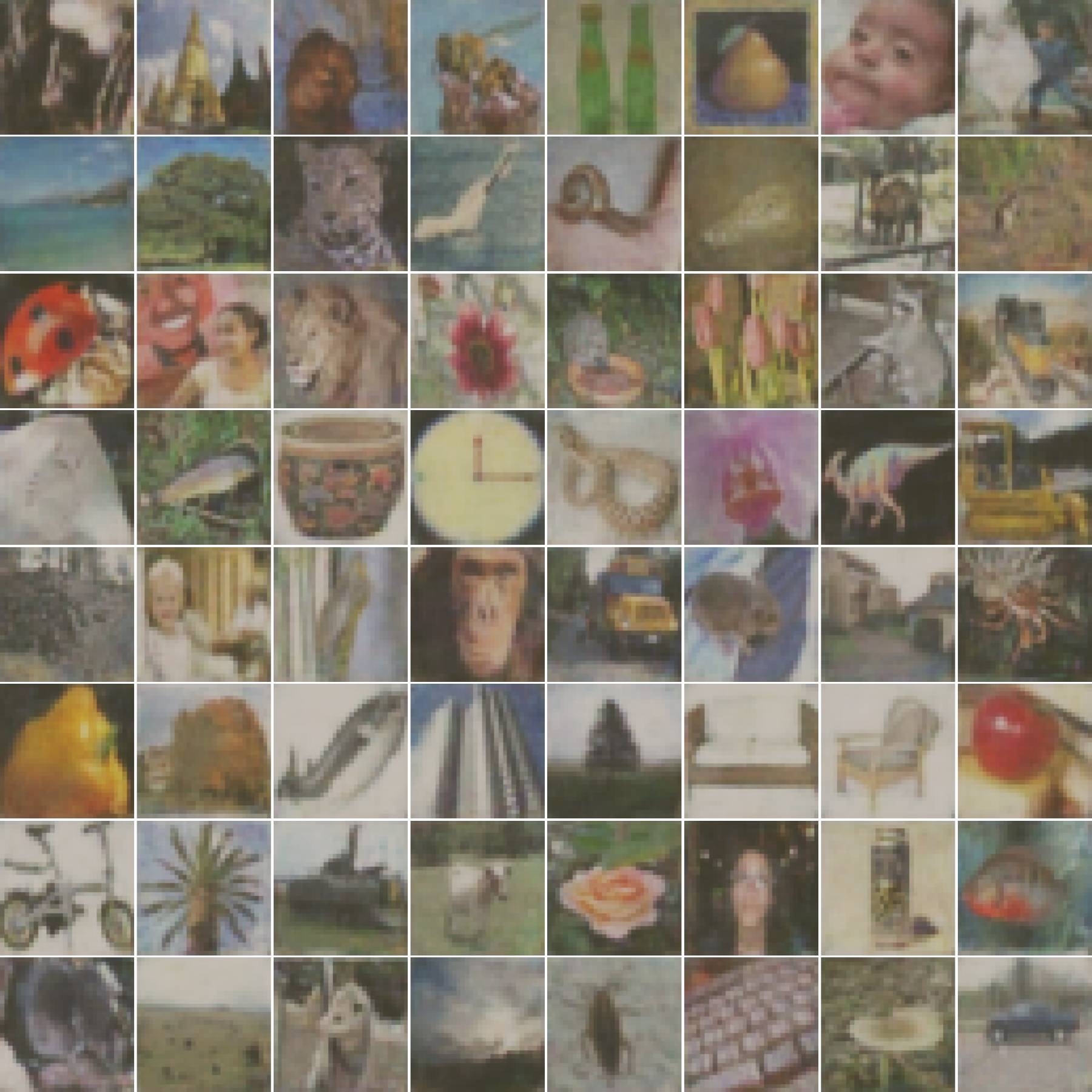}  
}   
\hfill
   \subfigure[Randomly selected. PSNR $\uparrow$: 17.87, SSIM $\uparrow$: 0.7365, LPIPS $\downarrow$: 0.0221.] {    
\includegraphics[width=0.29\linewidth]{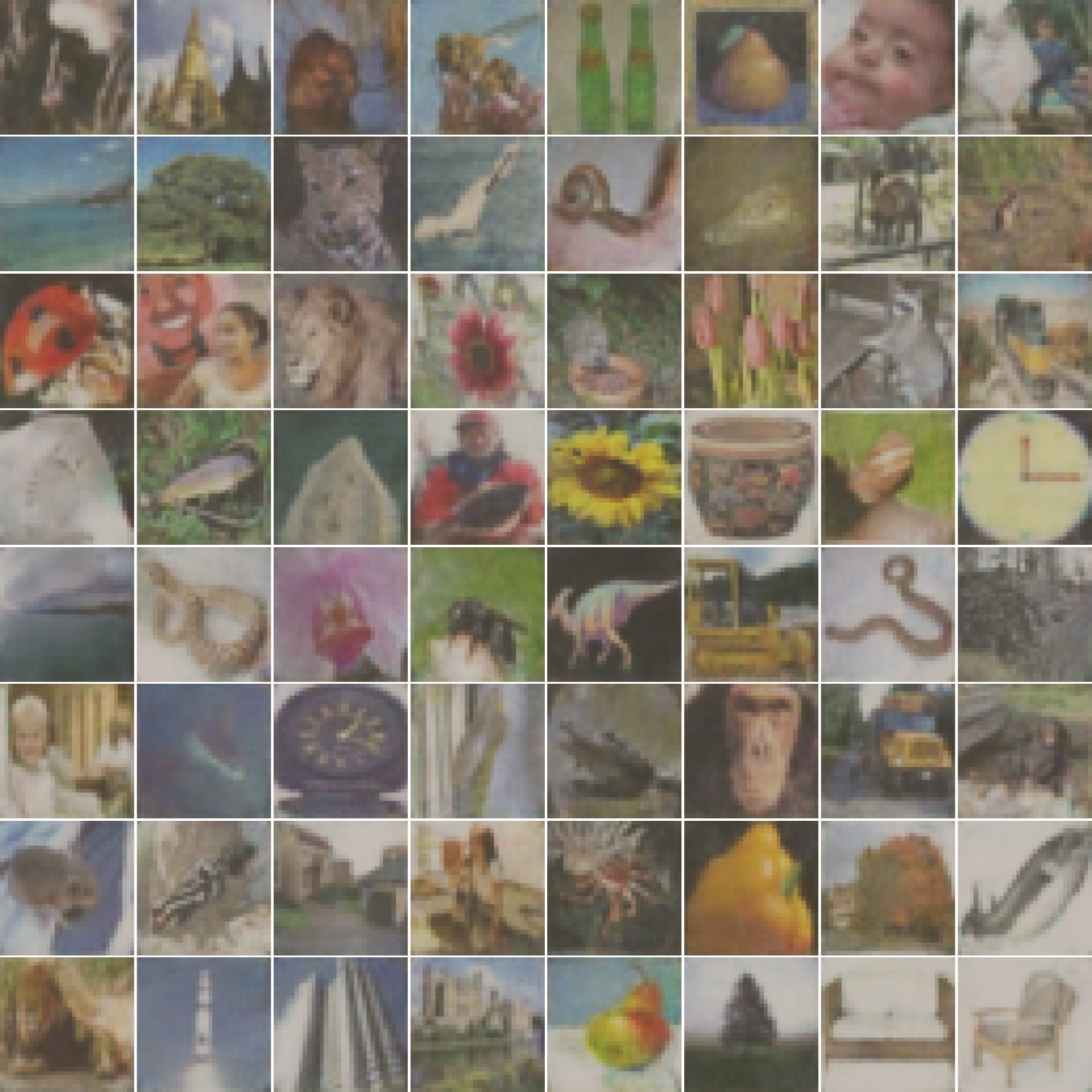}  
} 
\hfill
    \subfigure[All same label. PSNR $\uparrow$: 14.40, SSIM $\uparrow$: 0.4712, LPIPS $\downarrow$: 0.0769.] {    
\includegraphics[width=0.29\linewidth]{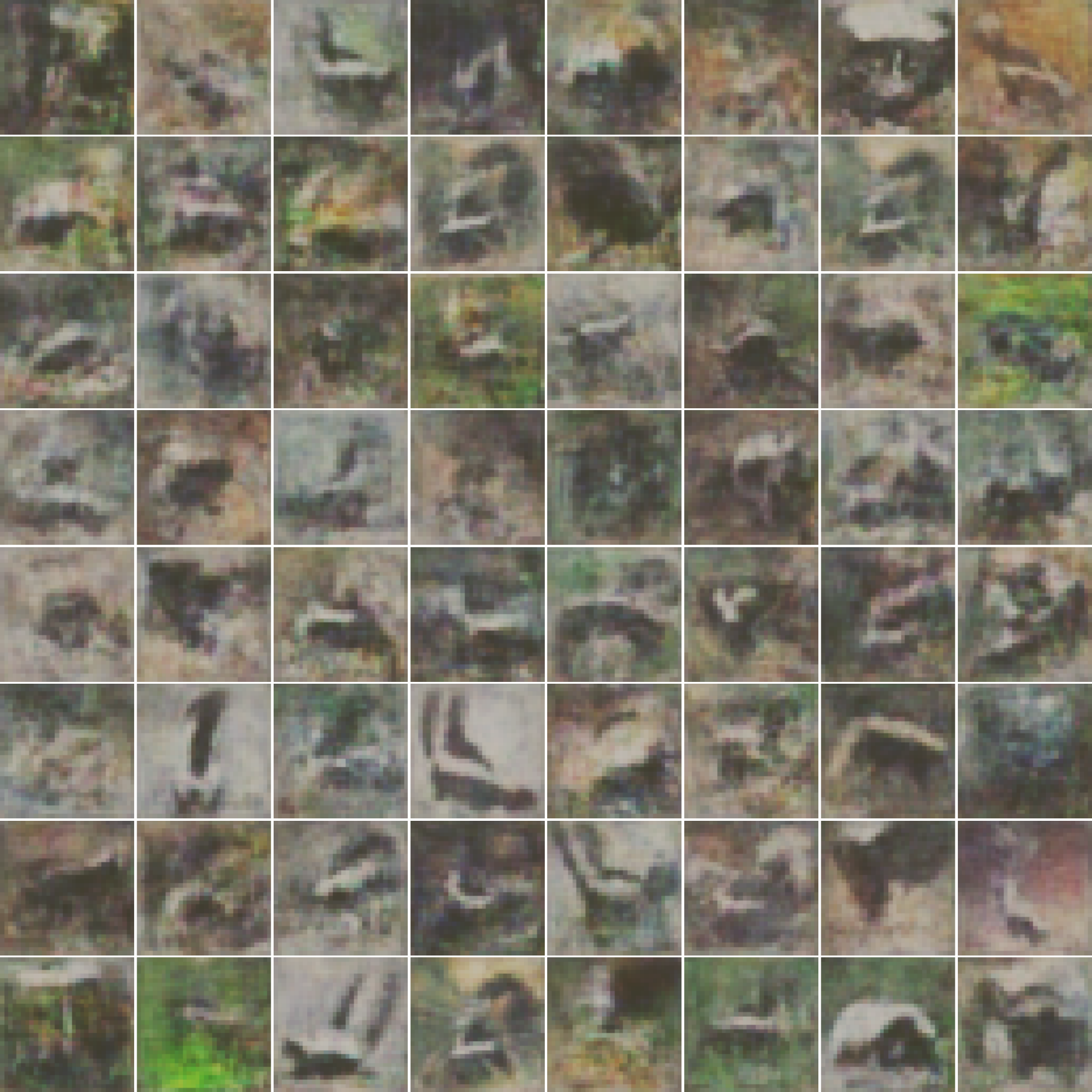}  
}   
  \caption{(a)-(c): Original images on CIFAR-100 with varying numbers of samples that share the same label. (d)-(f): Reconstruction results of CI-Net on CIFAR-100 with varying numbers of samples that share the same label in batches of size 64. These results indicate that more same labels in one batch lead to worse CI-Net performance.}
  \label{app_fig:CI-Net-same-label}
\end{figure}

The reconstruction results of CI-Net with all evaluation metrics are shown in Figures \ref{app_fig:CI-Net-all}, \ref{app_fig:CI-Net-bs-all}. Reconstruction results of CI-Net on the CIFAR-100 dataset with a batch size of 64 under different numbers of images with the same label within one batch are shown in Figure \ref{app_fig:CI-Net-same-label}. Reconstruction results of CI-Net on the CIFAR-100 dataset under various activation functions (e.g., Sigmoid, ReLU, Tanh, Leaky-ReLU, RReLU, and GeLU) are shown in Figure \ref{app_fig:CI-Net-acts-all}. These results show that when optimizing the generator's parameters $\bm{W}$, GEN-GIA can achieve pixel-level attacks, but is affected by the factors that influence OP-GIA. Moreover, it only works when the target model adopts the Sigmoid activation function and fails with other activation functions.

\subsubsection{Training an Inversion Generation Model} 
\begin{figure}[h]
  \centering
  \subfigure[PSNR $\uparrow$.] {     
\includegraphics[width=0.29\linewidth]{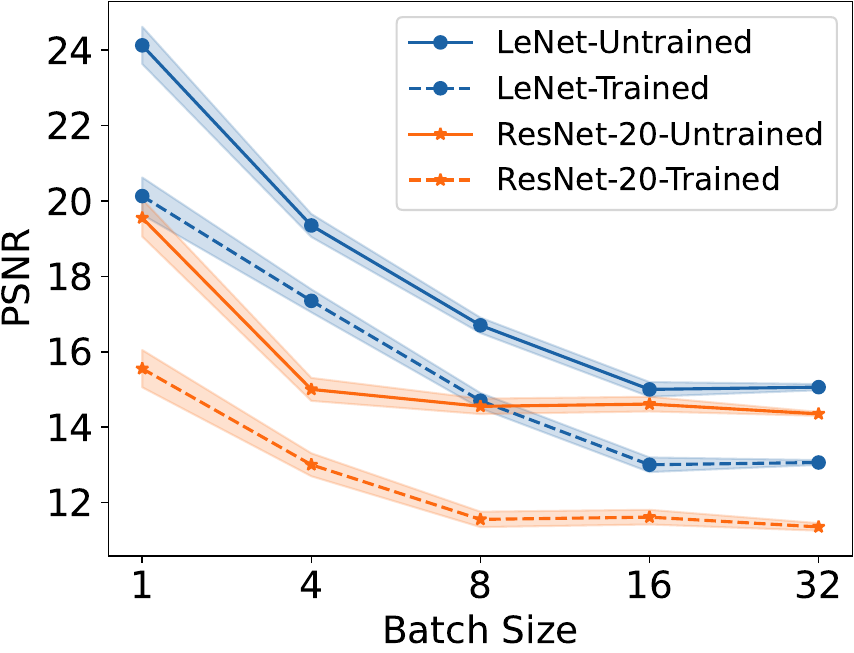}  
}   
\hfill
   \subfigure[SSIM $\uparrow$.] {    
\includegraphics[width=0.29\linewidth]{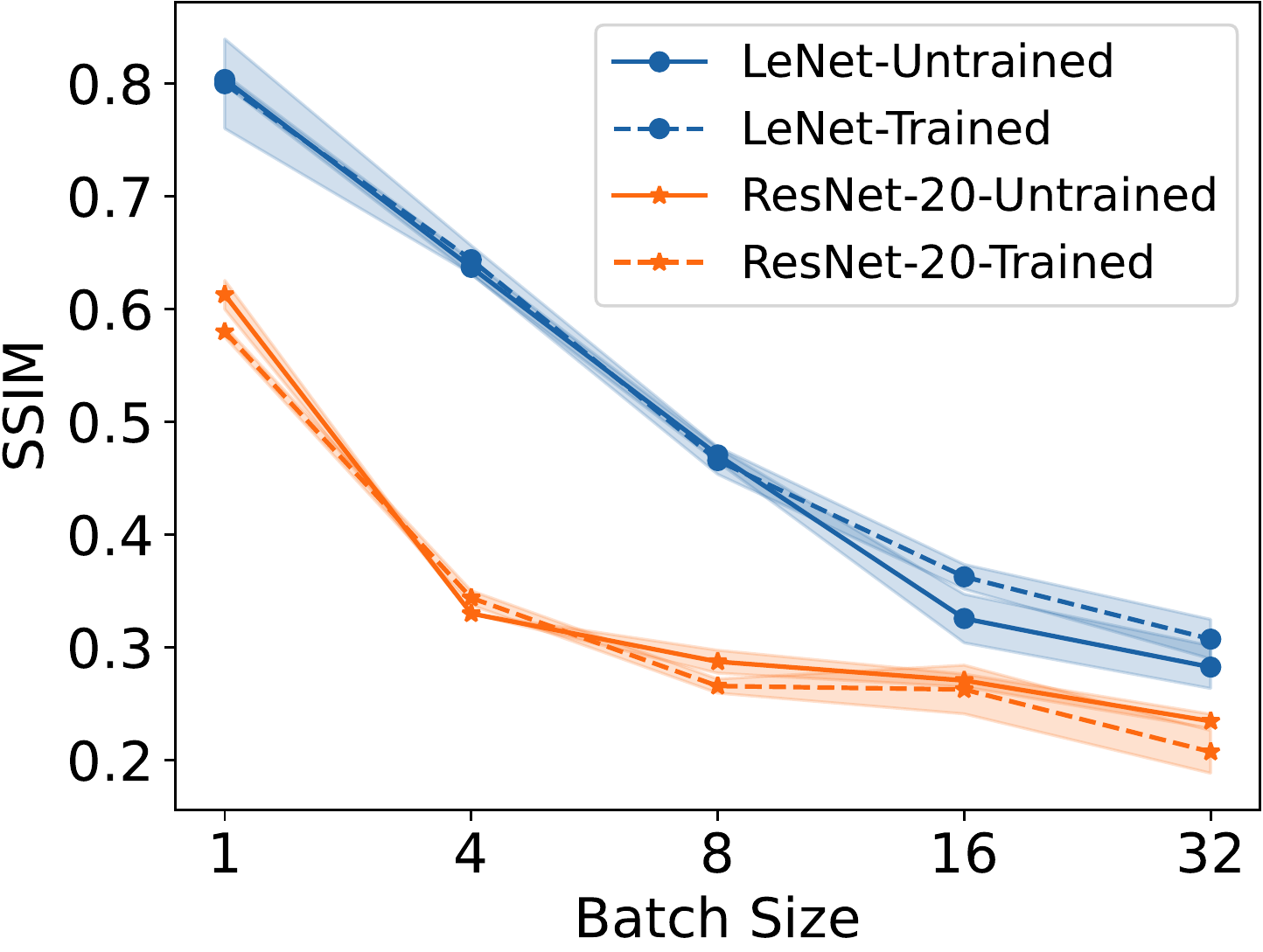}  
} 
\hfill
    \subfigure[LPIPS $\downarrow$.] {    
\includegraphics[width=0.29\linewidth]{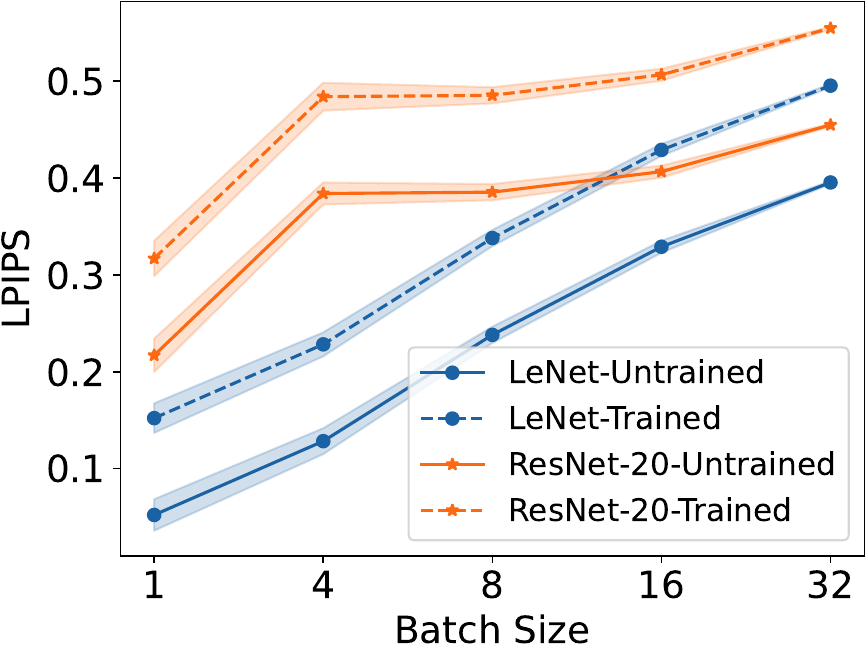}  
}   
\hfill 
\revise{
    \subfigure[Jaccard $\uparrow$.] {    
\includegraphics[width=0.29\linewidth]{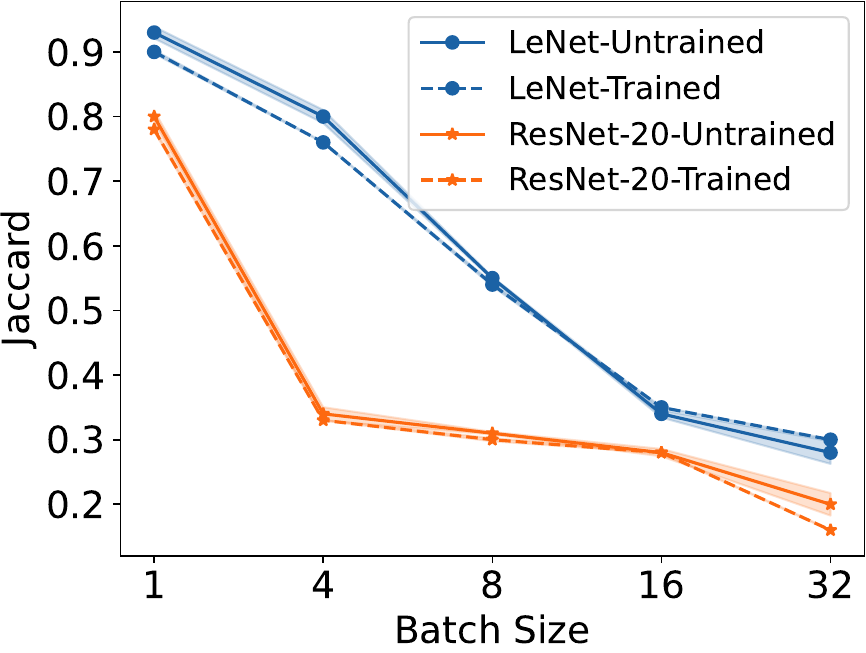}  
}   
    \subfigure[RDLV $\uparrow$.] {    
\includegraphics[width=0.29\linewidth]{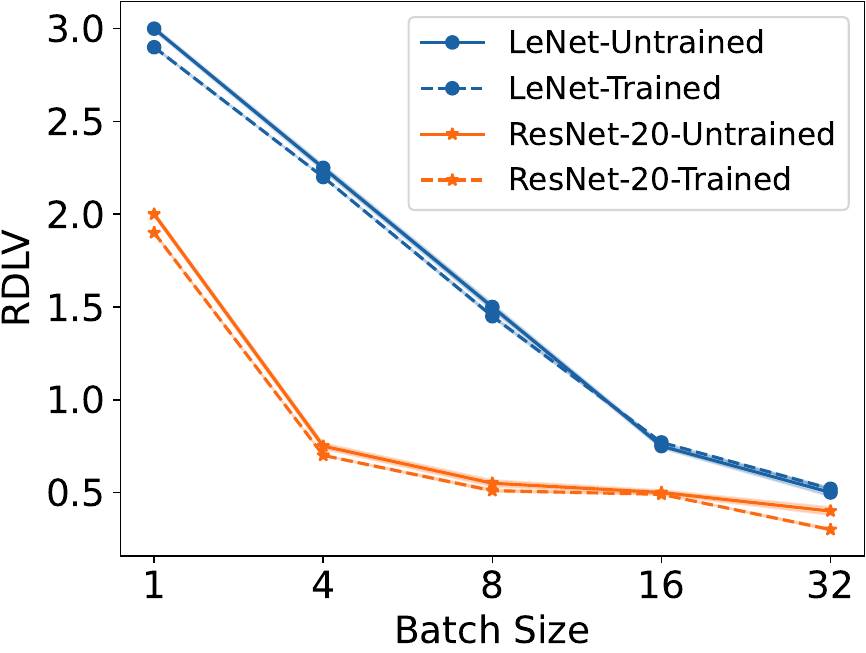}  
}
}
  \caption{Reconstruction results of LTI on CIFAR-10 and different model architectures under different model training states. These results show that a more complicated model architecture and better training state lead to worse LTI performance.
  }
  \label{app_fig:LTI-model-all}
\end{figure}

\begin{figure}[h]
  \centering
  \subfigure[PSNR $\uparrow$.] {     
\includegraphics[width=0.29\linewidth]{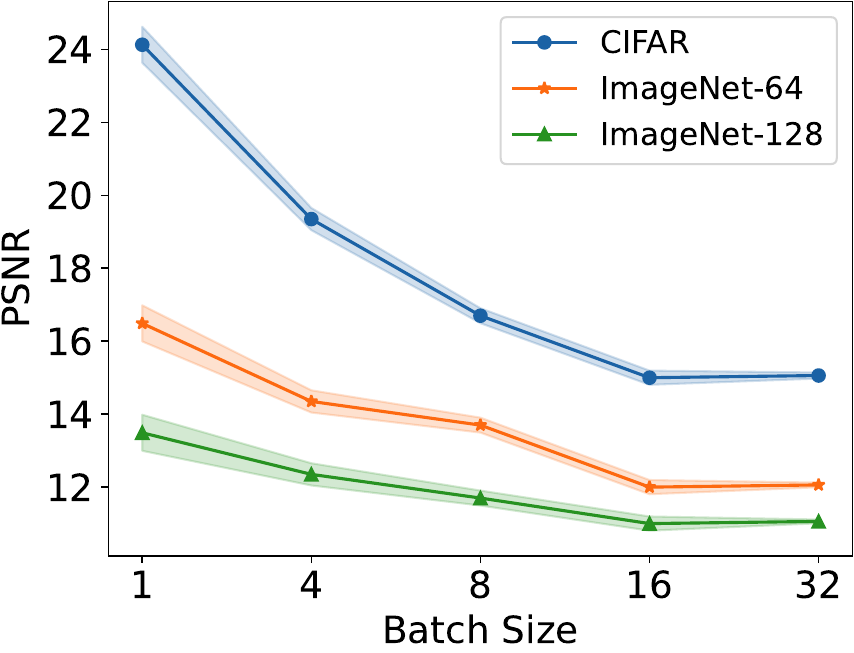}  
}   
\hfill
   \subfigure[SSIM $\uparrow$.] {    
\includegraphics[width=0.29\linewidth]{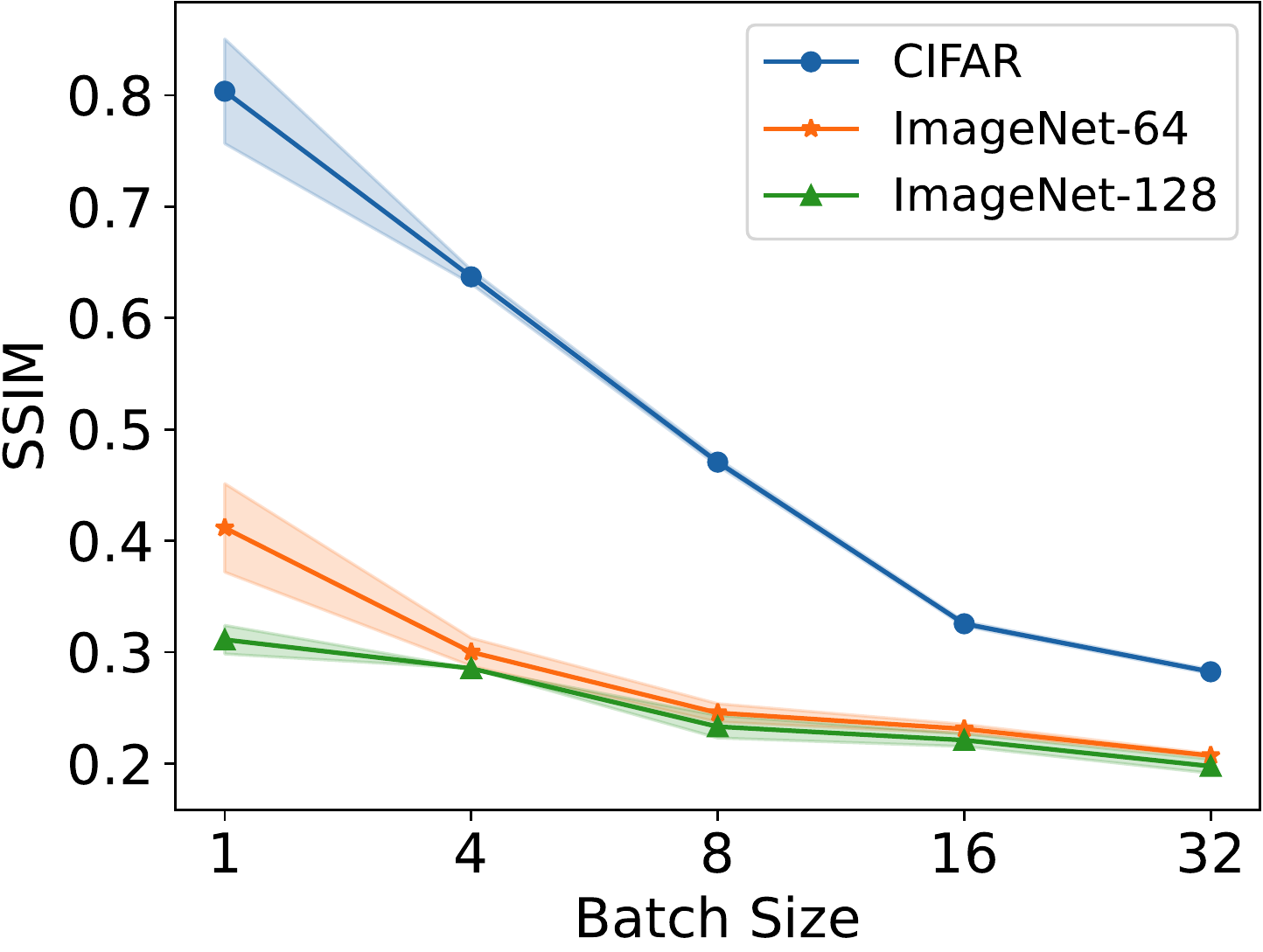}  
} 
\hfill
    \subfigure[LPIPS $\downarrow$.] {    
\includegraphics[width=0.29\linewidth]{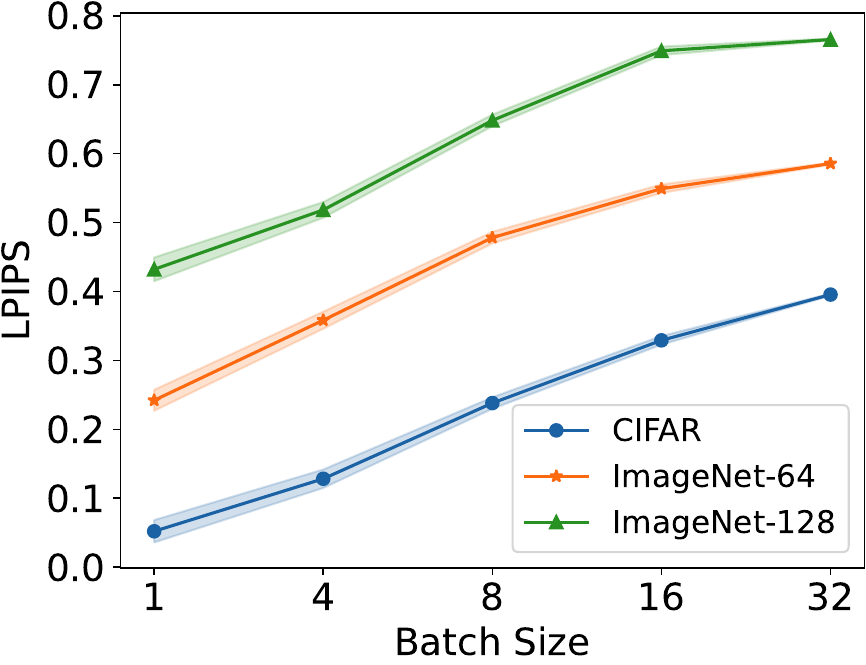}  
}   
\hfill 
\revise{
    \subfigure[Jaccard $\uparrow$.] {    
\includegraphics[width=0.29\linewidth]{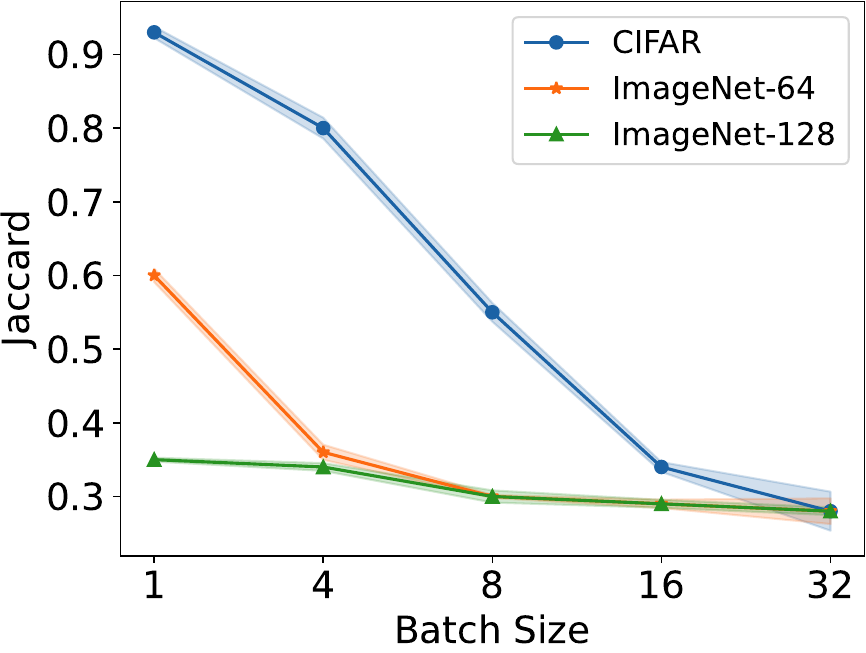}  
}   
    \subfigure[RDLV $\uparrow$.] {    
\includegraphics[width=0.29\linewidth]{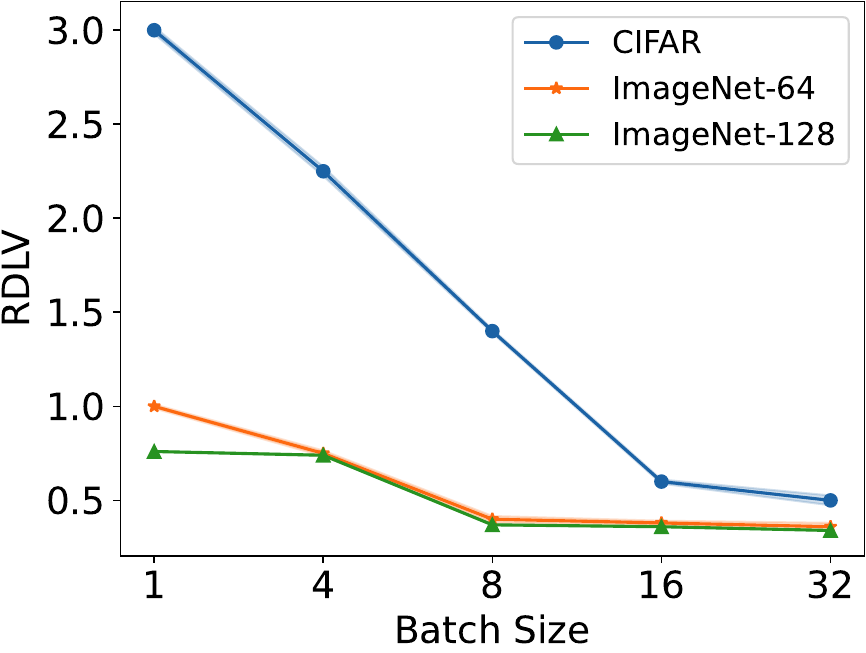}  
}
}
  \caption{Reconstruction results of LTI on various datasets with different resolutions. These results show that larger resolutions lead to worse LTI performance.
  }
  \label{app_fig:LTI-resolution-all}
\end{figure}

\begin{figure*}[t]
    \centering
    \includegraphics[width=1\linewidth]{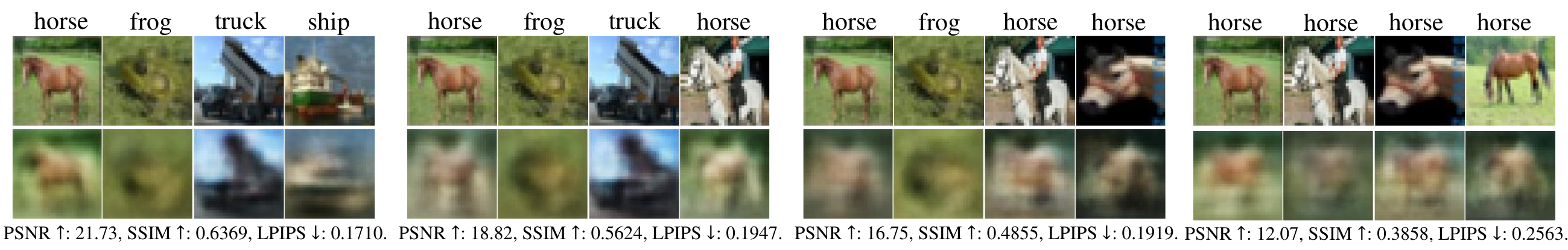}
    \caption{Reconstruction results of LTI on the CIFAR-10 dataset with a batch size of 4. From left to right, the number of images with the same label are 0, 2, 3, and 4. The first row represents the ground truth, while the second row shows the reconstruction results. These results indicate that more same labels in one batch lead to worse LTI performance.}
\label{app_fig:compare_label_number_LTI}
\end{figure*}

The reconstruction results of LTI with all evaluation metrics are shown in Figures \ref{app_fig:LTI-model-all} and \ref{app_fig:LTI-resolution-all}. Reconstruction results of LTI for the varying numbers of samples that share the same label in one batch are provided in Figure \ref{app_fig:compare_label_number_LTI}. These results show that when training an inversion generation model, GEN-GIA can achieve pixel-level attacks but is influenced by most of the factors that affect OP-GIA, except for the model's training state. Furthermore, such a paradigm relies on an auxiliary dataset where the data distribution is similar to the local data to train the inversion model, which is difficult to satisfy in real applications.

\subsection{Analytics-based GIA} 

\subsubsection{Manipulating Model Architecture}

\begin{figure*}[t]
  \centering
  \includegraphics[width=1\linewidth]{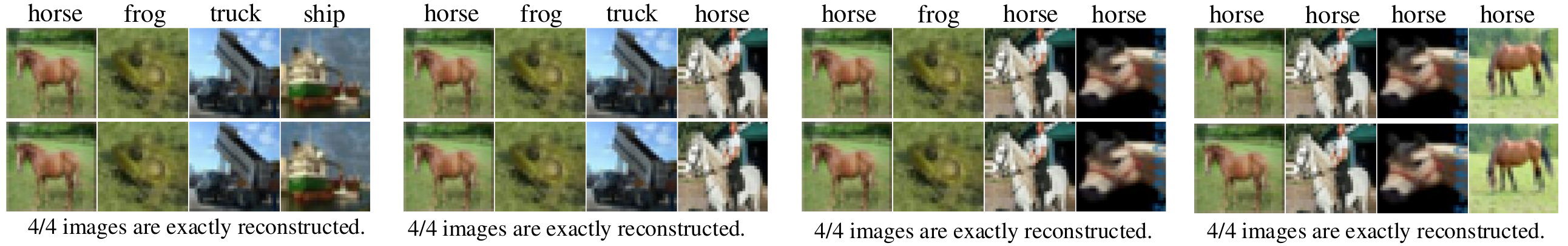}
  \caption{Reconstruction results of Robbing the Fed on the CIFAR-10 dataset with a batch size of 4. From left to right, the number of images with same label is 0, 2, 3, and 4. The first row represents the ground truth, while the second row shows the reconstruction results. These results show that the reconstruction results of ANA-GIA with manipulating model architecture are now affected by the number of same labels in one batch.}
\label{app_fig:compare_label_number_robbing_the_fed}
\end{figure*}

Reconstruction results of Robbing the Fed for the varying numbers of samples that share the same label in one batch are provided in Figure \ref{app_fig:compare_label_number_robbing_the_fed}. These results show that the reconstruction results of ANA-GIA with manipulating model architecture are now affected by the number of same labels in one batch.

\subsubsection{Manipulating Model Parameters} 
\begin{figure}[h]
  \centering
  \subfigure[PSNR $\uparrow$.] {     
\includegraphics[width=0.29\linewidth]{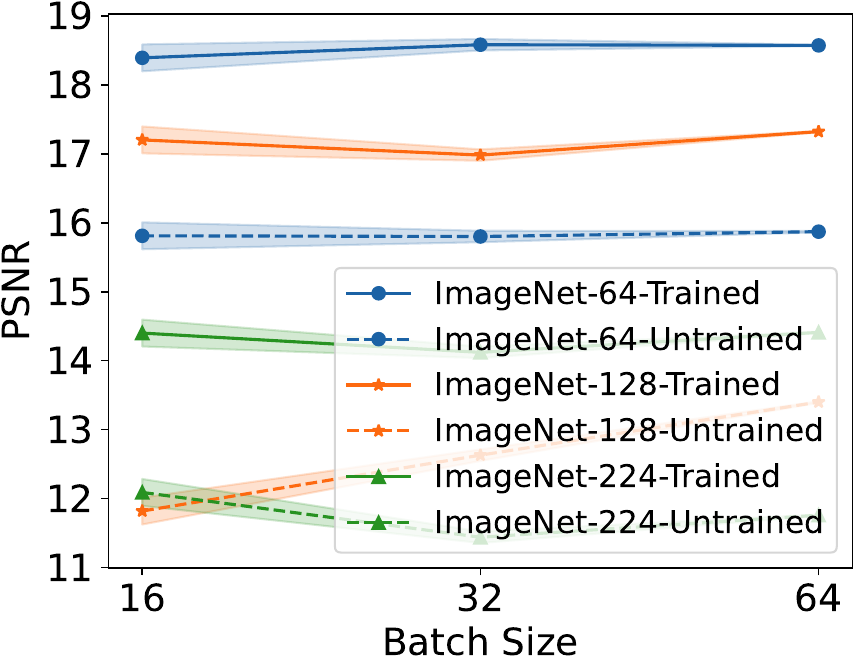}  
}   
\hfill
   \subfigure[SSIM $\uparrow$.] {    
\includegraphics[width=0.29\linewidth]{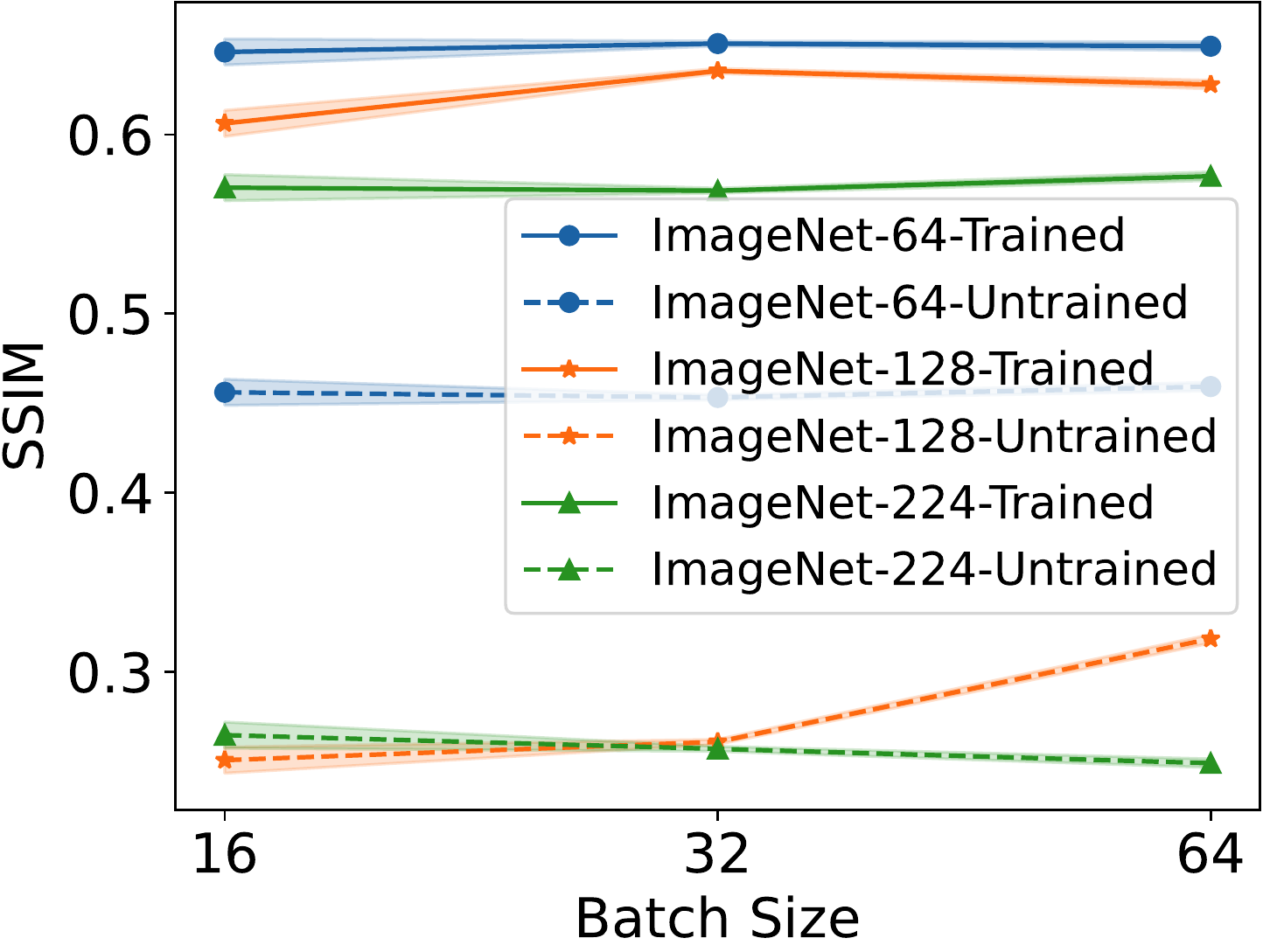}  
} 
\hfill
    \subfigure[LPIPS $\downarrow$.] {    
\includegraphics[width=0.29\linewidth]{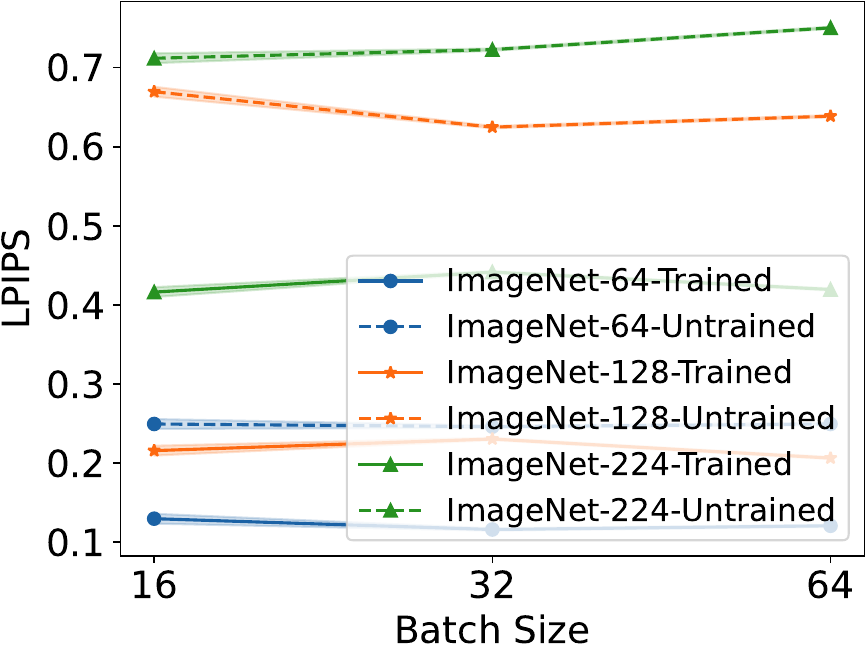}  
}   
\hfill 
\revise{
    \subfigure[Jaccard $\uparrow$.] {    
\includegraphics[width=0.29\linewidth]{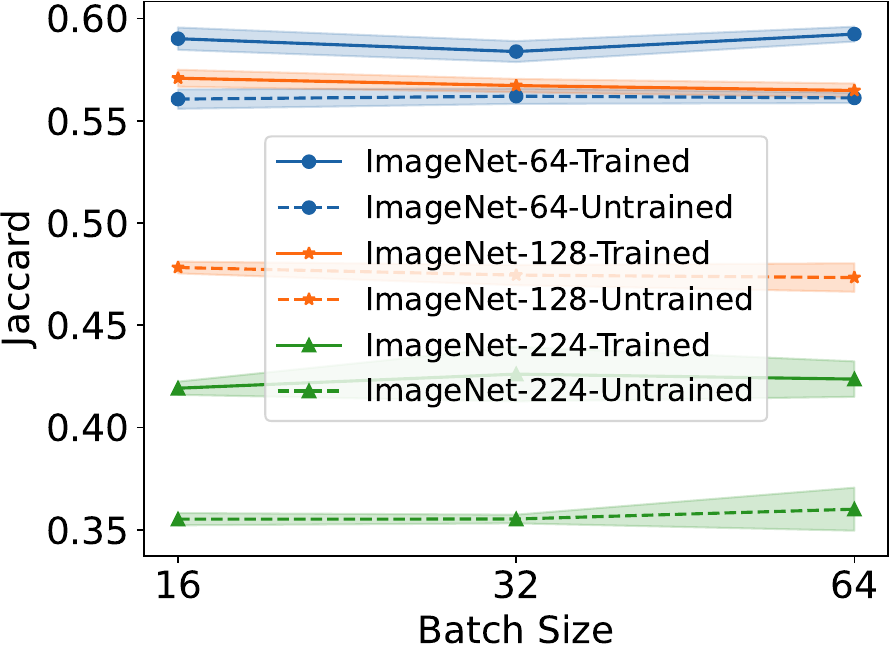}  
}   
    \subfigure[RDLV $\uparrow$.] {    
\includegraphics[width=0.29\linewidth]{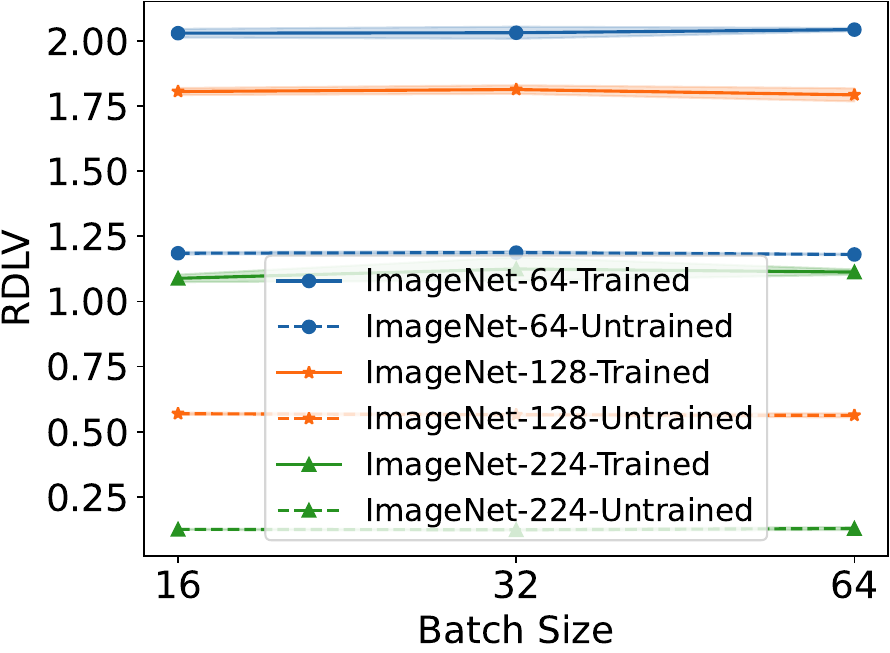}  
}
}
  \caption{Reconstruction results of Fishing on ImageNet with different image resolutions and model training states. These results show that the attack performance of ANA-GIA, which manipulates model parameters, is not affected by batch size but worsens with larger image resolutions and worse model training states.
  }
  \label{app_fig:fish_all}
\end{figure}

\begin{figure}[h]
  \centering
  \subfigure[PSNR $\uparrow$.] {     
\includegraphics[width=0.29\linewidth]{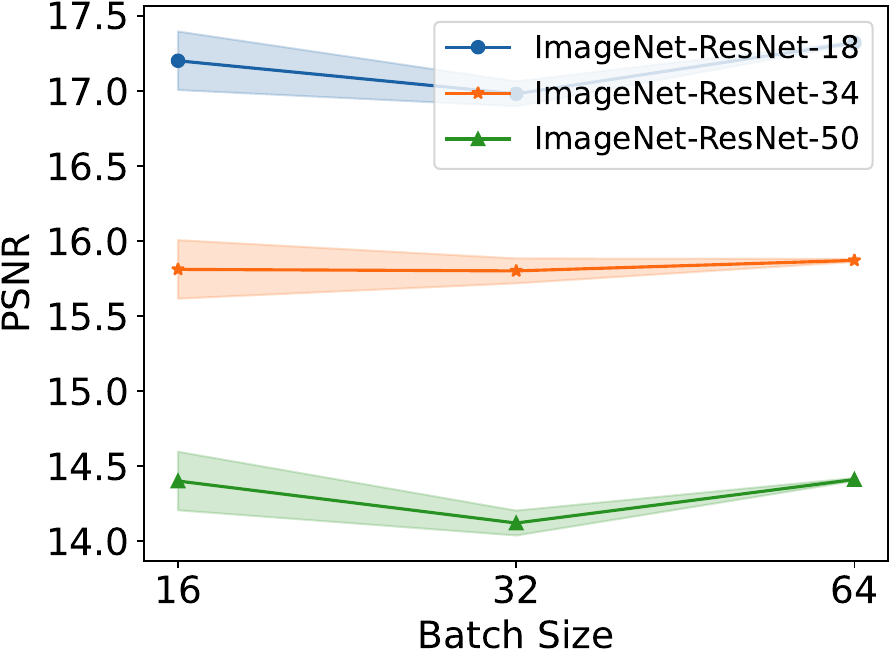}  
}   
\hfill
   \subfigure[SSIM $\uparrow$.] {    
\includegraphics[width=0.29\linewidth]{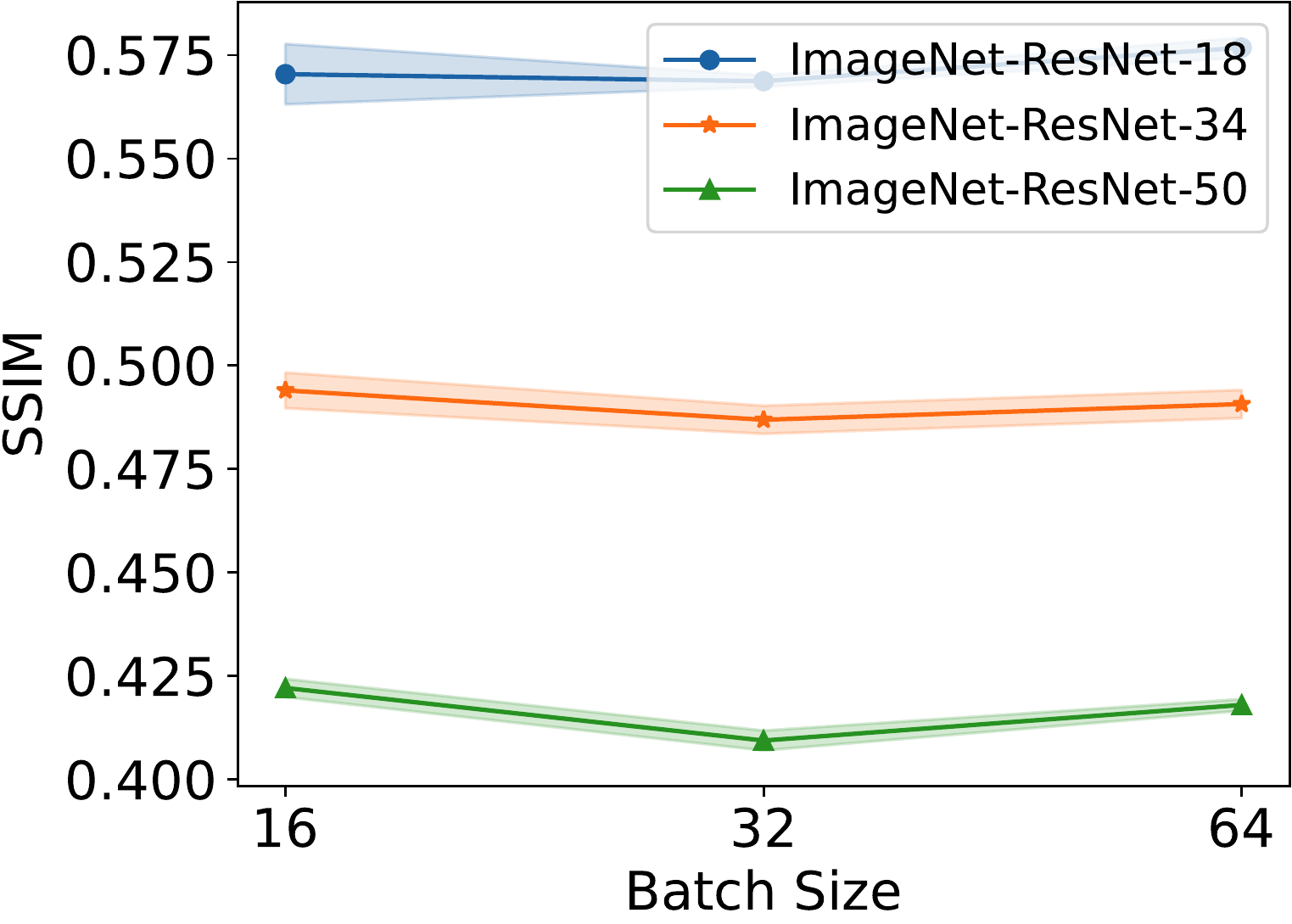}  
} 
\hfill
    \subfigure[LPIPS $\downarrow$.] {    
\includegraphics[width=0.29\linewidth]{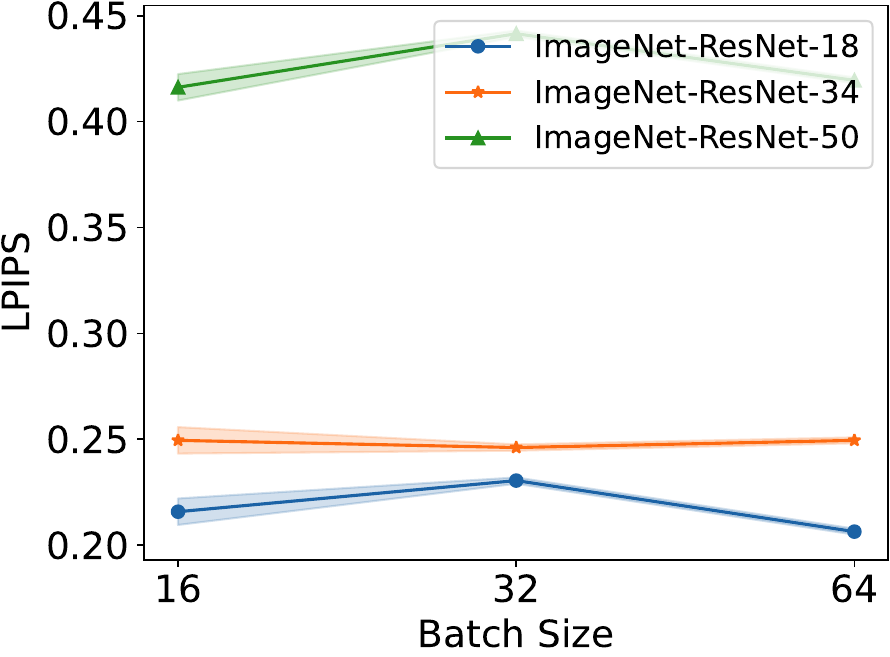}  
}   
\hfill 
\revise{
    \subfigure[Jaccard $\uparrow$.] {    
\includegraphics[width=0.29\linewidth]{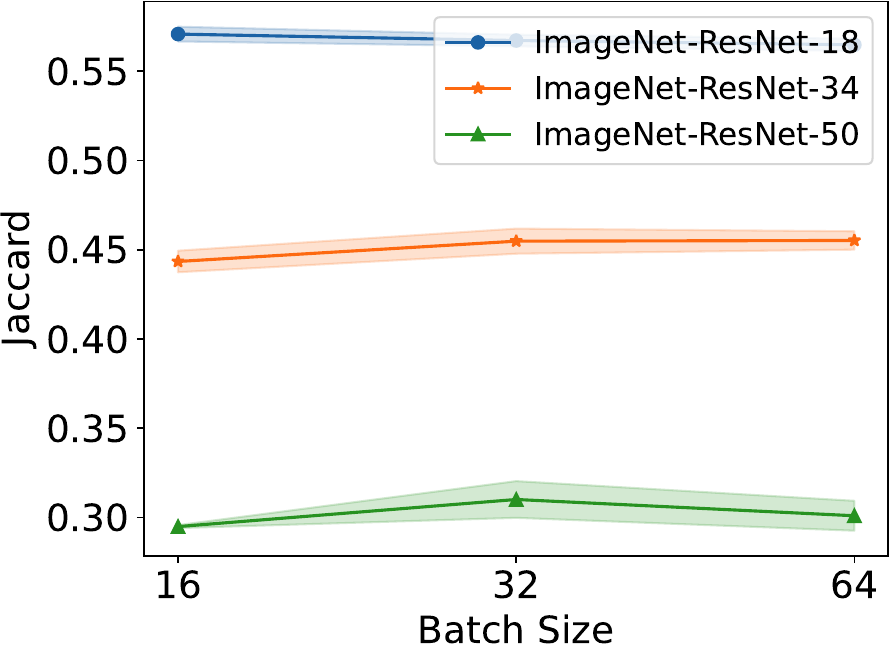}  
}   
    \subfigure[RDLV $\uparrow$.] {    
\includegraphics[width=0.29\linewidth]{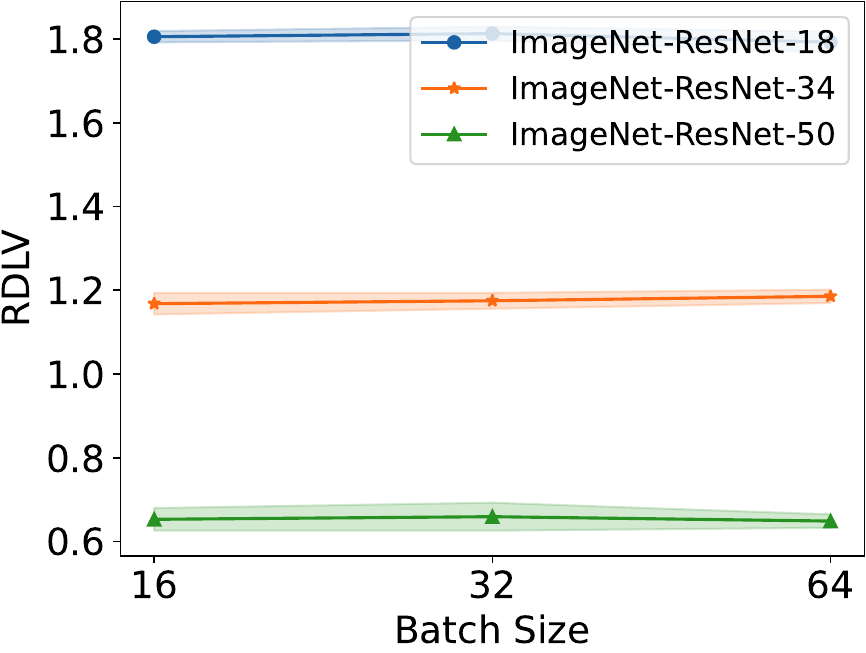}  
}
}
  \caption{Reconstruction results of Fishing on ImageNet with different network architectures. These results show that the attack performance of ANA-GIA, which manipulates model parameters, is not affected by batch size but worsens with more complicated model architecture.
  }
  \label{app_fig:fish_arch_all}
\end{figure}

The reconstruction results of Fishing with all evaluation metrics are shown in Figures \ref{app_fig:fish_all} and \ref{app_fig:fish_arch_all}. These results show that ANA-GIA which manipulates model parameters can achieve satisfactory attack performance regardless of batch size. However, performance decreases with increasing image resolution, from trained to untrained models, and complicated network architectures.

\subsection{Attacks under Parameter-Efficient Fine-Tuning} \label{app_sec:exp_attack_peft}

\begin{figure}[h]
  \centering
  \subfigure[PSNR $\uparrow$.] {     
\includegraphics[width=0.29\linewidth]{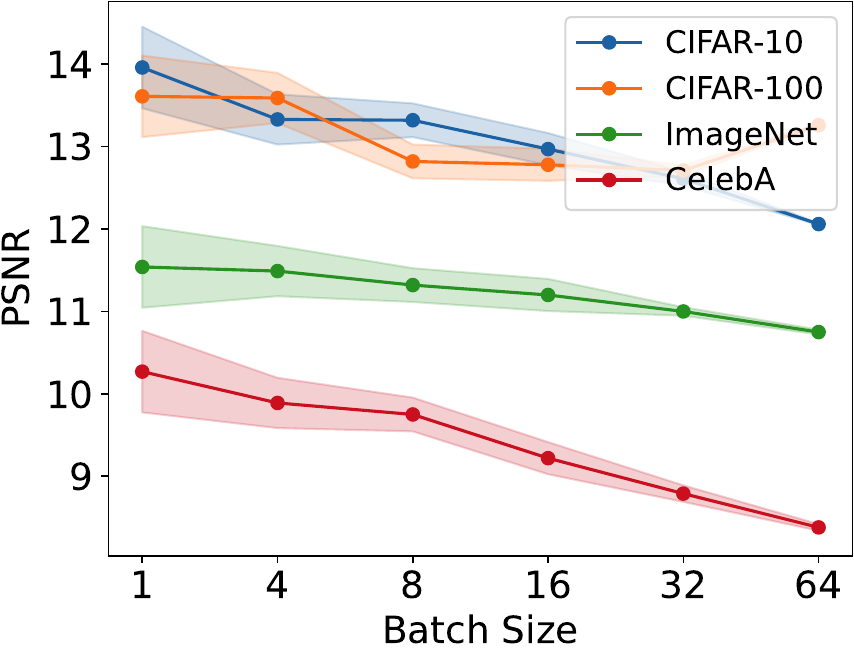}  
}   
\hfill
   \subfigure[SSIM $\uparrow$.] {    
\includegraphics[width=0.29\linewidth]{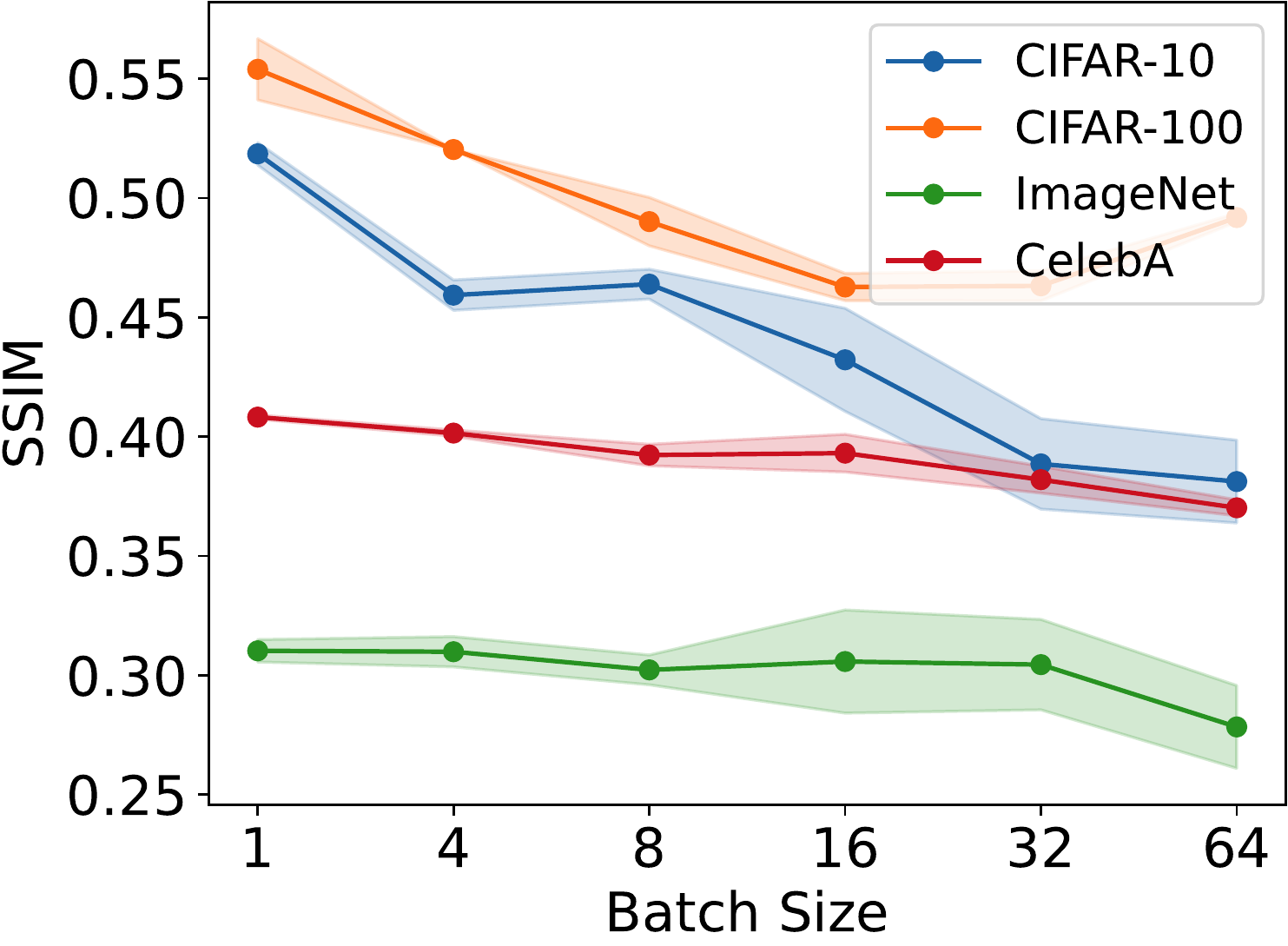}  
} 
\hfill
    \subfigure[LPIPS $\downarrow$.] {    
\includegraphics[width=0.29\linewidth]{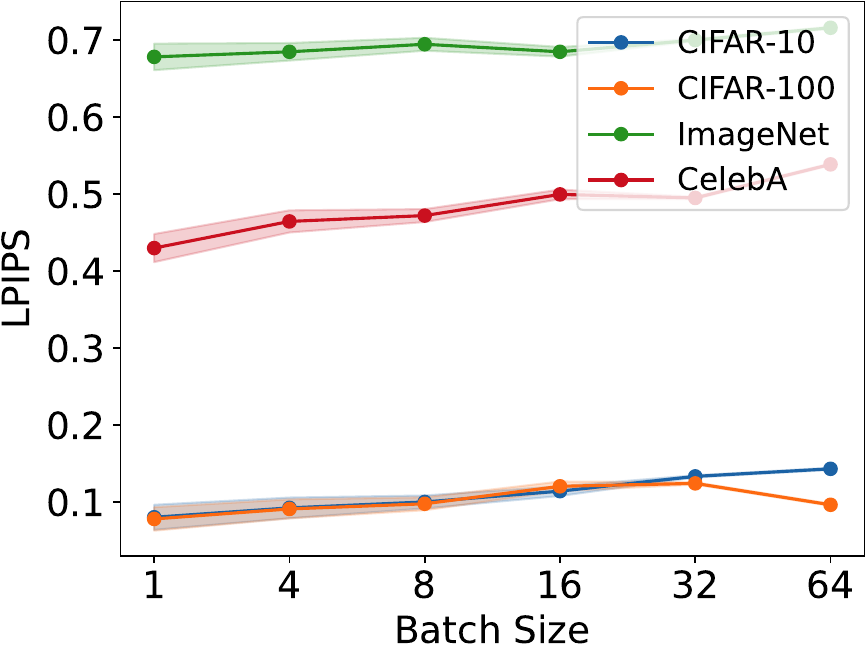}  
}   
\hfill 
\revise{
    \subfigure[Jaccard $\uparrow$.] {    
\includegraphics[width=0.29\linewidth]{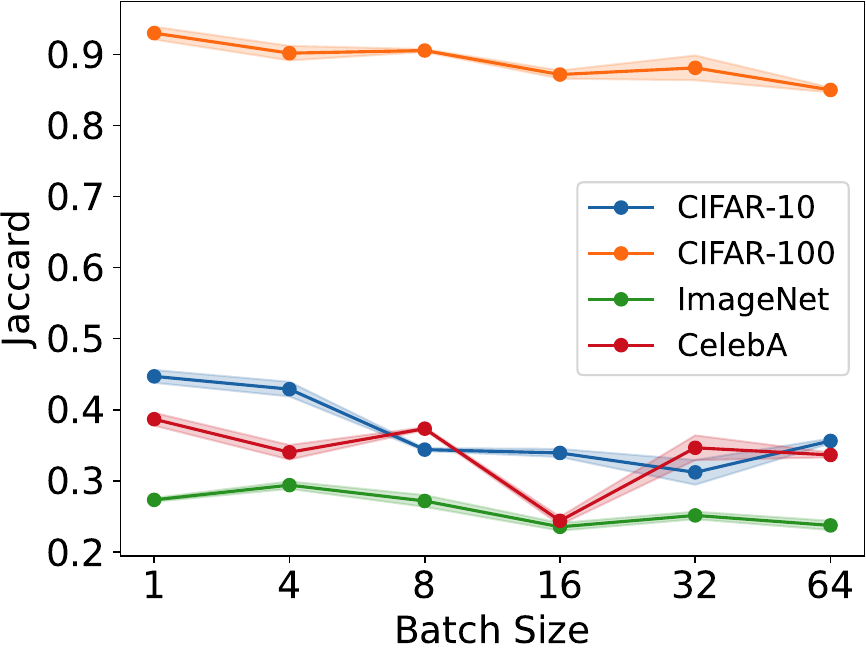}  
}   
    \subfigure[RDLV $\uparrow$.] {    
\includegraphics[width=0.29\linewidth]{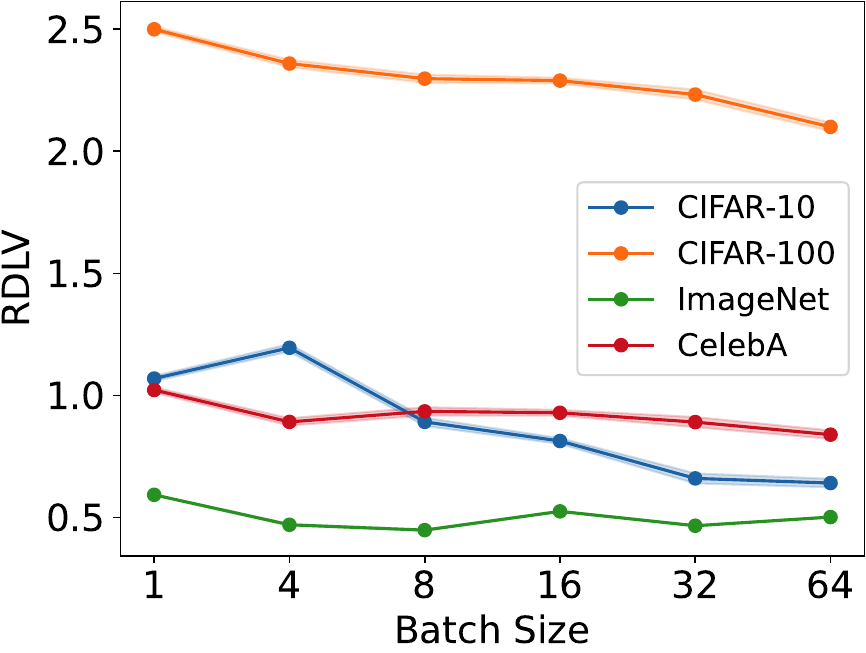}  
}
}
  \caption{The reconstruction results using Eq. (7) evaluated on ViT-base fine-tuned with LoRA on different datasets with all evaluation metrics. These results show that attackers can breach privacy on low-resolution images but fail with high-resolution ones under PEFT.
  }
  \label{app_fig:LoRA-dataset-all}
\end{figure}

\begin{figure}[h]
  \centering
  \subfigure[PSNR $\uparrow$.] {     
\includegraphics[width=0.29\linewidth]{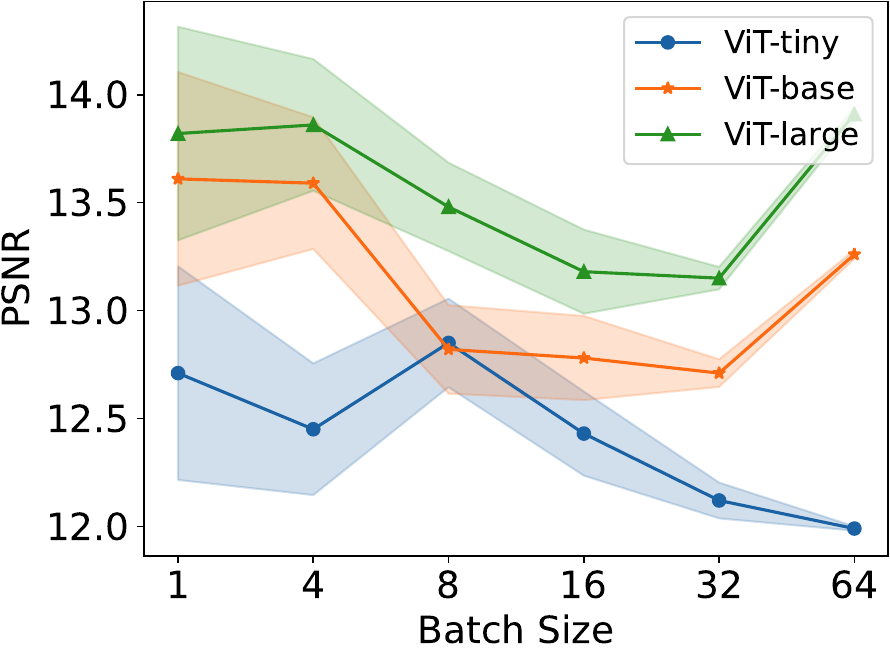}  
}   
\hfill
   \subfigure[SSIM $\uparrow$.] {    
\includegraphics[width=0.29\linewidth]{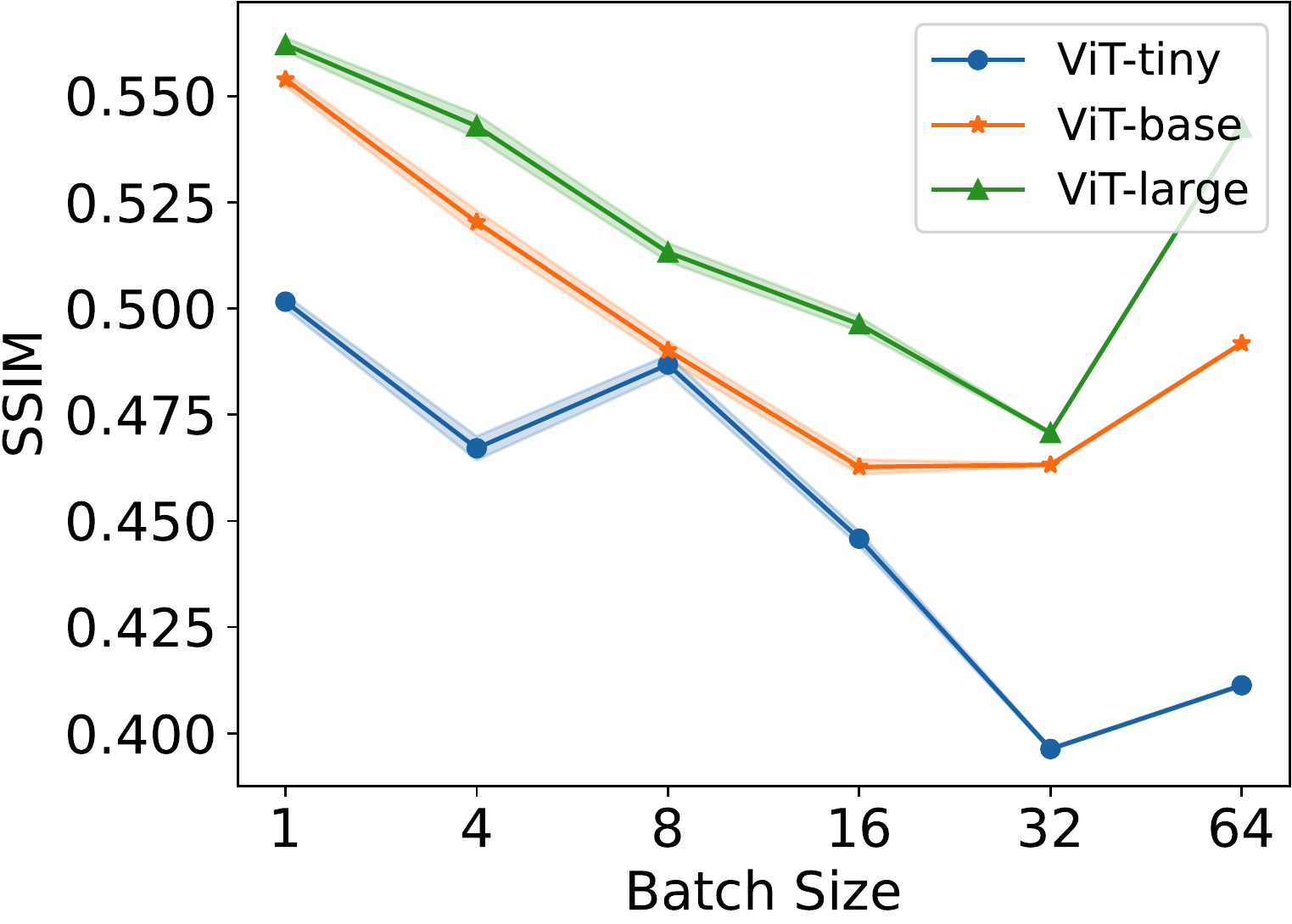}  
} 
\hfill
    \subfigure[LPIPS $\downarrow$.] {    
\includegraphics[width=0.29\linewidth]{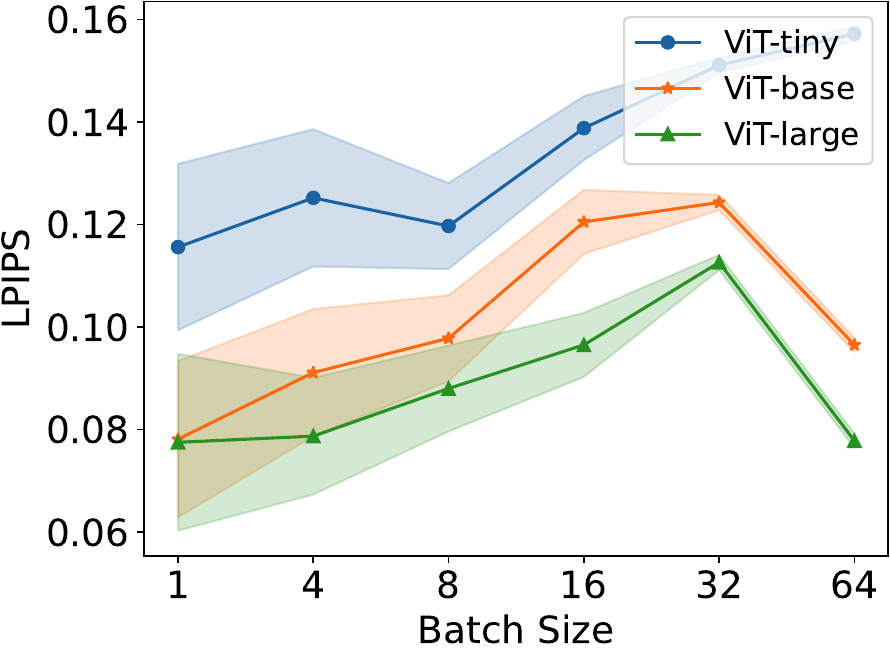}  
}   
\hfill 
\revise{
    \subfigure[Jaccard $\uparrow$.] {    
\includegraphics[width=0.29\linewidth]{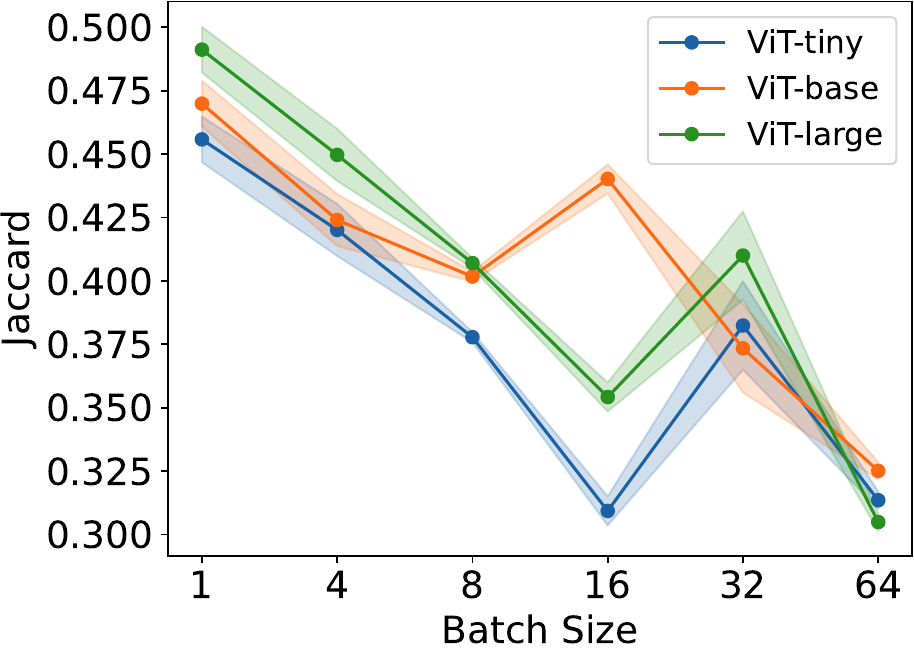}  
}   
    \subfigure[RDLV $\uparrow$.] {    
\includegraphics[width=0.29\linewidth]{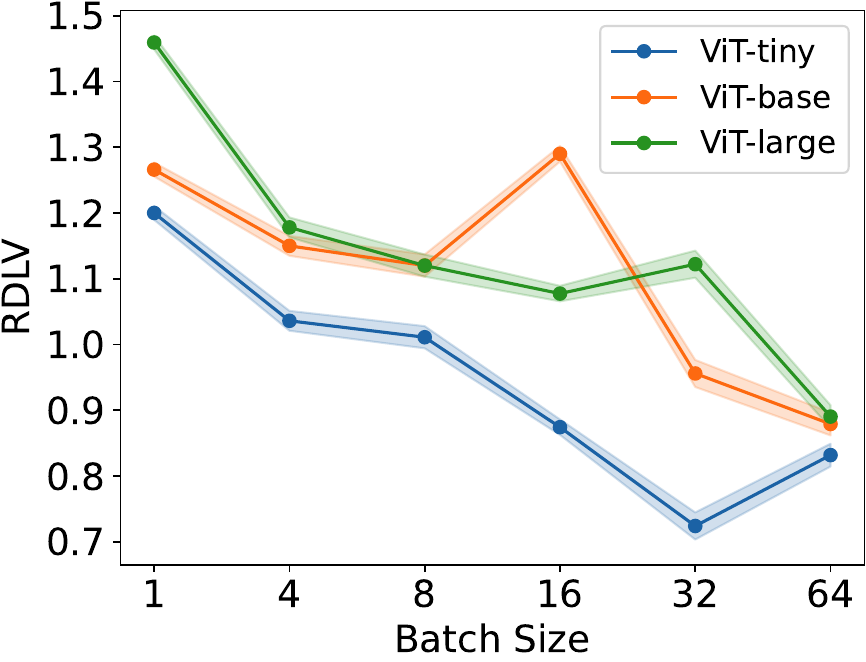}  
}
}
  \caption{The reconstruction results using Eq. (7) evaluated on different ViT architectures fine-tuned with LoRA on CIFAR-100 with all evaluation metrics. These results show that smaller pre-trained models are better at protecting privacy.
  }
  \label{app_fig:LoRA-model-all}
\end{figure}

\begin{figure}[h]
  \centering
  \subfigure[PSNR $\uparrow$.] {     
\includegraphics[width=0.29\linewidth]{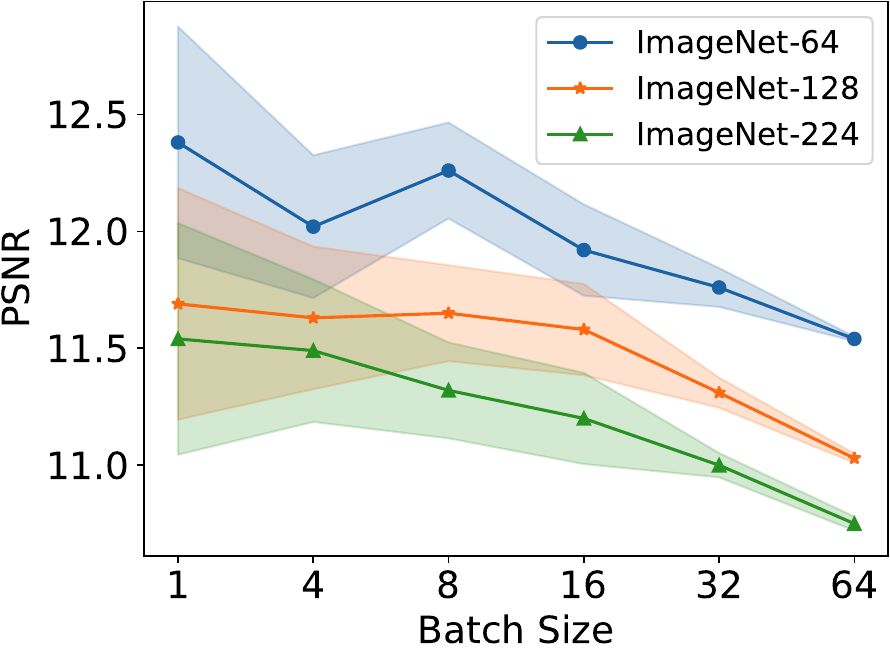}  
}   
\hfill
   \subfigure[SSIM $\uparrow$.] {    
\includegraphics[width=0.29\linewidth]{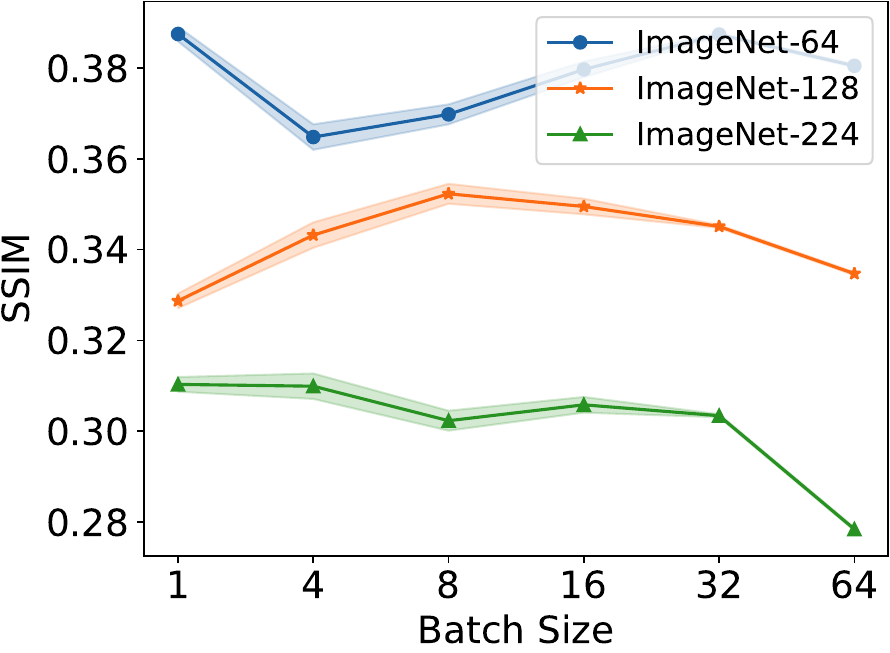}  
} 
\hfill
    \subfigure[LPIPS $\downarrow$.] {    
\includegraphics[width=0.29\linewidth]{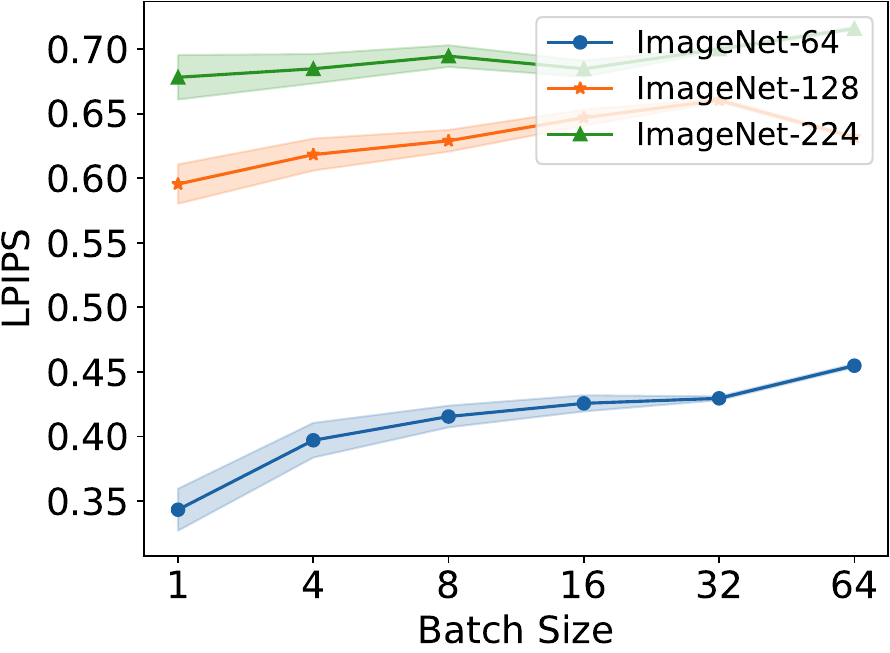}  
}   
\hfill 
\revise{
    \subfigure[Jaccard $\uparrow$.] {    
\includegraphics[width=0.29\linewidth]{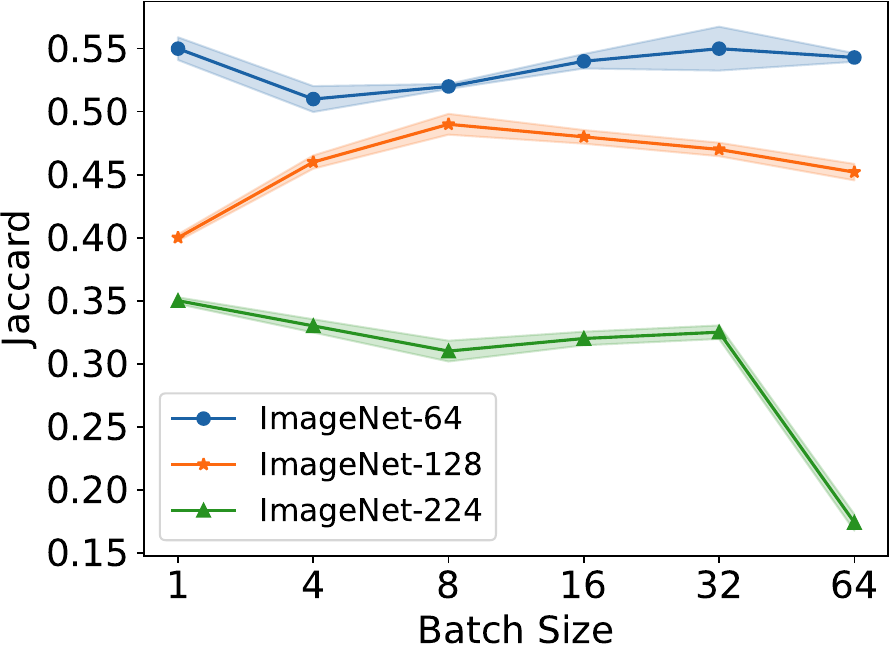}  
}   
    \subfigure[RDLV $\uparrow$.] {    
\includegraphics[width=0.29\linewidth]{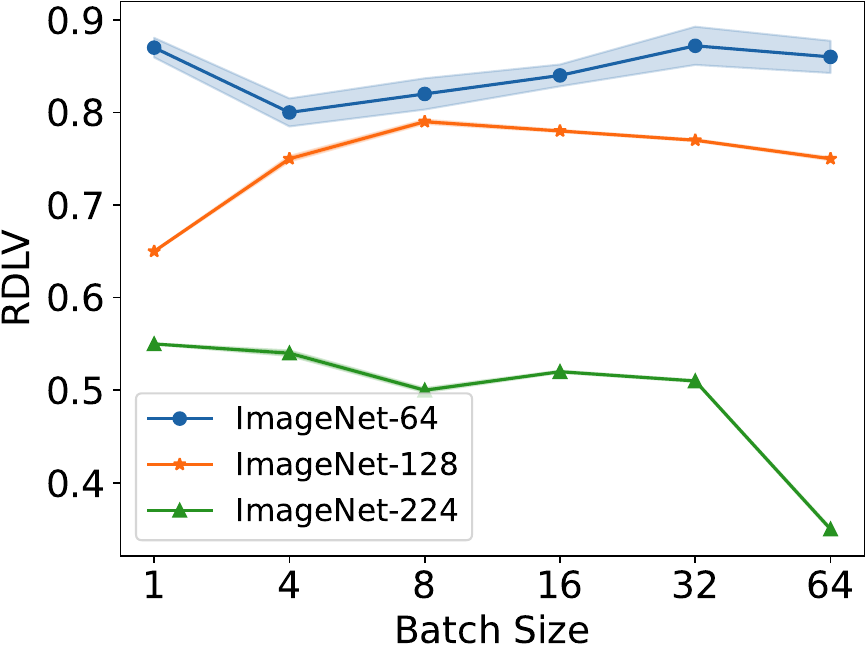}  
}
}
  \caption{The reconstruction results using Eq. (7) evaluated on different ViT-base fine-tuned with LoRA on ImageNet with different resolutions with all evaluation metrics. These results show that attackers can breach privacy on low-resolution images but fail with high-resolution ones under PEFT.
  }
  \label{app_fig:LoRA-resolution-all}
\end{figure}

The reconstruction results using Eq. (7) evaluated on different ViT architectures fine-tuned with LoRA on different datasets with all evaluation metrics are shown in Figures \ref{app_fig:LoRA-dataset-all} and \ref{app_fig:LoRA-model-all}. Reconstruction results on ImageNet with different image resolutions are shown in Figure \ref{app_fig:LoRA-resolution-all}. These results show that attackers can breach privacy on low-resolution images but fail with high-resolution ones under PEFT. Moreover, smaller pre-trained models are better at protecting privacy.

\subsubsection{Visualization}
\label{app_sec:exp_attack_peft_vis}

\begin{figure}[h]
\centering
\subfigure[Batch size = 1. PSNR $\uparrow$: 13.96, SSIM $\uparrow$: 0.5185, LPIPS $\downarrow$: 0.0803.] {    
\includegraphics[width=0.29\linewidth]{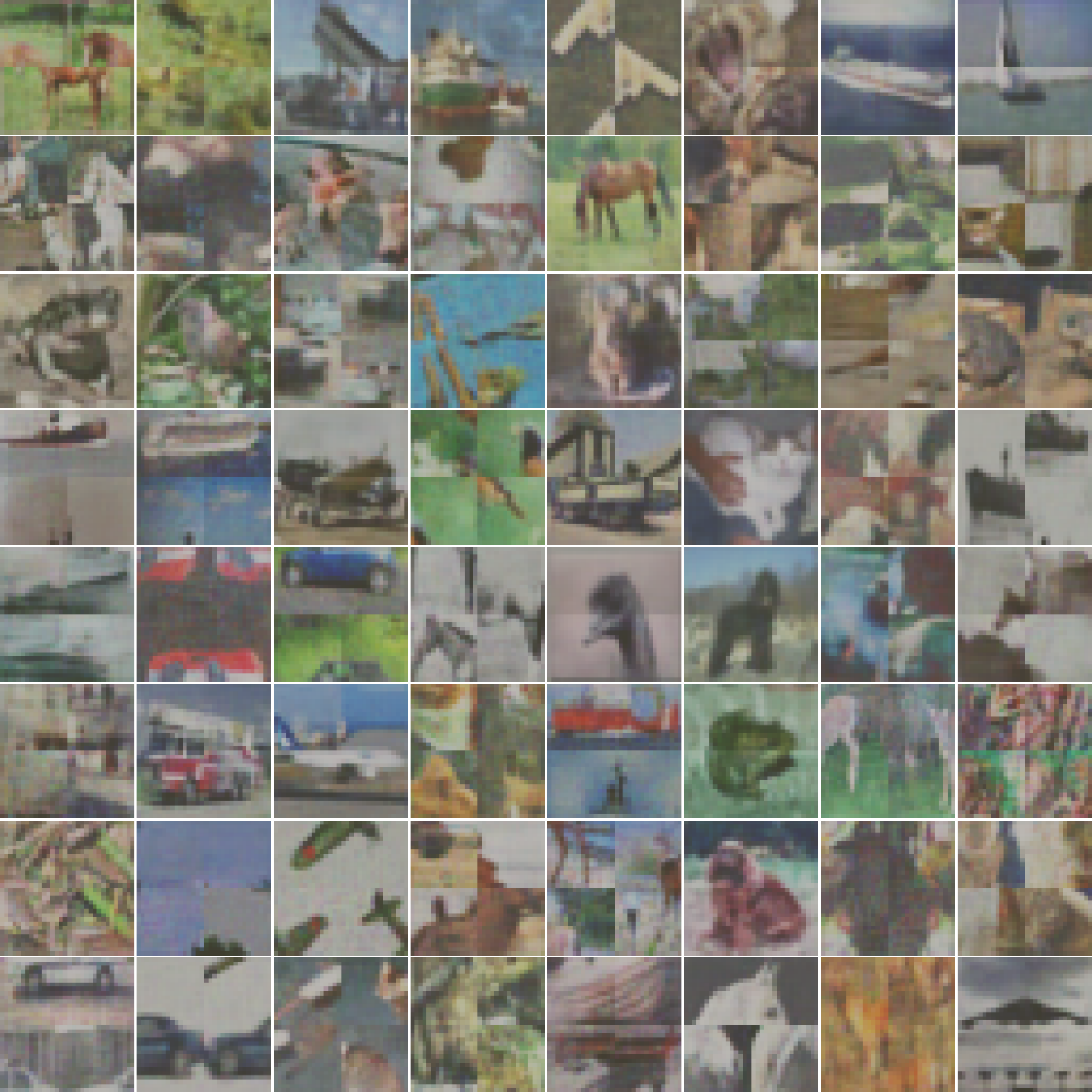}  
}   
\subfigure[Batch size = 32. PSNR $\uparrow$: 12.61, SSIM $\uparrow$: 0.3886, LPIPS $\downarrow$: 0.1334.] {    
\includegraphics[width=0.29\linewidth]{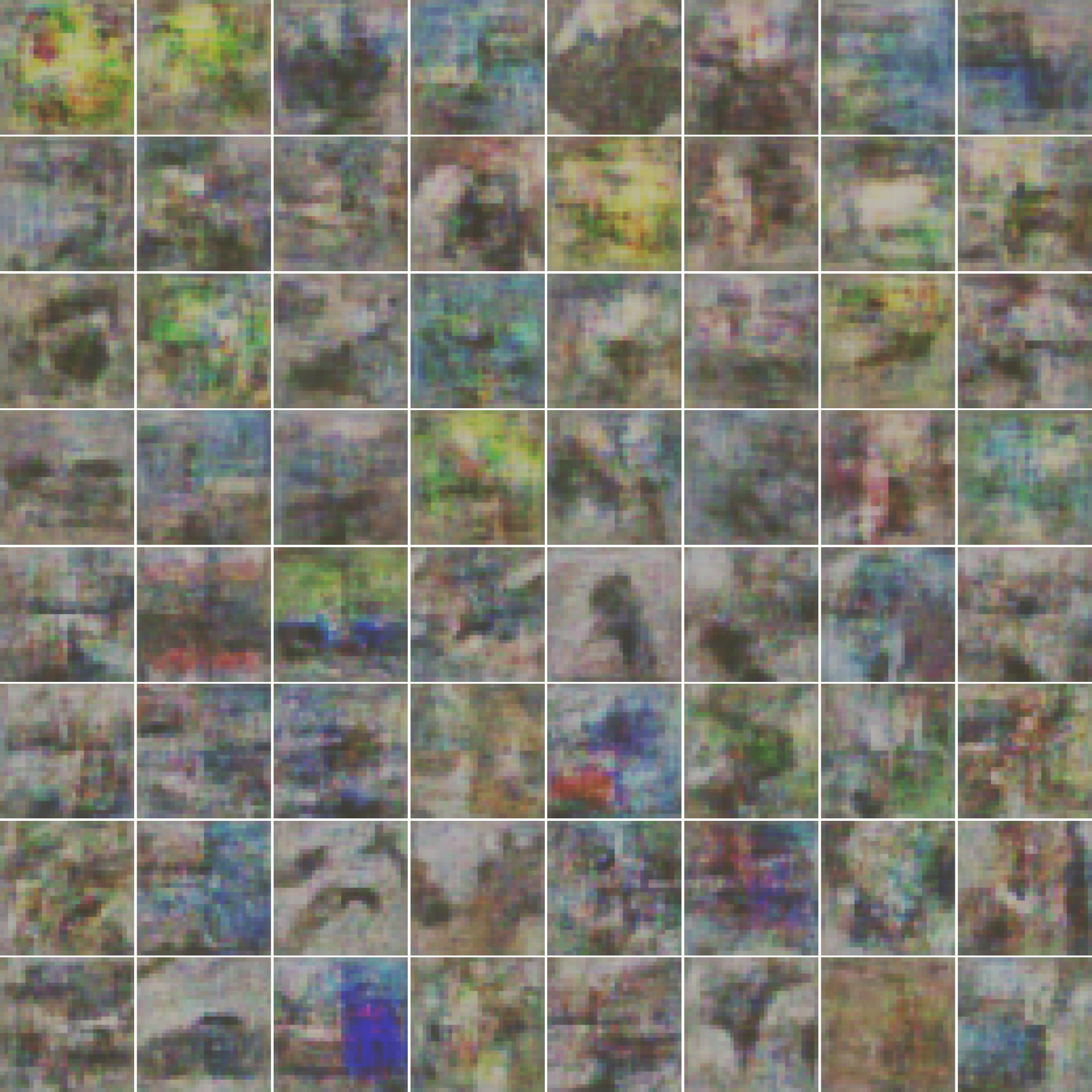}  
} 
\subfigure[Batch size = 64. PSNR $\uparrow$: 12.06, SSIM $\uparrow$: 0.3812, LPIPS $\downarrow$: 0.1432.] {     
\includegraphics[width=0.29\linewidth]{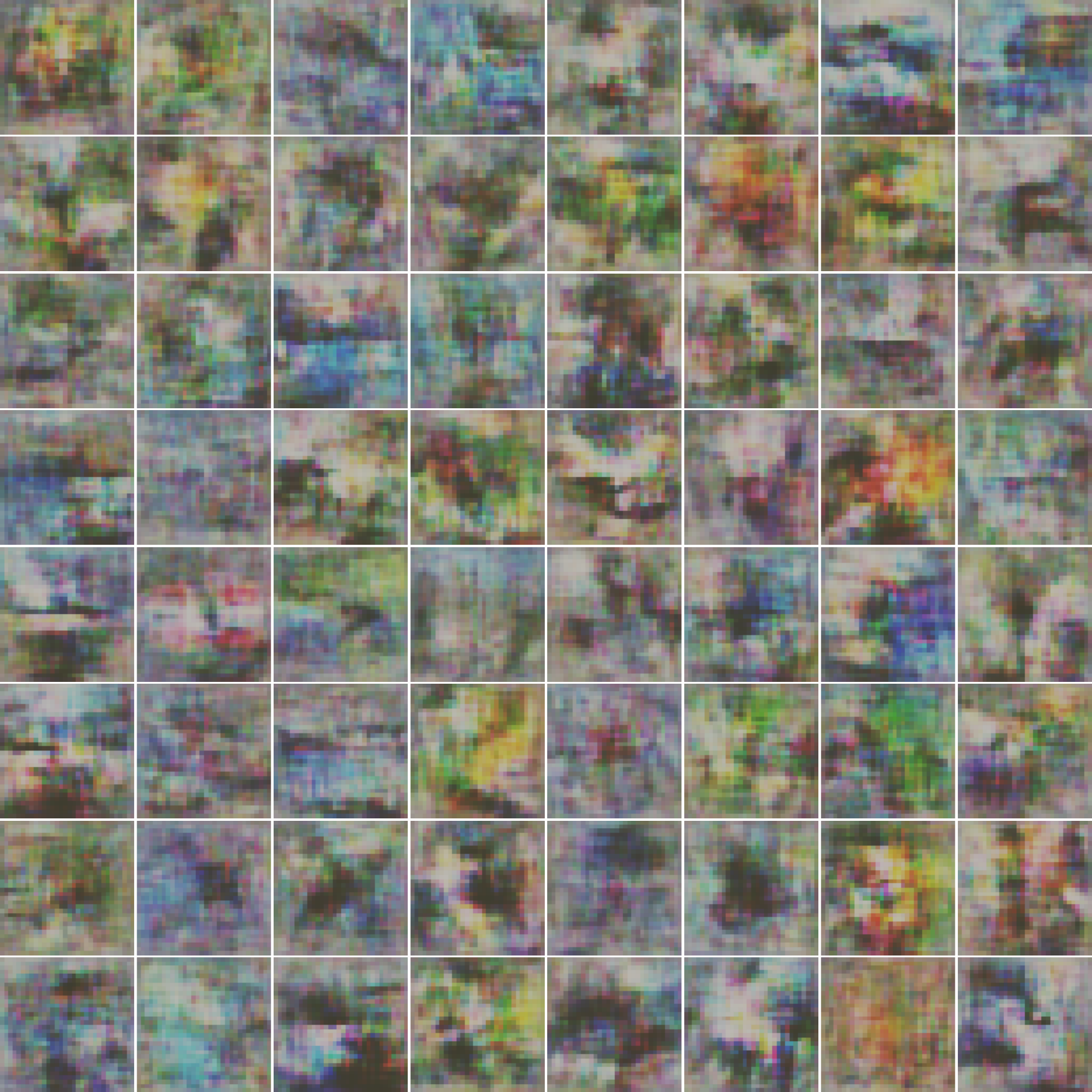}  
}   
\caption{Visualization of reconstruction results using Eq. (7) evaluated on the ViT-base fine-tuned with LoRA on the CIFAR-10 dataset.}
  \label{app_fig:vis_lora_cifar10}
\end{figure}

\begin{figure}[h]
\centering
\subfigure[Batch size = 1. PSNR $\uparrow$: 13.61, SSIM $\uparrow$: 0.5539, LPIPS $\downarrow$: 0.0781.] {    
\includegraphics[width=0.29\linewidth]{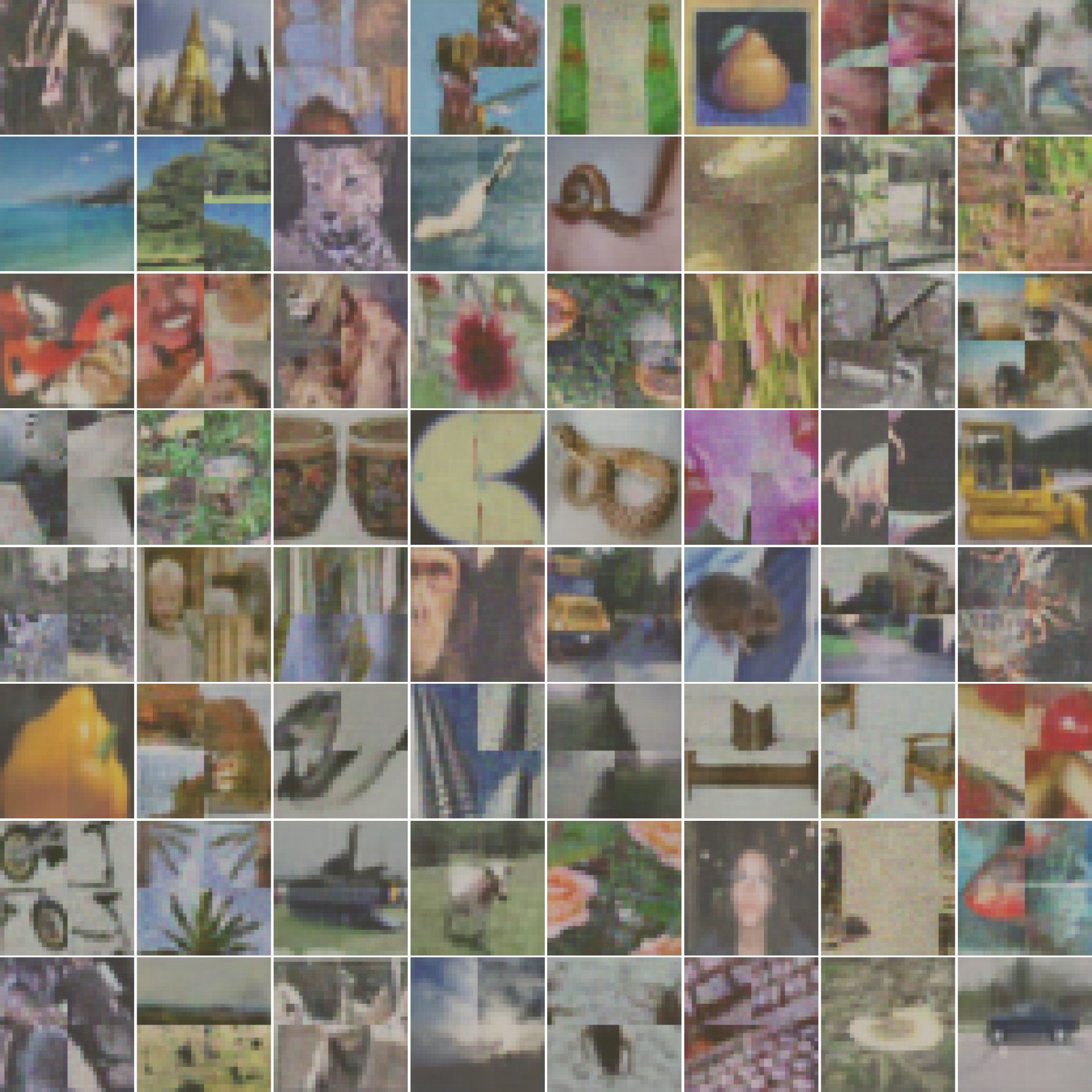}  
}   
\subfigure[Batch size = 32. PSNR $\uparrow$: 12.71, SSIM $\uparrow$: 0.4632, LPIPS $\downarrow$: 0.1243.] {    
\includegraphics[width=0.29\linewidth]{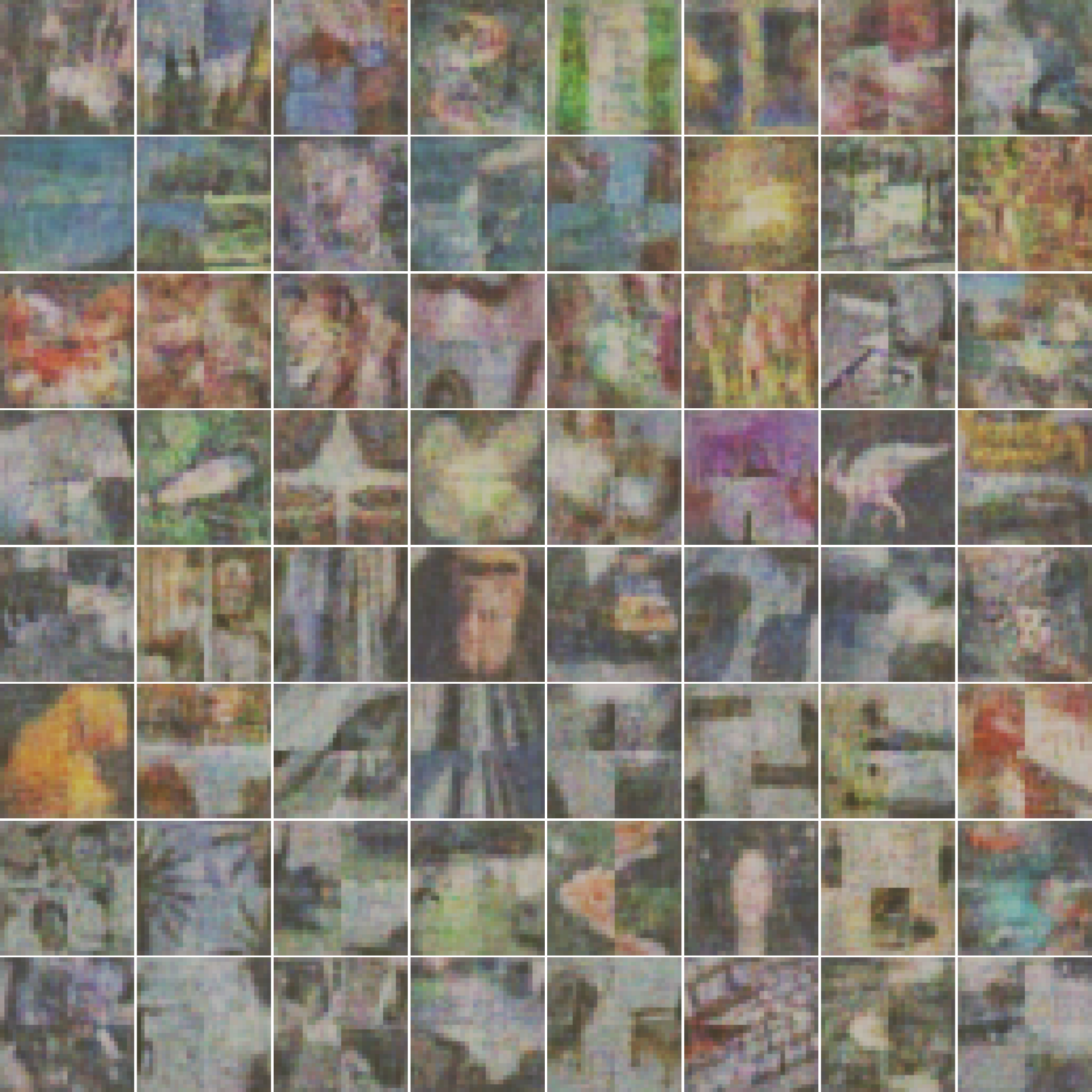}  
} 
\subfigure[Batch size = 64. PSNR $\uparrow$: 13.26, SSIM $\uparrow$: 0.4918, LPIPS $\downarrow$:  0.0965.] {     
\includegraphics[width=0.29\linewidth]{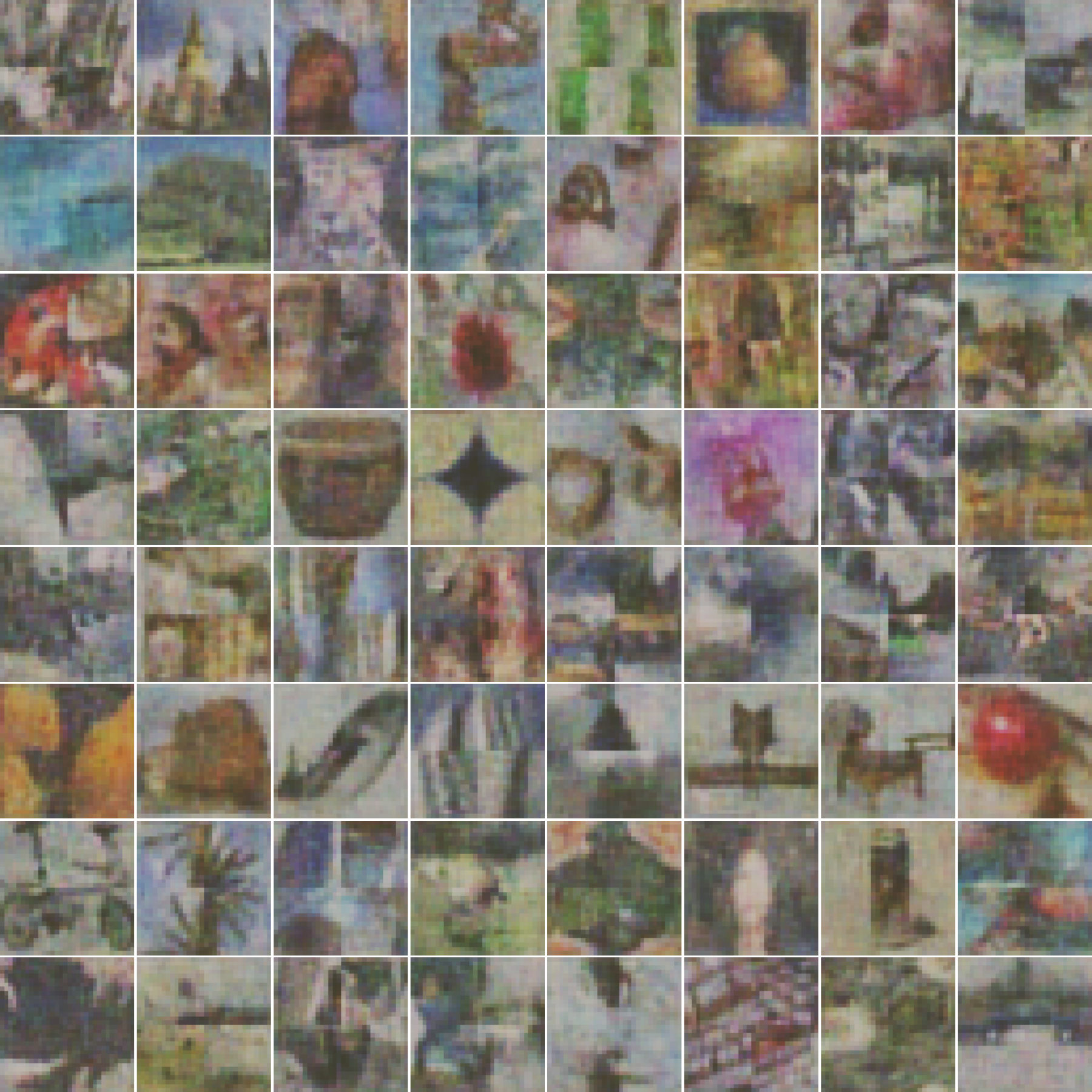}  
}   
\caption{Visualization of reconstruction results using Eq. (7) evaluated on the ViT-base fine-tuned with LoRA on the CIFAR-100 dataset.}
  \label{app_fig:vis_lora_cifar100}
\end{figure}

\begin{figure}[h]
\centering
\subfigure[Batch size = 1. PSNR $\uparrow$: 12.38, SSIM $\uparrow$: 0.3875, LPIPS $\downarrow$: 0.3432.] {    
\includegraphics[width=0.29\linewidth]{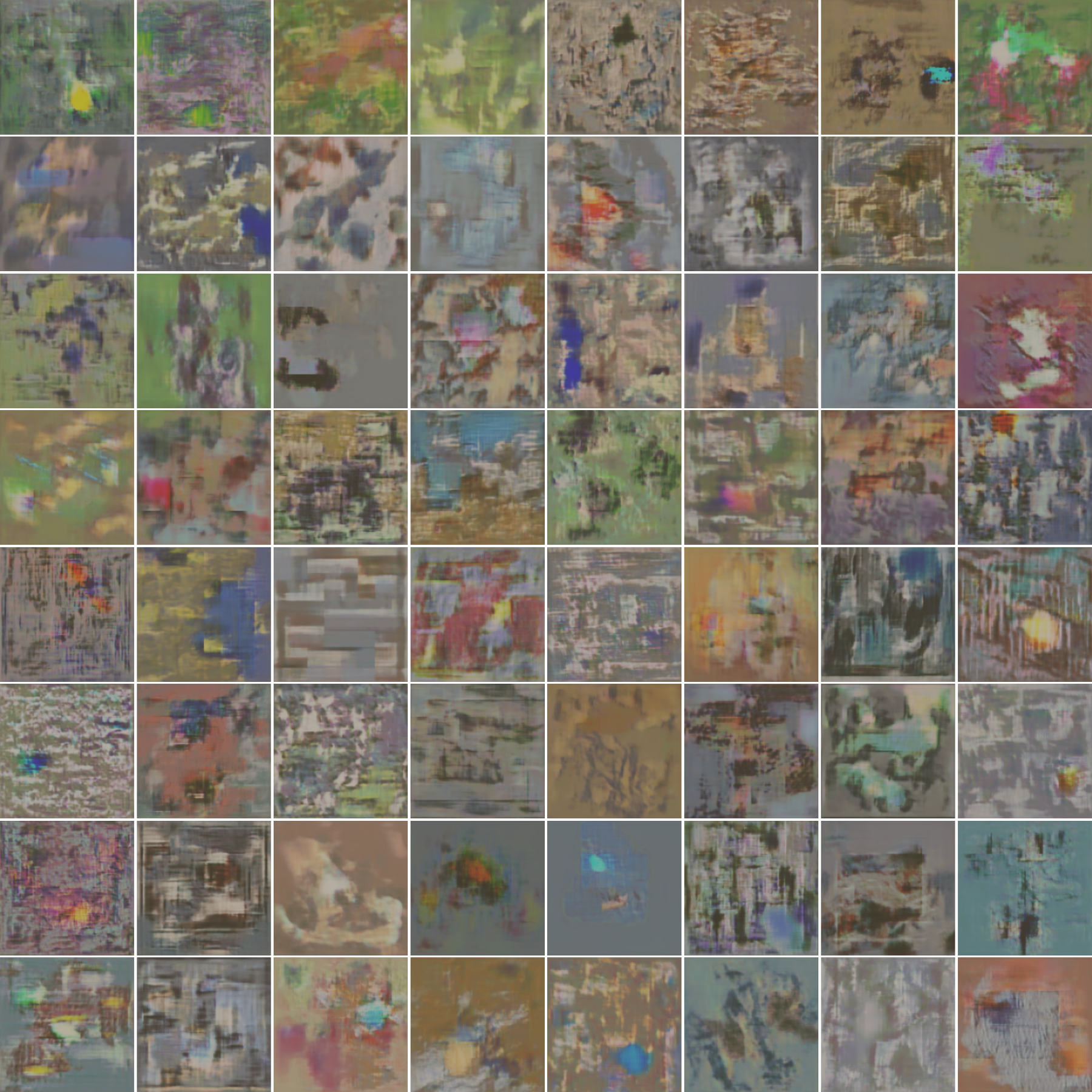}  
}   
\subfigure[Batch size = 32. PSNR $\uparrow$: 11.76, SSIM $\uparrow$: 0.3874, LPIPS $\downarrow$: 0.4295.] {    
\includegraphics[width=0.29\linewidth]{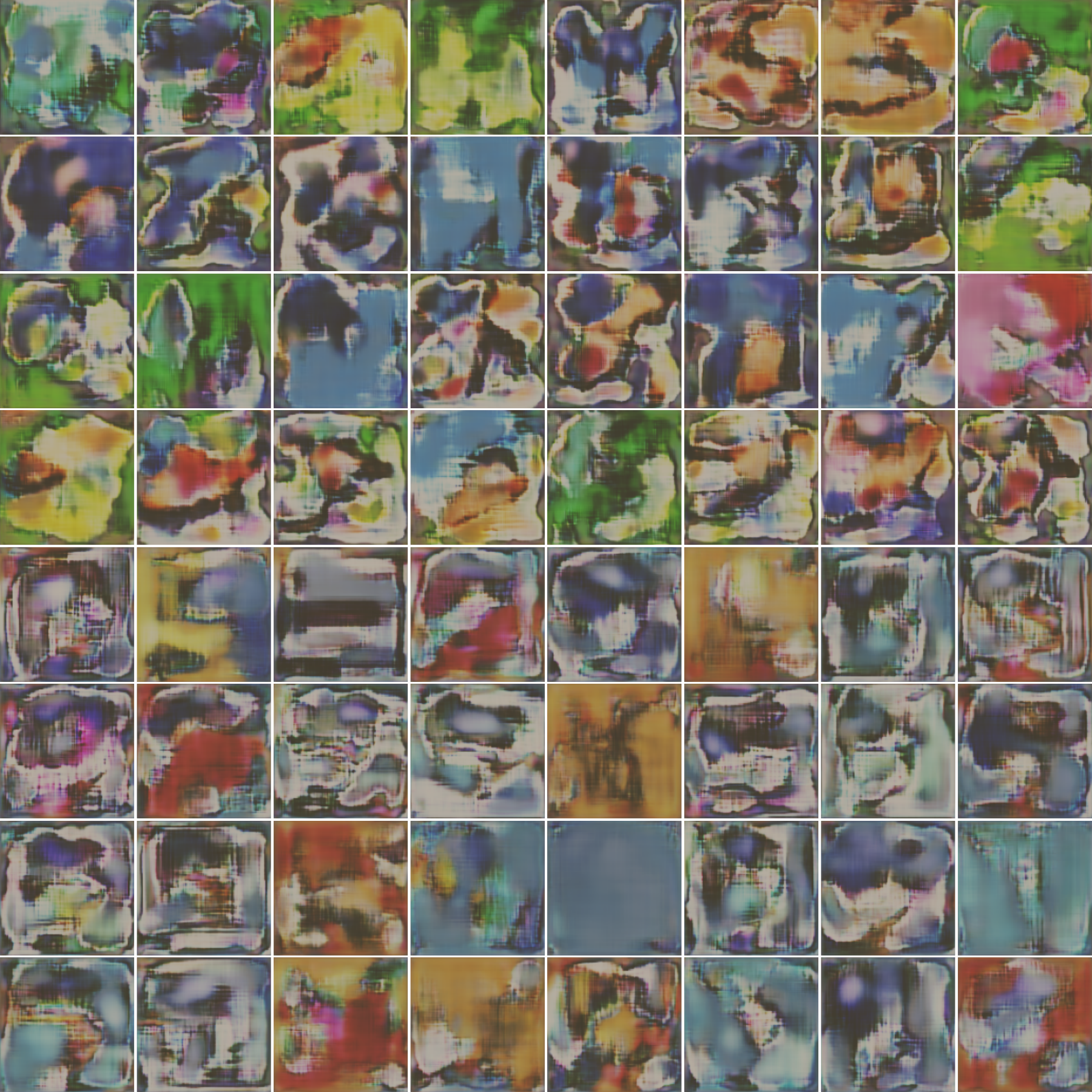}  
} 
\subfigure[Batch size = 64. PSNR $\uparrow$: 11.54, SSIM $\uparrow$: 0.3805, LPIPS $\downarrow$: 0.4548.] {     
\includegraphics[width=0.29\linewidth]{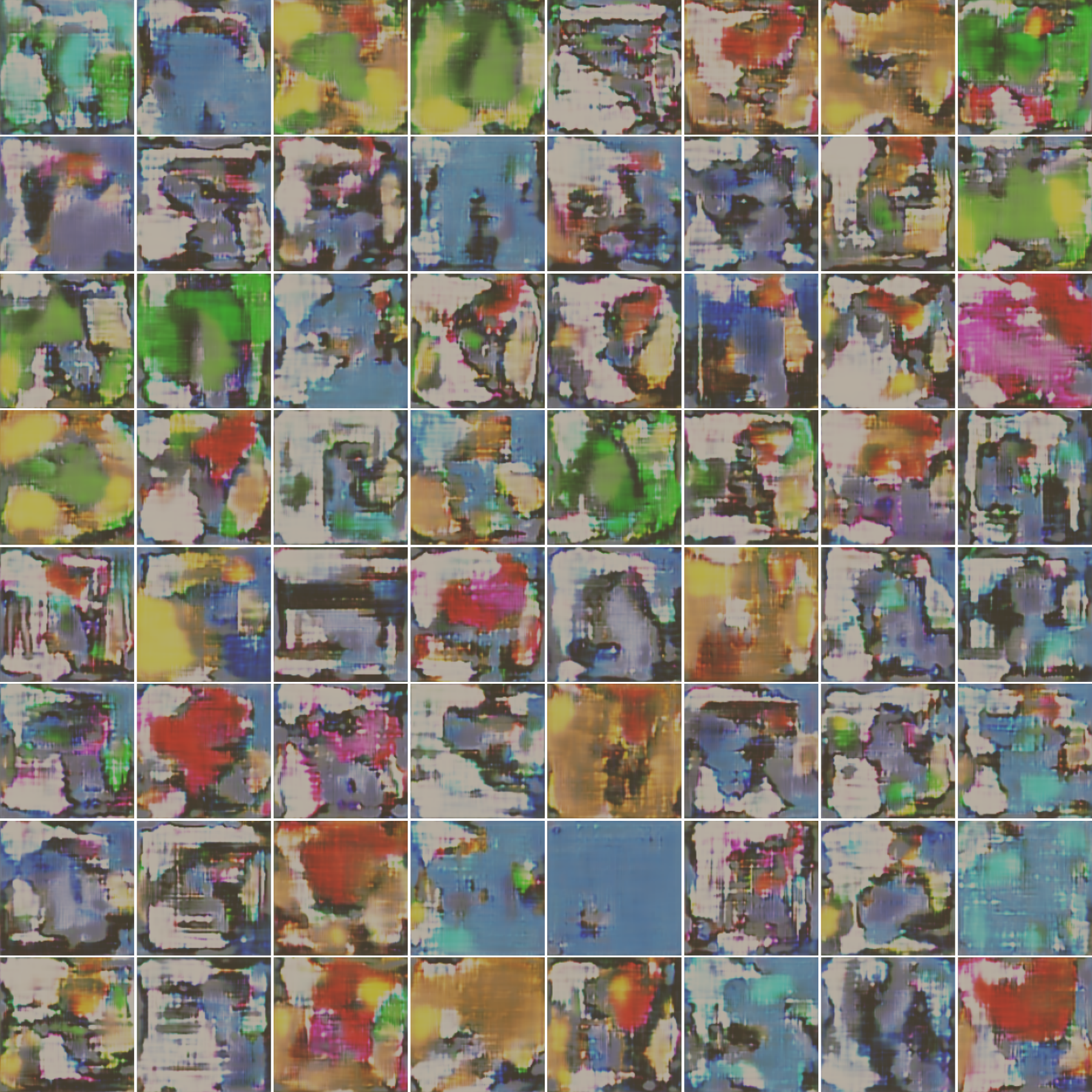}  
}   
\caption{Visualization of reconstruction results using Eq. (7) evaluated on the ViT-base fine-tuned with LoRA on the ImageNet dataset.}
  \label{app_fig:vis_lora_imagenet}
\end{figure}

\begin{figure}[h]
\centering
\subfigure[Batch size = 1. PSNR $\uparrow$: 10.27, SSIM $\uparrow$: 0.4082, LPIPS $\downarrow$: 0.4299.] {    
\includegraphics[width=0.29\linewidth]{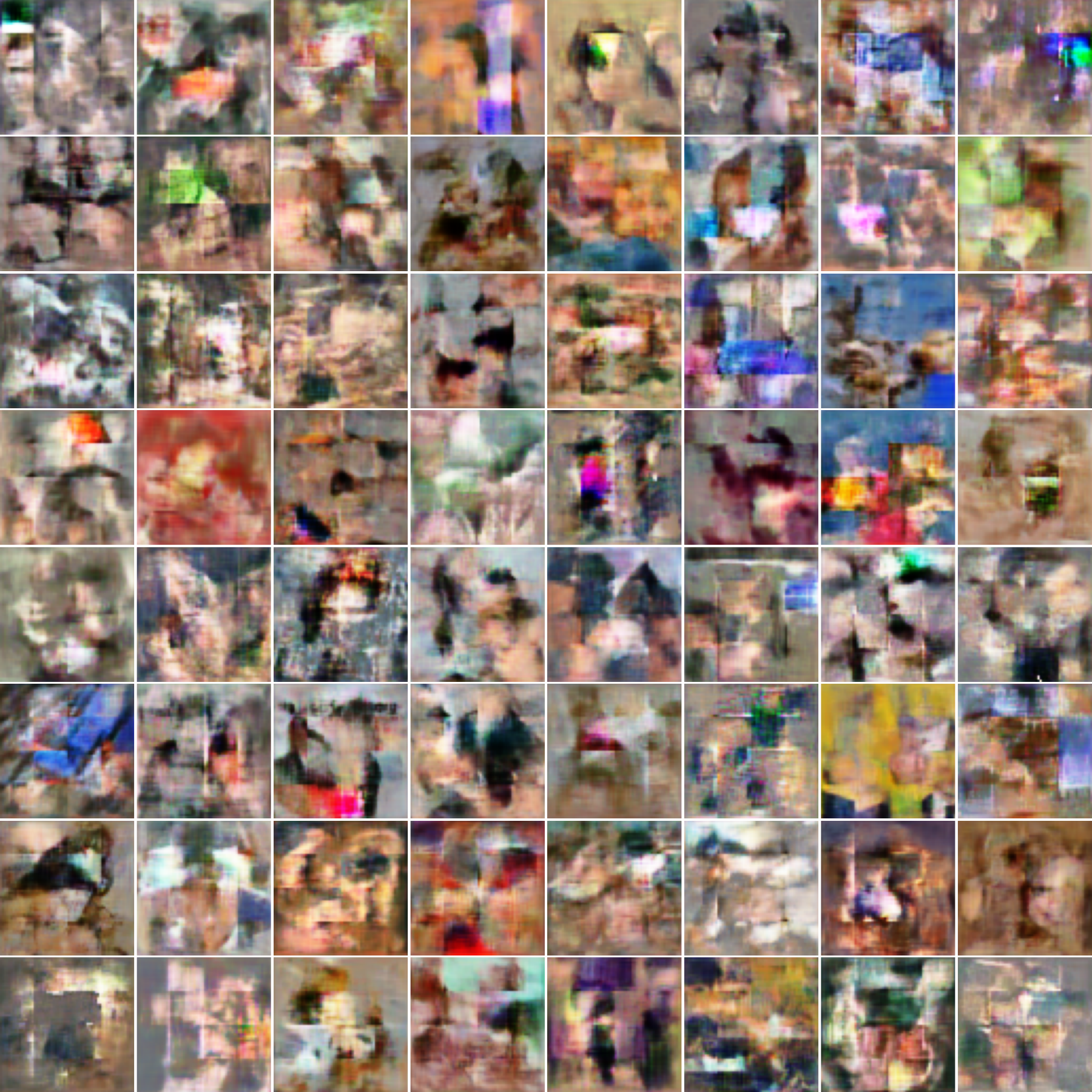}  
}   
\subfigure[Batch size = 32. PSNR $\uparrow$: 8.79, SSIM $\uparrow$: 0.3820, LPIPS $\downarrow$: 0.4950.] {    
\includegraphics[width=0.29\linewidth]{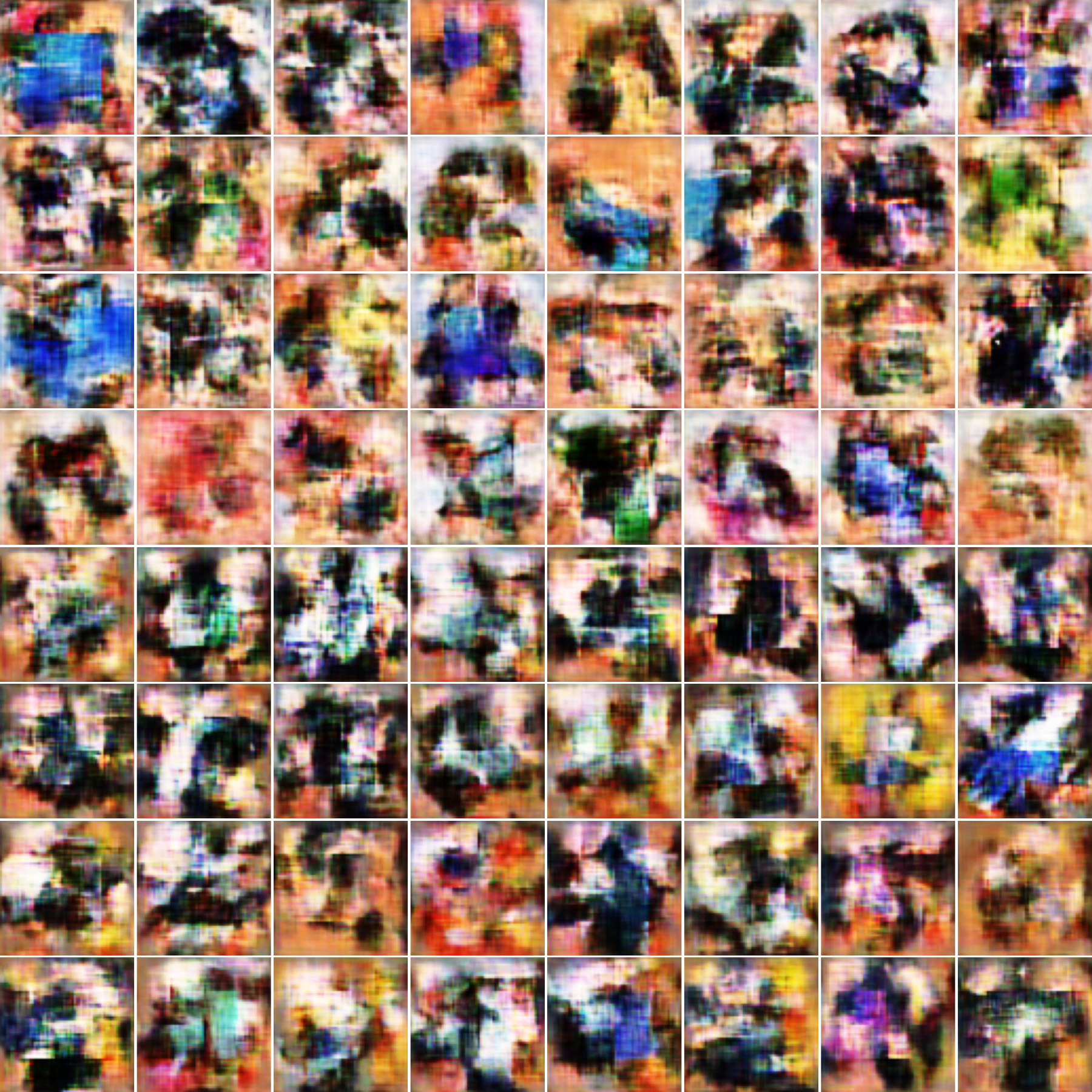}  
} 
\subfigure[Batch size = 64. PSNR $\uparrow$: 8.38, SSIM $\uparrow$: 0.3702, LPIPS $\downarrow$: 0.5386.] {     
\includegraphics[width=0.29\linewidth]{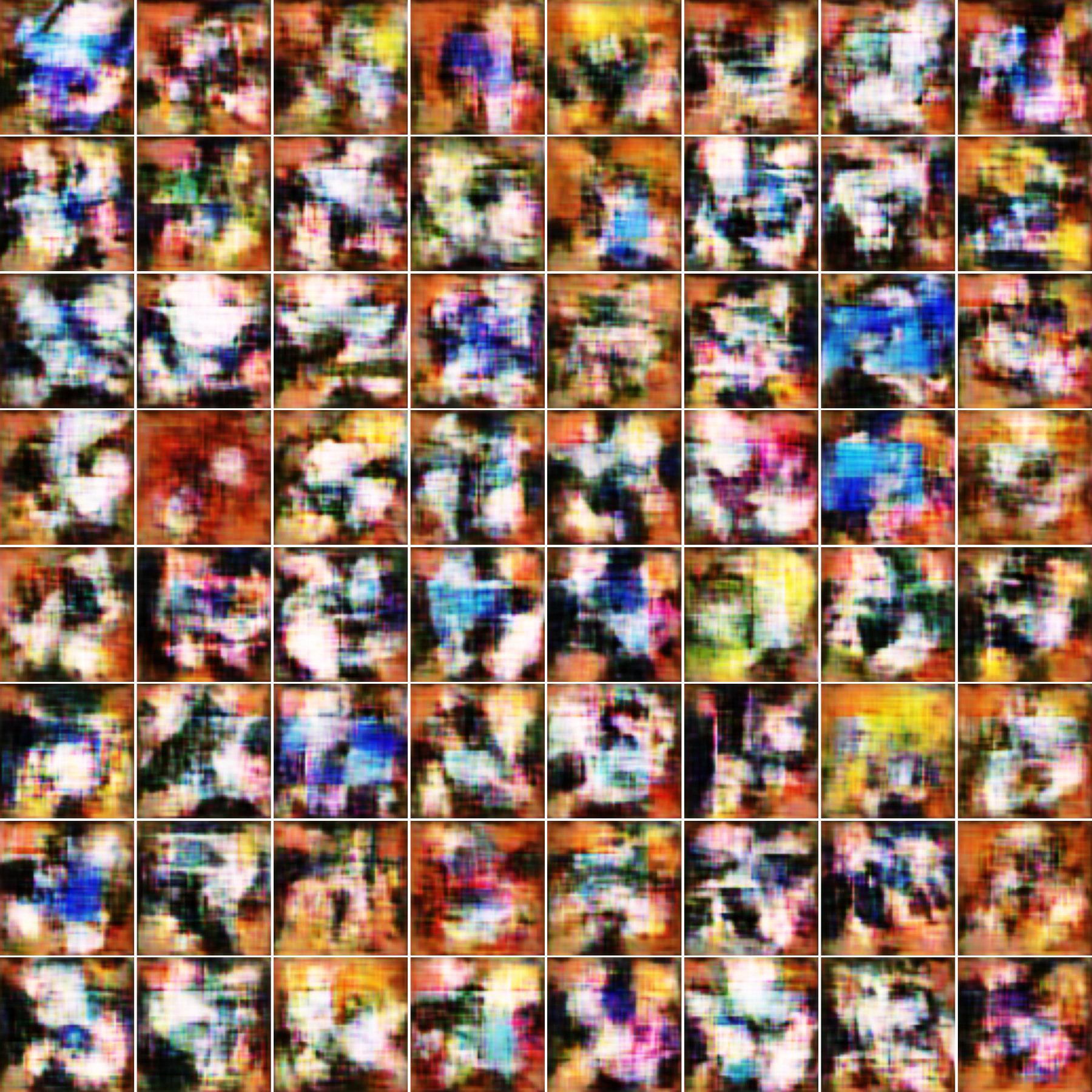}  
}   
\caption{Visualization of reconstruction results using Eq. (7) evaluated on the ViT-base fine-tuned with LoRA on the CelebA dataset.}
  \label{app_fig:vis_lora_celeba}
\end{figure}

The visualization of reconstruction results using Eq. (7) is shown in Figures \ref{app_fig:vis_lora_cifar10}, \ref{app_fig:vis_lora_cifar100}, \ref{app_fig:vis_lora_imagenet}, and \ref{app_fig:vis_lora_celeba}.

{
\section{Evaluation on Other Backbones}
\label{app_sec:exp_other_backbones}
We further evaluate the attack performance of various GIA methods on other backbones, such as LeNet \cite{lecun2002gradient}, AlexNet \cite{krizhevsky2012imagenet}, VGG \cite{simonyan2014very}, and GoogLeNet \cite{szegedy2015going}, to demonstrate the generalizability of our findings.

\subsection{Optimization-based GIA}

The reconstruction results of IG with all evaluation metrics on other backbones are shown in Figure \ref{app_fig:gia_all_other_backbones}. These results show a similar phenomenon to that observed with ResNet-18, which further demonstrates the generalizability of our findings.

\begin{figure}[h]
  \centering
\revise{
  \subfigure[PSNR $\uparrow$.] {     
\includegraphics[width=0.29\linewidth]{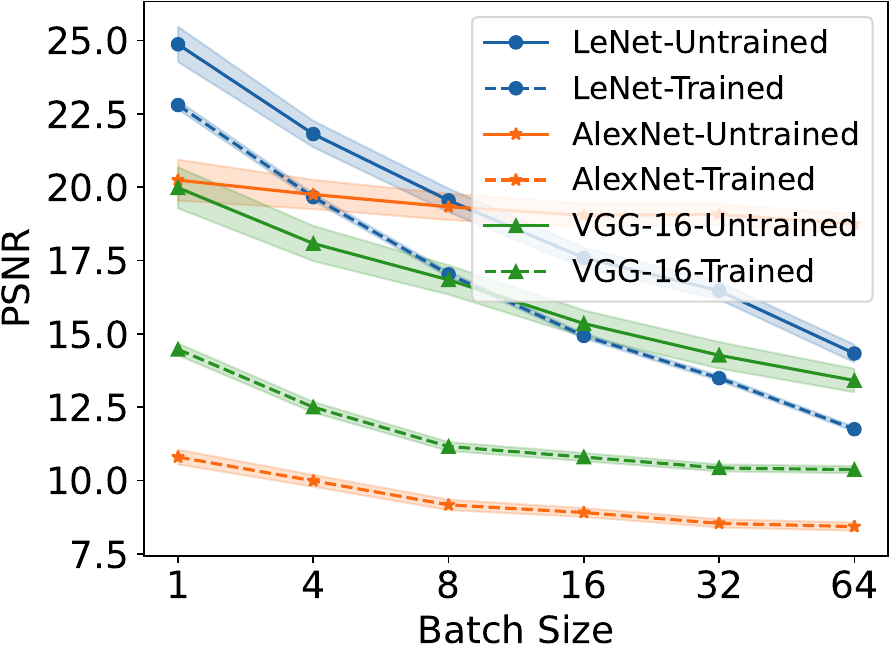}  
}   
\hfill
   \subfigure[SSIM $\uparrow$.] {    
\includegraphics[width=0.29\linewidth]{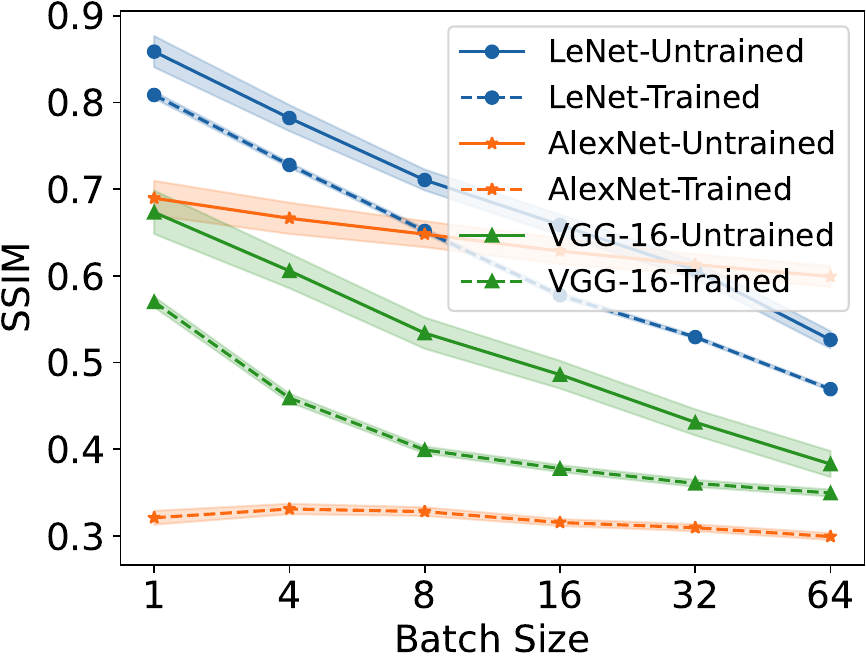}  
} 
\hfill
    \subfigure[LPIPS $\downarrow$.] {    
\includegraphics[width=0.29\linewidth]{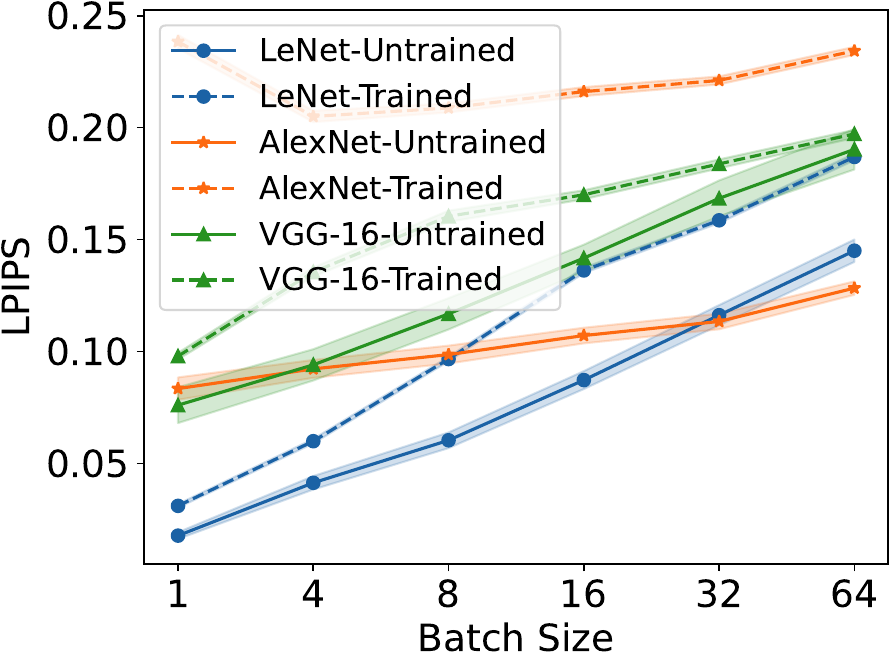} 
} 
\hfill
    \subfigure[Jaccard $\uparrow$.] {    
\includegraphics[width=0.29\linewidth]{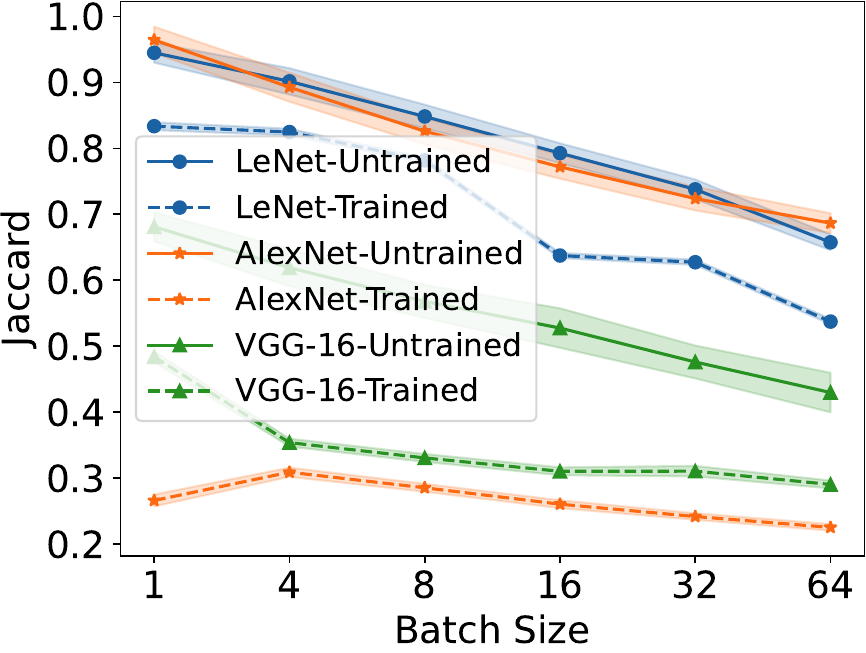} 
} 
    \subfigure[RDLV $\uparrow$.] {    
\includegraphics[width=0.29\linewidth]{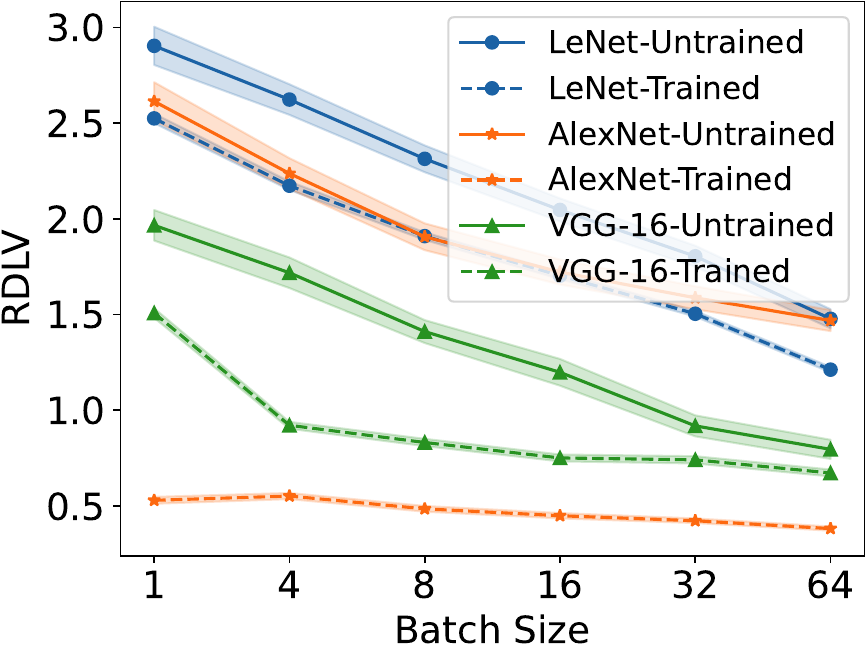} 
} 
}
\revise{
  \caption{Reconstruction results of IG evaluated on different models with different training states on CIFAR-100 with different batch sizes, where the shaded region represents the standard deviation. These results show that a larger batch size and better model training state lead to worse OP-GIA performance.
  }
  \label{app_fig:gia_all_other_backbones}
}
\end{figure}

\subsection{Generation-based GIA} 

Since the performance of GEN-GIA, when optimizing the latent vector $\bm{z}$, is unaffected by the leaked gradients, it indicates that the results are not influenced by the backbones used in the local setup. Additionally, experiments involving the training of an inversion generation model have utilized LeNet. Therefore, we will only present the results of GEN-GIA with optimizing generator parameters $\bm{W}$ on other backbones in this section.

\subsubsection{Optimizing Generator's Parameters $W$}

The reconstruction results of CI-Net with all evaluation metrics using LeNet as backbone are shown in Figure \ref{app_fig:CI-Net-all-other-backbones}. These results exhibit a phenomenon similar to that observed with ResNet-18, further highlighting the generalizability of our findings.

\begin{figure}[h]
  \centering
\revise{
  \subfigure[PSNR $\uparrow$.] {     
\includegraphics[width=0.29\linewidth]{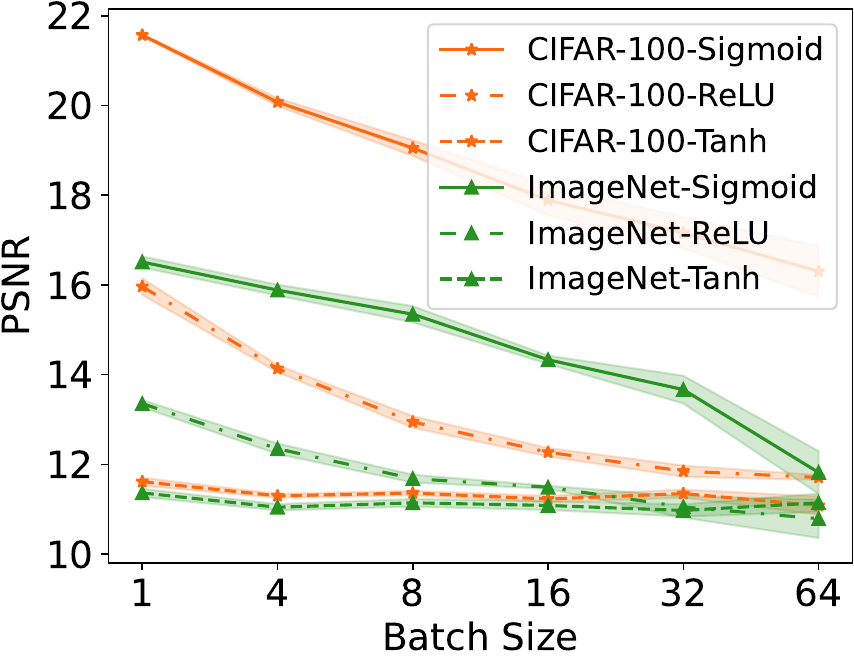}  
}   
\hfill
   \subfigure[SSIM $\uparrow$.] {    
\includegraphics[width=0.29\linewidth]{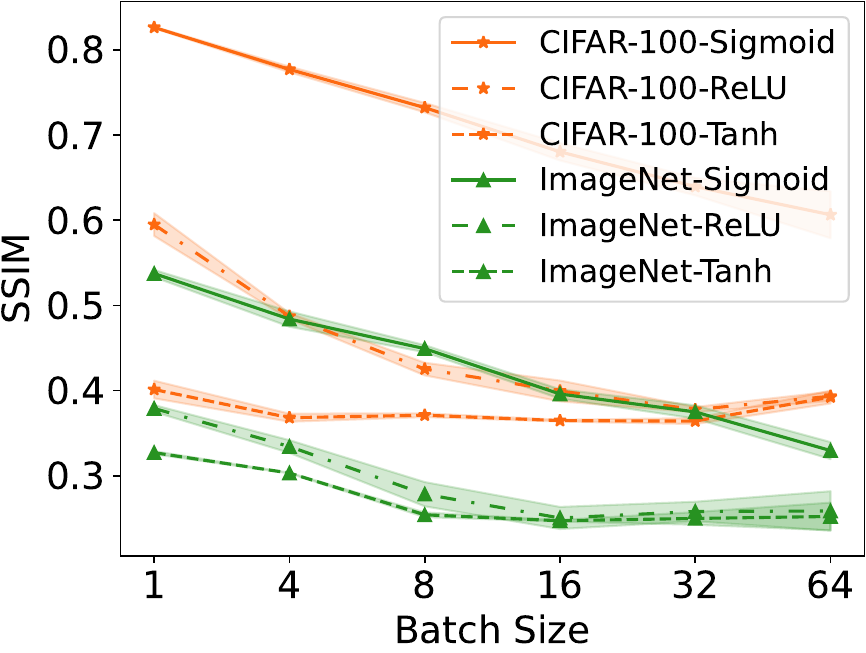}  
} 
\hfill
    \subfigure[LPIPS $\downarrow$.] {    
\includegraphics[width=0.29\linewidth]{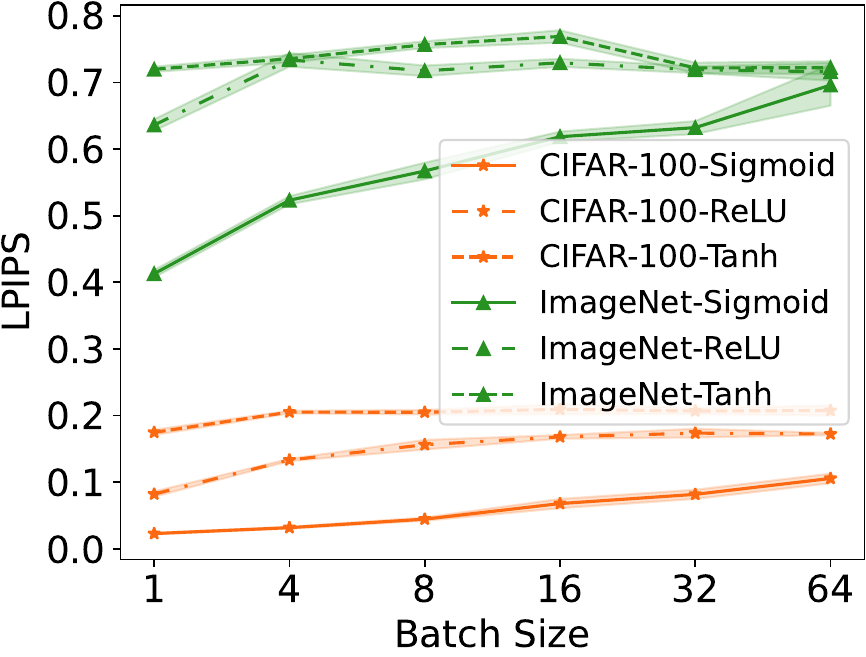}  
}   
\hfill 
    \subfigure[Jaccard $\uparrow$.] {    
\includegraphics[width=0.29\linewidth]{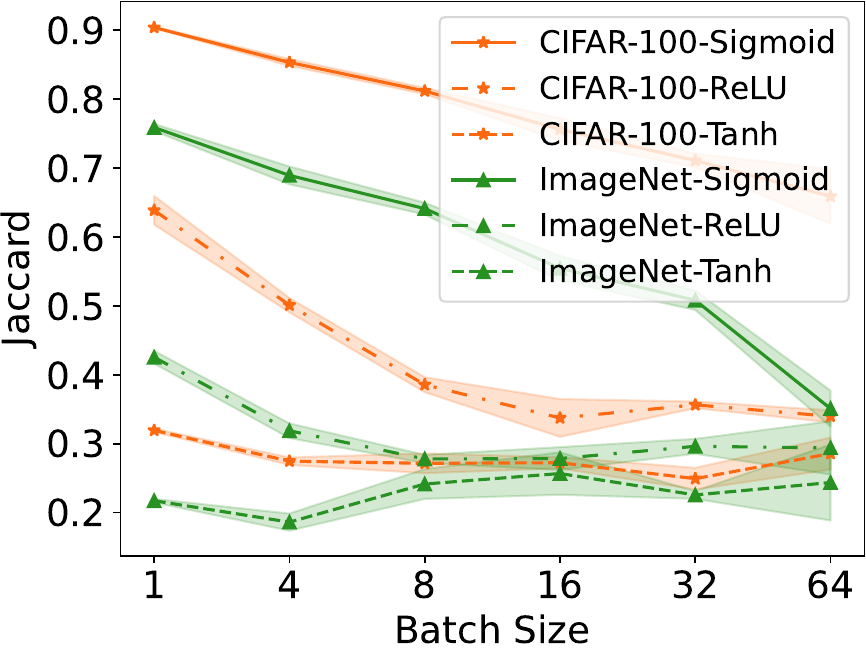}  
}   
    \subfigure[RDLV $\uparrow$.] {    
\includegraphics[width=0.29\linewidth]{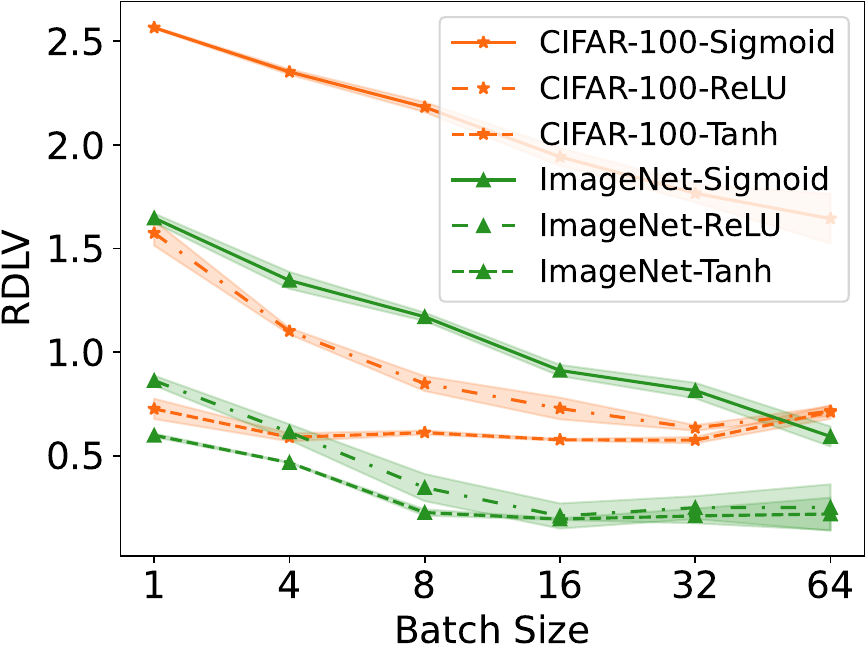}  
}
}
\revise{
  \caption{Reconstruction results of CI-Net evaluated on LeNet with different activation functions on various datasets with different batch sizes. These results show that GEN-GIA with optimizing the generator's parameters $\bm{W}$ is affected by the factors that influence OP-GIA. Moreover, it only works when the target model adopts the Sigmoid activation function and fails with other activation functions.
  }
  \label{app_fig:CI-Net-all-other-backbones}
}
\end{figure}

\subsection{Analytics-based GIA}

Since the performance of ANA-GIA, when manipulating the model architecture, is not affected by the local model—given that it directly adds a linear layer before the local model—we will only present the results of ANA-GIA with manipulated model parameters on other backbones in this section.

\subsubsection{Manipulating Model Parameters}

The reconstruction results of Fishing with all evaluation metrics using GoogLeNet as backbone are shown in Figures \ref{app_fig:fish_all_other_backbones}. These results reveal a comparable phenomenon to what was seen with ResNet-18, further emphasizing the generalizability of our findings.

\begin{figure}[!h]
  \centering
\revise{
  \subfigure[PSNR $\uparrow$.] {     
\includegraphics[width=0.29\linewidth]{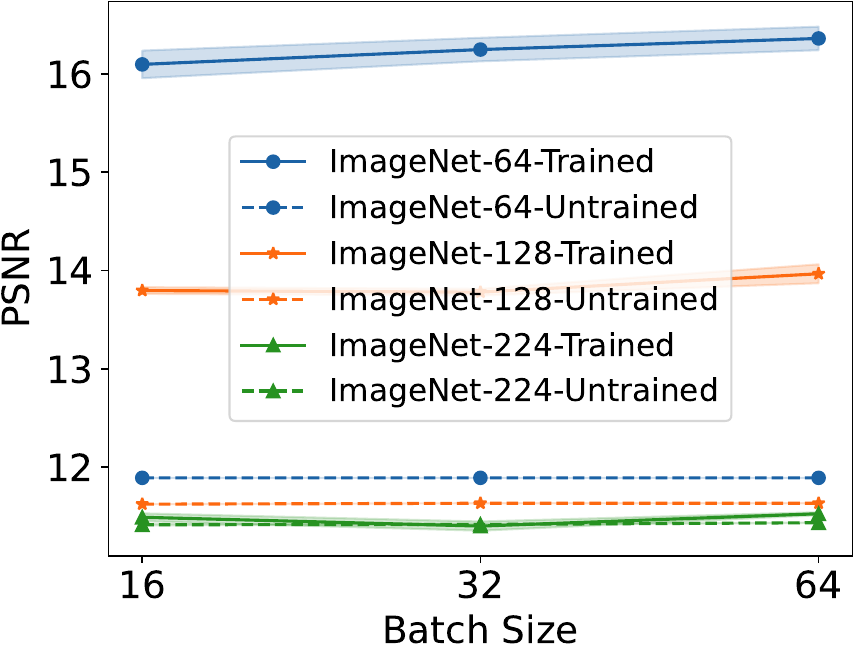}  
}   
\hfill
   \subfigure[SSIM $\uparrow$.] {    
\includegraphics[width=0.29\linewidth]{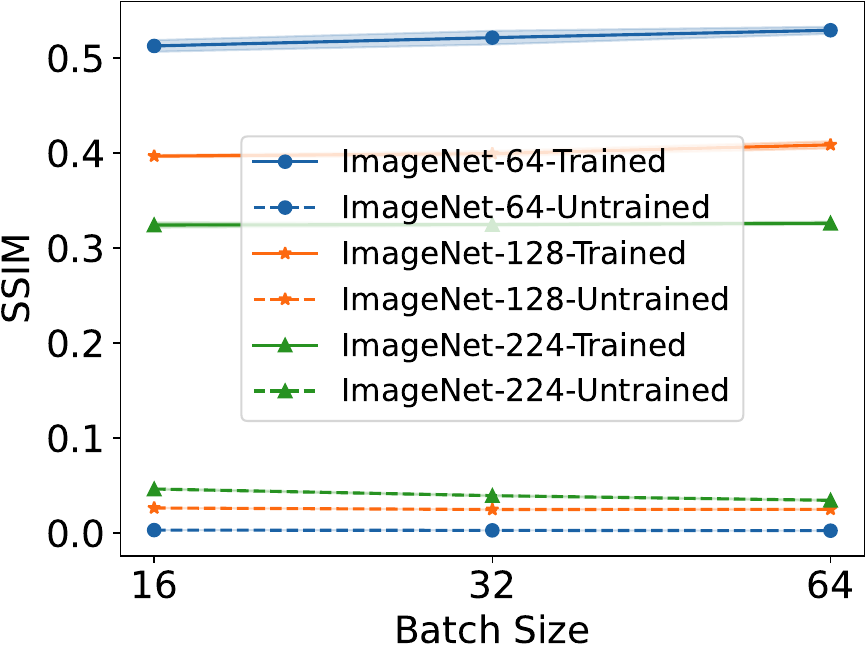}  
} 
\hfill
    \subfigure[LPIPS $\downarrow$.] {    
\includegraphics[width=0.29\linewidth]{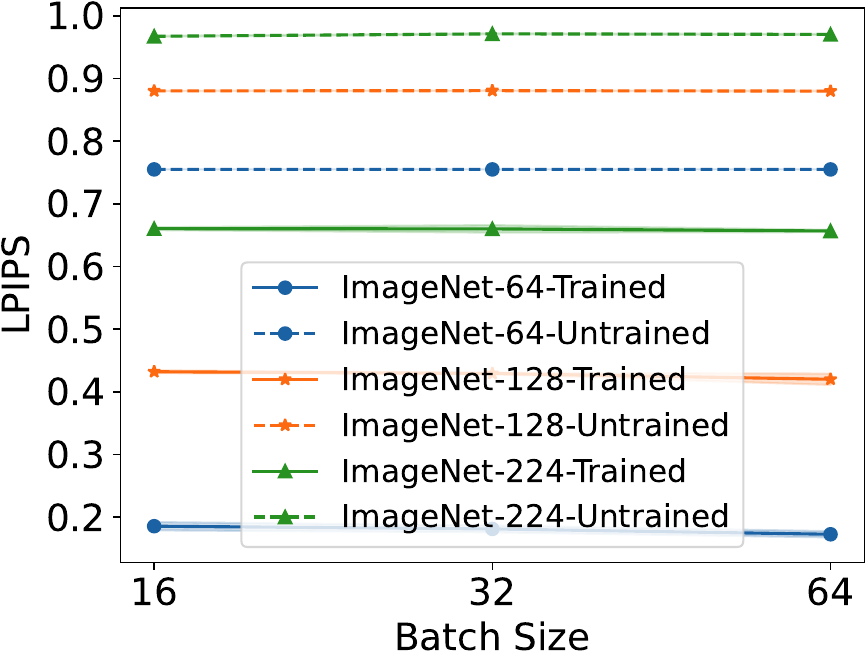}  
}   
\hfill 
    \subfigure[Jaccard $\uparrow$.] {    
\includegraphics[width=0.29\linewidth]{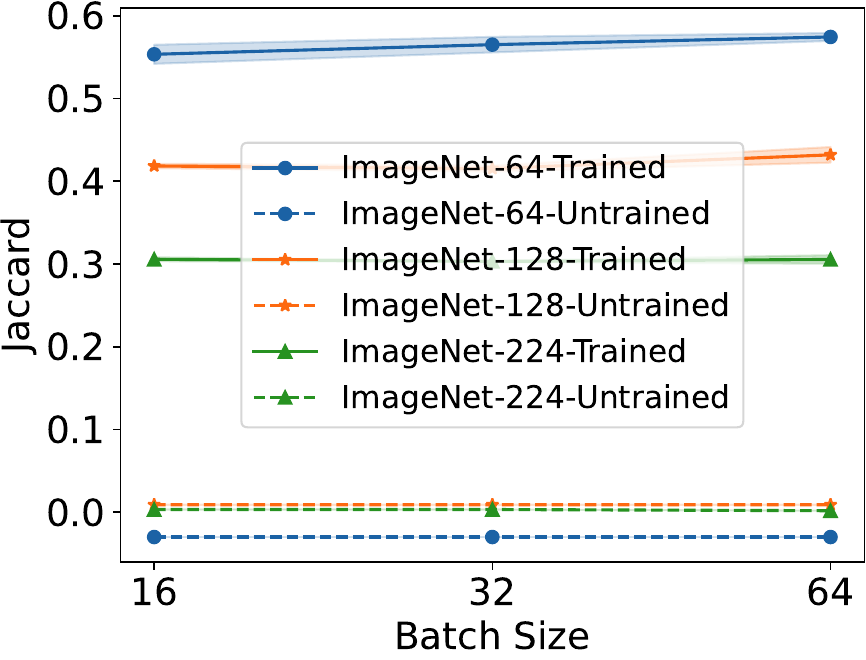}  
}   
    \subfigure[RDLV $\uparrow$.] {    
\includegraphics[width=0.29\linewidth]{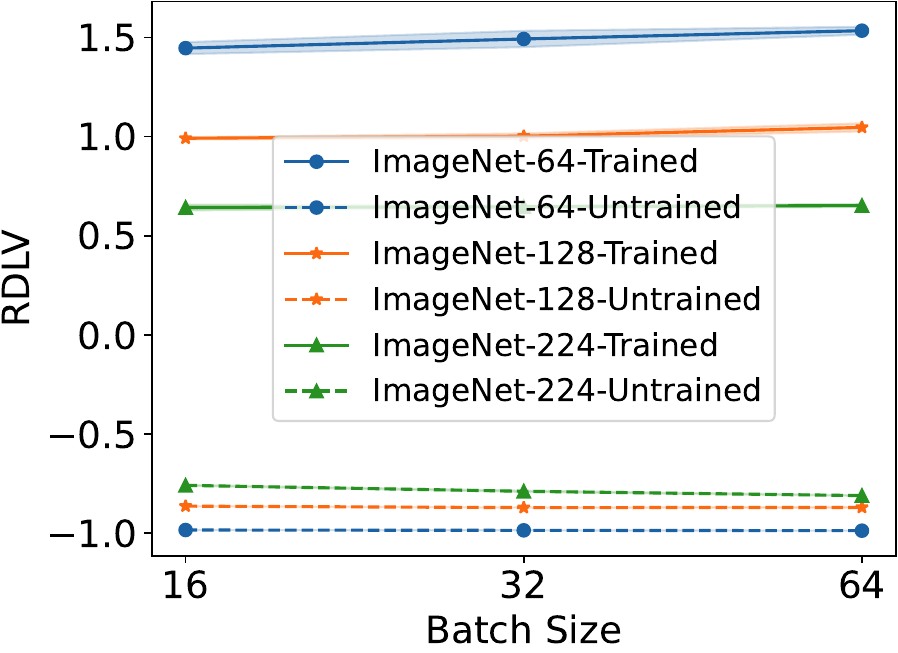}  
}
}
\revise{
  \caption{Reconstruction results of Fishing evaluated on GoogLeNet on ImageNet with different image resolutions and model training states. These results show that the attack performance of ANA-GIA, which manipulates model parameters, is not affected by batch size but worsens with larger image resolutions and worse model training states.
  }
  \label{app_fig:fish_all_other_backbones}
}
\end{figure}

}

\end{document}